\documentclass[11pt,letterpaper]{article}
\usepackage[margin=1in]{geometry} 
\usepackage[utf8]{inputenc}
\usepackage{mathtools} \usepackage{amssymb}

\usepackage{microtype}
\usepackage{enumerate}
\usepackage{accents} \usepackage{graphicx}
\usepackage{float}
\usepackage{placeins}
\usepackage[table,dvipsnames]{xcolor}
\definecolor{reviewcolor}{HTML}{8E24AA}
\definecolor{reviewcolorb}{HTML}{00897B}
\colorlet{reviewcolorb}{black}
\usepackage{tikz}
\usepackage{booktabs}

\usepackage{authblk}

\usepackage[colorlinks, pagebackref=true]{hyperref}
\hypersetup{
  linkcolor=[rgb]{0.3,0.3,0.6},
  citecolor=[rgb]{0.2, 0.6, 0.2},
  urlcolor=[rgb]{0.6, 0.2, 0.2}
}
\hypersetup{breaklinks=true}

\allowdisplaybreaks[1]

\usepackage{commath}
\usepackage{amsthm}
\usepackage{thmtools, thm-restate}
\usepackage[capitalize, nameinlink]{cleveref}

\declaretheorem[name=Theorem, parent=section]{theorem}
\declaretheorem[name=Corollary, sibling=theorem]{corollary}
\declaretheorem[name=Proposition, sibling=theorem]{proposition}
\declaretheorem[name=Lemma, sibling=theorem]{lemma}

\declaretheorem[name=Definition, sibling=theorem, style=definition]{definition}
\declaretheorem[name=Remark, sibling=theorem, style=definition]{remark}

\declaretheorem[name=Example, sibling=theorem, style=definition]{example}

\crefname{theorem}{Theorem}{Theorems}
\crefname{lemma}{Lemma}{Lemmas}
\crefname{proposition}{Proposition}{Propositions}
\crefname{corollary}{Corollary}{Corollaries}
\crefname{conjecture}{Conjecture}{Conjectures}
\crefname{observation}{Observation}{Observations}
\crefname{fact}{Fact}{Facts}
\crefname{definition}{Definition}{Definitions}

\crefname{remark}{Remark}{Remarks}
\crefname{problem}{Problem}{Problems}
\crefname{example}{Example}{Examples}
\crefname{question}{Question}{Questions}
\crefname{claim}{Claim}{Claims}
\crefname{section}{Section}{Sections}
\crefname{subsection}{Section}{Sections}
\crefname{subsubsection}{Section}{Sections}
\crefname{figure}{Figure}{Figures}

\DeclareMathOperator{\supp}{supp}
\DeclareMathOperator{\subrank}{Q}

\DeclareMathAccent{\wtilde}{\mathord}{largesymbols}{"65}
\DeclareMathOperator{\asympsubrank}{\underaccent{\wtilde}{Q}}

\DeclareMathOperator{\asymprank}{\underaccent{\wtilde}{R}}
\DeclareMathOperator{\declength}{L}
\DeclareMathOperator{\asympdeclength}{\underaccent{\wtilde}{L}}

\DeclareMathOperator{\tensorrank}{R}
\DeclareMathOperator{\merge}{merge}
\newcommand{\parts}{\mathsf{part}}
\newcommand{\bipart}{\mathsf{bipart}}
\newcommand{\polymult}{\mathsf{P}}
\newcommand{\FF}{\mathbb{F}}

\newcommand{\RR}{\mathbb{R}}
\newcommand{\CC}{\mathbb{C}}
\newcommand{\NN}{\mathbb{N}}
\newcommand{\ZZ}{\mathbb{Z}}

\def\<#1>{\left\langle\ignorespaces#1\unskip\right\rangle}

\newcommand{\asympleq}{\lesssim}

\usepackage{scalerel}

\DeclareMathOperator{\GL}{GL}

\DeclareMathOperator{\Id}{Id}
\DeclareMathOperator{\End}{End}

\DeclareMathOperator{\rank}{rank}  

\DeclareMathOperator{\Sym}{Sym}

\DeclareMathOperator{\conv}{conv}

\DeclarePairedDelimiterX\braket[2]{\langle}{\rangle}{#1\,\delimsize\vert\,\mathopen{}#2}

\title{The asymptotic spectrum of tensors is connected}

\author[1]{P\'eter Vrana}
\author[2]{Jeroen~Zuiddam}
\affil[1]{Budapest University of Technology and Economics}
\affil[2]{University of Amsterdam}
\date{November 2, 2026}

\begin{document}

\maketitle

\begin{abstract}
We prove path-connectedness of the asymptotic spectrum of complex tensors of every fixed order $k \geq 3$: every spectral point is connected by a path to every ``one leg versus the rest'' flattening rank. In particular, the asymptotic spectrum evaluated at any fixed tensor is a closed interval, answering a question of Strassen. 
This was previously known only on specific families of tensors (Strassen, FOCS 1986, Crelle 1987--1991; Wigderson and Zuiddam, Bull.~AMS 2026) or for specific families of spectral points (Christandl, Vrana and Zuiddam, STOC 2018; Sakabe, Do\u{g}an and Walter, FOCS 2026; Alman, Li and Pratt, FOCS 2026). 

Determining the topological and geometric properties of the asymptotic spectrum of tensors is central in Strassen's development of it, as it leads to new routes for upper bounding asymptotic rank (and, e.g.,~the matrix multiplication exponent $\omega$) beyond standard asymptotic methods, or for obtaining barriers for such bounds. Indeed, path-connectedness naturally leads to such routes. 

To obtain our path-connectedness theorem we develop a general method for ``interpolating'' between spectral points, which applies to  Strassen semirings that are appropriately decomposable (in the sense of an information-theoretic ``method of types'' on the semiring) and whose spectral points satisfy a cover property (inspired by slice rank). This covers tensors and asymptotic rank (with ``types'' given by decomposition into symmetric group irreps), as well as other objects and asymptotic problems, like graphs and Shannon capacity, where it matches the interpolation of Vrana~(2019).

Using our interpolation methods we furthermore obtain the following new results and simpler proofs of previous results: \begin{enumerate}[(1)]
\item We give a~simpler proof of Strassen's log-star-convexity theorem for the asymptotic spectrum of rectangular matrix multiplication, and of its extension to graph tensors by Wigderson and Zuiddam, using a notion of irreducibility for tensors.
\item We generalise the log-star-convexity theorem to the asymptotic spectrum of polynomial multiplication tensors and their permutations, which are central in the recent~$\omega \leq 9/4$ proof.
\item We give a simpler proof that the singleton quantum functionals of Christandl, Vrana and Zuiddam~(STOC 2018) are spectral points, by interpolating between the singleton flattening ranks.
\item By alternating interpolation between flattening ranks and merging (partial flattening), we prove that the non-crossing (laminar) upper quantum functionals are spectral points on complex tensors of every fixed order $k \geq 3$.
\end{enumerate}

\end{abstract}

\newpage
{\small\tableofcontents}
\newpage

\section{Introduction}
\suppressfloats[t]
\begin{figure}[b]
\centering
\begin{tikzpicture}[font=\small,scale=1.05]
\fill[gray!18] (1.710,-0.250) -- (1.716,-0.203) -- (1.715,-0.155) -- (1.707,-0.108) -- (1.693,-0.061) -- (1.672,-0.016) -- (1.646,0.028) -- (1.615,0.071) -- (1.579,0.111) -- (1.540,0.150) -- (1.498,0.187) -- (1.453,0.222) -- (1.406,0.254) -- (1.357,0.285) -- (1.308,0.314) -- (1.258,0.342) -- (1.207,0.368) -- (1.157,0.392) -- (1.106,0.415) -- (1.055,0.437) -- (1.005,0.458) -- (0.954,0.477) -- (0.903,0.495) -- (0.853,0.512) -- (0.802,0.527) -- (0.751,0.541) -- (0.699,0.554) -- (0.648,0.565) -- (0.596,0.574) -- (0.544,0.582) -- (0.493,0.588) -- (0.441,0.592) -- (0.390,0.595) -- (0.339,0.596) -- (0.289,0.597) -- (0.239,0.596) -- (0.190,0.595) -- (0.142,0.593) -- (0.094,0.591) -- (0.047,0.588) -- (0.000,0.587) -- (-0.047,0.585) -- (-0.093,0.585) -- (-0.140,0.585) -- (-0.188,0.586) -- (-0.237,0.588) -- (-0.287,0.591) -- (-0.338,0.594) -- (-0.391,0.598) -- (-0.446,0.601) -- (-0.502,0.604) -- (-0.561,0.606) -- (-0.620,0.607) -- (-0.681,0.607) -- (-0.743,0.604) -- (-0.805,0.598) -- (-0.867,0.590) -- (-0.928,0.579) -- (-0.987,0.564) -- (-1.045,0.546) -- (-1.100,0.524) -- (-1.151,0.499) -- (-1.199,0.471) -- (-1.242,0.439) -- (-1.280,0.405) -- (-1.314,0.368) -- (-1.343,0.330) -- (-1.367,0.289) -- (-1.386,0.247) -- (-1.400,0.205) -- (-1.410,0.161) -- (-1.417,0.118) -- (-1.421,0.075) -- (-1.422,0.032) -- (-1.421,-0.010) -- (-1.418,-0.051) -- (-1.415,-0.092) -- (-1.411,-0.132) -- (-1.408,-0.172) -- (-1.404,-0.211) -- (-1.401,-0.250) -- (-1.399,-0.289) -- (-1.397,-0.327) -- (-1.395,-0.366) -- (-1.393,-0.405) -- (-1.391,-0.445) -- (-1.388,-0.485) -- (-1.383,-0.525) -- (-1.377,-0.565) -- (-1.369,-0.606) -- (-1.358,-0.646) -- (-1.344,-0.686) -- (-1.327,-0.726) -- (-1.307,-0.765) -- (-1.283,-0.803) -- (-1.255,-0.840) -- (-1.223,-0.876) -- (-1.188,-0.909) -- (-1.149,-0.941) -- (-1.107,-0.971) -- (-1.063,-0.998) -- (-1.016,-1.024) -- (-0.967,-1.048) -- (-0.917,-1.069) -- (-0.865,-1.089) -- (-0.813,-1.106) -- (-0.759,-1.123) -- (-0.706,-1.137) -- (-0.652,-1.151) -- (-0.598,-1.163) -- (-0.543,-1.174) -- (-0.489,-1.184) -- (-0.435,-1.193) -- (-0.381,-1.201) -- (-0.327,-1.209) -- (-0.272,-1.215) -- (-0.218,-1.219) -- (-0.163,-1.223) -- (-0.109,-1.224) -- (-0.054,-1.224) -- (-0.000,-1.222) -- (0.054,-1.218) -- (0.107,-1.212) -- (0.160,-1.203) -- (0.212,-1.191) -- (0.262,-1.178) -- (0.311,-1.162) -- (0.358,-1.144) -- (0.403,-1.124) -- (0.447,-1.103) -- (0.488,-1.080) -- (0.528,-1.056) -- (0.566,-1.032) -- (0.602,-1.007) -- (0.638,-0.983) -- (0.672,-0.958) -- (0.706,-0.935) -- (0.741,-0.912) -- (0.776,-0.889) -- (0.812,-0.868) -- (0.849,-0.848) -- (0.889,-0.828) -- (0.930,-0.809) -- (0.974,-0.791) -- (1.020,-0.772) -- (1.069,-0.753) -- (1.120,-0.733) -- (1.173,-0.712) -- (1.226,-0.690) -- (1.281,-0.666) -- (1.336,-0.640) -- (1.390,-0.611) -- (1.442,-0.580) -- (1.492,-0.546) -- (1.539,-0.510) -- (1.581,-0.472) -- (1.619,-0.431) -- (1.651,-0.388) -- (1.678,-0.343) -- (1.697,-0.297) -- cycle;
\draw[gray!70!black] (1.710,-0.250) -- (1.716,-0.203) -- (1.715,-0.155) -- (1.707,-0.108) -- (1.693,-0.061) -- (1.672,-0.016) -- (1.646,0.028) -- (1.615,0.071) -- (1.579,0.111) -- (1.540,0.150) -- (1.498,0.187) -- (1.453,0.222) -- (1.406,0.254) -- (1.357,0.285) -- (1.308,0.314) -- (1.258,0.342) -- (1.207,0.368) -- (1.157,0.392) -- (1.106,0.415) -- (1.055,0.437) -- (1.005,0.458) -- (0.954,0.477) -- (0.903,0.495) -- (0.853,0.512) -- (0.802,0.527) -- (0.751,0.541) -- (0.699,0.554) -- (0.648,0.565) -- (0.596,0.574) -- (0.544,0.582) -- (0.493,0.588) -- (0.441,0.592) -- (0.390,0.595) -- (0.339,0.596) -- (0.289,0.597) -- (0.239,0.596) -- (0.190,0.595) -- (0.142,0.593) -- (0.094,0.591) -- (0.047,0.588) -- (0.000,0.587) -- (-0.047,0.585) -- (-0.093,0.585) -- (-0.140,0.585) -- (-0.188,0.586) -- (-0.237,0.588) -- (-0.287,0.591) -- (-0.338,0.594) -- (-0.391,0.598) -- (-0.446,0.601) -- (-0.502,0.604) -- (-0.561,0.606) -- (-0.620,0.607) -- (-0.681,0.607) -- (-0.743,0.604) -- (-0.805,0.598) -- (-0.867,0.590) -- (-0.928,0.579) -- (-0.987,0.564) -- (-1.045,0.546) -- (-1.100,0.524) -- (-1.151,0.499) -- (-1.199,0.471) -- (-1.242,0.439) -- (-1.280,0.405) -- (-1.314,0.368) -- (-1.343,0.330) -- (-1.367,0.289) -- (-1.386,0.247) -- (-1.400,0.205) -- (-1.410,0.161) -- (-1.417,0.118) -- (-1.421,0.075) -- (-1.422,0.032) -- (-1.421,-0.010) -- (-1.418,-0.051) -- (-1.415,-0.092) -- (-1.411,-0.132) -- (-1.408,-0.172) -- (-1.404,-0.211) -- (-1.401,-0.250) -- (-1.399,-0.289) -- (-1.397,-0.327) -- (-1.395,-0.366) -- (-1.393,-0.405) -- (-1.391,-0.445) -- (-1.388,-0.485) -- (-1.383,-0.525) -- (-1.377,-0.565) -- (-1.369,-0.606) -- (-1.358,-0.646) -- (-1.344,-0.686) -- (-1.327,-0.726) -- (-1.307,-0.765) -- (-1.283,-0.803) -- (-1.255,-0.840) -- (-1.223,-0.876) -- (-1.188,-0.909) -- (-1.149,-0.941) -- (-1.107,-0.971) -- (-1.063,-0.998) -- (-1.016,-1.024) -- (-0.967,-1.048) -- (-0.917,-1.069) -- (-0.865,-1.089) -- (-0.813,-1.106) -- (-0.759,-1.123) -- (-0.706,-1.137) -- (-0.652,-1.151) -- (-0.598,-1.163) -- (-0.543,-1.174) -- (-0.489,-1.184) -- (-0.435,-1.193) -- (-0.381,-1.201) -- (-0.327,-1.209) -- (-0.272,-1.215) -- (-0.218,-1.219) -- (-0.163,-1.223) -- (-0.109,-1.224) -- (-0.054,-1.224) -- (-0.000,-1.222) -- (0.054,-1.218) -- (0.107,-1.212) -- (0.160,-1.203) -- (0.212,-1.191) -- (0.262,-1.178) -- (0.311,-1.162) -- (0.358,-1.144) -- (0.403,-1.124) -- (0.447,-1.103) -- (0.488,-1.080) -- (0.528,-1.056) -- (0.566,-1.032) -- (0.602,-1.007) -- (0.638,-0.983) -- (0.672,-0.958) -- (0.706,-0.935) -- (0.741,-0.912) -- (0.776,-0.889) -- (0.812,-0.868) -- (0.849,-0.848) -- (0.889,-0.828) -- (0.930,-0.809) -- (0.974,-0.791) -- (1.020,-0.772) -- (1.069,-0.753) -- (1.120,-0.733) -- (1.173,-0.712) -- (1.226,-0.690) -- (1.281,-0.666) -- (1.336,-0.640) -- (1.390,-0.611) -- (1.442,-0.580) -- (1.492,-0.546) -- (1.539,-0.510) -- (1.581,-0.472) -- (1.619,-0.431) -- (1.651,-0.388) -- (1.678,-0.343) -- (1.697,-0.297) -- cycle;
\draw[gray!70!black,dashed] (1.330,-0.126) -- (1.328,-0.093) -- (1.322,-0.060) -- (1.311,-0.028) -- (1.296,0.002) -- (1.277,0.031) -- (1.254,0.058) -- (1.227,0.084) -- (1.198,0.108) -- (1.166,0.130) -- (1.133,0.151) -- (1.098,0.170) -- (1.062,0.187) -- (1.025,0.203) -- (0.987,0.217) -- (0.949,0.230) -- (0.912,0.242) -- (0.874,0.253) -- (0.836,0.263) -- (0.799,0.273) -- (0.761,0.281) -- (0.724,0.289) -- (0.687,0.295) -- (0.650,0.301) -- (0.613,0.306) -- (0.576,0.310) -- (0.540,0.312) -- (0.503,0.314) -- (0.466,0.315) -- (0.429,0.314) -- (0.393,0.312) -- (0.357,0.309) -- (0.321,0.305) -- (0.286,0.300) -- (0.251,0.295) -- (0.217,0.289) -- (0.183,0.282) -- (0.150,0.275) -- (0.117,0.268) -- (0.085,0.261) -- (0.053,0.255) -- (0.021,0.249) -- (-0.011,0.243) -- (-0.044,0.238) -- (-0.077,0.233) -- (-0.111,0.229) -- (-0.146,0.225) -- (-0.181,0.221) -- (-0.219,0.217) -- (-0.257,0.213) -- (-0.296,0.209) -- (-0.337,0.204) -- (-0.378,0.198) -- (-0.420,0.190) -- (-0.462,0.181) -- (-0.504,0.170) -- (-0.546,0.158) -- (-0.587,0.143) -- (-0.626,0.126) -- (-0.664,0.107) -- (-0.699,0.085) -- (-0.732,0.062) -- (-0.762,0.037) -- (-0.788,0.010) -- (-0.810,-0.018) -- (-0.830,-0.047) -- (-0.845,-0.077) -- (-0.857,-0.108) -- (-0.865,-0.139) -- (-0.870,-0.170) -- (-0.872,-0.201) -- (-0.872,-0.232) -- (-0.869,-0.262) -- (-0.865,-0.291) -- (-0.860,-0.320) -- (-0.853,-0.349) -- (-0.846,-0.377) -- (-0.839,-0.404) -- (-0.832,-0.431) -- (-0.825,-0.458) -- (-0.819,-0.484) -- (-0.812,-0.511) -- (-0.806,-0.537) -- (-0.801,-0.564) -- (-0.795,-0.590) -- (-0.789,-0.617) -- (-0.782,-0.644) -- (-0.775,-0.672) -- (-0.766,-0.699) -- (-0.755,-0.726) -- (-0.743,-0.753) -- (-0.729,-0.779) -- (-0.713,-0.804) -- (-0.694,-0.829) -- (-0.673,-0.853) -- (-0.649,-0.875) -- (-0.623,-0.896) -- (-0.595,-0.915) -- (-0.565,-0.932) -- (-0.533,-0.948) -- (-0.499,-0.962) -- (-0.464,-0.974) -- (-0.427,-0.985) -- (-0.390,-0.994) -- (-0.352,-1.001) -- (-0.314,-1.008) -- (-0.275,-1.013) -- (-0.236,-1.017) -- (-0.197,-1.020) -- (-0.159,-1.022) -- (-0.120,-1.023) -- (-0.081,-1.024) -- (-0.043,-1.024) -- (-0.005,-1.024) -- (0.034,-1.022) -- (0.072,-1.020) -- (0.110,-1.017) -- (0.148,-1.013) -- (0.186,-1.008) -- (0.223,-1.002) -- (0.261,-0.994) -- (0.298,-0.985) -- (0.334,-0.975) -- (0.369,-0.962) -- (0.403,-0.949) -- (0.437,-0.934) -- (0.469,-0.917) -- (0.499,-0.899) -- (0.528,-0.880) -- (0.556,-0.861) -- (0.582,-0.840) -- (0.606,-0.819) -- (0.630,-0.798) -- (0.652,-0.777) -- (0.673,-0.756) -- (0.695,-0.735) -- (0.715,-0.714) -- (0.736,-0.695) -- (0.758,-0.675) -- (0.780,-0.657) -- (0.804,-0.638) -- (0.829,-0.620) -- (0.856,-0.602) -- (0.884,-0.584) -- (0.914,-0.566) -- (0.945,-0.547) -- (0.978,-0.528) -- (1.012,-0.507) -- (1.046,-0.486) -- (1.081,-0.463) -- (1.116,-0.438) -- (1.150,-0.412) -- (1.183,-0.385) -- (1.213,-0.356) -- (1.242,-0.326) -- (1.266,-0.294) -- (1.288,-0.261) -- (1.305,-0.228) -- (1.318,-0.194) -- (1.326,-0.160) -- cycle;
\draw[->,>=stealth] (1.696,-0.300) -- (1.666,-0.284) -- (1.636,-0.269) -- (1.605,-0.254) -- (1.575,-0.241) -- (1.544,-0.228) -- (1.513,-0.216) -- (1.482,-0.205) -- (1.451,-0.195) -- (1.420,-0.186) -- (1.388,-0.177) -- (1.357,-0.169) -- (1.326,-0.162) -- (1.294,-0.156) -- (1.263,-0.151) -- (1.232,-0.147) -- (1.200,-0.143) -- (1.169,-0.141) -- (1.138,-0.139) -- (1.107,-0.138) -- (1.075,-0.138) -- (1.045,-0.138) -- (1.014,-0.140) -- (0.983,-0.142) -- (0.953,-0.145) -- (0.922,-0.149) -- (0.892,-0.154) -- (0.862,-0.159) -- (0.833,-0.165);
\draw (0.833,-0.165) -- (0.803,-0.172) -- (0.774,-0.180) -- (0.745,-0.189) -- (0.716,-0.198) -- (0.688,-0.208) -- (0.660,-0.219) -- (0.633,-0.231) -- (0.605,-0.243) -- (0.578,-0.256) -- (0.552,-0.270) -- (0.526,-0.285) -- (0.500,-0.300);
\draw[->,>=stealth] (1.102,0.417) -- (1.077,0.407) -- (1.053,0.397) -- (1.029,0.386) -- (1.005,0.374) -- (0.982,0.362) -- (0.959,0.350) -- (0.937,0.337) -- (0.915,0.323) -- (0.894,0.309) -- (0.873,0.295) -- (0.853,0.280) -- (0.833,0.264) -- (0.814,0.248) -- (0.795,0.232) -- (0.776,0.216) -- (0.758,0.199) -- (0.741,0.181) -- (0.724,0.163) -- (0.708,0.145) -- (0.692,0.127) -- (0.677,0.108) -- (0.662,0.089) -- (0.648,0.069) -- (0.635,0.049) -- (0.622,0.029) -- (0.609,0.009) -- (0.598,-0.012) -- (0.586,-0.033);
\draw (0.586,-0.033) -- (0.576,-0.054) -- (0.566,-0.075) -- (0.557,-0.097) -- (0.548,-0.119) -- (0.540,-0.141) -- (0.532,-0.163) -- (0.525,-0.186) -- (0.519,-0.208) -- (0.513,-0.231) -- (0.508,-0.254) -- (0.504,-0.277) -- (0.500,-0.300);
\draw[->,>=stealth] (-1.398,-0.300) -- (-1.351,-0.325) -- (-1.303,-0.350) -- (-1.254,-0.372) -- (-1.206,-0.394) -- (-1.157,-0.414) -- (-1.108,-0.433) -- (-1.059,-0.450) -- (-1.009,-0.467) -- (-0.960,-0.482) -- (-0.910,-0.495) -- (-0.860,-0.507) -- (-0.811,-0.518) -- (-0.761,-0.528) -- (-0.711,-0.536) -- (-0.661,-0.543) -- (-0.611,-0.549) -- (-0.562,-0.553) -- (-0.512,-0.556) -- (-0.463,-0.557) -- (-0.413,-0.558) -- (-0.364,-0.557) -- (-0.315,-0.554) -- (-0.267,-0.551) -- (-0.218,-0.546) -- (-0.170,-0.540) -- (-0.122,-0.532) -- (-0.075,-0.524) -- (-0.028,-0.514);
\draw (-0.028,-0.514) -- (0.019,-0.503) -- (0.065,-0.490) -- (0.111,-0.477) -- (0.157,-0.462) -- (0.201,-0.446) -- (0.246,-0.428) -- (0.290,-0.410) -- (0.333,-0.390) -- (0.376,-0.369) -- (0.418,-0.347) -- (0.459,-0.324) -- (0.500,-0.300);
\draw[->,>=stealth] (-0.033,-1.224) -- (-0.007,-1.207) -- (0.018,-1.191) -- (0.042,-1.174) -- (0.067,-1.156) -- (0.090,-1.138) -- (0.113,-1.120) -- (0.135,-1.101) -- (0.157,-1.081) -- (0.178,-1.061) -- (0.199,-1.041) -- (0.219,-1.020) -- (0.238,-0.999) -- (0.257,-0.977) -- (0.275,-0.955) -- (0.292,-0.933) -- (0.309,-0.911) -- (0.325,-0.888) -- (0.340,-0.864) -- (0.355,-0.841) -- (0.369,-0.817) -- (0.382,-0.793) -- (0.395,-0.768) -- (0.407,-0.743) -- (0.418,-0.719) -- (0.428,-0.693) -- (0.438,-0.668) -- (0.447,-0.643) -- (0.456,-0.617);
\draw (0.456,-0.617) -- (0.463,-0.591) -- (0.470,-0.565) -- (0.477,-0.539) -- (0.482,-0.513) -- (0.487,-0.486) -- (0.491,-0.460) -- (0.494,-0.433) -- (0.497,-0.407) -- (0.499,-0.380) -- (0.500,-0.353) -- (0.500,-0.327) -- (0.500,-0.300);
\draw[->,>=stealth] (0.824,-0.861) -- (0.824,-0.843) -- (0.823,-0.825) -- (0.821,-0.807) -- (0.819,-0.789) -- (0.817,-0.771) -- (0.814,-0.753) -- (0.811,-0.735) -- (0.807,-0.718) -- (0.803,-0.701) -- (0.799,-0.684) -- (0.794,-0.667) -- (0.788,-0.650) -- (0.783,-0.634) -- (0.777,-0.618) -- (0.770,-0.602) -- (0.763,-0.586) -- (0.756,-0.571) -- (0.749,-0.556) -- (0.741,-0.541) -- (0.732,-0.526) -- (0.724,-0.512) -- (0.715,-0.498) -- (0.705,-0.484) -- (0.695,-0.470) -- (0.685,-0.457) -- (0.675,-0.444) -- (0.664,-0.432) -- (0.653,-0.420);
\draw (0.653,-0.420) -- (0.642,-0.408) -- (0.631,-0.396) -- (0.619,-0.385) -- (0.606,-0.374) -- (0.594,-0.363) -- (0.581,-0.353) -- (0.568,-0.343) -- (0.555,-0.334) -- (0.542,-0.325) -- (0.528,-0.316) -- (0.514,-0.308) -- (0.500,-0.300);
\fill (0.500,-0.300) circle (1.5pt);
\node[above left=-2pt] at (0.500,-0.300) {\(\tensorrank_1\)};
\node at (-1.020,0.250) {\(\Delta\)};
\node[above,inner sep=1pt] at (0.000,1.037) {(a) contractible \(\Delta\)};
\fill[gray!18] (7.077,0.000) -- (7.083,0.047) -- (7.082,0.095) -- (7.074,0.142) -- (7.059,0.189) -- (7.039,0.234) -- (7.013,0.278) -- (6.982,0.321) -- (6.946,0.361) -- (6.907,0.400) -- (6.864,0.437) -- (6.819,0.472) -- (6.772,0.504) -- (6.724,0.535) -- (6.674,0.564) -- (6.624,0.592) -- (6.574,0.618) -- (6.523,0.642) -- (6.473,0.665) -- (6.422,0.687) -- (6.371,0.708) -- (6.321,0.727) -- (6.270,0.745) -- (6.219,0.762) -- (6.168,0.777) -- (6.117,0.791) -- (6.066,0.804) -- (6.015,0.815) -- (5.963,0.824) -- (5.911,0.832) -- (5.859,0.838) -- (5.808,0.842) -- (5.756,0.845) -- (5.706,0.846) -- (5.655,0.847) -- (5.606,0.846) -- (5.557,0.845) -- (5.508,0.843) -- (5.461,0.841) -- (5.413,0.838) -- (5.367,0.837) -- (5.320,0.835) -- (5.273,0.835) -- (5.226,0.835) -- (5.179,0.836) -- (5.130,0.838) -- (5.080,0.841) -- (5.029,0.844) -- (4.976,0.848) -- (4.921,0.851) -- (4.864,0.854) -- (4.806,0.856) -- (4.746,0.857) -- (4.685,0.857) -- (4.624,0.854) -- (4.562,0.848) -- (4.500,0.840) -- (4.439,0.829) -- (4.379,0.814) -- (4.322,0.796) -- (4.267,0.774) -- (4.216,0.749) -- (4.168,0.721) -- (4.125,0.689) -- (4.086,0.655) -- (4.052,0.618) -- (4.024,0.580) -- (4.000,0.539) -- (3.981,0.497) -- (3.967,0.455) -- (3.956,0.411) -- (3.950,0.368) -- (3.946,0.325) -- (3.945,0.282) -- (3.946,0.240) -- (3.948,0.199) -- (3.952,0.158) -- (3.955,0.118) -- (3.959,0.078) -- (3.962,0.039) -- (3.965,0.000) -- (3.968,-0.039) -- (3.970,-0.077) -- (3.972,-0.116) -- (3.974,-0.155) -- (3.976,-0.195) -- (3.979,-0.235) -- (3.983,-0.275) -- (3.989,-0.315) -- (3.998,-0.356) -- (4.008,-0.396) -- (4.022,-0.436) -- (4.039,-0.476) -- (4.060,-0.515) -- (4.084,-0.553) -- (4.112,-0.590) -- (4.144,-0.626) -- (4.179,-0.659) -- (4.218,-0.691) -- (4.259,-0.721) -- (4.304,-0.748) -- (4.351,-0.774) -- (4.399,-0.798) -- (4.450,-0.819) -- (4.502,-0.839) -- (4.554,-0.856) -- (4.607,-0.873) -- (4.661,-0.887) -- (4.715,-0.901) -- (4.769,-0.913) -- (4.823,-0.924) -- (4.877,-0.934) -- (4.931,-0.943) -- (4.986,-0.951) -- (5.040,-0.959) -- (5.094,-0.965) -- (5.149,-0.969) -- (5.203,-0.973) -- (5.258,-0.974) -- (5.312,-0.974) -- (5.367,-0.972) -- (5.421,-0.968) -- (5.474,-0.962) -- (5.527,-0.953) -- (5.578,-0.941) -- (5.629,-0.928) -- (5.678,-0.912) -- (5.725,-0.894) -- (5.770,-0.874) -- (5.814,-0.853) -- (5.855,-0.830) -- (5.895,-0.806) -- (5.933,-0.782) -- (5.969,-0.757) -- (6.004,-0.733) -- (6.039,-0.708) -- (6.073,-0.685) -- (6.107,-0.662) -- (6.142,-0.639) -- (6.178,-0.618) -- (6.216,-0.598) -- (6.255,-0.578) -- (6.297,-0.559) -- (6.341,-0.541) -- (6.387,-0.522) -- (6.436,-0.503) -- (6.487,-0.483) -- (6.539,-0.462) -- (6.593,-0.440) -- (6.648,-0.416) -- (6.703,-0.390) -- (6.756,-0.361) -- (6.809,-0.330) -- (6.859,-0.296) -- (6.906,-0.260) -- (6.948,-0.222) -- (6.986,-0.181) -- (7.018,-0.138) -- (7.044,-0.093) -- (7.064,-0.047) -- cycle;
\draw[gray!70!black] (7.077,0.000) -- (7.083,0.047) -- (7.082,0.095) -- (7.074,0.142) -- (7.059,0.189) -- (7.039,0.234) -- (7.013,0.278) -- (6.982,0.321) -- (6.946,0.361) -- (6.907,0.400) -- (6.864,0.437) -- (6.819,0.472) -- (6.772,0.504) -- (6.724,0.535) -- (6.674,0.564) -- (6.624,0.592) -- (6.574,0.618) -- (6.523,0.642) -- (6.473,0.665) -- (6.422,0.687) -- (6.371,0.708) -- (6.321,0.727) -- (6.270,0.745) -- (6.219,0.762) -- (6.168,0.777) -- (6.117,0.791) -- (6.066,0.804) -- (6.015,0.815) -- (5.963,0.824) -- (5.911,0.832) -- (5.859,0.838) -- (5.808,0.842) -- (5.756,0.845) -- (5.706,0.846) -- (5.655,0.847) -- (5.606,0.846) -- (5.557,0.845) -- (5.508,0.843) -- (5.461,0.841) -- (5.413,0.838) -- (5.367,0.837) -- (5.320,0.835) -- (5.273,0.835) -- (5.226,0.835) -- (5.179,0.836) -- (5.130,0.838) -- (5.080,0.841) -- (5.029,0.844) -- (4.976,0.848) -- (4.921,0.851) -- (4.864,0.854) -- (4.806,0.856) -- (4.746,0.857) -- (4.685,0.857) -- (4.624,0.854) -- (4.562,0.848) -- (4.500,0.840) -- (4.439,0.829) -- (4.379,0.814) -- (4.322,0.796) -- (4.267,0.774) -- (4.216,0.749) -- (4.168,0.721) -- (4.125,0.689) -- (4.086,0.655) -- (4.052,0.618) -- (4.024,0.580) -- (4.000,0.539) -- (3.981,0.497) -- (3.967,0.455) -- (3.956,0.411) -- (3.950,0.368) -- (3.946,0.325) -- (3.945,0.282) -- (3.946,0.240) -- (3.948,0.199) -- (3.952,0.158) -- (3.955,0.118) -- (3.959,0.078) -- (3.962,0.039) -- (3.965,0.000) -- (3.968,-0.039) -- (3.970,-0.077) -- (3.972,-0.116) -- (3.974,-0.155) -- (3.976,-0.195) -- (3.979,-0.235) -- (3.983,-0.275) -- (3.989,-0.315) -- (3.998,-0.356) -- (4.008,-0.396) -- (4.022,-0.436) -- (4.039,-0.476) -- (4.060,-0.515) -- (4.084,-0.553) -- (4.112,-0.590) -- (4.144,-0.626) -- (4.179,-0.659) -- (4.218,-0.691) -- (4.259,-0.721) -- (4.304,-0.748) -- (4.351,-0.774) -- (4.399,-0.798) -- (4.450,-0.819) -- (4.502,-0.839) -- (4.554,-0.856) -- (4.607,-0.873) -- (4.661,-0.887) -- (4.715,-0.901) -- (4.769,-0.913) -- (4.823,-0.924) -- (4.877,-0.934) -- (4.931,-0.943) -- (4.986,-0.951) -- (5.040,-0.959) -- (5.094,-0.965) -- (5.149,-0.969) -- (5.203,-0.973) -- (5.258,-0.974) -- (5.312,-0.974) -- (5.367,-0.972) -- (5.421,-0.968) -- (5.474,-0.962) -- (5.527,-0.953) -- (5.578,-0.941) -- (5.629,-0.928) -- (5.678,-0.912) -- (5.725,-0.894) -- (5.770,-0.874) -- (5.814,-0.853) -- (5.855,-0.830) -- (5.895,-0.806) -- (5.933,-0.782) -- (5.969,-0.757) -- (6.004,-0.733) -- (6.039,-0.708) -- (6.073,-0.685) -- (6.107,-0.662) -- (6.142,-0.639) -- (6.178,-0.618) -- (6.216,-0.598) -- (6.255,-0.578) -- (6.297,-0.559) -- (6.341,-0.541) -- (6.387,-0.522) -- (6.436,-0.503) -- (6.487,-0.483) -- (6.539,-0.462) -- (6.593,-0.440) -- (6.648,-0.416) -- (6.703,-0.390) -- (6.756,-0.361) -- (6.809,-0.330) -- (6.859,-0.296) -- (6.906,-0.260) -- (6.948,-0.222) -- (6.986,-0.181) -- (7.018,-0.138) -- (7.044,-0.093) -- (7.064,-0.047) -- cycle;
\draw[->] (3.695,-1.294) -- (7.383,-1.294);
\draw[very thick] (3.945,-1.294) -- (7.083,-1.294);
\draw[densely dotted] (3.945,0.282) -- (3.945,-1.294);
\draw (3.945,-1.294) ++(0,-3.5pt) -- ++(0,7pt);
\draw[densely dotted] (7.083,0.047) -- (7.083,-1.294);
\draw (7.083,-1.294) ++(0,-3.5pt) -- ++(0,7pt);
\draw[densely dotted] (4.817,-0.450) -- (4.817,-1.294);
\draw[thick] (4.817,-0.450) -- (4.842,-0.469) -- (4.867,-0.487) -- (4.892,-0.503) -- (4.917,-0.518) -- (4.942,-0.531) -- (4.968,-0.543) -- (4.993,-0.553) -- (5.019,-0.562) -- (5.044,-0.569) -- (5.070,-0.575) -- (5.095,-0.579) -- (5.121,-0.582) -- (5.147,-0.583) -- (5.173,-0.583) -- (5.199,-0.581) -- (5.225,-0.578) -- (5.251,-0.573) -- (5.277,-0.567) -- (5.303,-0.559) -- (5.329,-0.550) -- (5.355,-0.539) -- (5.382,-0.527) -- (5.408,-0.513) -- (5.435,-0.498) -- (5.461,-0.481) -- (5.488,-0.463) -- (5.514,-0.443) -- (5.541,-0.422) -- (5.568,-0.399) -- (5.595,-0.375) -- (5.622,-0.349) -- (5.649,-0.322) -- (5.676,-0.293) -- (5.703,-0.263) -- (5.730,-0.231) -- (5.757,-0.198) -- (5.784,-0.163) -- (5.812,-0.127) -- (5.839,-0.089) -- (5.867,-0.050);
\fill (4.817,-0.450) circle (1.5pt) node[left] {\(f\)};
\fill (4.817,-1.294) circle (1.5pt) node[below=4pt] {\(f(s)\)};
\draw[densely dotted] (6.367,-0.420) -- (6.367,-1.294);
\draw[thick] (6.367,-0.420) -- (6.364,-0.394) -- (6.361,-0.370) -- (6.357,-0.346) -- (6.353,-0.323) -- (6.348,-0.300) -- (6.343,-0.279) -- (6.337,-0.259) -- (6.331,-0.239) -- (6.324,-0.220) -- (6.317,-0.202) -- (6.309,-0.185) -- (6.301,-0.168) -- (6.292,-0.153) -- (6.283,-0.138) -- (6.273,-0.124) -- (6.263,-0.111) -- (6.252,-0.099) -- (6.241,-0.088) -- (6.229,-0.077) -- (6.217,-0.067) -- (6.204,-0.059) -- (6.191,-0.051) -- (6.177,-0.044) -- (6.163,-0.037) -- (6.148,-0.032) -- (6.133,-0.027) -- (6.117,-0.023) -- (6.101,-0.020) -- (6.084,-0.018) -- (6.067,-0.017) -- (6.049,-0.016) -- (6.031,-0.017) -- (6.012,-0.018) -- (5.993,-0.020) -- (5.973,-0.023) -- (5.953,-0.027) -- (5.932,-0.031) -- (5.911,-0.037) -- (5.889,-0.043) -- (5.867,-0.050);
\fill (6.367,-0.420) circle (1.5pt) node[left] {\(g\)};
\fill (6.367,-1.294) circle (1.5pt) node[below=4pt] {\(g(s)\)};
\fill (5.867,-0.050) circle (1.5pt) node[above] {\(\tensorrank_1\)};
\node[below=4pt] at (3.945,-1.294) {\(\asympsubrank(s)\)};
\node[below=4pt] at (7.083,-1.294) {\(\asymprank(s)\)};
\node at (4.347,0.500) {\(\Delta\)};
\node[above,inner sep=1pt] at (5.367,1.037) {(b) path-connected \(\Delta\), closed interval \(\Delta(s)\)};
\fill[gray!18] (11.846,-0.018) -- (11.167,0.695) -- (10.921,0.106) -- (10.207,0.246) -- (10.508,-0.338) -- (9.750,-0.994) -- (10.978,-1.404) -- (11.497,-1.780) -- (11.829,-1.002) -- (12.870,-0.556) -- cycle;
\draw[gray!70!black] (11.846,-0.018) -- (11.167,0.695) -- (10.921,0.106) -- (10.207,0.246) -- (10.508,-0.338) -- (9.750,-0.994) -- (10.978,-1.404) -- (11.497,-1.780) -- (11.829,-1.002) -- (12.870,-0.556) -- cycle;
\draw[thick] (10.207,0.246) -- (11.271,-0.500);
\draw[thick] (10.991,-0.900) -- (11.271,-0.500);
\fill (10.207,0.246) circle (1.5pt) node[left] {\(\log f\)};
\fill (10.991,-0.900) circle (1.5pt) node[left] {\(\log g\)};
\fill (11.271,-0.500) circle (1.5pt) node[right] {\(\log\tensorrank_1\)};
\node[above,inner sep=1pt] at (11.171,1.037) {(c) log-star-convex \(\Delta(b_1,b_2)\)};
\end{tikzpicture}
\setlength{\abovecaptionskip}{4pt}
\caption{We prove that the asymptotic spectrum of complex $k$-tensors~$\Delta$ is contractible (to every flattening rank $\tensorrank_i$), hence path-connected, so for every complex $k$-tensor $s$, \(\Delta(s)=\{h(s):h\in\Delta\}\) is one closed interval $[\asympsubrank(s), \asymprank(s)]$. For any (permuted) polynomial multiplication tensors  $b_1,\ldots,  b_\ell$ we moreover prove that \(\Delta(b_1,\ldots, b_\ell)=\{(h(b_1),\ldots, h(b_\ell)):h\in\Delta\}\) is log-star-convex.}\label{fig:intro-connectedness}
\end{figure}
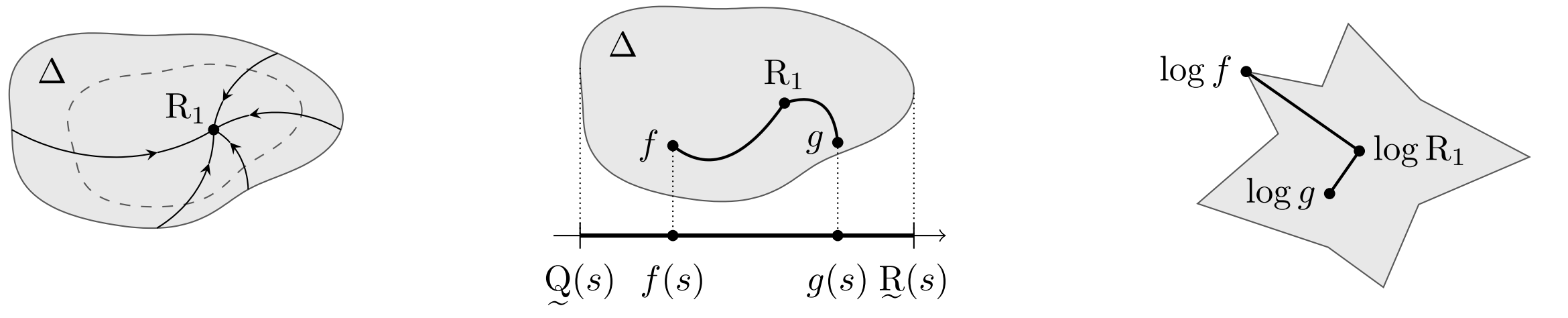

Since Strassen's 1969 ``Gaussian elimination is not optimal'' \cite{strassen1969gaussian}, which bounded the matrix multiplication exponent by $\omega \leq \log_2 7$, we have come a long way; decades of development of new methods and ideas have resulted not only in the remarkable~2026 bound $\omega \leq 9/4$ \cite{OAI:Matrix-Multiplication-Nine-Fourths-October-2-2026}, but in a flourishing asymptotic theory of tensors \cite{burgisser1997algebraic, blaser2013fast, wigderson2022asymptotic}, with many connections beyond algebraic complexity theory, via the asymptotic rank conjecture \cite{MR1341854,MR4204580,kaski2024universalsequencetensorsasymptotic,alman2026asymptoticrankspeeduptheorems} (with strong ties to other problems in complexity theory \cite{10.1145/3618260.3649656, 10.1145/3618260.3649620,björklund2024chromaticnumber19999ntime,MR5048041}), applications of asymptotic subrank and asymptotic slice rank in combinatorics \cite{MR3583358,tao,MR3631613,naslund_sawin_2017}, and study of quantum entanglement \cite{MR4495838,vrana2015transformation}, to name a few. 

Binding it all together is Strassen's own 1986--1991 theory of asymptotic spectrum duality \cite{DBLP:conf/focs/Strassen86,MR882307,strassen1988asymptotic,strassen1991degeneration}, which collects and explains the many techniques that arose in this area in one mathematical ``dual object'', from upper bounds on $\omega$ to barriers for it \cite{alman2018limits,v017a002,MR4870821,amanov2026multiplyingmatricespolynomialsexposition,shpilka2026exposition}. The duality has found applications to totally different areas \cite{wigderson2022asymptotic}, from catalytic computation \cite{MR4399732}, to circuits for Kronecker power matrices \cite{MR5048041}, to Shannon capacity \cite{MR4039606,MR4357434,MR4231964,deboer2024asymptoticspectrumdistancegraph}, to hypothesis testing \cite{MR4286265,MR4395416}.

This theory shifts from the study of a single tensor, say the matrix multiplication tensor, to the relations between tensors, the asymptotic restrictions. It introduces a ``complete'' set of monotone functions on tensors, called spectral points, that characterizes these relations, and thus the original problem, via a strong duality theorem (in essence a semiring version of linear programming duality). The spectral points form a compact topological space~$\Delta$, the asymptotic spectrum of tensors. For instance, the duality describes the matrix multiplication exponent, which is naturally a minimization problem, as a maximization over the asymptotic spectrum. The asymptotic tensor rank of a tensor is the largest value of the spectral points at it, and its asymptotic subrank is the smallest. 

While families of spectral points have been constructed  \cite{strassen1991degeneration,MR4495838,botero2026quantumfunctionalshigherordertensors,sakabe2026strassenssupportfunctionalscoincide,alman2026edgeasymptoticspectrumtensors}, it is not known how to construct all. Indeed, if the current constructions gave all spectral points of 3-tensors, then \(\omega=2\) and the asymptotic rank conjecture would hold. Even partial knowledge of topological and geometric structure of~$\Delta$, such as connectedness and convexity, leads to new routes for the original problem, but little is known about this structure.

We prove as our main result that the asymptotic spectrum~\(\Delta\) of complex $k$-tensors is path-connected, for every order \(k\geq3\). Every spectral point is joined by a path to each flattening rank~\(\tensorrank_i\). In fact, \(\Delta\) is contractible. In particular, for every complex tensor~\(s\), the values of the spectral points at~\(s\) form the closed interval $
[\asympsubrank(s),\asymprank(s)]$
from the asymptotic subrank to the asymptotic rank of~\(s\), which answers a question of Strassen \cite{DBLP:conf/issac/Strassen12}. We picture these properties in \cref{fig:intro-connectedness}~(a),(b). 

With Strassen's duality theorem, path-connectedness implies that asymptotic restrictions among elements of the $\oplus$-$\otimes$-semiring generated by any complex tensor~$s$ are decided by polynomial inequalities on this interval. This leads to new routes to upper bounds on asymptotic rank, which we will explain below.  

We obtain our results from a general interpolation method for abstract Strassen semirings, which combines information-theoretic ``method of types'' ideas with an abstract notion of slice rank. It applies not only to tensors (via decomposition into irreducible representations of symmetric groups and subspace averaging), but also to other asymptotic problems, which we will illustrate. 

Our interpolation methods moreover give a simpler proof of Strassen's theorem that the asymptotic spectrum of rectangular matrix multiplication is log-star-convex; an extension of this theorem to all (permuted) polynomial multiplication tensors (\cref{fig:intro-connectedness}~(c)), central in the recent $\omega \leq 9/4$ proof; a simpler proof that the quantum functionals \cite{MR4495838} are spectral points; and a proof that the non-crossing upper quantum functionals \cite{botero2026quantumfunctionalshigherordertensors} are spectral points.

Before stating our results in detail, we discuss their context.

\subsubsection*{Matrix multiplication and the asymptotic spectrum}

The matrix multiplication exponent~\(\omega\) is the infimum of the real numbers~\(\tau\) such that two \(n\times n\) matrices can be multiplied with \(O(n^\tau)\) arithmetic operations. 

Strassen's approach \cite{DBLP:conf/focs/Strassen86,strassen1988asymptotic} is to put matrix multiplication and all bilinear problems in a common framework of tensors in which computational reductions are encoded by a restriction preorder, pointwise multiplication is the basic measure, and problems can be combined in a semiring fashion.   Let~\(\FF\) be a field and let \(k\geq2\) be an integer. A \(k\)-tensor over~\(\FF\) is an element~\(t\) of a tensor product \(V_1\otimes\cdots\otimes V_k\) of finite-dimensional vector spaces over~\(\FF\). A \(k\)-tensor \(s\in W_1\otimes\cdots\otimes W_k\) is a \emph{restriction} of~\(t\), written \(s\leq t\), if \(s=(A_1\otimes\cdots\otimes A_k)t\) for some linear maps \(A_i:V_i\to W_i\). Two tensors are \emph{equivalent} if each is a restriction of the other. The direct sum \(s\oplus t\) places~\(s\) and~\(t\) in blocks of \((W_1\oplus V_1)\otimes\cdots\otimes(W_k\oplus V_k)\). The Kronecker product \(s\otimes t\) is the \(k\)-tensor in \((W_1\otimes V_1)\otimes\cdots\otimes(W_k\otimes V_k)\) obtained from the tensor product of~\(s\) and~\(t\) by grouping their \(i\)th legs, for every~\(i\). For every integer \(r\geq0\), the unit tensor is \(\langle r\rangle=\sum_{j=1}^r e_j\otimes\cdots\otimes e_j\).

The tensor rank~\(\tensorrank(t)\) is the smallest~\(r\) with \(t\leq\langle r\rangle\), and the subrank~\(\subrank(t)\) is the largest~\(r\) with \(\langle r\rangle\leq t\). The asymptotic tensor rank and the asymptotic subrank of~\(t\) are
\begin{align*}
\asymprank(t)&\coloneqq\lim_{n\to\infty}\tensorrank(t^{\otimes n})^{1/n},\\
\asympsubrank(t)&\coloneqq\lim_{n\to\infty}\subrank(t^{\otimes n})^{1/n}.
\end{align*}
For positive integers \(n_1,n_2,n_3\), let \(\langle n_1,n_2,n_3\rangle\) be the tensor of the bilinear map that multiplies an \(n_1\times n_2\) matrix with an \(n_2\times n_3\) matrix. Strassen \cite[(1.3)]{strassen1988asymptotic} showed that the exponent~\(\omega\) satisfies
\[
2^\omega=\asymprank(\langle2,2,2\rangle).
\]
The asymptotic subrank of every tensor is an upper bound on the asymptotic induced matching number of its support, and it is used for cap sets~\cite{MR3583358,tao,sawin}, sunflowers~\cite{naslund_sawin_2017} and barriers for upper bounds on~\(\omega\)~\cite{MR3631613,MR3899623,v017a002}.

Strassen defined the \emph{asymptotic spectrum}~\(\Delta_\FF\) of \(k\)-tensors over~\(\FF\). It consists of the functions~\(f\) from \(k\)-tensors over~\(\FF\) to~\(\RR_{\geq0}\) with \(f(\langle1\rangle)=1\) that are additive under direct sum, multiplicative under Kronecker product and monotone under restriction. Its elements are the \emph{spectral points}. With the topology of pointwise convergence, \(\Delta_\FF\) is compact. 

A tensor~\(s\) is an \emph{asymptotic restriction} of~\(t\), written \(s\asympleq t\), if \(s^{\otimes n}\leq t^{\smash{\otimes(n+o(n))}}\). Strassen's duality theorem says that \(s\asympleq t\) if and only if \(f(s)\leq f(t)\) for every \(f\in\Delta_\FF\). For tensors \(s_1,\ldots,s_q\) over~\(\FF\), let \(\Delta_\FF(s_1,\ldots,s_q)=\{(f(s_1),\ldots,f(s_q)):f\in\Delta_\FF\}\). By the duality theorem, this subset of~\(\RR^q\) characterizes all asymptotic restrictions among the tensors generated by \(s_1,\ldots,s_q\) under direct sum and Kronecker product. A consequence of the duality is that every tensor~\(t\) satisfies
\begin{align*}
\asymprank(t)&=\max\Delta_\FF(t),\\
\asympsubrank(t)&=\min\Delta_\FF(t).
\end{align*}
We note that for \(k\geq3\), asymptotic rank and asymptotic subrank are themselves not in~\(\Delta_\FF\), since they are not multiplicative (use $\langle 2,2,2\rangle = \langle 2,1,1\rangle \otimes\langle 1,2,1\rangle \otimes\langle 1,1,2\rangle$ and $\asymprank(\langle 2,2,2\rangle) < 8$, $\asympsubrank(\langle 2,2,2\rangle) = 4$, $\asymprank(\langle 2,1,1\rangle) = 2$, $\asympsubrank(\langle 2,1,1\rangle) = 1$ and same for permutations).

What elements in~$\Delta_\FF$ do we know? For \(i\in[k]=\{1,\ldots,k\}\), the flattening rank~\(\tensorrank_i(t)\) is the dimension of the smallest subspace \(U\subseteq V_i\) with \(t\in V_1\otimes\cdots\otimes U\otimes\cdots\otimes V_k\). Every flattening rank is a spectral point. The asymptotic rank conjecture \cite{MR1341854}, in the general form of \cite[Problem~15.5]{burgisser1997algebraic}, says that every complex \(3\)-tensor~\(t\) satisfies \(\asymprank(t)=\max_i\tensorrank_i(t)\): the largest value of the spectral points at~\(t\) would be attained by a flattening rank. 
In the direction of the asymptotic subrank, the situation is much richer already, as it is not integer-valued. Strassen \cite{strassen1991degeneration} constructed the support functionals, which are spectral points on oblique tensors (tensors whose support in suitable bases is an antichain in the product order), such as the matrix multiplication tensors. For instance, \(W=e_1\otimes e_1\otimes e_2+e_1\otimes e_2\otimes e_1+e_2\otimes e_1\otimes e_1\) is oblique with \(\tensorrank_i(W)=2\) for every~\(i\), and the smallest value of the support functionals at~\(W\) is its asymptotic subrank \(3\cdot2^{-2/3}\approx1.89\) \cite[Theorem~6.7]{strassen1991degeneration}. 
Christandl, Vrana and Zuiddam \cite{MR4495838} constructed the quantum functionals, a continuous family of spectral points of complex tensors. Sakabe, Do\u{g}an and Walter \cite{sakabe2026strassenssupportfunctionalscoincide} proved that they coincide with the upper support functionals. 

The goal of obtaining better bounds on asymptotic rank (for matrix multiplication or other tensors) and asymptotic subrank, and the asymptotic spectrum characterization, raises several natural questions: How can we make more spectral points from known ones? What structure do the spectral points have? What consequences does that structure have?

\subsubsection*{Connectedness and convexity properties and their implications}

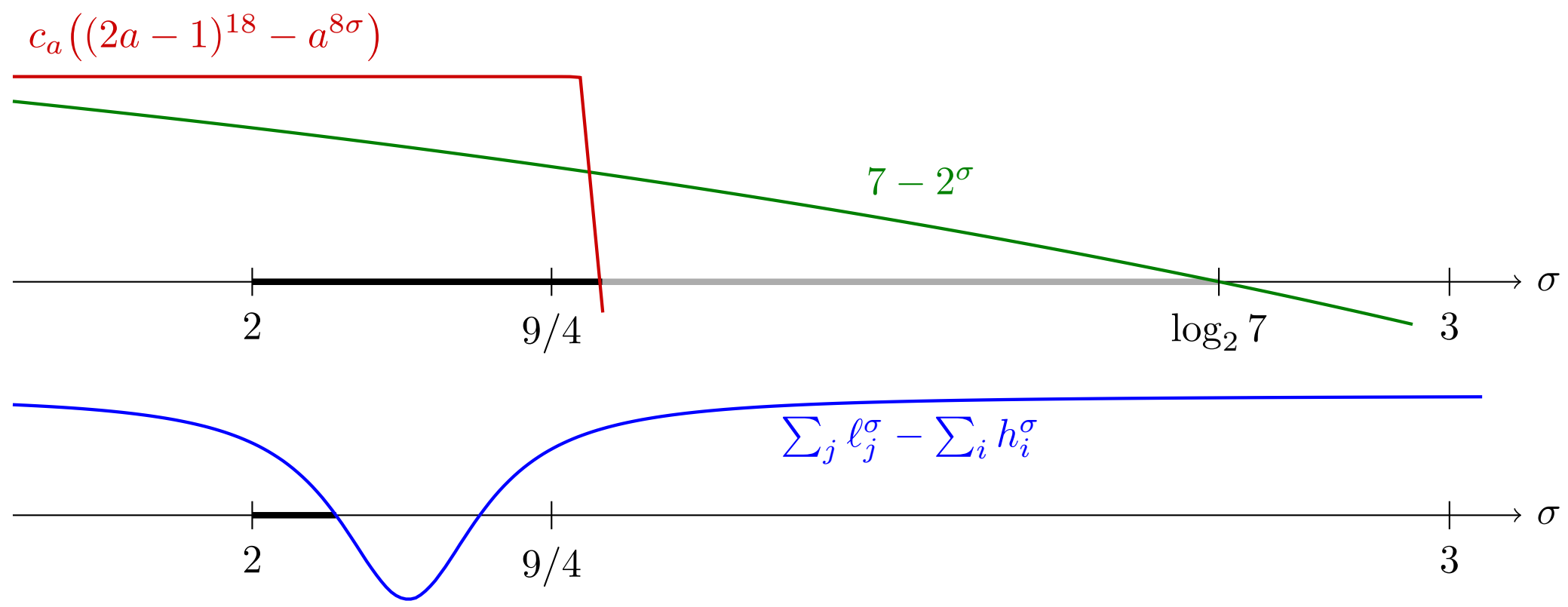
\begin{figure}[t]
\centering
\begin{tikzpicture}[font=\small]
\begin{scope}[x=10.5cm,y=0.45cm,shift={(-1.8,0)}]
\draw[->] (1.8,0) -- (3.06,0) node[right] {\(\sigma\)};
\draw[line width=1.6pt,gray!65] (2,0) -- (2.8074,0);
\draw[line width=1.6pt] (2,0) -- (2.292332,0);
\draw[green!50!black,thick] plot[domain=1.8:2.97,samples=100] (\x,{7-2^\x});
\foreach \t/\l in {2/2,2.8074/\log_27,3/3}{\draw (\t,0) ++(0,-3.5pt) -- ++(0,7pt); \node[below=4pt] at (\t,0) {\(\l\)};}
\draw (2.25,0) ++(0,-3.5pt) -- ++(0,7pt);
\node[below=4pt] at (2.25,0) {\(9/4\)};
\node[green!50!black,above right] at (2.5,1.3431) {\(7-2^\sigma\)};
\end{scope}
\begin{scope}[x=10.5cm,y=0.45cm,shift={(-1.8,0)}]
\draw[red!80!black,thick] plot[domain=1.8:2.274,samples=60] (\x,{4*(1-exp(294.7309*(\x-2.292332)))})
  -- plot[domain=2.274:2.2928,samples=200] (\x,{4*(1-exp(294.7309*(\x-2.292332)))});
\node[above right,red!80!black] at (1.8,4) {\(c_a\big((2a-1)^{18}-a^{8\sigma}\big)\)};
\end{scope}
\begin{scope}[x=10.5cm,y=1.05cm,shift={(-1.8,-1.95)}]
\draw[->] (1.8,0) -- (3.06,0) node[right] {\(\sigma\)};
\draw[line width=1.6pt] (2,0) -- (2.07,0);
\draw[blue,thick] plot[domain=1.8:3.03,samples=300] (\x,{((\x-2.13)*(\x-2.13)-0.0036)/((\x-2.13)*(\x-2.13)+0.00514)});
\foreach \t/\l in {2/2,3/3}{\draw (\t,0) ++(0,-3.5pt) -- ++(0,7pt); \node[below=4pt] at (\t,0) {\(\l\)};}
\draw (2.25,0) ++(0,-3.5pt) -- ++(0,7pt);
\node[below=4pt] at (2.25,0) {\(9/4\)};
\node[blue,below] at (2.55,0.9519) {\(\sum_j\ell_j^\sigma-\sum_ih_i^\sigma\)};
\end{scope}
\end{tikzpicture}
\caption{Differences of asymptotic restrictions as functions of~\(\sigma\), with the largest intervals~\([2,x]\) on which they are nonnegative as thick segments. Top: ``traditional'', monotone differences, for \(\langle2,2,2\rangle\asympleq\langle7\rangle\) \cite{strassen1969gaussian} and \(\langle a^8,a^8,a^8\rangle\asympleq\langle(2a-1)^{18}\rangle\) \cite{amanov2026multiplyingmatricespolynomialsexposition}, scaled by \(c_a=4(2a-1)^{-18}\) and drawn for \(a=10^{16}\) (\(a\to\infty\) gives \(\omega\leq9/4\)). Bottom: a potential ``non-traditional'', non-monotone difference, which would give \(\omega<9/4\) using connectedness.}\label{fig:intro-differences}
\end{figure}

Schönhage's \(\tau\)-theorem \cite{MR623057}, also called the asymptotic sum inequality, is the basic method used in every upper bound on~\(\omega\). It implies that for integers \(h_1,\ldots,h_q\geq2\) and \(r\geq1\), if \(\bigoplus_{i=1}^q\langle h_i,h_i,h_i\rangle\asympleq\langle r\rangle\), then \(\sum_{i=1}^qh_i^\omega\leq r\), that is, the difference \(r-\sum_{i=1}^qh_i^\sigma\) of such an asymptotic restriction is nonnegative at~\(\sigma=\omega\). Since this difference is decreasing in~\(\sigma\), \(\omega\) is at most its zero. For instance, the proof of \(\omega\leq9/4\) \cite{OAI:Matrix-Multiplication-Nine-Fourths-October-2-2026} gives, for every integer \(a\geq2\), the asymptotic restriction \(\langle a^8,a^8,a^8\rangle\asympleq\langle(2a-1)^{18}\rangle\) \cite{amanov2026multiplyingmatricespolynomialsexposition}. The zero \((9/4)\log(2a-1)/\log a\) of its difference \((2a-1)^{18}-a^{8\sigma}\) tends to~\(9/4\) as \(a\to\infty\) (\cref{fig:intro-differences}, top).

More generally, an asymptotic restriction can have a direct sum of matrix multiplication tensors on the right (see, e.g., the group-theoretic approach~\cite{cohn2003group,cohn2005group}),
\[
\bigoplus_{i=1}^q\langle h_i,h_i,h_i\rangle\asympleq\bigoplus_{j=1}^r\langle \ell_j,\ell_j,\ell_j\rangle.
\]
Does this give an upper bound on $\omega$?
Its difference \(\sum_{j=1}^r\ell_j^\sigma-\sum_{i=1}^qh_i^\sigma\) is still nonnegative at~\(\sigma=\omega\), but it need not be decreasing in~\(\sigma\), so this alone does not bound~\(\omega\). Here connectedness helps, since it turns the duality theorem into inequalities on intervals. Strassen \cite[Corollaries~6.7 and~6.8]{strassen1988asymptotic} proved that the values of the spectral points at~\(\langle2,2,2\rangle\) form the interval \([4,2^\omega]\), and hence that such an asymptotic restriction holds if and only if its difference is nonnegative on~\([2,\omega]\). In particular, the difference gives an upper bound on~\(\omega\) as soon as it is negative at some \(\sigma>2\), even if it is nonnegative again for larger~\(\sigma\). All upper bounds on~\(\omega\) so far have been obtained from asymptotic restrictions with a monotone difference (``traditional'' constructions, \cref{fig:intro-differences}, top). Connectedness allows a new route, which uses non-monotone differences more effectively (\cref{fig:intro-differences}, bottom). Our results imply that this route applies to every complex~\(k\)-tensor (\cref{cor:intro-polynomials}).

Path-connectedness was known only for a few families of tensors, and for two of them in the stronger form of log-star-convexity. A set \(Y\subseteq\RR^r\) is star-convex with respect to a point \(c\in Y\) if, for every \(y\in Y\), the segment from~\(y\) to~\(c\) lies in~\(Y\). Such a set is path-connected. Logarithms are to base~\(2\), and we apply~\(\log\) to vectors and to sets of vectors coordinatewise. A set~\(Y\subseteq\RR_{>0}^r\) is log-star-convex if \(\log Y\) is star-convex. Strassen \cite[Corollary~6.6]{strassen1988asymptotic} proved that the asymptotic spectrum of rectangular matrix multiplication is log-star-convex (\cref{th:intro-mm}). Wigderson and Zuiddam \cite[Corollary~9.11]{wigderson2022asymptotic} used this to bound~\(\omega\) beyond Schönhage's \(\tau\)-theorem. For positive integers \(a_i,b_i,c_i\) and \(u_j,v_j,w_j\), if \(\bigoplus_{i=1}^q\langle a_i,b_i,c_i\rangle\asympleq\bigoplus_{j=1}^r\langle u_j,v_j,w_j\rangle\), then \(\sum_{i=1}^q(a_ib_ic_i)^\kappa\leq\sum_{j=1}^r(u_jv_jw_j)^\kappa\) with \(\kappa=(4-\omega)/(7-2\omega)\). 

Wigderson and Zuiddam \cite[Theorem~12.23]{wigderson2022asymptotic} proved log-star-convexity for graph tensors (tensor networks). For a graph with vertex set~\([k]\) and a positive integer~\(n_{\mathsf{e}}\) for every edge~\(\mathsf{e}=\{i,j\}\), the graph tensor is the Kronecker product over the edges~\(\mathsf{e}\) of the \(k\)-tensors \(\sum_{x=1}^{n_{\mathsf{e}}}e_x\otimes e_x\) on the legs~\(i\) and~\(j\), with one-dimensional other legs. For the triangle graph with varying~$n_{\mathsf{e}}$, the graph tensors are the matrix multiplication tensors. 

More restrictive forms of connectedness were known, namely for~\(\Delta_\FF(t)\) for specific tensors~\(t\) and for specific constructions of spectral points. Strassen \cite[Section~6]{strassen1991degeneration} determined the asymptotic spectra of the multiplication tensors of the null algebras and of the algebras \(\FF[T]/(A)\) of univariate polynomials modulo a nonconstant polynomial~\(A\). Each of them is a closed interval. Over~\(\CC\), the quantum functionals of Christandl, Vrana and Zuiddam \cite{MR4495838} depend continuously on their parameter, and they give paths between the flattening ranks. Vrana \cite{MR4357434} constructed an interpolation between spectral points of graphs. Alman, Li and Pratt \cite[Corollary~8.7]{alman2026edgeasymptoticspectrumtensors} proved that the upper support functionals with weight zero on the third leg are spectral points of \(3\)-tensors over every field. These functionals form a path from~\(\tensorrank_2\) to~\(\tensorrank_1\).

\FloatBarrier
\subsubsection*{Other work on algebraic, geometric, computational structure of spectral points}

Recent work has proved many different structural properties of spectral points, asymptotic rank and asymptotic subrank, including algebraic, geometric and computational ones. 

Kaski and Micha\l{}ek \cite{kaski2024universalsequencetensorsasymptotic} constructed explicit sequences of zero-one-valued tensors that are universal for the asymptotic rank conjecture, and they related upper bounds on asymptotic rank to the absence of low-degree polynomials vanishing on the tensors of low asymptotic rank. Christandl, Hoeberechts, Nieuwboer, Vrana and Zuiddam \cite{MR4928468} proved that asymptotic tensor rank is characterized by polynomials. For every real number~\(r\), the tensors of a fixed format with asymptotic rank at most~\(r\) are the common zeros of finitely many polynomials. Thus asymptotic rank is semicontinuous. Over every computable field, this gives, for every~\(r\), an algorithm that decides whether a given tensor has asymptotic rank at most~\(r\). They proved the same results for every spectral point~\(f\). 

Several results concern the set of values that such a parameter takes on all tensors. Christandl, Hoeberechts, Nieuwboer, Vrana and Zuiddam \cite{MR4928468} also proved that this set is well-ordered for every spectral point. Blatter, Draisma and Rupniewski \cite{bdr} proved that over a finite field, the values of a class of parameters that includes asymptotic rank and asymptotic subrank form a well-ordered set, and that over~\(\CC\) such parameters take countably many values \cite{https://doi.org/10.48550/arxiv.2212.12219}. Christandl, Vrana and Zuiddam \cite{MR4495838} proved that over~\(\CC\), asymptotic slice rank takes only finitely many values on the tensors of each format. Bri\"et, Christandl, Leigh, Shpilka and Zuiddam \cite{briet_et_al:LIPIcs.ITCS.2024.20} proved that asymptotic subrank has no accumulation points on the \(3\)-tensors with coefficients in a fixed finite set. Costa and Dalai \cite{DBLP:journals/jcta/CostaD21}, Christandl, Gesmundo and Zuiddam \cite{cgz}, and Gesmundo and Zuiddam \cite{gesmundo2023gap} found explicit gaps in the smallest values of such parameters.

\FloatBarrier
\subsubsection*{New results}

In this paper:
\begin{itemize}
\item We prove that the asymptotic spectrum of complex \(k\)-tensors is contractible (to every flattening rank) and thus path-connected. In particular, from every spectral point there is a continuous path in the asymptotic spectrum to every flattening rank. It follows that the asymptotic spectrum evaluated on any complex tensor is a closed interval. This answers a question of Strassen (ISSAC 2012). \item Connectedness enables new routes (not following from standard asymptotic techniques) to upper bound the asymptotic rank of a tensor. These transform seemingly weak asymptotic restrictions among $\oplus$-$\otimes$-polynomial expressions of the tensor into upper bounds on its asymptotic rank, by letting asymptotic spectrum duality cut out (in the sense of semi-algebraic geometry) an outer approximation of the asymptotic spectrum, of which only the connected component of the known points (flattening ranks) matters. \item As the core ingredient for these results, we prove a general interpolation theorem for Strassen semirings, the abstract setting of Strassen's duality theorem. Under a decomposability condition and a cover property, it joins every spectral point to a fixed spectral point by a path that depends continuously on its starting point, which gives contractibility. 
Besides tensors, it applies to other asymptotic problems, for instance Shannon capacity of graphs. \end{itemize}
Using our interpolation methods, we prove several further results:
\begin{itemize}
\item  We give a new proof of Strassen's theorem that the asymptotic spectrum of rectangular matrix multiplication is log-star-convex \cite[Corollary~6.6]{strassen1988asymptotic}, and of the analogous theorem of Wigderson and Zuiddam for tensor networks \cite[Theorem~12.23]{wigderson2022asymptotic}. As a new ingredient we use a notion of irreducibility for tensors. We moreover extend Strassen's theorem to polynomial multiplication tensors and their permutations, which are central in the recent proof of~\(\omega\leq9/4\) \cite{OAI:Matrix-Multiplication-Nine-Fourths-October-2-2026} (see, e.g.,~\cite{amanov2026multiplyingmatricespolynomialsexposition,shpilka2026exposition}). \item By repeated interpolation between the flattening ranks, we obtain the quantum functionals, which Christandl, Vrana and Zuiddam \cite{MR4495838} defined through representation theory (moment polytopes) and quantum information theory (marginal spectra). This gives a new, self-contained proof that they are spectral points \cite[Corollary~3.31]{MR4495838}.
\item By more intricate repeated interpolation (alternated with merging) between flattening ranks of groups of legs, we prove that the non-crossing upper quantum functionals \cite[Definition~3.14]{MR4495838} are spectral points. Previously, this was known only on a subsemiring of the complex tensors, which is strictly contained in the semiring of all complex tensors for \(k\geq4\) \cite[Proposition~4.2 and Theorem~3.1]{botero2026quantumfunctionalshigherordertensors}.
\end{itemize}
In the rest of the introduction we will discuss our results in more detail.

\subsection{Path-connectedness and contractibility for complex tensors}

We prove that the asymptotic spectrum of complex tensors is contractible and thus path-connected.

\begin{theorem}[{\cref{th:tensor-contractible}}]\label{th:intro-complex}
Let \(k\geq2\), and let~\(\Delta\) be the asymptotic spectrum of complex \(k\)-tensors. For every \(i\in[k]\) and every \(f\in\Delta\), there is a path in~\(\Delta\) from~\(f\) to~\(\tensorrank_i\). For every complex \(k\)-tensor~\(s\), the set \(\{f(s):f\in\Delta\}\) is a closed interval. Moreover, \(\Delta\) is contractible.
\end{theorem}

By Strassen's duality theorem \cite[Theorem~3.8]{strassen1988asymptotic}, the interval in \cref{th:intro-complex} is
\[
[\asympsubrank(s),\asymprank(s)].
\]
The paths of \cref{th:intro-complex} give paths in \(\Delta_\CC(s_1,\ldots,s_q)\) to the point \((\tensorrank_i(s_1),\ldots,\tensorrank_i(s_q))\) (\cref{cor:tensor-paths}).

For a complex tensor~\(t\) and a polynomial \(p(x)=c_0+c_1x+\cdots+c_dx^d\) with nonnegative integer coefficients, let
\[
p(t)\coloneqq\langle c_0\rangle\oplus\langle c_1\rangle\otimes t\oplus\cdots\oplus\langle c_d\rangle\otimes t^{\otimes d}.
\]
Every \(f\in\Delta\) satisfies \(f(p(t))=p(f(t))\). Thus \cref{th:intro-complex} and the duality theorem give the following corollary.

\begin{corollary}\label{cor:intro-polynomials}
Let \(k\geq3\), and let~\(t\) be a complex \(k\)-tensor. For all polynomials \(p\) and~\(q\) with nonnegative integer coefficients, \(p(t)\asympleq q(t)\) if and only if \(p(x)\leq q(x)\) for every \(x\in[\asympsubrank(t),\asymprank(t)]\).
\end{corollary}

As for matrix multiplication, \cref{cor:intro-polynomials} gives a new route to upper bounds on asymptotic rank. Let \(t\), \(p\) and~\(q\) be as in \cref{cor:intro-polynomials}, with \(p(t)\asympleq q(t)\). For any $x\in \RR$, if \(p(x)>q(x)\) and \(x>\tensorrank_1(t)\), then \(\asymprank(t)<x\). Without connectedness, Strassen's duality theorem gives \(p(f(t))\leq q(f(t))\) for every spectral point~\(f\), so \(p(x)>q(x)\) excludes only the value~\(x\). To conclude \(\asymprank(t)<x\), we would moreover need \(p(y)>q(y)\) for every~\(y\) between~\(x\) and a known upper bound on~\(\asymprank(t)\).

\subsubsection{General interpolation theorem}

We will prove \cref{th:intro-complex} as an application of a general interpolation theorem, for which we introduce new notions of decomposability and of a cover property. 

Let~\(\mathcal{S}\) be a Strassen semiring, for instance the semiring of tensors under $\oplus$ and $\otimes$ with $\leq$ the restriction order on tensors. Let~\(\Delta\) be its asymptotic spectrum. Let \(\mathcal{A}\subseteq\mathcal{S}\setminus\{0\}\). For \(s\in\mathcal{S}\), we call a list of elements \(a_1,\ldots,a_q\in\mathcal{A}\) with \(a_j\leq s\) for every \(j\in[q]\) and \(s\leq a_1+\cdots+a_q\) an \emph{\(\mathcal{A}\)-decomposition} of~\(s\) of length~\(q\). We say that~\(\mathcal{S}\) is \emph{\(\mathcal{A}\)-decomposable} if~\(\mathcal{A}\) is multiplicatively closed and contains every positive integer, and if the powers~\(s^n\) of every nonzero \(s\in\mathcal{S}\) have \(\mathcal{A}\)-decompositions of length \(2^{\smash{o(n)}}\). For such a decomposition \(a_1,\ldots,a_q\) of~\(s^n\), every \(f\in\Delta\) satisfies the sandwich \(\max_jf(a_j)\leq f(s)^n\leq q\max_jf(a_j)\). By taking $n$th roots and applying the sandwich for $n\to\infty$, we find that if~\(\mathcal{S}\) is \(\mathcal{A}\)-decomposable, every spectral point is determined by its values on~\(\mathcal{A}\).
We say that spectral points \(f_0\) and~\(f_1\) have the \emph{\(\mathcal{A}\)-cover property} if \(a^n\leq s_0+s_1\) with \(a\in\mathcal{A}\) implies that \(f_0(s_0)/f_0(a)^n+f_1(s_1)/f_1(a)^n\) is at least \(2^{\smash{-o(n)}}\), where the \(o(n)\) depends only on~\(a\).

For \(f_0,f_1\in\Delta\), \(t\in[0,1]\) and \(s\in\mathcal{S}\), let
\[
F^{\smash[t]{f_0}}_t(s)\coloneqq\sup\bigl\{(f_0(a)^{1-t}f_1(a)^t)^{1/n}:n\geq1,\ a\in\mathcal{A},\ a\leq s^n\bigr\}.
\]

\begin{theorem}[{\cref{th:interpolation}}]\label{th:intro-interpolation}
Let~\(\mathcal{S}\) be \(\mathcal{A}\)-decomposable, and let \(f_1\in\Delta\).
\begin{enumerate}[\upshape(i)]
\item\label{it:intro-interp-path} For every \(f_0\in\Delta\) such that the pair \(f_0,f_1\) has the \(\mathcal{A}\)-cover property, the function \(t\mapsto F^{\smash[t]{f_0}}_t\) is a path in~\(\Delta\) from~\(f_0\) to~\(f_1\), and for every \(t\in[0,1]\) and \(a\in\mathcal{A}\), we have \(F^{\smash[t]{f_0}}_t(a)=f_0(a)^{1-t}f_1(a)^t\).
\item\label{it:intro-interp-cont} If for every \(f_0\in\Delta\) the pair \(f_0,f_1\) has the \(\mathcal{A}\)-cover property, then the map \(\Delta\times[0,1]\to\Delta:(f_0,t)\mapsto F^{\smash[t]{f_0}}_t\) is continuous.
\end{enumerate}
\end{theorem}

If for every \(f_0\in\Delta\) the pair \(f_0,f_1\) has the \(\mathcal{A}\)-cover property, then the joint continuity in~\((f_0,t)\) makes the asymptotic spectrum~\(\Delta\) contractible (\cref{cor:contractible}).

We will prove \cref{th:intro-interpolation} from two ingredients. First, let~\(\mathcal{S}\) be \(\mathcal{A}\)-decomposable, and let \(w:\mathcal{A}\to\RR_{\geq0}\) be multiplicative, with \(w(m)=m\) for every positive integer~\(m\), and such that for all \(a,a_1,\ldots,a_q\in\mathcal{A}\), if \(a\leq a_1+\cdots+a_q\), then \(w(a)\leq w(a_1)+\cdots+w(a_q)\). Then the function that maps \(s\in\mathcal{S}\) to the supremum of \(w(a)^{\smash{1/n}}\) over all \(n\geq1\) and \(a\in\mathcal{A}\) with \(a\leq s^n\) is a spectral point, and it is the only spectral point that agrees with~\(w\) on~\(\mathcal{A}\). Second, the \(\mathcal{A}\)-cover property of \(f_0,f_1\) implies that \(w=f_0^{1-t}f_1^t\) has these three properties. For the last property, we will expand \((a_1+\cdots+a_q)^n\) and assign each of its terms to whichever of~\(f_0\) and~\(f_1\) gives it the smaller value relative to~\(a^n\).

Besides tensors, as illustration, we will apply \cref{th:intro-interpolation} to graphs, where it coincides with the interpolation of Vrana \cite{MR4357434} (\cref{sec:graphs}), and to directed graphs and uniform hypergraphs (\cref{sec:digraphs} and~\cref{sec:hypergraphs}).

\subsubsection{Interpolation towards a flattening rank}

To prove \cref{th:intro-complex}, we will fix \(i\in[k]\) and apply \cref{th:intro-interpolation} to the complex \(k\)-tensors with \(f_1=\tensorrank_i\). We say that a tensor is \emph{\(i\)-irreducible} if it is nonzero and equivalent to a tensor that is invariant under an action of a finite group on its legs that is irreducible on the \(i\)th leg. Let~\(\mathcal{A}_i\) be the set of \(i\)-irreducible tensors. For every nonzero complex \(k\)-tensor~\(s\), decomposing the legs of~\(s^{\otimes n}\) into irreducible representations of the symmetric group gives \(\mathcal{A}_i\)-decompositions of~\(s^{\otimes n}\) of length polynomial in~\(n\). Thus the semiring of complex \(k\)-tensors is \(\mathcal{A}_i\)-decomposable. We will prove the \(\mathcal{A}_i\)-cover property of every pair \(f,\tensorrank_i\). Then \cref{th:intro-interpolation} will give, for every \(f\in\Delta\), a path \(t\mapsto F^{\smash[t]{f}}_t\) from~\(f\) to~\(\tensorrank_i\) with \(F^{\smash[t]{f}}_t=f^{1-t}\tensorrank_i^t\) on the \(i\)-irreducible tensors.

As crucial ingredients for the proof of the \(\mathcal{A}_i\)-cover property, we will prove the following two lemmas, which may be of independent interest.

\begin{lemma}[{\cref{lem:dim-averaging}}]\label{lem:intro-averaging}
Let~\(\FF\) be a field, let a finite group~\(G\) act irreducibly on a nonzero \(\FF\)-vector space~\(V\), and let \(U,W\subseteq V\) be subspaces. Then
\[
\frac{1}{|G|}\sum_{g\in G}\dim(U\cap gW)\leq\frac{\dim W}{\dim V}\dim U.
\]
\end{lemma}

Applied repeatedly, \cref{lem:intro-averaging} gives, for every proper subspace \(W\subsetneq V\), elements \(g_1,\ldots,g_q\in G\) with \(g_1W\cap\cdots\cap g_qW=0\) and \(q\leq\frac{\dim V}{\dim V-\dim W}(1+\ln\dim V)\). For~\(W\) the kernel of a linear map on the first leg of a tensor that~\(G\) fixes, this will lead to the following lemma, which says that ``removing'' a small part of a tensor does not make the tensor much smaller. 

\begin{lemma}[{\cref{lem:restriction}}]\label{lem:intro-reverse}
Let \(a\in V_1\otimes\cdots\otimes V_k\) be a \(1\)-irreducible complex tensor, let \(\pi:V_1\to W\) be linear, and let \(b=(\pi\otimes\Id\otimes\cdots\otimes\Id)a\). Then every \(f\in\Delta\) satisfies
\[
\frac{\tensorrank_1(b)}{\tensorrank_1(a)}f(a)\leq f(b).
\]
\end{lemma}

\Cref{lem:intro-reverse} will give the \(\mathcal{A}_1\)-cover property of every pair \(f,\tensorrank_1\), and relabelling the legs will give the \(\mathcal{A}_i\)-cover property of every pair \(f,\tensorrank_i\). Let \(a\in V_1\otimes\cdots\otimes V_k\) be \(1\)-irreducible, and let~\(s_0\) and~\(s_1\) be tensors with \(a\leq s_0\oplus s_1\). Then there is a subspace \(U\subseteq V_1\) of dimension at most~\(\tensorrank_1(s_1)\) such that~\(a\) is the sum of a restriction of~\(s_0\) and a restriction of~\(s_1\) in \(U\otimes V_2\otimes\cdots\otimes V_k\). Let \(\pi:V_1\to V_1/U\) be the quotient map, and let \(b=(\pi\otimes\Id\otimes\cdots\otimes\Id)a\). Since~\(\pi\) vanishes on~\(U\), the tensor~\(b\) is a restriction of~\(s_0\). Since \(\dim U\leq\tensorrank_1(s_1)\), we have \(\tensorrank_1(b)\geq\tensorrank_1(a)-\tensorrank_1(s_1)\). By \cref{lem:intro-reverse}, every \(f\in\Delta\) satisfies \(f(s_0)\geq f(b)\geq(1-\tensorrank_1(s_1)/\tensorrank_1(a))f(a)\). Applied to the powers~\(a^n\), which lie in~\(\mathcal{A}_1\), this inequality gives the \(\mathcal{A}_1\)-cover property.

\subsection{Log-star-convexity for matrix and polynomial multiplication}

We give a new proof of Strassen's theorem that the asymptotic spectrum of rectangular matrix multiplication is log-star-convex (\cref{fig:intro-connectedness}~(c)). Let \(b_1=\langle2,1,1\rangle\), \(b_2=\langle1,2,1\rangle\) and \(b_3=\langle1,1,2\rangle\), and let \(\Gamma=\conv\{(1,1,0),(0,1,1),(1,0,1)\}\).

\begin{theorem}[{\cite[Corollary~6.6]{strassen1988asymptotic}, \cref{th:star-convex}}]\label{th:intro-mm}
The set \(\log\Delta_\CC(b_1,b_2,b_3)\) contains~\(\Gamma\), and it is star-convex with respect to every point of~\(\Gamma\).
\end{theorem}

We will obtain \cref{th:intro-mm} by applying \cref{th:intro-interpolation} to the semiring generated by~\(b_1\), \(b_2\) and~\(b_3\), with~\(\mathcal{A}\) the set of tensors \(\langle m\rangle\otimes b_1^{\otimes u_1}\otimes b_2^{\otimes u_2}\otimes b_3^{\otimes u_3}\) for positive integers~\(m\) and nonnegative integers \(u_1,u_2,u_3\). The products \(b_1^{\otimes u_1}\otimes b_2^{\otimes u_2}\otimes b_3^{\otimes u_3}\) are the matrix multiplication tensors \(\langle2^{u_1},2^{u_2},2^{u_3}\rangle\). The multinomial expansion of powers shows that this semiring is \(\mathcal{A}\)-decomposable. Every element of~\(\mathcal{A}\) is \(i\)-irreducible for every \(i\in[3]\), since the unit tensors and the matrix multiplication tensors are invariant under groups of signed permutation matrices that act irreducibly on every leg, and Kronecker products of \(i\)-irreducible tensors are \(i\)-irreducible. Thus \cref{lem:intro-reverse} gives the \(\mathcal{A}\)-cover property of every pair \(f,\tensorrank_i\). Hence the path from~\(f\) to~\(\tensorrank_i\) of \cref{th:intro-interpolation} satisfies \(F^{\smash[t]{f}}_t(b_j)=f(b_j)^{1-t}\tensorrank_i(b_j)^t\) for every \(j\in[3]\). After taking logarithms, its values at \(b_1\), \(b_2\) and~\(b_3\) form the segment from \((\log f(b_j))_{j\in[3]}\) to the vertex \((\log\tensorrank_i(b_j))_{j\in[3]}\) of~\(\Gamma\), and combining these segments for \(i=1,2,3\) will give \cref{th:intro-mm}.

We extend \cref{th:intro-mm} to polynomial multiplication tensors.
For positive integers \(m,n\), the polynomial multiplication tensor~\(\polymult_{m,n}\) is the tensor of the bilinear map that multiplies a polynomial of degree at most \(m-1\) with a polynomial of degree at most \(n-1\). For a permutation~\(\pi\) in the symmetric group~\(\mathfrak{S}_3\) and a \(3\)-tensor~\(t\), let~\(\pi t\) be the tensor obtained from~\(t\) by moving its \(i\)th leg to position~\(\pi(i)\), for every \(i\in[3]\). For every positive integer~\(a\), the tensors \(\pi\polymult_{1,a}\) with \(\pi\in\mathfrak{S}_3\) are, up to equivalence, the matrix multiplication tensors \(\langle a,1,1\rangle\), \(\langle1,a,1\rangle\) and~\(\langle1,1,a\rangle\). Since \(\langle n_1,n_2,n_3\rangle=\langle n_1,1,1\rangle\otimes\langle1,n_2,1\rangle\otimes\langle1,1,n_3\rangle\) for all positive integers \(n_1,n_2,n_3\), the semiring generated by the tensors~\(\pi\polymult_{m,n}\) contains all matrix multiplication tensors. The proof of~\(\omega\leq9/4\) relies heavily on polynomial multiplication tensors \cite{OAI:Matrix-Multiplication-Nine-Fourths-October-2-2026,amanov2026multiplyingmatricespolynomialsexposition}.

\begin{theorem}[{\cref{th:poly-star-convex}}]\label{th:intro-poly}
Let \(b_1,\ldots,b_r\) be tensors of the form~\(\pi\polymult_{m,n}\) with \(\pi\in\mathfrak{S}_3\) and positive integers \(m,n\). Then \(\log\Delta_\CC(b_1,\ldots,b_r)\) is star-convex with respect to every point of \(\conv\{(\log\tensorrank_i(b_j))_{j\in[r]}:i\in[3]\}\).
\end{theorem}

To prove \cref{th:intro-poly}, we will replace the finite groups by an action of the compact group \(\mathrm{SU}(2)\) that fixes \(\polymult_{m,n}\) and is irreducible on every leg, together with a version of \cref{lem:intro-averaging} for compact groups acting on complex vector spaces.

Besides extending \cref{th:intro-mm} to polynomial multiplication tensors, we prove it over every field~\(\FF\), and more generally for graph tensors. Both follow from log-star-convexity under an irreducibility condition. Let \(b_1,\ldots,b_r\) be nonzero \(k\)-tensors over~\(\FF\), and let \(I\subseteq[k]\) be such that for every \(i\in I\), every Kronecker product of copies of \(b_1,\ldots,b_r\) is \(i\)-irreducible over~\(\FF\). Then \(\log\Delta_\FF(b_1,\ldots,b_r)\) is star-convex with respect to every point of \(\conv\{(\log\tensorrank_i(b_j))_{j\in[r]}:i\in I\}\) (\cref{th:star-convex-general}). The condition holds for the matrix multiplication tensors \(b_1,b_2,b_3\) of \cref{th:intro-mm} and for graph tensors. This gives new proofs, over every field, of the theorems of Strassen \cite[Theorem~3.10 and Corollary~6.6]{strassen1988asymptotic} and of Wigderson and Zuiddam \cite[Theorem~12.23]{wigderson2022asymptotic} (\cref{th:star-convex-any-field} and~\cref{th:graph-tensors}).

\subsection{Quantum functionals}

For every probability distribution~\(\theta\) on~\([k]\), Christandl, Vrana and Zuiddam \cite{MR4495838} defined the quantum functional~\(F^\theta\) of complex tensors through moment polytopes and quantum marginal spectra. They proved that~\(F^\theta\) is a spectral point \cite[Corollary~3.31]{MR4495838}. Algorithms for moment polytopes and for the quantum functionals were developed in \cite{MR3899650,MR4228202,MR4849336,MR4928469,doğan2026convexoptimizationmomentpolytopes}. 

We will obtain the quantum functionals by repeated interpolation between the flattening ranks \(\tensorrank_1,\ldots,\tensorrank_k\) with \cref{th:intro-interpolation}, which gives a new proof that they are spectral points. To do this, we say that a nonzero tensor is \emph{jointly irreducible} if it is invariant under an action of a finite group on its legs that is irreducible on every leg. On these tensors, the interpolating spectral points are weighted geometric means of the flattening ranks, and this characterizes the quantum functionals.

\begin{theorem}[{\cref{cor:qf} and~\cref{th:cvz}}]\label{th:intro-qf}
Let~\(\theta\) be a probability distribution on~\([k]\). Then the quantum functional~\(F^\theta\) is the unique spectral point of complex \(k\)-tensors that agrees with \(\tensorrank_1^{\theta_1}\cdots\tensorrank_k^{\theta_k}\) on all jointly irreducible tensors.
\end{theorem}

Our proof that these interpolating spectral points exist is elementary. Their identification with the quantum functionals as defined by Christandl, Vrana and Zuiddam \cite{MR4495838} requires some representation theory.

\subsection{Non-crossing upper quantum functionals}

We prove that the upper quantum functionals of non-crossing families of bipartitions are spectral points. A bipartition of~\([k]\) is a set \(\{X_1,X_2\}\) of two disjoint nonempty sets with \(X_1\cup X_2=[k]\). For a bipartition \(b=\{X_1,X_2\}\), the flattening rank~\(\tensorrank_b(t)\) is the rank of the flattening of~\(t\) that groups the legs in~\(X_1\) and the legs in~\(X_2\). Two bipartitions \(\{X_1,X_2\}\) and \(\{Y_1,Y_2\}\) are non-crossing if \(X_i\subseteq Y_j\) for some \(i,j\in\{1,2\}\). Let~\(B\) be a set of pairwise non-crossing bipartitions of~\([k]\). Christandl, Vrana and Zuiddam \cite[Definition~3.14]{MR4495838} defined the upper quantum functional~\(F^\theta\) for every probability distribution~\(\theta\) on~\(B\). The quantum functionals are those with~\(B\) the set of bipartitions that separate one leg from the others.

\begin{theorem}[{\cref{th:upper-spectral}}]\label{th:intro-noncrossing}
Let~\(B\) be a set of pairwise non-crossing bipartitions of~\([k]\), and let~\(\theta\) be a probability distribution on~\(B\). Then the upper quantum functional~\(F^\theta\) is a spectral point of complex \(k\)-tensors. It is the largest spectral point~\(f\) with \(f\leq\prod_{b\in B}\tensorrank_b^{\smash{\theta(b)}}\).
\end{theorem}

Unlike in \cref{th:intro-qf}, for \(k\geq5\) there need not be a spectral point that agrees with \(\prod_{b\in B}\tensorrank_b^{\smash{\theta(b)}}\) on all jointly irreducible tensors (\cref{ex:not-B-good}). We will therefore interpolate towards the flattening ranks~\(\tensorrank_b\) with \(b\in B\) one at a time, by decreasing size of the part~\(S_b\) of~\(b\) that does not contain~\(k\). In the step for~\(b\), we merge into one leg every group of legs that neither~\(b\) nor an earlier bipartition separates. Then~\(\tensorrank_b\) is the flattening rank of a single leg of the merged tensors, and \cref{th:intro-interpolation} applies (\cref{sec:upper-interp}). We will then identify the resulting spectral point with~\(F^\theta\) (\cref{sec:upper-identify}).

Botero, Christandl, Fraser, Leigh and Nieuwboer \cite[Proposition~4.2 and Theorem~3.1]{botero2026quantumfunctionalshigherordertensors} proved before that~\(F^\theta\) is a spectral point on the subsemiring of the complex tensors at which every upper quantum functional of a non-crossing family agrees with the corresponding lower quantum functional \cite[Definition~3.3]{MR4495838}. This subsemiring is strict for \(k\geq4\).

\section{Decomposable semirings and path-connectedness}\label{sec:dec}

In this section we will construct paths between spectral points of a Strassen semiring, using decompositions of powers of its elements and a cover property of the spectral points.

We write $\NN = \{0, 1, 2, \ldots\}$ and $[n] = \{1, \ldots, n\}$, use $0^0 = 1$, $\sup \emptyset = 0$ in $[0, \infty]$ and $\log = \log_2$, and all vector spaces are finite-dimensional. For any set~$I$ and $X \subseteq \RR_{>0}^I$, let $\log X = \{(\log x_i)_{i \in I} : x \in X\}$. A set $Y \subseteq \RR^I$ is \emph{star-convex} with respect to a point $c \in Y$ if for every $y \in Y$ the segment $\{(1-t)y + tc : t \in [0,1]\}$ lies in~$Y$, and \emph{star-convex} if it is star-convex with respect to some point. We call the set of points with respect to which~$Y$ is star-convex the \emph{kernel} of~$Y$. The kernel is convex. We call~$X$ \emph{log-convex} if $\log X$ is convex, and \emph{log-star-convex} if $\log X$ is star-convex.

Throughout, $\mathcal{S}$ is a commutative semiring with additive unit 0, multiplicative unit 1, and a preorder $\leq$ such that for every $s,t,u \in \mathcal{S}$, $s\leq t$ implies $s + u \leq t + u$ and $su \leq tu$. We identify~$\NN$ with its image in~$\mathcal{S}$ and assume that its induced order is the usual one. In particular $0 \leq 1$, so every $s \in \mathcal{S}$ satisfies $0 = 0 \cdot s \leq s$. Finally, we assume that the preorder is \emph{Archimedean}: for all nonzero $s, t \in \mathcal{S}$ there is an integer $m \geq 1$ with $s \leq mt$, as in \cite[(2.5)]{strassen1988asymptotic}. Consequently, products of nonzero elements are nonzero, and no nonzero element is at most~$0$: for nonzero $s, t \in \mathcal{S}$, integers $m_1, m_2 \geq 1$ with $1 \leq m_1 s$ and $1 \leq m_2 t$ give $1 \leq m_1 m_2 st$, so $st \not\leq 0$. We call the preordered semiring $\mathcal{S} = (\mathcal{S}, \leq)$ a Strassen semiring.

Let~$\Delta$ denote the set of all semiring homomorphisms $f : \mathcal{S} \to \RR_{\geq0}$ such that $f(1) = 1$ and such that for every $s,t \in \mathcal{S}$, if $s \leq t$, then $f(s) \leq f(t)$. We call~$\Delta$ the \emph{asymptotic spectrum} of~$\mathcal{S}$ and its elements \emph{spectral points}. We give~$\Delta$ the topology of pointwise convergence, that is, the coarsest topology for which $f \mapsto f(s)$ is continuous for every $s \in \mathcal{S}$. Then~$\Delta$ is compact \cite[Corollary~2.6]{strassen1988asymptotic}. Let $s \in \mathcal{S}$ be nonzero. The Archimedean property, applied to~$s$ and~$1$ and to~$1$ and~$s$, gives integers $m, n \geq 1$ with $s \leq m$ and $1 \leq ns$. So every $f \in \Delta$ satisfies $1/n \leq f(s) \leq m$. In particular, every spectral point is positive at every nonzero element of~$\mathcal{S}$.

\subsection{\texorpdfstring{$\mathcal{A}$-length,}{A-length,} \texorpdfstring{$\mathcal{A}$}{A}-decomposable semirings and \texorpdfstring{$\mathcal{A}$}{A}-good functions}

Given a Strassen semiring~$\mathcal{S}$ we are looking for a special subset $\mathcal{A}\subseteq \mathcal{S}$ such that for every element $s \in \mathcal{S}$ and every $n\geq1$ we can upper bound~$s^n$ by a sum of, subexponentially in~$n$ many, elements $a_1, \ldots, a_q \in \mathcal{A}$ each of which is at most~$s^n$, thus getting an ``asymptotic sandwich''
\begin{equation}\label{eq:sand}
a_j \leq s^n \leq a_1 + \cdots + a_q
\end{equation}
for every $j \in [q]$, which will provide for  $\mathcal{S}$ an information-theoretic ``method of types''. The following notions make this precise.

Let $\mathcal{A} \subseteq \mathcal{S} \setminus \{0\}$ and $s \in \mathcal{S}$. An \emph{$\mathcal{A}$-decomposition} of~$s$ is a list $a_1, \ldots, a_q$ of elements of~$\mathcal{A}$, with $q \in \NN$ and possibly with repetitions, such that $a_j \leq s$ for every $j \in [q]$ and $s \leq a_1 + \cdots + a_q$. Its \emph{length} is~$q$. The \emph{$\mathcal{A}$-length} $\declength_{\mathcal{A}}(s)$ is the smallest length of an $\mathcal{A}$-decomposition of~$s$, and $\declength_{\mathcal{A}}(s) = \infty$ if~$s$ has no $\mathcal{A}$-decomposition. The $\mathcal{A}$-length need not be monotone: for $\mathcal{S} = \RR_{\geq0}$ with its usual order and $\mathcal{A} = \{1, 2, \ldots\}$, we have $3/2 \leq 2$ but $\declength_{\mathcal{A}}(3/2) = 2 > 1 = \declength_{\mathcal{A}}(2)$. The \emph{asymptotic $\mathcal{A}$-length} of~$s$ is $\asympdeclength_{\mathcal{A}}(s) \coloneqq \inf_{n \geq 1} \declength_{\mathcal{A}}(s^n)^{1/n}$.

\begin{definition}\label{def:decomposable}
Let $\mathcal{A} \subseteq \mathcal{S} \setminus \{0\}$.
We say~$\mathcal{S}$ is $\mathcal{A}$-decomposable if the following properties hold:
\begin{enumerate}[(a)]
\item\label{it:dec-mult} $\mathcal{A}$ is multiplicatively closed: if $a,b \in \mathcal{A}$, then $ab \in \mathcal{A}$.
\item\label{it:dec-nat} Every positive integer lies in~$\mathcal{A}$.
\item\label{it:dec-finite} For every $s \in \mathcal{S} \setminus \{0\}$ there is $a \in \mathcal{A}$ with $a \leq s$.
\item\label{it:dec-count} For every $s \in \mathcal{S} \setminus \{0\}$, we have $\asympdeclength_{\mathcal{A}}(s) = 1$.
\end{enumerate}
\end{definition}

Property~(\ref{it:dec-finite}) implies that every $s \in \mathcal{S}$ has finite $\mathcal{A}$-length. For $s = 0$ the empty list is an $\mathcal{A}$-decomposition. For nonzero~$s$, let $a \in \mathcal{A}$ with $a \leq s$. Since $a \neq 0$, the Archimedean property gives an integer $m \geq 1$ with $s \leq ma$. So the list of~$m$ copies of~$a$ is an $\mathcal{A}$-decomposition of~$s$. If $1 \leq s$ for every nonzero $s \in \mathcal{S}$, then (\ref{it:dec-finite}) follows from~(\ref{it:dec-nat}), with $a = 1$.

If~$\mathcal{A}$ is multiplicatively closed, then $\declength_{\mathcal{A}}(st) \leq \declength_{\mathcal{A}}(s) \declength_{\mathcal{A}}(t)$ for all $s, t \in \mathcal{S}$ of finite $\mathcal{A}$-length. Indeed, let $a_1, \ldots, a_q$ be an $\mathcal{A}$-decomposition of~$s$ of length $q = \declength_{\mathcal{A}}(s)$ and let $b_1, \ldots, b_r$ be an $\mathcal{A}$-decomposition of~$t$ of length $r = \declength_{\mathcal{A}}(t)$. Every product $a_i b_j$ lies in~$\mathcal{A}$ and satisfies $a_i b_j \leq s b_j \leq st$. Moreover $st \leq (\sum_i a_i)(\sum_j b_j) = \sum_{i,j} a_i b_j$. So the $qr$ products $a_i b_j$ are an $\mathcal{A}$-decomposition of $st$.

Now suppose~$\mathcal{S}$ is $\mathcal{A}$-decomposable, and let $s \in \mathcal{S} \setminus \{0\}$. For every $n \geq 1$, an~$\mathcal{A}$-decomposition of~$s^n$ cannot have length~$0$, since an empty decomposition would give $s^n \leq 0$. Therefore $1 \leq \declength_{\mathcal{A}}(s^n) < \infty$, using \cref{def:decomposable}~(\ref{it:dec-finite}). The sequence $\log \declength_{\mathcal{A}}(s^n)$ is real and subadditive. Fekete's lemma and \cref{def:decomposable}~(\ref{it:dec-count}) give
\begin{equation}\label{eq:length-limit}
\lim_{n \to \infty} \declength_{\mathcal{A}}(s^n)^{1/n} = \asympdeclength_{\mathcal{A}}(s) = 1.
\end{equation}

Given $\mathcal{A}\subseteq \mathcal{S}$ such that~$\mathcal{S}$ is $\mathcal{A}$-decomposable as in \cref{def:decomposable}, we consider a special kind of real-valued functions $w: \mathcal{A} \to \RR_{\geq0}$. We will use them to produce spectral points via the asymptotic sandwich \eqref{eq:sand}.

\begin{definition}\label{def:w}
Let $\mathcal{A} \subseteq \mathcal{S} \setminus \{0\}$ be multiplicatively closed and contain every positive integer, and let $w : \mathcal{A} \to \RR_{\geq0}$. We say~$w$ is $\mathcal{A}$-good if the following properties hold:
\begin{enumerate}[(a)]
\item\label{it:good-mult} For every $a,b \in \mathcal{A}$, we have $w(ab) = w(a)w(b)$.
\item\label{it:good-norm} For every positive integer~$m$, we have $w(m) = m$.
\item\label{it:good-sub} For every $q \in \NN$ and every $a, a_1, \ldots, a_q \in \mathcal{A}$, if $a \leq \sum_{j=1}^q a_j$, then $w(a) \leq \sum_{j=1}^q w(a_j)$.
\end{enumerate}
\end{definition}

\subsection{Spectral points from \texorpdfstring{$\mathcal{A}$}{A}-good functions}

We now construct from every $\mathcal{A}$-good function~$w$ a spectral point $F_w : \mathcal{S} \to \RR_{\geq0}$. The construction is a simple normalised supremum of~$w$ over elements in~$\mathcal{A}$ that are at most~$s^n$.

\begin{definition}\label{def:Fw}
For $w : \mathcal{A} \to \RR_{\geq0}$ and $s \in \mathcal{S}$, let
\[
F_w(s) \coloneqq \sup \{ w(a)^{1/n} : n \geq 1, a \in \mathcal{A}, a \leq s^n \} \in [0, \infty].
\]
\end{definition}

For every $\mathcal{A}$-good~$w$ and every $s \in \mathcal{S}$, the supremum~$F_w(s)$ is finite. Indeed, let $m \geq 1$ be an integer with $s \leq m$ (by the Archimedean property if $s \neq 0$, and $m = 1$ if $s = 0$). If $n \geq 1$, $a \in \mathcal{A}$ and $a \leq s^n$, then $a \leq m^n$ and $m^n \in \mathcal{A}$. So \cref{def:w}~(\ref{it:good-sub}) with a single summand, together with (\ref{it:good-norm}), gives $w(a) \leq w(m^n) = m^n$. Hence $F_w(s) \leq m$. Moreover $F_w(0) = 0$, since no element of~$\mathcal{A}$ is at most~$0$.

\cref{def:w} and \cref{def:Fw} are stated for functions $w : \mathcal{A} \to \RR_{\geq0}$. We will later also apply them to functions $w : \mathcal{S} \to \RR_{\geq0}$, through their restriction to~$\mathcal{A}$. 

\begin{theorem}\label{th:w-good-spec}
Let~$\mathcal{S}$ be $\mathcal{A}$-decomposable and let $w : \mathcal{A} \to \RR_{\geq0}$ be $\mathcal{A}$-good. Then
\begin{enumerate}[\upshape(i)]
\item\label{it:Fw-spectral} $F_w \in \Delta$.
\item\label{it:Fw-limit} For every $s \in \mathcal{S} \setminus \{0\}$, every $n \geq 1$ and every $\mathcal{A}$-decomposition $a_1, \ldots, a_q$ of~$s^n$, we have $\max_{j \in [q]} w(a_j) \leq F_w(s)^n \leq \sum_{j=1}^q w(a_j)$.
\item\label{it:Fw-family} $F_w = w$ on~$\mathcal{A}$.
\item\label{it:Fw-largest} Every $g \in \Delta$ with $g \leq w$ on~$\mathcal{A}$ satisfies $g \leq F_w$.
\item\label{it:Fw-below} Let $v : \mathcal{S} \to \RR_{\geq0}$ be such that $w \leq v$ on~$\mathcal{A}$ and such that for every $s, t \in \mathcal{S}$, we have $v(st) = v(s)v(t)$, and if $s \leq t$, then $v(s) \leq v(t)$. Then $F_w \leq v$.
\item\label{it:Fw-unique} $F_w$ is the unique element of~$\Delta$ that agrees with~$w$ on~$\mathcal{A}$.
\end{enumerate}
\end{theorem}

Suppose~$\mathcal{S}$ is $\mathcal{A}$-decomposable. For every $g \in \Delta$, the restriction of~$g$ to~$\mathcal{A}$ is $\mathcal{A}$-good, because~$g$ is multiplicative, monotone and additive, with $g(1) = 1$. So by \cref{th:w-good-spec}~(\ref{it:Fw-spectral}), (\ref{it:Fw-family}) and (\ref{it:Fw-unique}), restriction to~$\mathcal{A}$ is a bijection from~$\Delta$ to the set of $\mathcal{A}$-good functions $\mathcal{A} \to \RR_{\geq0}$, with inverse $w \mapsto F_w$.

We will prove \cref{th:w-good-spec} by comparing~$F_w$ with two functions on~$\mathcal{S}$ that are built from~$w$. Let $\mathcal{A} \subseteq \mathcal{S} \setminus \{0\}$ be multiplicatively closed and contain every positive integer, and let $w : \mathcal{A} \to \RR_{\geq0}$ be $\mathcal{A}$-good. For $s \in \mathcal{S}$, the \emph{$w$-subrank} of~$s$ is
\[
\subrank_w(s) \coloneqq \sup \{ w(a) : a \in \mathcal{A},\ a \leq s \},
\]
and the \emph{$w$-rank} of~$s$ is
\[
\tensorrank_w(s) \coloneqq \inf \Bigl\{ \sum_{j=1}^q w(a_j) : q \in \NN,\ a_1, \ldots, a_q \in \mathcal{A},\ s \leq a_1 + \cdots + a_q \Bigr\}.
\]
By \cref{def:Fw}, $F_w(s) = \sup_{n \geq 1} \subrank_w(s^n)^{1/n}$.

\begin{lemma}\label{lem:QR}
Let $\mathcal{A} \subseteq \mathcal{S} \setminus \{0\}$ be multiplicatively closed and contain every positive integer, let $w : \mathcal{A} \to \RR_{\geq0}$ be $\mathcal{A}$-good, and let $s, t \in \mathcal{S}$.
\begin{enumerate}[\upshape(i)]
\item\label{it:QR-finite} $0 \leq \subrank_w(s) \leq \tensorrank_w(s) < \infty$.
\item\label{it:QR-mono} If $s \leq t$, then $\subrank_w(s) \leq \subrank_w(t)$ and $\tensorrank_w(s) \leq \tensorrank_w(t)$.
\item\label{it:QR-family} If $s \in \mathcal{A}$, then $\subrank_w(s) = \tensorrank_w(s) = w(s)$.
\item\label{it:QR-mult} $\subrank_w(st) \geq \subrank_w(s) \subrank_w(t)$, $\tensorrank_w(st) \leq \tensorrank_w(s) \tensorrank_w(t)$ and $\tensorrank_w(s + t) \leq \tensorrank_w(s) + \tensorrank_w(t)$.
\item\label{it:QR-length} If $\declength_{\mathcal{A}}(s) < \infty$, then $\tensorrank_w(s) \leq \declength_{\mathcal{A}}(s) \subrank_w(s)$.
\end{enumerate}
\end{lemma}
\begin{proof}
(i) Since~$w$ takes nonnegative values, $\subrank_w(s) \geq 0$. If $a \in \mathcal{A}$ and $a \leq s \leq a_1 + \cdots + a_q$ with $a_1, \ldots, a_q \in \mathcal{A}$, then $w(a) \leq \sum_{j=1}^q w(a_j)$ by \cref{def:w}~(\ref{it:good-sub}), so $\subrank_w(s) \leq \tensorrank_w(s)$. The empty list gives $\tensorrank_w(0) = 0$. For nonzero~$s$, the Archimedean property gives an integer $m \geq 1$ with $s \leq m$, and $m \in \mathcal{A}$, so $\tensorrank_w(s) \leq w(m) = m$.

(ii) If $s \leq t$, then every $a \leq s$ satisfies $a \leq t$, and all $a_1, \ldots, a_q$ with $t \leq a_1 + \cdots + a_q$ satisfy $s \leq a_1 + \cdots + a_q$.

(iii) Taking~$s$ itself in both definitions, and using~(\ref{it:QR-finite}), gives $w(s) \leq \subrank_w(s) \leq \tensorrank_w(s) \leq w(s)$.

(iv) Let $a, b \in \mathcal{A}$ with $a \leq s$ and $b \leq t$. Then $ab \in \mathcal{A}$, $ab \leq st$ and $w(ab) = w(a)w(b)$, which gives the first inequality. Let $a_1, \ldots, a_q, b_1, \ldots, b_r \in \mathcal{A}$ with $s \leq \sum_i a_i$ and $t \leq \sum_\ell b_\ell$. Then $st \leq \sum_{i,\ell} a_i b_\ell$, every $a_i b_\ell$ lies in~$\mathcal{A}$, and $\sum_{i,\ell} w(a_i b_\ell) = \sum_i w(a_i) \sum_\ell w(b_\ell)$, which gives the second inequality. Also $s + t \leq \sum_i a_i + \sum_\ell b_\ell$, which gives the third.

(v) Let $a_1, \ldots, a_q$ be an $\mathcal{A}$-decomposition of~$s$ of length $q = \declength_{\mathcal{A}}(s)$. Every~$a_j$ satisfies $a_j \leq s$, so $w(a_j) \leq \subrank_w(s)$. Since $s \leq a_1 + \cdots + a_q$, we get $\tensorrank_w(s) \leq \sum_{j=1}^q w(a_j) \leq q \subrank_w(s)$.
\end{proof}

The $w$-subrank and the $w$-rank of~$s^n$ bound the two sides of the asymptotic sandwich~\eqref{eq:sand}: the values of~$w$ on its left side are at most~$\subrank_w(s^n)$, and the sum of the values of~$w$ on its right side is at least~$\tensorrank_w(s^n)$. Both grow at the exponential rate~$F_w(s)$, as follows.

\begin{lemma}\label{lem:MZ}
Suppose~$\mathcal{S}$ is $\mathcal{A}$-decomposable and~$w$ is $\mathcal{A}$-good, and let $s \in \mathcal{S} \setminus \{0\}$.
\begin{enumerate}[\upshape(i)]
\item\label{it:MZ-bounds} For every $n \geq 1$, we have $0 < \subrank_w(s^n) \leq F_w(s)^n \leq \tensorrank_w(s^n) \leq \declength_{\mathcal{A}}(s^n) \subrank_w(s^n)$.
\item\label{it:MZ-limit} $F_w(s) = \lim_{n\to\infty} \subrank_w(s^n)^{1/n} = \lim_{n\to\infty} \tensorrank_w(s^n)^{1/n}$.
\end{enumerate}
\end{lemma}
\begin{proof}
(i) We first prove positivity. Since $s^n \neq 0$, the Archimedean property gives an integer $\ell \geq 1$ with $1 \leq \ell s^n$. By \cref{lem:QR},
\[
1 = \tensorrank_w(1) \leq \tensorrank_w(\ell s^n) \leq \ell \tensorrank_w(s^n) \leq \ell \declength_{\mathcal{A}}(s^n) \subrank_w(s^n),
\]
and $\declength_{\mathcal{A}}(s^n) < \infty$ by \cref{def:decomposable}~(\ref{it:dec-finite}). Hence $\subrank_w(s^n) > 0$.

The second inequality holds by \cref{def:Fw}. For the third, let $m \geq 1$ and let $b \in \mathcal{A}$ with $b \leq s^m$. Then $b^n \in \mathcal{A}$ and $b^n \leq (s^n)^m$, so \cref{lem:QR} gives
\[
w(b)^n = w(b^n) \leq \subrank_w((s^n)^m) \leq \tensorrank_w((s^n)^m) \leq \tensorrank_w(s^n)^m.
\]
Taking $nm$th roots and then the supremum over~$m$ and~$b$ gives $F_w(s) \leq \tensorrank_w(s^n)^{1/n}$. The fourth inequality is \cref{lem:QR}~(\ref{it:QR-length}), since $\declength_{\mathcal{A}}(s^n) < \infty$ by \cref{def:decomposable}~(\ref{it:dec-finite}).

(ii) Take~$n$th roots of (i) and use $\declength_{\mathcal{A}}(s^n)^{\smash{1/n}} \to 1$, by \eqref{eq:length-limit}.
\end{proof}

\begin{proof}[Proof of \cref{th:w-good-spec}]
Write $F = F_w$. We first prove~(\ref{it:Fw-family}), since the proof of~(\ref{it:Fw-spectral}) uses it.

(iii) Let $a \in \mathcal{A}$. For every $n \geq 1$ we have $a^n \in \mathcal{A}$, so $\subrank_w(a^n) = w(a^n) = w(a)^n$ by \cref{lem:QR}~(\ref{it:QR-family}). Hence $F(a) = \sup_{n \geq 1} \subrank_w(a^n)^{1/n} = w(a)$.

(i)
Monotonicity of~$F$ follows directly from its definition as a supremum over $a \leq s^n$. Normalisation $F(m) = m$ for every positive integer~$m$ follows from (iii) and \cref{def:w}~(\ref{it:good-norm}), and $F(0) = 0$.

For multiplicativity, suppose $s,t \neq 0$ (if $s = 0$ or $t = 0$, then $st = 0$ and the claim follows from $F(0) = 0$). For every $n \geq 1$, \cref{lem:QR}~(\ref{it:QR-mult}) gives
\[
\subrank_w(s^n) \subrank_w(t^n) \leq \subrank_w((st)^n)
\quad\text{and}\quad
\tensorrank_w((st)^n) \leq \tensorrank_w(s^n) \tensorrank_w(t^n).
\]
Taking~$n$th roots and letting $n \to \infty$, \cref{lem:MZ}~(\ref{it:MZ-limit}), applied to~$s$, $t$ and $st \neq 0$, gives $F(s)F(t) \leq F(st) \leq F(s)F(t)$.

For additivity, suppose $s,t \neq 0$ (if $s = 0$ or $t = 0$, the claim follows from $F(0) = 0$), and let $n \geq 1$. For every $0 \leq j \leq n$ we have $\binom{n}{j} s^j t^{n-j} \leq (s+t)^n$, so monotonicity, multiplicativity and normalisation of~$F$ give
\[
\binom{n}{j} F(s)^j F(t)^{n-j} \leq F(s+t)^n.
\]
Summing over~$j$ gives $(F(s) + F(t))^n \leq (n+1) F(s+t)^n$. Taking~$n$th roots and letting $n \to \infty$ gives $F(s) + F(t) \leq F(s+t)$.

For the reverse inequality, take $x,y \in \RR$ such that $F(s) < x$ and $F(t) < y$. By \cref{lem:MZ}~(\ref{it:MZ-limit}) there are constants $C_s, C_t > 0$ such that $\tensorrank_w(s^j) \leq C_s x^j$ and $\tensorrank_w(t^j) \leq C_t y^j$ for every $j \geq 0$. Since $s \leq s + t$, also $s + t \neq 0$. By \cref{lem:MZ}~(\ref{it:MZ-bounds}), the binomial expansion of~$(s+t)^n$, and \cref{lem:QR}~(\ref{it:QR-family}) and (\ref{it:QR-mult}),
\[
F(s+t)^n \leq \tensorrank_w((s+t)^n) \leq \sum_{j=0}^n \binom{n}{j} \tensorrank_w(s^j) \tensorrank_w(t^{n-j}) \leq C_s C_t (x+y)^n.
\]
Taking~$n$th roots and letting $n \to \infty$ gives $F(s+t) \leq x + y$. Letting $x \to F(s)$ and $y \to F(t)$ gives the claim.
We have thus proved $F \in \Delta$.

(ii) Apply~$F$ to the inequalities $a_j \leq s^n \leq a_1 + \cdots + a_q$, using~(\ref{it:Fw-spectral}) and~(\ref{it:Fw-family}).

(iv) Let $g \in \Delta$ with $g \leq w$ on~$\mathcal{A}$, and let $s \in \mathcal{S}$. For all $a_1, \ldots, a_q \in \mathcal{A}$ with $s \leq a_1 + \cdots + a_q$, since~$g$ is monotone and additive, $g(s) \leq \sum_{j=1}^q g(a_j) \leq \sum_{j=1}^q w(a_j)$. So $g(s) \leq \tensorrank_w(s)$. Applied to~$s^n$, and since~$g$ is multiplicative, this gives $g(s)^n \leq \tensorrank_w(s^n)$ for every $n \geq 1$. So $g(s) \leq F(s)$ by \cref{lem:MZ}~(\ref{it:MZ-limit}) if $s \neq 0$, and $g(0) = 0 = F(0)$.

(v) Let $s \in \mathcal{S}$. For every $n \geq 1$ and $a \in \mathcal{A}$ with $a \leq s^n$, we have $w(a) \leq v(a) \leq v(s^n) = v(s)^n$. So $F(s) \leq v(s)$ by definition of~$F$.

(vi) By (i) and (iii), $F$ is an element of~$\Delta$ that agrees with~$w$ on~$\mathcal{A}$. Let $g \in \Delta$ agree with~$w$ on~$\mathcal{A}$. Then $g \leq F$ by (iv). Since~$g$ is monotone and multiplicative, (v) with $v = g$ gives $F \leq g$. Hence $g = F$.
\end{proof}

\subsection{\texorpdfstring{$\mathcal{A}$}{A}-cover property for spectral points}

We will now construct $\mathcal{A}$-good functions from spectral points, simply by taking their weighted geometric average, assuming that these spectral points satisfy a special ``cover property''.

Let $g_1, \ldots, g_r \in \Delta$.
For a finite set~$X$ write $\mathcal{P}(X) \coloneqq \{\theta \in \RR_{\geq0}^X : \sum_{x \in X} \theta_x = 1\}$.
For $\theta \in \mathcal{P}([r])$ and $s \in \mathcal{S}$ let
\[
w_\theta(s) \coloneqq \prod_{i=1}^r g_i(s)^{\theta_i}
\]
be the $\theta$-weighted geometric average of the~$g_i$.
We see directly that
$w_\theta(st) = w_\theta(s)w_\theta(t)$ for every $s,t \in \mathcal{S}$ and $w_\theta(m) = m$ for every $m \in \NN$. 

We aim to prove that~$w_\theta$ is $\mathcal{A}$-good. To get there, we will need a cover property that says that if some power of an element in~$\mathcal{A}$ is at most a sum of elements $s_1, \ldots, s_r \in \mathcal{S}$ (note that we take as many elements as we have chosen spectral points $g_1, \ldots, g_r$), then these elements cannot all be small, as measured respectively by $g_1, \ldots, g_r$.

\begin{definition}\label{def:cover}
We say $g_1, \ldots, g_r \in \Delta$ have the $\mathcal{A}$-\emph{cover property} if for every $a \in \mathcal{A}$, there are numbers $1 \leq \kappa_a(n)$ such that $\kappa_a(n)^{1/n} \to 1$ and for every $n \in \NN$, $s_1, \ldots, s_r \in \mathcal{S}$, if $a^n \leq s_1 + \cdots + s_r$, then 
\[
\kappa_a(n)^{-1} \leq \sum_{i=1}^r \frac{g_i(s_i)}{g_i(a)^n}.
\]
\end{definition}
If $g_1 = \cdots = g_r = g \in \Delta$, then $w_\theta = g$, and \cref{def:cover} holds with $\kappa_a(n) = 1$, because $a^n \leq s_1 + \cdots + s_r$ implies $g(a)^n \leq g(s_1) + \cdots + g(s_r)$. In our application however the spectral points $g_1, \ldots, g_r$ will not all be the same, and it is this mixed and asymptotic kind of subadditivity that we will need.

\begin{remark}\label{rem:cover-lambda}
For $a \in \mathcal{S} \setminus \{0\}$ let $\mu(a)$ be the infimum of $\max_{i \in [r]} g_i(s_i)/g_i(a)$ over all $s_1, \ldots, s_r \in \mathcal{S}$ with $a \leq s_1 + \cdots + s_r$. Then $\mu(a) \leq 1$, and $g_1, \ldots, g_r$ have the $\mathcal{A}$-cover property if and only if, for every $a \in \mathcal{A}$, every $\mu(a^n)$ is positive and $\mu(a^n)^{1/n} \to 1$. Indeed, $\kappa_a(n) = 1/\mu(a^n)$ satisfies the inequality of \cref{def:cover}, since the sum there is at least its largest term. Conversely, every $\kappa_a(n)$ that satisfies it is at least $1/(r\mu(a^n))$, since the sum is at most~$r$ times its largest term.
\end{remark}

\begin{lemma}\label{lem:A-cover}
Let $\mathcal{A} \subseteq \mathcal{S} \setminus \{0\}$ be multiplicatively closed and contain every positive integer. If $g_1, \ldots, g_r \in \Delta$ have the $\mathcal{A}$-cover property, then for every $\theta \in \mathcal{P}([r])$, the function~$w_\theta$ is $\mathcal{A}$-good.
\end{lemma}
\begin{proof}
Multiplicativity and normalisation we already noted follow directly from the definition of~$w_\theta$ as a weighted geometric average of spectral points, so it remains to prove \cref{def:w}~(\ref{it:good-sub}). Suppose $a \leq a_1 + \cdots + a_q$ for some $a, a_1, \ldots, a_q \in \mathcal{A}$, and let $n \geq 1$. For every word $J = (j_1, \ldots, j_n) \in [q]^n$ let $a_J = a_{j_1} \cdots a_{j_n}$, and choose $i(J) \in [r]$ minimising $g_i(a_J)/g_i(a)^n$ over $i \in [r]$. For $i \in [r]$ let~$b_i$ be the sum of the~$a_J$ with $i(J) = i$, allowing an empty sum. Expanding the $n$th power gives
\[
a^n \leq (a_1 + \cdots + a_q)^n = \sum_{J \in [q]^n} a_J = \sum_{i=1}^r b_i,
\]
so by the $\mathcal{A}$-cover property,
\begin{align*}
\kappa_a(n)^{-1} &\leq \sum_i \frac{g_i(b_i)}{g_i(a)^n} = \sum_{J \in [q]^n} \min_i \frac{g_i(a_J)}{g_i(a)^n} \\
&\leq \sum_{J \in [q]^n} \frac{w_\theta(a_J)}{w_\theta(a)^n} = \biggl( \frac{\sum_{j=1}^q w_\theta(a_j)}{w_\theta(a)} \biggr)^n,
\end{align*}
where we use that the minimum is at most the $\theta$-weighted geometric mean. Taking~$n$th roots gives $w_\theta(a) \leq \kappa_a(n)^{1/n} \sum_{j=1}^q w_\theta(a_j)$, and letting $n \to \infty$ gives \cref{def:w}~(\ref{it:good-sub}).
\end{proof}

\subsection{Path-connectedness of the asymptotic spectrum}

Recall from \cref{def:Fw} that $F_w(s) \coloneqq \sup \{ w(a)^{1/n} : n \geq 1, a \in \mathcal{A}, a \leq s^n \}$ for $s \in \mathcal{S}$. \cref{th:w-good-spec} says that~$F_w$ is a spectral point if~$\mathcal{S}$ is $\mathcal{A}$-decomposable and~$w$ is $\mathcal{A}$-good, and \cref{lem:A-cover} says that the geometric average $w_\theta(a) \coloneqq \prod_i g_i(a)^{\theta_i}$ is $\mathcal{A}$-good if $g_1, \ldots, g_r \in \Delta$ have the $\mathcal{A}$-cover property. 

We will now combine these ingredients to construct paths between spectral points. To do this we keep one endpoint $f_1 \in \Delta$ of our paths fixed. For $f_0 \in \Delta$ and $t \in [0,1]$, let~$w_t$ be the geometric average of~$f_0$ and~$f_1$ given by $w_t(s) \coloneqq f_0(s)^{1-t} f_1(s)^t$ for $s \in \mathcal{S}$. Let $F_t = F^{\smash[t]{f_0}}_t \coloneqq F_{w_t}$. Suppose that~$\mathcal{S}$ is $\mathcal{A}$-decomposable and that the pair $f_0, f_1$ has the $\mathcal{A}$-cover property. Then $t \mapsto F_t$ will be a path in~$\Delta$ from~$f_0$ to~$f_1$, the ``interpolation'' between them. If every pair $g, f_1$ with $g \in \Delta$ has the $\mathcal{A}$-cover property, then every point of~$\Delta$ will be joined to~$f_1$ by a path, so~$\Delta$ will be path-connected.

\begin{theorem}\label{th:interpolation}
Suppose~$\mathcal{S}$ is $\mathcal{A}$-decomposable, and let $f_1 \in \Delta$. For every $f_0 \in \Delta$ such that the pair $f_0, f_1$ has the $\mathcal{A}$-cover property, the following hold.
\begin{enumerate}[\upshape(i)]
\item\label{it:int-spectral} $F_t \in \Delta$ for every $t \in [0,1]$.
\item\label{it:int-start} $F_0 = f_0$.
\item\label{it:int-end} $F_1 = f_1$.
\item\label{it:int-family} For every $t \in [0,1]$ and $a \in \mathcal{A}$, we have $F_t(a) = f_0(a)^{1-t} f_1(a)^t$.
\item\label{it:int-cont} The map $[0,1] \to \Delta : t \mapsto F_t$ is continuous.
\end{enumerate}
If for every $f_0 \in \Delta$ the pair $f_0, f_1$ has the $\mathcal{A}$-cover property, then moreover the following holds.
\begin{enumerate}[\upshape(i)]\setcounter{enumi}{5}
\item\label{it:int-joint} The map $\Delta \times [0,1] \to \Delta : (f_0, t) \mapsto F^{f_0}_t$ is continuous.
\end{enumerate}
\end{theorem}
\begin{proof}
(i) Let $t \in [0,1]$. By \cref{lem:A-cover}, applied to the pair $f_0, f_1$ with $\theta = (1-t,t)$, the function~$w_t$ is $\mathcal{A}$-good. Since~$\mathcal{S}$ is $\mathcal{A}$-decomposable, \cref{th:w-good-spec}~(\ref{it:Fw-spectral}) gives the claim.

(ii) and (iii). For $t = 0$ we have $w_0 = f_0$, and for $t = 1$ we have $w_1 = f_1$. Since $f_0, f_1 \in \Delta$, \cref{th:w-good-spec}~(\ref{it:Fw-unique}) gives the claims.

(iv) Since~$w_t$ is $\mathcal{A}$-good, \cref{th:w-good-spec}~(\ref{it:Fw-family}) gives $F_t = w_t$ on~$\mathcal{A}$, which is the claim.

(v) and (vi). Give~$\RR^{\mathcal{A}}$ the product topology. Let $\rho : \Delta \to \RR^{\mathcal{A}} : f \mapsto (f(a))_{a \in \mathcal{A}}$. This map is continuous by the definition of the topology on~$\Delta$. It is injective by \cref{th:w-good-spec}~(\ref{it:Fw-unique}). Since~$\Delta$ is compact and $\RR^{\mathcal{A}}$ is Hausdorff, $\rho$ is a homeomorphism onto its image.

For $a \in \mathcal{A}$, the value
\[
w_t(a)
=
2^{(1-t)\log f_0(a)+t\log f_1(a)}
\]
is jointly continuous in $(f_0,t)\in\Delta\times[0,1]$. Hence the map
\[
\Delta\times[0,1]\to\RR^{\mathcal{A}}:(f_0,t)\mapsto(w_t(a))_{a\in\mathcal{A}}
\]
is continuous. Let $f_0 \in \Delta$ be such that the pair $f_0, f_1$ has the $\mathcal{A}$-cover property. By~(\ref{it:int-family}), the value of this map at~$(f_0,t)$ is $\rho(F^{\smash[t]{f_0}}_t)$ for every $t \in [0,1]$. Restricting this map to $\{f_0\} \times [0,1]$ and composing with~$\rho^{-1}$ proves~(\ref{it:int-cont}). Now suppose that every pair $g, f_1$ with $g \in \Delta$ has the $\mathcal{A}$-cover property. Then the value of this map at every $(g,t) \in \Delta \times [0,1]$ is $\rho(F^{\smash[t]{g}}_t)$. Composing with~$\rho^{-1}$ proves~(\ref{it:int-joint}).
\end{proof}

The proof of \cref{th:interpolation} gives more generally: if~$\mathcal{S}$ is $\mathcal{A}$-decomposable and $C \subseteq \Delta^r$ is a set of $r$-tuples with the $\mathcal{A}$-cover property, then $(g_1, \ldots, g_r, \theta) \mapsto F_{w_\theta}$ is a continuous map from $C \times \mathcal{P}([r])$ to~$\Delta$, where $F_{w_\theta}$ agrees with~$w_\theta$ on~$\mathcal{A}$ and equals~$g_i$ when~$\theta$ is the $i$th standard basis vector.

\begin{corollary}\label{cor:path-connected}
Fix $f_1 \in \Delta$. Suppose~$\mathcal{S}$ is $\mathcal{A}$-decomposable and that for every $f_0 \in \Delta$ the pair $f_0, f_1$ has the $\mathcal{A}$-cover property. Let $s_1, \ldots, s_q \in \mathcal{S}$ and
\[
\Delta(s_1, \ldots, s_q) \coloneqq \{(g(s_1), \ldots, g(s_q)) : g \in \Delta\} \subseteq \RR^q.
\]
Then every point of $\Delta(s_1, \ldots, s_q)$ is joined by a path to $(f_1(s_1), \ldots, f_1(s_q))$. In particular $\Delta(s_1, \ldots, s_q)$ is path-connected, and for every $s \in \mathcal{S}$, $\Delta(s) \subseteq \RR$ is a closed interval.
\end{corollary}
\begin{proof}
Let $f \in \Delta$. The function $[0,1] \to \Delta : t \mapsto F^f_t$ takes values in~$\Delta$ by \cref{th:interpolation}~(\ref{it:int-spectral}), is continuous by (\ref{it:int-cont}), and runs from $F^f_0 = f$ to $F^f_1 = f_1$ by (\ref{it:int-start}) and (\ref{it:int-end}). So it is a path from~$f$ to~$f_1$.

The function $\Delta \to \RR^q : g \mapsto (g(s_1), \ldots, g(s_q))$ is continuous by definition of the topology on~$\Delta$, and $\Delta(s_1, \ldots, s_q)$ is its image. So $[0,1] \to \RR^q : t \mapsto (F^{\smash[t]{f}}_t(s_1), \ldots, F^{\smash[t]{f}}_t(s_q))$ is a path in $\Delta(s_1, \ldots, s_q)$ from $(f(s_1), \ldots, f(s_q))$ to $(f_1(s_1), \ldots, f_1(s_q))$. This proves the first claim, and path-connectedness follows.

Let $q = 1$. Then~$\Delta(s)$ is the image of the compact space~$\Delta$ under a continuous function, hence compact, and it is nonempty because $f_1 \in \Delta$. A nonempty compact connected subset of~$\RR$ is a closed interval.
\end{proof}

A topological space~$X$ is \emph{contractible} if there are a point $c \in X$ and a continuous map $H : X \times [0,1] \to X$ such that $H(x,0) = x$ and $H(x,1) = c$ for every $x \in X$. Then $t \mapsto H(x,t)$ is a path from~$x$ to~$c$, so a contractible space is path-connected, by paths that depend continuously on their starting point. 

Suppose that~$\mathcal{S}$ is $\mathcal{A}$-decomposable and that every pair $f_0, f_1$ with $f_0 \in \Delta$ has the $\mathcal{A}$-cover property. Then the interpolation of \cref{th:interpolation} gives such a map~$H$ for $X = \Delta$ and $c = f_1$.

\begin{corollary}\label{cor:contractible}
Fix $f_1 \in \Delta$. Suppose~$\mathcal{S}$ is $\mathcal{A}$-decomposable and that for every $f_0 \in \Delta$ the pair $f_0, f_1$ has the $\mathcal{A}$-cover property. Then
\[
H : \Delta \times [0,1] \to \Delta : (f,t) \mapsto F^f_t
\]
is continuous. It satisfies $H(f,0) = f$ and $H(f,1) = f_1$ for every $f \in \Delta$, and $H(f_1,t) = f_1$ for every $t \in [0,1]$. In particular~$\Delta$ is contractible.
\end{corollary}
\begin{proof}
The map~$H$ takes values in~$\Delta$ by \cref{th:interpolation}~(\ref{it:int-spectral}). It is continuous by (\ref{it:int-joint}). The first two identities are (\ref{it:int-start}) and (\ref{it:int-end}). For $f = f_1$ the function $w_t(s) = f_1(s)^{1-t} f_1(s)^t = f_1(s)$ does not depend on~$t$, so $H(f_1,t) = H(f_1,1) = f_1$ for every $t \in [0,1]$. So~$H$ satisfies the definition of contractibility with $c = f_1$.
\end{proof}

Contractibility does not pass from~$\Delta$ to the sets $\Delta(s_1, \ldots, s_q)$ of \cref{cor:path-connected}, which are continuous images of~$\Delta$: in \cref{sec:mm-complex} we will give elements $s_1, s_2, s_3$ for which $\Delta(s_1, s_2, s_3)$ is homotopy equivalent to a circle.

When every pair of spectral points has the $\mathcal{A}$-cover property, \cref{th:interpolation} interpolates between any two spectral points, and the values on~$\mathcal{A}$ interpolate geometrically.
\begin{corollary}\label{cor:convex}
Suppose~$\mathcal{S}$ is $\mathcal{A}$-decomposable and that every pair $f_0, f_1 \in \Delta$ has the $\mathcal{A}$-cover property. Give~$\RR^{\mathcal{A}}$ the product topology, and let
\[
\Delta(\mathcal{A}) \coloneqq \{ (f(a))_{a \in \mathcal{A}} : f \in \Delta \} \subseteq \RR^{\mathcal{A}}.
\]
Then $f \mapsto (f(a))_{a \in \mathcal{A}}$ is a homeomorphism from~$\Delta$ onto $\Delta(\mathcal{A})$, which is compact and log-convex.
\end{corollary}
\begin{proof}
The map $f \mapsto (f(a))_{a \in \mathcal{A}}$ is continuous, because~$\Delta$ has the topology of pointwise convergence. It is injective: if $f(a) = g(a)$ for every $a \in \mathcal{A}$, then $f = g$ by \cref{th:w-good-spec}~(\ref{it:Fw-unique}). So it is a continuous injection from the compact space~$\Delta$ into a Hausdorff space, hence a homeomorphism onto its image $\Delta(\mathcal{A})$, and this image is compact.

Let $f, g \in \Delta$ and $t \in [0,1]$. By \cref{th:interpolation}~(\ref{it:int-spectral}) and (\ref{it:int-family}) with $f_0 = f$ and $f_1 = g$, the spectral point~$F^{\smash[t]{f}}_t$ satisfies $\log F^{\smash[t]{f}}_t(a) = (1-t)\log f(a) + t\log g(a)$ for every $a \in \mathcal{A}$. So $\log \Delta(\mathcal{A})$ is convex.
\end{proof}

\subsection{Asymptotic rank and asymptotic subrank}\label{sec:asymptotic-rank}

In this subsection we will relate the values of the spectral points at an element to its asymptotic rank and asymptotic subrank. Throughout this subsection only, we additionally assume that $1 \leq s$ for every nonzero $s \in \mathcal{S}$. Together with the Archimedean property, this is the stronger assumption that for every nonzero $s \in \mathcal{S}$ there is an $m \in \NN$ with $1 \leq s \leq m$. The rank and subrank limits, the duality statements, and the assertions about gapped elements in this subsection use this stronger assumption.

For $s \in \mathcal{S}$ the \emph{subrank} is defined as $
\subrank(s) \coloneqq \max\{m \in \NN : m \leq s\}$ and the \emph{rank} is defined as  $\tensorrank(s) \coloneqq \min\{m \in \NN : s \leq m\}$.
Their asymptotic counterparts are $\asympsubrank(s) \coloneqq \sup_{n\geq1} \subrank(s^n)^{1/n}$ and $\asymprank(s) \coloneqq \inf_{n\geq1} \tensorrank(s^n)^{\smash{1/n}}$, which can each be replaced by a limit by Fekete's lemma.

Following Wigderson and Zuiddam \cite[Definition~3.27]{wigderson2022asymptotic}, we call $s \in \mathcal{S}$ \emph{gapped} if there is an $n \geq 1$ such that $2 \leq s^n$, or there is an $f \in \Delta$ such that $f(s) = 1$. Asymptotic spectrum duality gives
\[
\max_{f \in \Delta} f(s) = \asymprank(s)
\]
for every $s \in \mathcal{S}$ \cite[Corollary~3.30]{wigderson2022asymptotic}, and
\[
\min_{f \in \Delta} f(s) = \asympsubrank(s)
\]
for every gapped $s \in \mathcal{S}$ \cite[Corollary~3.31]{wigderson2022asymptotic}. For tensors, both identities are due to Strassen \cite[Theorem~3.8]{strassen1988asymptotic}. Combining these identities with \cref{cor:path-connected} gives the following.

\begin{corollary}\label{cor:interval}
Fix $f_1 \in \Delta$. Suppose~$\mathcal{S}$ is $\mathcal{A}$-decomposable and that for every $f_0 \in \Delta$ the pair $f_0, f_1$ has the $\mathcal{A}$-cover property. Then for every gapped $s \in \mathcal{S}$,
\[
\Delta(s) = [\asympsubrank(s), \asymprank(s)].
\]
\end{corollary}
\begin{proof}
By \cref{cor:path-connected} the set~$\Delta(s)$ is a closed interval. By duality its minimum is~$\asympsubrank(s)$ and its maximum is~$\asymprank(s)$.
\end{proof}

In the setting of \cref{th:interpolation}, let $s \in \mathcal{S}$ be nonzero. By \cref{def:Fw}, the function $t \mapsto \log F_t(s)$ is a supremum of affine functions of~$t$, so it is convex. With \cref{th:interpolation}~(\ref{it:int-start}) and~(\ref{it:int-end}), convexity gives
\[
F_t(s) \leq f_0(s)^{1-t} f_1(s)^t \leq \max\{f_0(s), f_1(s)\}
\]
for every $t \in [0,1]$. The same holds for the interpolation between $r$-tuples described after \cref{th:interpolation}. Every $f \in \Delta$ gives the bounds
\[
\asympsubrank(s) \leq f(s) \leq \asymprank(s).
\]
As lower bounds on the asymptotic rank, interpolations are therefore no better than their end points. For $s \in \mathcal{A}$ the same holds for upper bounds on the asymptotic subrank, since $\log F_t(s)$ is affine in~$t$ by \cref{th:interpolation}~(\ref{it:int-family}). For other~$s$, interpolations can give better upper bounds, as an example in \cref{sec:qf-identify} will show.

\subsection{Extension of spectral points}\label{sec:extension}

In this subsection we state a general extension theorem for spectral points of subsemirings, which we will use later.

Let $\mathcal{T} \subseteq \mathcal{S}$ be a subsemiring, with the preorder induced from~$\mathcal{S}$. Then~$\mathcal{T}$ is a Strassen semiring, since it contains~$\NN$ and since the relations $s \leq mt$ of the Archimedean property, for nonzero $s, t \in \mathcal{T}$, are relations between elements of~$\mathcal{T}$. For every $f \in \Delta$, the restriction~$f|_{\mathcal{T}}$ of~$f$ to~$\mathcal{T}$ is a spectral point of~$\mathcal{T}$. Conversely, every spectral point of~$\mathcal{T}$ arises in this way, by the following proposition of Strassen.

\begin{proposition}[{\cite[Proposition~3.3]{strassen1988asymptotic}}]\label{prop:extension}
Let~$\mathcal{S}$ be a Strassen semiring and let $\mathcal{T} \subseteq \mathcal{S}$ be a subsemiring, with the preorder induced from~$\mathcal{S}$. Then for every spectral point~$g$ of~$\mathcal{T}$ there is a spectral point~$f$ of~$\mathcal{S}$ with $f|_{\mathcal{T}} = g$.
\end{proposition}

\section{Path-connectedness and contractibility for complex tensors}\label{sec:tensors}

In this section we will prove that the asymptotic spectrum of complex tensors is path-connected and contractible.

Fix $k \geq 2$ and let~$\FF$ be a field. All vector spaces are over~$\FF$. For $k$-tensors $s \in W_1 \otimes \cdots \otimes W_k$ and $t \in V_1 \otimes \cdots \otimes V_k$ we write $s \leq t$, and say that~$s$ is a \emph{restriction} of~$t$, if $s = (A_1 \otimes \cdots \otimes A_k)t$ for some linear maps $A_i : V_i \to W_i$. We call~$s$ and~$t$ \emph{equivalent} if $s \leq t$ and $t \leq s$. The direct sum $s \oplus t$ is the $k$-tensor in $(W_1 \oplus V_1) \otimes \cdots \otimes (W_k \oplus V_k)$ obtained by embedding~$s$ and~$t$ and adding them, and the Kronecker product $s \otimes t$ is the $k$-tensor in $(W_1 \otimes V_1) \otimes \cdots \otimes (W_k \otimes V_k)$ obtained from the tensor product of~$s$ and~$t$ by regrouping the factors. Let~$\mathcal{S}_\FF$ be the set of $k$-tensors over~$\FF$ up to equivalence, with~$\oplus$, $\otimes$ and~$\leq$, all of which are well defined on equivalence classes. For $m \in \NN$, the element~$m$ of~$\mathcal{S}_\FF$ is the class of the unit tensor $\langle m \rangle = \sum_{j=1}^m e_j \otimes \cdots \otimes e_j$. This is a Strassen semiring, and we write~$\Delta_\FF$ for its asymptotic spectrum. When we need to indicate the order~$k$, we write~$\mathcal{S}_{\FF,k}$ and~$\Delta_{\FF,k}$.

Let $t \in V_1 \otimes \cdots \otimes V_k$ and let $i \in [k]$. The \emph{$i$-support} of~$t$ is the smallest subspace $U \subseteq V_i$ such that $t \in V_1 \otimes \cdots \otimes U \otimes \cdots \otimes V_k$. The \emph{$i$-flattening rank} $\tensorrank_i(t)$ is the dimension of the $i$-support. We say~$t$ is \emph{$i$-concise} if its $i$-support equals~$V_i$, that is, if $\tensorrank_i(t) = \dim V_i$. We say~$t$ is \emph{concise} if~$t$ is $i$-concise for every $i \in [k]$. For every $i \in [k]$ we have $\tensorrank_i \in \Delta_\FF$. Every tensor is equivalent to a concise one. Conciseness is closed under direct sum and Kronecker product.

In the rest of this section, we will only use complex tensors: $\FF = \CC$,  $\mathcal{S} = \mathcal{S}_\CC$ and $\Delta = \Delta_\CC$. 

The main result of this section is the following theorem.

\begin{theorem}\label{th:tensor-contractible}
The asymptotic spectrum~$\Delta$ of complex $k$-tensors is path-connected: for every $i \in [k]$ and every $f \in \Delta$ there is a path in~$\Delta$ from~$f$ to~$\tensorrank_i$. For every $s \in \mathcal{S}$, the set~$\Delta(s)$ is a closed interval. In fact, $\Delta$ is contractible.
\end{theorem}

We will prove \cref{th:tensor-contractible} in \cref{sec:tensor-paths}, by applying \cref{th:interpolation} to a set~$\mathcal{A}_i$ that we define in the next subsection. In \cref{sec:mm-complex} we will apply \cref{th:interpolation} to rectangular matrix multiplication, and recover Strassen's theorem that its asymptotic spectrum is log-star-convex (\cref{th:star-convex}). In \cref{sec:poly-complex} we will extend this to polynomial multiplication (\cref{th:poly-star-convex}), using compact groups in place of finite groups.

\subsection{\texorpdfstring{$\mathcal{A}$}{A}-decomposability}

To instantiate the method of \cref{sec:dec} for the complex tensor semiring~$\mathcal{S}$ we need an $\mathcal{A}\subseteq \mathcal{S}$ such that~$\mathcal{S}$ is $\mathcal{A}$-decomposable (\cref{def:decomposable}).

Let $i \in [k]$. An element of~$\mathcal{S}$ is \emph{$i$-irreducible} if it has a concise representative $a \in V_1 \otimes \cdots \otimes V_k$ with the following property: there are a finite group~$G$ and representations $\rho_j : G \to \GL(V_j)$ for $j \in [k]$ such that~$\rho_i$ is irreducible and, for every $g \in G$,
\[
(\rho_1(g) \otimes \cdots \otimes \rho_k(g)) a = a.
\]

Let $\mathcal{A}_i \subseteq \mathcal{S}$ be the set of $i$-irreducible tensors. 

We will prove that~$\mathcal{S}$ is decomposable with respect to the following smaller set~$\mathcal{B}$, which \cref{sec:qf} will use when it interpolates towards several flattening ranks in turn, and deduce that~$\mathcal{S}$ is $\mathcal{A}_i$-decomposable for every $i \in [k]$. An element of~$\mathcal{S}$ is \emph{jointly irreducible} if it has a concise representative $a \in V_1 \otimes \cdots \otimes V_k$ for which there are a finite group~$G$ and, for every $i \in [k]$, a representation $\rho_i : G \to \GL(V_i)$, such that $(\rho_1(g) \otimes \cdots \otimes \rho_k(g)) a = a$ for every $g \in G$, and such that~$\rho_i$ is irreducible for every $i \in [k]$. Let $\mathcal{B} \subseteq \mathcal{S}$ be the set of jointly irreducible tensors. Thus $\mathcal{B} \subseteq \bigcap_{i=1}^k \mathcal{A}_i$. Joint irreducibility asks for one group serving every leg, while the intersection allows a different group for each leg. Over a finite field, with the same definitions, the inclusion is an equality: the groups for the different legs can be moved to one concise representative, since equivalent concise tensors are related by invertible linear maps on the legs, and there they generate a finite group that acts irreducibly on every leg. Over~$\CC$ the generated group can be infinite, but its Zariski closure is reductive, as every leg is an irreducible representation of it, and a compact real form of the closure (see Adams and Ta\"{\i}bi \cite[Definition~3.3, Lemma~3.6 and Theorem~3.7]{MR3786301}) is a Zariski-dense compact subgroup that still fixes the representative and acts irreducibly on every leg; we do not know whether the inclusion is an equality over~$\CC$.

We first prove two lemmas about representations of finite groups, which we will use to show that~$\mathcal{S}$ is $\mathcal{B}$-decomposable. For $m \geq 1$ and $r \geq 2$, let $G_{m,r} \subseteq \GL(\CC^m)$ be the group of monomial matrices whose nonzero entries are~$r$th roots of unity, that is, the matrices of the maps $e_x \mapsto \zeta_x e_{\pi(x)}$ with $\pi \in \mathfrak{S}_m$ and $\zeta_1, \ldots, \zeta_m$ $r$th roots of unity. For $r = 2$ these are the signed permutation matrices.

\begin{lemma}\label{lem:monomial-irred}
For all $m \geq 1$ and $r \geq 2$, the representation of $G_{m,r}$ on~$\CC^m$ is irreducible.
\end{lemma}
\begin{proof}
Let $U \subseteq \CC^m$ be a nonzero $G_{m,r}$-invariant subspace. Let $v \in U$ be nonzero and let $j \in [m]$ with $v_j \neq 0$. Let $\omega \neq 1$ be an~$r$th root of unity and let $D_j \in G_{m,r}$ be the diagonal matrix with~$\omega$ in position~$j$ and~$1$ elsewhere. Then $(D_j - \Id)v = (\omega - 1) v_j e_j$ lies in~$U$, so $e_j \in U$. The permutation matrices lie in $G_{m,r}$ and map~$e_j$ to every~$e_\ell$ with $\ell \in [m]$, so $U = \CC^m$.
\end{proof}

The following count is standard, and we include a proof.

\begin{lemma}\label{lem:irrep-count}
Let~$V$ be a complex vector space of dimension $d \geq 1$, let $n \geq 1$, and let~$\mathfrak{S}_n$ act on $V^{\otimes n}$ by permuting the factors. Then every decomposition of $V^{\otimes n}$ into irreducible $\mathfrak{S}_n$-subrepresentations has at most $(n+1)^{d^2}$ summands.
\end{lemma}
\begin{proof}
Let $V^{\otimes n} = \bigoplus_{\smash[b]{j=1}}^q U_j$ be a decomposition into irreducible $\mathfrak{S}_n$-subrepresentations, and let $\End_{\mathfrak{S}_n}(V^{\otimes n})$ be the space of linear maps $V^{\otimes n} \to V^{\otimes n}$ that commute with the action of~$\mathfrak{S}_n$. The projections onto the~$U_j$ along the other summands lie in this space and are linearly independent, so $q \leq \dim \End_{\mathfrak{S}_n}(V^{\otimes n})$. Under the isomorphism $\End(V^{\otimes n}) \cong \End(V)^{\otimes n}$, the space $\End_{\mathfrak{S}_n}(V^{\otimes n})$ corresponds to the tensors fixed by all permutations of the~$n$ factors. For a basis of $\End(V)$, which has dimension~$d^2$, the sums over the orbits of the corresponding product basis of $\End(V)^{\otimes n}$ are a basis of the space of these fixed tensors, and the orbits correspond to the multisets of size~$n$ of basis elements. Hence
\[
\dim \End_{\mathfrak{S}_n}(V^{\otimes n}) = \binom{n+d^2-1}{d^2-1} = \prod_{j=1}^{d^2-1} \frac{n+j}{j} \leq (n+1)^{d^2-1},
\]
since $n + j \leq j(n+1)$ for every $j \geq 1$.
\end{proof}

The next lemma decomposes the Kronecker powers of every tensor explicitly into elements of~$\mathcal{B}$. \Cref{sec:qf-identify} will use this explicit form.

\begin{lemma}\label{lem:qf-explicit}
Let $t \in V_1 \otimes \cdots \otimes V_k$ be nonzero and let $n \geq 1$. For every $i \in [k]$ let~$\mathfrak{S}_n$ act on~$V_i^{\otimes n}$ by permuting the factors, let
\[
V_i^{\otimes n} = \bigoplus_{j=1}^{q_i} U^{(i)}_j
\]
be a decomposition into irreducible $\mathfrak{S}_n$-subrepresentations, and let~$p^{\smash{(i)}}_{\!j}$ be the projection onto~$U^{\smash{(i)}}_{\!j}$ along the other summands. Then the nonzero tensors 
\[
a_{j_1, \ldots, j_k} \coloneqq (p^{(1)}_{\!j_1} \otimes \cdots \otimes p^{(k)}_{\!j_k}) t^{\otimes n}
\]
give a $\mathcal{B}$-decomposition of~$t^{\otimes n}$ of length at most $\prod_{i=1}^k (n+1)^{(\dim V_i)^2}$. Every nonzero $a_{j_1, \ldots, j_k}$ is concise in $U^{\smash{(1)}}_{\!j_1} \otimes \cdots \otimes U^{\smash{(k)}}_{\!j_k}$.
\end{lemma}
\begin{proof}
The simultaneous action of~$\mathfrak{S}_n$ on $V_1^{\otimes n} \otimes \cdots \otimes V_k^{\otimes n}$ fixes~$t^{\otimes n}$. The~$p^{\smash{(i)}}_{\!j}$ are projections along $\mathfrak{S}_n$-invariant summands, and hence $\mathfrak{S}_n$-equivariant. So every nonzero $a_{j_1, \ldots, j_k}$ is invariant under the simultaneous action of~$\mathfrak{S}_n$, and for every $i \in [k]$ its $i$-support is a nonzero invariant subspace of~$U^{\smash{(i)}}_{\!j_i}$, which equals~$U^{\smash{(i)}}_{\!j_i}$ by irreducibility. So it is concise in $U^{\smash{(1)}}_{\!j_1} \otimes \cdots \otimes U^{\smash{(k)}}_{\!j_k}$, where it is fixed by~$\mathfrak{S}_n$ acting irreducibly on every leg. Hence it lies in~$\mathcal{B}$, and it restricts from~$t^{\smash{\otimes n}}$. For every $i \in [k]$ the projections satisfy $\sum_{j=1}^{q_i} p^{\smash{(i)}}_{\!j} = \Id$, so $t^{\smash{\otimes n}}$ is the sum of the nonzero $a_{j_1, \ldots, j_k}$. A sum of tensors in one space is a restriction of their direct sum. So~$t^{\smash{\otimes n}}$ is at most the direct sum of the nonzero $a_{j_1, \ldots, j_k}$, and these give a $\mathcal{B}$-decomposition of~$t^{\otimes n}$. Its length is at most $\prod_{i=1}^{\smash[t]{k}} q_i$, and $q_i \leq (n+1)^{\smash{(\dim V_i)^2}}$ for every $i \in [k]$ by \cref{lem:irrep-count}.
\end{proof}

\begin{lemma}\label{lem:tensor-decomp}
The semiring~$\mathcal{S}$ is $\mathcal{B}$-decomposable, and $\mathcal{A}_i$-decomposable for every $i \in [k]$.
\end{lemma}
\begin{proof}
We check the properties of \cref{def:decomposable}, for~$\mathcal{B}$ and for~$\mathcal{A}_i$ with $i \in [k]$.

(\ref{it:dec-mult}) We prove that~$\mathcal{B}$ and~$\mathcal{A}_i$ are multiplicatively closed. Let $a \in V_1 \otimes \cdots \otimes V_k$ and $b \in W_1 \otimes \cdots \otimes W_k$ be concise representatives of elements of~$\mathcal{B}$, or of~$\mathcal{A}_i$, with finite groups~$G$ and~$H$ and representations $\rho_\ell : G \to \GL(V_\ell)$ and $\sigma_\ell : H \to \GL(W_\ell)$ for $\ell \in [k]$, as in the definition. The finite group $G \times H$ has the representations $\rho_\ell \otimes \sigma_\ell : G \times H \to \GL(V_\ell \otimes W_\ell)$, whose simultaneous action fixes $a \otimes b$. If~$\rho_\ell$ and~$\sigma_\ell$ are irreducible, then so is $\rho_\ell \otimes \sigma_\ell$, because the tensor product of irreducible complex representations is irreducible for the product group. Also $a \otimes b$ is concise. Hence $a \otimes b$ lies in~$\mathcal{B}$, or in~$\mathcal{A}_i$, respectively.

(\ref{it:dec-nat}) We prove that every positive integer lies in~$\mathcal{B}$, and hence in~$\mathcal{A}_i$. Let $m \geq 1$, let $G = G_{m,k}$, let $\rho : G \to \GL(\CC^m)$ be the inclusion and let $\rho_1 = \cdots = \rho_k = \rho$. The simultaneous action fixes $\langle m \rangle$, because the nonzero entries of the matrices in~$G$ are~$k$th roots of unity, and $\langle m \rangle$ is concise. Moreover~$\rho$ is irreducible by \cref{lem:monomial-irred}. Hence $\langle m \rangle \in \mathcal{B}$.

(\ref{it:dec-finite}) We prove that every nonzero tensor~$t$ satisfies $a \leq t$ for some $a \in \mathcal{B}$, and hence for some $a \in \mathcal{A}_i$. Every nonzero tensor~$t$ satisfies $\langle 1 \rangle \leq t$, so this follows from (\ref{it:dec-nat}).

(\ref{it:dec-count}) We prove that every nonzero tensor has asymptotic $\mathcal{B}$-length~$1$ and asymptotic $\mathcal{A}_i$-length~$1$. Let $t \in \mathcal{S} \setminus \{0\}$, represented by a tensor in $V_1 \otimes \cdots \otimes V_k$, and let $n \geq 1$. By Maschke's theorem, every~$V_\ell^{\otimes n}$ is a direct sum of irreducible $\mathfrak{S}_n$-subrepresentations. So \cref{lem:qf-explicit} gives $1 \leq \declength_{\mathcal{B}}(t^{\otimes n}) \leq \prod_{\ell=1}^{\smash[t]{k}} (n+1)^{\smash{(\dim V_\ell)^2}}$, which is a polynomial in~$n$. So $\declength_{\mathcal{B}}(t^{\otimes n})^{1/n} \to 1$, that is, $\asympdeclength_{\mathcal{B}}(t) = 1$. Since $\mathcal{B} \subseteq \mathcal{A}_i$, every $\mathcal{B}$-decomposition is an $\mathcal{A}_i$-decomposition, so $1 \leq \declength_{\mathcal{A}_i}(t^{\otimes n}) \leq \declength_{\mathcal{B}}(t^{\otimes n})$. Hence also $\asympdeclength_{\mathcal{A}_i}(t) = 1$.
\end{proof}

\subsection{Subspace translate averaging and intersection}

The following lemma upper bounds the average overlap of $G$-translates of a subspace $W\subseteq V$ with some fixed subspace $U \subseteq V$ when~$V$ is an irreducible $G$-representation for a finite group~$G$.

\begin{lemma}\label{lem:dim-averaging}
Let~$\FF$ be any field, let a finite group~$G$ act irreducibly on a nonzero $\FF$-vector space~$V$, and let $U, W \subseteq V$ be subspaces. Then
\[
\frac{1}{|G|} \sum_{g \in G} \dim(U \cap gW) \leq \frac{\dim W}{\dim V} \dim U.
\]
\end{lemma}
\begin{proof}
Let $d \coloneqq \dim V$. For a subspace $X \subseteq V$ let
\[
\Psi(X) \coloneqq d \sum_{g \in G} \dim(X \cap gW) - |G| \dim W \dim X \in \ZZ.
\]
We show $\Psi \leq 0$, which proves the claim after dividing by $|G| d$. 

First, we claim~$\Psi$ is supermodular: for all subspaces $X, Y \subseteq V$,
\begin{equation}\label{eq:psi-super}
\Psi(X) + \Psi(Y) \leq \Psi(X \cap Y) + \Psi(X + Y).
\end{equation}
Indeed, every pair of subspaces $A, B \subseteq V$ satisfies $\dim A + \dim B = \dim(A \cap B) + \dim(A+B)$. Let $C \subseteq V$ be any subspace. Applying this with $A = X \cap C$ and $B = Y \cap C$, and using $(X \cap C) \cap (Y \cap C) = (X \cap Y) \cap C$ and $(X \cap C) + (Y \cap C) \subseteq (X+Y) \cap C$, gives
\begin{equation}\label{eq:cap-super}
\dim(X \cap C) + \dim(Y \cap C) \leq \dim\bigl((X \cap Y) \cap C\bigr) + \dim\bigl((X+Y) \cap C\bigr).
\end{equation}
Taking $C = gW$ in \eqref{eq:cap-super} and summing over all $g \in G$ gives
\begin{multline}\label{eq:cap-super-sum}
\sum_{g \in G} \dim(X \cap gW) + \sum_{g \in G} \dim(Y \cap gW)\\[-0.5em]
\leq \sum_{g \in G} \dim\bigl((X \cap Y) \cap gW\bigr) + \sum_{g \in G} \dim\bigl((X+Y) \cap gW\bigr).
\end{multline}
Next we multiply the inequality \eqref{eq:cap-super-sum} by~$d$ and then subtract $|G| \dim W$ times the equality $\dim X + \dim Y = \dim(X \cap Y) + \dim(X+Y)$ to get the claim. 

Second, $\Psi$ is $G$-invariant, since $\dim(hX \cap gW) = \dim(X \cap h^{-1}gW)$ and $g \mapsto hg$ permutes~$G$ for every $h \in G$.

Third, $\Psi(0) = 0$, and $\Psi(V) = d \sum_g \dim(gW) - |G| \dim W \, d = 0$. 

The function~$\Psi$ takes values in~$\ZZ$ and satisfies $\Psi(X) \leq d |G| \dim W$ for every~$X$, so it attains its maximum. Among the subspaces at which~$\Psi$ attains its maximum, let~$Z$ be one of largest dimension. Let $g \in G$. By $G$-invariance $\Psi(gZ) = \Psi(Z)$, so supermodularity \eqref{eq:psi-super} with $X = Z$ and $Y = gZ$ gives
\[
2 \Psi(Z) \leq \Psi(Z \cap gZ) + \Psi(Z + gZ).
\]
Since~$Z$ maximises~$\Psi$, both terms on the right are at most~$\Psi(Z)$, so both equal~$\Psi(Z)$. So $Z + gZ$ is also a maximiser, and the choice of~$Z$ gives $\dim(Z + gZ) \leq \dim Z$. Hence $gZ \subseteq Z$, and $\dim gZ = \dim Z$ gives $gZ = Z$. As~$g$ was arbitrary, $Z$ is $G$-invariant. By irreducibility of~$V$, we have $Z = 0$ or $Z = V$. In both cases $\Psi(Z) = 0$, so the maximum of~$\Psi$ is~$0$ and $\Psi(U) \leq 0$.
\end{proof}

We now apply the subspace averaging \cref{lem:dim-averaging} iteratively to show existence of group elements $g_1, \ldots, g_q$ such that the translates $g_1 W, \ldots, g_q W$ have intersection equal to~$0$.

\begin{lemma}\label{lem:spanning}
Let~$\FF$ be any field, let a finite group~$G$ act irreducibly on a nonzero $\FF$-vector space~$V$, and let $W \subsetneq V$ be a proper subspace. Then there is $q \in \NN$ with 
\[
q \leq \frac{\dim V}{\dim V - \dim W}(1 + \ln \dim V)
\]
and $g_1, \ldots, g_q \in G$ such that 
\[
g_1 W \cap \cdots \cap g_q W = 0.
\]
\end{lemma}
\begin{proof}
Let $d \coloneqq \dim V$ and $c \coloneqq \dim W / d$, so that $c < 1$. If $W = 0$ then $q = 1$ works, so let $W \neq 0$. We construct $g_1, g_2, \ldots \in G$ and subspaces $V = U_0 \supseteq U_1 \supseteq \cdots$ recursively, as follows. Given~$U_j$, \cref{lem:dim-averaging} gives a $g_{j+1} \in G$ with $\dim(U_j \cap g_{j+1} W) \leq c \dim U_j$, and we let $U_{j+1} \coloneqq U_j \cap g_{j+1} W$. Then $U_j = g_1 W \cap \cdots \cap g_j W$ and $\dim U_j \leq d c^j$ for every $j \geq 0$.
Let $q \coloneqq \lfloor \ln d / \ln(1/c) \rfloor + 1$, so that $d c^q < 1$ and hence $U_q = 0$. Since $\ln(1/c) \geq 1 - c$, we have
\[
q \leq \frac{\ln d}{1-c} + 1 \leq \frac{1 + \ln d}{1-c} = \frac{d}{d - \dim W}(1 + \ln d). \qedhere
\]
\end{proof}

We note the following equivalent form of \cref{lem:spanning} in terms of spanning a space by translates of a subspace: For every nonzero subspace $U \subseteq V$ with a finite group~$G$ acting irreducibly on~$V$, applying \cref{lem:spanning} to the dual action of~$G$ on~$V^*$, which is again irreducible, and to the annihilator $U^\perp \subseteq V^*$ of~$U$ gives $g_1, \ldots, g_q \in G$ with $g_1 U + \cdots + g_q U = V$ and $q \leq \frac{\dim V}{\dim U}(1 + \ln \dim V)$. The intersection version \cref{lem:spanning} will be more natural for us.

\subsection{Reverse monotonicity}

Let $a \in V_1 \otimes \cdots \otimes V_k$, let $\pi : V_1 \to W$ be linear and let $b = (\pi \otimes \Id \otimes \cdots \otimes \Id)a$. Then $b \leq a$, so by monotonicity every $f \in \Delta$ satisfies
\[
f(b) \leq f(a).
\]
The following lemma provides a reverse inequality in terms of the flattening rank~$\tensorrank_1$, and thus bounds how much spectral points can drop under restriction.

\begin{lemma}\label{lem:restriction}
Let $a \in \mathcal{A}_1$ lie in $V_1 \otimes \cdots \otimes V_k$. Let $b \coloneqq (\pi \otimes \Id \otimes \cdots \otimes \Id)a$ for some linear map $\pi : V_1 \to W$. Then for every $f \in \Delta$ we have
\[
\frac{\tensorrank_1(b)}{\tensorrank_1(a)} f(a) \leq f(b).
\]
\end{lemma}
\begin{proof}
Without loss of generality~$a$ is concise: replace every~$V_j$ by the $j$-support of~$a$, and~$\pi$ by its restriction to the $1$-support, which changes neither~$b$ nor~$\tensorrank_1(a)$. Then $d = \tensorrank_1(a) = \dim V_1$ and $r = \tensorrank_1(b) = \rank \pi$. The case $r = 0$ is immediate, so let $r > 0$. Equivalent concise tensors are related by invertible maps on their legs, since the maps of a restriction between them are surjective on every leg and the dimensions agree. So the group and representations of the definition of~$\mathcal{A}_1$ can be transported to~$a$: there are a finite group~$G$ and representations $\rho_j : G \to \GL(V_j)$ whose simultaneous action fixes~$a$, with~$\rho_1$ irreducible. Applying \cref{lem:spanning} to~$\rho_1$ and $\ker \pi$ gives $g_1, \ldots, g_q \in G$ with $q \leq \frac{d}{r}(1 + \ln d)$ such that $g_1 \ker\pi \cap \cdots \cap g_q \ker\pi = 0$. Let $\pi_j \coloneqq \pi \rho_1(g_j)^{-1}$ and $b_j \coloneqq (\pi_j \otimes \Id \otimes \cdots \otimes \Id)a$, so that $\ker \pi_j = g_j \ker \pi$. Then the map
\[
P : V_1 \rightarrow \bigoplus_{j=1}^q W : v \mapsto (\pi_1 v, \ldots, \pi_q v)
\]
is injective, so there is a linear map $L : \bigoplus_{j=1}^q W \to V_1$ with $LP = \Id$. Applying~$L$ on the first leg of $\bigoplus_{j=1}^q b_j$ and adding up the components on each other leg gives $(LP \otimes \Id \otimes \cdots \otimes \Id)a = a$, so $a \leq \bigoplus_{j=1}^q b_j$. Since~$a$ is invariant under~$g_j$, we have $b_j = (\Id_W \otimes \rho_2(g_j) \otimes \cdots \otimes \rho_k(g_j))b$, so~$b_j$ is equivalent to~$b$. Hence
\begin{equation}\label{eq:fabound}
f(a) \leq \sum_{j=1}^q f(b_j) = q \, f(b) \leq \frac{d}{r} (1 + \ln d) \, f(b).
\end{equation}
We remove the logarithmic factor by applying \eqref{eq:fabound} to Kronecker powers. Now $a^{\otimes n} \in \mathcal{A}_1$, since~$\mathcal{A}_1$ is multiplicatively closed by \cref{lem:tensor-decomp}. The map $\pi^{\otimes n}$ has rank~$r^n$, and $(\pi^{\otimes n} \otimes \Id \otimes \cdots \otimes \Id) a^{\otimes n} = b^{\otimes n}$. Applying inequality \eqref{eq:fabound} to these gives $f(a)^n \leq (d/r)^n (1 + n \ln d) f(b)^n$. Then take~$n$th roots and let $n \to \infty$.
\end{proof}

\subsection{\texorpdfstring{$\mathcal{A}$}{A}-cover property}

Using \cref{lem:restriction} we will prove, for every $f\in \Delta$, that the pair $(f, \tensorrank_1)$ has the $\mathcal{A}_1$-cover property (\cref{def:cover}).

\begin{lemma}\label{lem:tensor-cover}
Let $f \in \Delta$.
\begin{enumerate}[\upshape(i)]
\item\label{it:cover-one} For every $a \in \mathcal{A}_1$ and all $s_1, s_2 \in \mathcal{S}$ with $a \leq s_1 \oplus s_2$,
\[
1 \leq \frac{f(s_1)}{f(a)} + \frac{\tensorrank_1(s_2)}{\tensorrank_1(a)}.
\]
\item\label{it:cover-all} The pair $f, \tensorrank_1$ has the $\mathcal{A}_1$-cover property with $\kappa_a(n) = 1$.
\end{enumerate}
\end{lemma}
\begin{proof}
(\ref{it:cover-one}) We may take $a \in V_1 \otimes \cdots \otimes V_k$, $s_1$ and~$s_2$ to be concise (replace each by a concise representative), and we let $d = \tensorrank_1(a) = \dim V_1$. If $\tensorrank_1(s_2) \geq d$ then the second term is at least~$1$ and there is nothing to prove, so let $\tensorrank_1(s_2) < d$. Write $a = (A_1 \otimes \cdots \otimes A_k)(s_1 \oplus s_2)$ with linear maps~$A_\ell$, let~$E_2$ be the first leg of~$s_2$, and let $U = A_1(E_2) \subseteq V_1$, so that $\dim U \leq \dim E_2 = \tensorrank_1(s_2)$. Let $\pi : V_1 \to V_1/U$ be the quotient map and let $b = (\pi \otimes \Id \otimes \cdots \otimes \Id)a$, so that $\tensorrank_1(b) = d - \dim U$. Since~$\pi$ vanishes on $U = A_1(E_2)$, applying~$\pi$ annihilates the contribution of~$s_2$. Hence $b \leq s_1$. By \cref{lem:restriction},
\[
f(s_1) \geq f(b) \geq \frac{d - \dim U}{d} f(a) \geq \Bigl( 1 - \frac{\tensorrank_1(s_2)}{d} \Bigr) f(a).
\]
Dividing by~$f(a)$ gives (\ref{it:cover-one}).

(\ref{it:cover-all}) Apply (\ref{it:cover-one}) to~$a^{\otimes n}$, which lies in~$\mathcal{A}_1$ because~$\mathcal{A}_1$ is multiplicatively closed and contains $a^{\otimes 0} = \langle 1 \rangle$, and use $f(a^{\otimes n}) = f(a)^n$ and $\tensorrank_1(a^{\otimes n}) = \tensorrank_1(a)^n$.
\end{proof}

\subsection{Path-connectedness and contractibility}\label{sec:tensor-paths}

We now prove \cref{th:tensor-contractible}. We first record the paths that \cref{cor:path-connected} gives, with the fixed spectral point $f_1 = \tensorrank_i$ and any $f_0 \in \Delta$.

\begin{corollary}\label{cor:tensor-paths}
Let $i \in [k]$ and $s_1, \ldots, s_q \in \mathcal{S}$. Then every point of $\Delta(s_1, \ldots, s_q)$ is joined by a path to $(\tensorrank_i(s_1), \ldots, \tensorrank_i(s_q))$. In particular $\Delta(s_1, \ldots, s_q)$ is path-connected, and for every $s \in \mathcal{S}$, $\Delta(s)$ is a closed interval.
\end{corollary}
\begin{proof}
By \cref{lem:tensor-decomp} and \cref{lem:tensor-cover}, the hypotheses of \cref{cor:path-connected} hold with $\mathcal{A} = \mathcal{A}_1$ and $f_1 = \tensorrank_1$. Relabelling the legs, they hold with~$\mathcal{A}_i$ and~$\tensorrank_i$ for every $i \in [k]$.
\end{proof}

\begin{proof}[Proof of \cref{th:tensor-contractible}]
Let $i \in [k]$. By \cref{lem:tensor-decomp} and \cref{lem:tensor-cover}, after relabelling the legs, the hypotheses of \cref{cor:contractible} hold with $\mathcal{A} = \mathcal{A}_i$ and $f_1 = \tensorrank_i$. So the map $H : \Delta \times [0,1] \to \Delta$ given by $H(f,t) = F^{\smash[t]{f}}_t$ is continuous, $H(f,0) = f$ and $H(f,1) = \tensorrank_i$ for every $f \in \Delta$, and $H(\tensorrank_i,t) = \tensorrank_i$ for every $t \in [0,1]$. For every $f \in \Delta$, the map $t \mapsto H(f,t)$ is a path in~$\Delta$ from~$f$ to~$\tensorrank_i$. So~$\Delta$ is path-connected, and~$H$ shows that~$\Delta$ is contractible. The assertion about~$\Delta(s)$ follows from \cref{cor:tensor-paths}.
\end{proof}

\subsection{Log-star-convexity for rectangular matrix multiplication}\label{sec:mm-complex}

In this subsection we will recover, as an application of \cref{th:interpolation} over the complex numbers, Strassen's theorem that the asymptotic spectrum of rectangular matrix multiplication is log-star-convex \cite[Corollary~6.6]{strassen1988asymptotic}. In \cref{sec:star-convex-any-field} we will do so over every field.

In this subsection $k = 3$. The \emph{matrix multiplication tensors}, for positive integers $n_1, n_2, n_3$, are defined as 
\begin{equation}\label{eq:mm-tensor}
\langle n_1, n_2, n_3 \rangle \coloneqq \sum_{x=1}^{n_1} \sum_{y=1}^{n_2} \sum_{z=1}^{n_3} e_{xy} \otimes e_{yz} \otimes e_{zx}  \in \CC^{n_1 n_2} \otimes \CC^{n_2 n_3} \otimes \CC^{n_3 n_1},
\end{equation}
where the~$e_{xy}$ with $x \in [n_1]$ and $y \in [n_2]$ are the standard basis of~$\CC^{n_1 n_2}$, and similarly for the other two legs. 
Each~$\langle n_1, n_2, n_3 \rangle$ is concise, so its flattening ranks are $n_1 n_2$, $n_2 n_3$ and $n_3 n_1$. 

Let $b_1 = \langle 2,1,1 \rangle$, $b_2 = \langle 1,2,1 \rangle$ and $b_3 = \langle 1,1,2 \rangle$. 
The set
\[
\Delta(b_1,b_2,b_3) = \{ (g(b_1), g(b_2), g(b_3)) : g \in \Delta \} \subseteq \RR^3
\]
is the asymptotic spectrum of rectangular matrix multiplication: by \cite[Proposition~4.3]{strassen1988asymptotic}, every $g \in \Delta$ satisfies
\[
g(\langle n_1, n_2, n_3 \rangle) = g(b_1)^{\log n_1} g(b_2)^{\log n_2} g(b_3)^{\log n_3}
\]
for all positive integers $n_1, n_2, n_3$. Evaluating the flattening ranks at~$b_1$, $b_2$ and~$b_3$ gives $(\log \tensorrank_1(b_j))_{j \in [3]} = (1,1,0)$, $(\log \tensorrank_2(b_j))_{j \in [3]} = (0,1,1)$ and $(\log \tensorrank_3(b_j))_{j \in [3]} = (1,0,1)$. We let $\Gamma \coloneqq \conv\{(1,1,0), (0,1,1), (1,0,1)\} \subseteq \RR^3$.

By \cref{cor:tensor-paths}, every point of $\Delta(b_1,b_2,b_3)$ is joined by a path to $(\tensorrank_i(b_1), \tensorrank_i(b_2), \tensorrank_i(b_3))$, for every $i \in [3]$. In particular $\Delta(b_1,b_2,b_3)$ is path-connected. Strassen's theorem says that, after taking logarithm, these paths may be taken to be line segments. 

\begin{theorem}[{\cite[Corollary~6.6]{strassen1988asymptotic}}]\label{th:star-convex}
The set~$\Gamma$ is contained in $\log \Delta(b_1,b_2,b_3)$, and $\log \Delta(b_1,b_2,b_3)$ is star-convex with respect to every point of~$\Gamma$.
\end{theorem}
\begin{proof}
We first check that $b_j \in \mathcal{A}_i$ for all $i, j \in [3]$. Two legs of~$b_j$ are~$\CC^2$ and one is~$\CC$, and on the two legs~$\CC^2$, with the basis of \eqref{eq:mm-tensor}, the tensor~$b_j$ is $e_1 \otimes e_1 + e_2 \otimes e_2$. Let~$G_{2,2}$ act by its inclusion on the two legs~$\CC^2$ and trivially on the leg~$\CC$. This action fixes~$b_j$, because every signed permutation matrix~$g$ maps $e_1 \otimes e_1 + e_2 \otimes e_2$ under $g \otimes g$ to itself. The two-dimensional representations are irreducible by \cref{lem:monomial-irred}, and the one-dimensional one is irreducible, so $b_j \in \mathcal{A}_i$.

Let $f \in \Delta$ and $i \in [3]$. By \cref{lem:tensor-decomp} and \cref{lem:tensor-cover}, after relabelling the legs, \cref{th:interpolation} applies with $\mathcal{A} = \mathcal{A}_i$ and $f_1 = \tensorrank_i$. Since $b_1, b_2, b_3 \in \mathcal{A}_i$, its parts~(\ref{it:int-spectral}) and~(\ref{it:int-family}) give $F^{\smash[t]{f}}_t \in \Delta$ and $F^{\smash[t]{f}}_t(b_j) = f(b_j)^{1-t} \tensorrank_i(b_j)^t$ for every $t \in [0,1]$ and $j \in [3]$.
So the segment from $(\log f(b_j))_{j \in [3]}$ to $(\log \tensorrank_i(b_j))_{j \in [3]}$ lies in $\log \Delta(b_1,b_2,b_3)$. Taking such convex combinations successively shows that the convex hull of $(\log f(b_j))_{j \in [3]}$, $(1,1,0)$, $(0,1,1)$ and $(1,0,1)$ is contained in $\log \Delta(b_1,b_2,b_3)$. Taking $f = \tensorrank_1$ gives $\Gamma \subseteq \log \Delta(b_1,b_2,b_3)$. For every $v \in \Gamma$ the segment from $(\log f(b_j))_{j \in [3]}$ to~$v$ lies in $\log \Delta(b_1,b_2,b_3)$, which is the star-convexity.
\end{proof}

Unlike~$\Delta$ (\cref{th:tensor-contractible}), the sets $\Delta(s_1, \ldots, s_q)$ of \cref{cor:tensor-paths} need not be contractible. Every $g \in \Delta$ satisfies $1 \leq g(b_j) \leq 2$ and $g(b_1) g(b_2) = g(\langle 2,2,1 \rangle) \geq g(\langle 2 \rangle) = 2$. These bounds and \cref{th:star-convex} give $\Delta(b_1, b_2) = D$, where $D = \{(u,v) \in [1,2]^2 : uv \geq 2\}$. Let $s_1 = b_1 \oplus b_2$, $s_2 = b_1 \otimes b_2$ and $s_3 = b_1^{\otimes 2} \otimes b_2 \oplus b_2^{\otimes 2} \oplus b_1$. Then $\Delta(s_1, s_2, s_3)$ is the image of~$D$ under the map $(u,v) \mapsto (u+v, uv, u^2v + v^2 + u)$. Two distinct points of~$D$ with the same image are of the form $(u,v)$ and $(v,u)$ with $(u-1)(v-1) = 0$, so they are $(1,2)$ and $(2,1)$. Hence $\Delta(s_1, s_2, s_3)$ is a disc with two boundary points identified, which is homotopy equivalent to a circle.

\subsection{Log-star-convexity for polynomial multiplication}\label{sec:poly-complex}

We extend \cref{th:star-convex} from rectangular matrix multiplication to polynomial multiplication. For positive integers $m, n$, the \emph{polynomial multiplication tensor} is the $3$-tensor
\begin{equation}\label{eq:poly-tensor}
\polymult_{m,n} \coloneqq \sum_{u=1}^{m} \sum_{v=1}^{n} e_u \otimes e_v \otimes e_{u+v-1} \in \CC^m \otimes \CC^n \otimes \CC^{m+n-1},
\end{equation}
where the $e_\ell$ are standard basis vectors. It is concise, so its flattening ranks are $m$, $n$ and $m+n-1$. For $\pi \in \mathfrak{S}_3$ and a $3$-tensor~$t$, let~$\pi t$ be the tensor obtained from~$t$ by moving its $i$th leg to position~$\pi(i)$, for every $i \in [3]$. For all positive integers $a, b, c$, the tensors $\pi\polymult_{1,a}$ with $\pi \in \mathfrak{S}_3$ are, up to equivalence, the matrix multiplication tensors $\langle a,1,1 \rangle$, $\langle 1,a,1 \rangle$ and $\langle 1,1,a \rangle$, and $\langle a,b,c \rangle = \langle a,1,1 \rangle \otimes \langle 1,b,1 \rangle \otimes \langle 1,1,c \rangle$. So the semiring generated by the tensors~$\pi\polymult_{m,n}$ contains all matrix multiplication tensors.

\begin{theorem}\label{th:poly-star-convex}
Let $k = 3$, and let $b_1, \ldots, b_r$ be tensors of the form~$\pi\polymult_{m,n}$ with $\pi \in \mathfrak{S}_3$ and positive integers $m, n$. Then $\log \Delta(b_1, \ldots, b_r)$ is star-convex with respect to every point of
\[
\Gamma_r \coloneqq \conv\bigl\{ (\log \tensorrank_i(b_j))_{j \in [r]} : i \in [3] \bigr\}.
\]
\end{theorem}

We prove \cref{th:poly-star-convex} after \cref{lem:poly-irred}. The tensor~$\polymult_{m,n}$ is fixed by an action of the compact group~$\mathrm{SU}(2)$ that is irreducible on every leg (\cref{lem:poly-irred}). To use this action, we replace the finite groups in the definition of~$\mathcal{A}_i$ by compact groups.

An element of~$\mathcal{S}$ is \emph{compactly $i$-irreducible} if it has a concise representative $a \in V_1 \otimes \cdots \otimes V_k$ for which there are a compact group~$K$ and, for every $j \in [k]$, a continuous representation $\rho_j : K \to \GL(V_j)$, such that $(\rho_1(g) \otimes \cdots \otimes \rho_k(g))a = a$ for every $g \in K$, and such that~$\rho_i$ is irreducible. Let $\mathcal{A}^{\mathrm{c}}_i \subseteq \mathcal{S}$ be the set of compactly $i$-irreducible tensors. A finite group with the discrete topology is compact, so $\mathcal{A}_i \subseteq \mathcal{A}^{\mathrm{c}}_i$.

To extend the proofs of \cref{lem:spanning}, \cref{lem:restriction} and \cref{lem:tensor-cover} to compact groups, we need the following replacement for \cref{lem:dim-averaging}.

\begin{lemma}\label{lem:dim-averaging-compact}
Let a compact group~$K$ act continuously and irreducibly on a nonzero complex vector space~$V$, and let $U, W \subseteq V$ be subspaces. Then there is $g \in K$ with
\[
\dim(U \cap gW) \leq \frac{\dim W}{\dim V} \dim U.
\]
\end{lemma}
\begin{proof}
Let~$\mu$ be the Haar probability measure on~$K$. Averaging a Hermitian inner product on~$V$ over~$K$ gives a $K$-invariant one, so we may assume that~$K$ acts by unitary maps. For a subspace $X \subseteq V$ let~$\Pi_X$ be the orthogonal projection onto~$X$, so that $\Pi_{gW} = g \Pi_W g^{-1}$ for $g \in K$. The operator $M \coloneqq \int_K g \Pi_W g^{-1} \, d\mu(g)$ commutes with every element of~$K$, by invariance of~$\mu$. By Schur's lemma, $M$ is a scalar multiple of~$\Id_V$, and taking traces gives $M = (\dim W / \dim V) \Id_V$. The function $\varphi(g) \coloneqq \operatorname{tr}(\Pi_U \Pi_{gW})$ is continuous on~$K$ and
\[
\int_K \varphi \, d\mu = \operatorname{tr}(\Pi_U M) = \frac{\dim W}{\dim V} \dim U,
\]
so there is $g \in K$ with $\varphi(g) \leq (\dim W/\dim V) \dim U$. Finally, let $u_1, \ldots, u_{\dim U}$ be an orthonormal basis of~$U$ whose first $\dim(U \cap gW)$ vectors span $U \cap gW$. Then
\[
\varphi(g) = \sum_{j=1}^{\dim U} \langle u_j, \Pi_{gW} u_j \rangle = \sum_{j=1}^{\dim U} \lVert \Pi_{gW} u_j \rVert^2 \geq \dim(U \cap gW). \qedhere
\]
\end{proof}

\begin{lemma}\label{lem:compact-decomp}
Let $i \in [k]$. Then~$\mathcal{S}$ is $\mathcal{A}^{\mathrm{c}}_i$-decomposable.
\end{lemma}
\begin{proof}
We first show that~$\mathcal{A}^{\mathrm{c}}_i$ is multiplicatively closed. Let $a \in V_1 \otimes \cdots \otimes V_k$ and $b \in W_1 \otimes \cdots \otimes W_k$ be concise representatives of elements of~$\mathcal{A}^{\mathrm{c}}_i$, with compact groups~$K$ and~$H$ and representations $\rho_\ell : K \to \GL(V_\ell)$ and $\sigma_\ell : H \to \GL(W_\ell)$ for $\ell \in [k]$, as in the definition. The compact group $K \times H$ acts continuously by $\rho_\ell \otimes \sigma_\ell$ on the legs $V_\ell \otimes W_\ell$ of the concise tensor $a \otimes b$ and fixes it. By Burnside's theorem, a representation of a group on a nonzero complex vector space~$E$ is irreducible if and only if the linear span of its image is~$\End(E)$. So the span of the image of $\rho_i \otimes \sigma_i$ is $\End(V_i) \otimes \End(W_i) = \End(V_i \otimes W_i)$, hence $\rho_i \otimes \sigma_i$ is irreducible, and $a \otimes b \in \mathcal{A}^{\mathrm{c}}_i$.

By \cref{lem:tensor-decomp}, $\mathcal{S}$ is $\mathcal{A}_i$-decomposable. Since $\mathcal{A}_i \subseteq \mathcal{A}^{\mathrm{c}}_i$, every positive integer lies in~$\mathcal{A}^{\mathrm{c}}_i$, and every $\mathcal{A}_i$-decomposition is an $\mathcal{A}^{\mathrm{c}}_i$-decomposition. So the remaining properties of \cref{def:decomposable} hold for~$\mathcal{A}^{\mathrm{c}}_i$.
\end{proof}

\begin{lemma}\label{lem:compact-cover}
Let $i \in [k]$ and $f \in \Delta$. Then the pair $f, \tensorrank_i$ has the $\mathcal{A}^{\mathrm{c}}_i$-cover property with $\kappa_a(n) = 1$.
\end{lemma}
\begin{proof}
After relabelling the legs, let $i = 1$. The proof of \cref{lem:spanning} uses \cref{lem:dim-averaging} only to choose, for a subspace~$U_j$, a group element~$g_{j+1}$ such that $\dim(U_j \cap g_{j+1}W)$ is at most $(\dim W/\dim V) \dim U_j$. \Cref{lem:dim-averaging-compact} provides such an element for a compact group acting continuously and irreducibly on a complex vector space, so \cref{lem:spanning} holds for such actions. Transporting such an action to another concise representative conjugates it by invertible linear maps, which keeps it continuous. All Kronecker powers of an element of~$\mathcal{A}^{\mathrm{c}}_1$, including~$\langle 1 \rangle$, lie in~$\mathcal{A}^{\mathrm{c}}_1$ by \cref{lem:compact-decomp}. So the proofs of \cref{lem:restriction} and \cref{lem:tensor-cover} apply verbatim with~$\mathcal{A}^{\mathrm{c}}_1$ in place of~$\mathcal{A}_1$, and give the claim.
\end{proof}

For the rest of this subsection $k = 3$. For an integer $d \geq 0$ let~$\CC[x,y]_d$ be the space of homogeneous polynomials of degree~$d$ in the coordinate functions~$x$ and~$y$ of~$\CC^2$, with the ordered basis of monomials $x^d, x^{d-1}y, \ldots, y^d$. The group $\mathrm{SU}(2)$ acts continuously on~$\CC[x,y]_d$ by $(g \cdot h)(w) = h(g^{-1} w)$ for $w \in \CC^2$. We write~$\rho_d(g)$ for the matrix of this action of $g \in \mathrm{SU}(2)$ in the monomial basis.

\begin{lemma}\label{lem:poly-irred}\hfill
\begin{enumerate}[\upshape(i)]
\item\label{it:poly-irred-sym} For every $d \geq 0$, the representation~$\rho_d$ of~$\mathrm{SU}(2)$ is irreducible.
\item\label{it:poly-irred-tensor} For all positive integers $m, n$ and every $\pi \in \mathfrak{S}_3$, the tensor~$\pi\polymult_{m,n}$ lies in~$\mathcal{A}^{\mathrm{c}}_i$ for every $i \in [3]$.
\end{enumerate}
\end{lemma}
\begin{proof}
(\ref{it:poly-irred-sym}) Let $X \subseteq \CC[x,y]_d$ be a nonzero invariant subspace. Choose $z \in \CC$ with $|z| = 1$ that is not a root of unity. The matrix with diagonal entries~$z$ and~$z^{-1}$ acts on the monomials by the distinct scalars $z^{-d}, z^{2-d}, \ldots, z^d$, so~$X$ is spanned by the monomials it contains. The rotation $g \in \mathrm{SU}(2)$ by the angle~$\pi/4$ maps~$x$ and~$y$ to linear forms with both coefficients nonzero. So~$g$ maps every monomial to a polynomial with a nonzero coefficient at~$x^d$, which gives $x^d \in X$, and it maps~$x^d$ to a polynomial with all coefficients nonzero. Hence every monomial lies in~$X$, and $X = \CC[x,y]_d$.

(\ref{it:poly-irred-tensor}) Let $g \in \mathrm{SU}(2)$. Substitution commutes with multiplication of polynomials: $g \cdot (h_1 h_2) = (g \cdot h_1)(g \cdot h_2)$ for $h_1 \in \CC[x,y]_{m-1}$ and $h_2 \in \CC[x,y]_{n-1}$. The product of the $u$th monomial of~$\CC[x,y]_{m-1}$ and the $v$th monomial of~$\CC[x,y]_{n-1}$ is the $(u+v-1)$th monomial of~$\CC[x,y]_{m+n-2}$. So the coefficient vector of~$h_1 h_2$ is obtained by contracting~$\polymult_{m,n}$ on its first two legs with the coefficient vectors of~$h_1$ and~$h_2$. Comparing coefficients in $g \cdot (h_1 h_2) = (g \cdot h_1)(g \cdot h_2)$ for all~$h_1$ and~$h_2$ gives
\[
\bigl(\rho_{m-1}(g^{-1})^{\mathsf{T}} \otimes \rho_{n-1}(g^{-1})^{\mathsf{T}} \otimes \rho_{m+n-2}(g)\bigr) \polymult_{m,n} = \polymult_{m,n}.
\]
The maps $g \mapsto \rho_{m-1}(g^{-1})^{\mathsf{T}}$ and $g \mapsto \rho_{n-1}(g^{-1})^{\mathsf{T}}$ are the duals of~$\rho_{m-1}$ and~$\rho_{n-1}$, and the dual of an irreducible representation is irreducible. So by~(\ref{it:poly-irred-sym}) all three representations are continuous and irreducible. Since~$\polymult_{m,n}$ is concise, $\polymult_{m,n} \in \mathcal{A}^{\mathrm{c}}_i$ for every $i \in [3]$. Permuting the legs of~$\polymult_{m,n}$ permutes the three representations accordingly.
\end{proof}

\begin{proof}[Proof of \cref{th:poly-star-convex}]
Let $i \in [3]$. By \cref{lem:compact-decomp} and \cref{lem:compact-cover}, \cref{th:interpolation} applies with $\mathcal{A} = \mathcal{A}^{\mathrm{c}}_i$ and $f_1 = \tensorrank_i$. By \cref{lem:poly-irred}, the tensors $b_1, \ldots, b_r$ lie in~$\mathcal{A}^{\mathrm{c}}_i$. So for every $f \in \Delta$ and $t \in [0,1]$, parts~(\ref{it:int-spectral}) and~(\ref{it:int-family}) of \cref{th:interpolation} give $F^{\smash[t]{f}}_t \in \Delta$ with $F^{\smash[t]{f}}_t(b_j) = f(b_j)^{1-t} \tensorrank_i(b_j)^t$ for every $j \in [r]$. Hence the segment from $(\log f(b_j))_{j \in [r]}$ to $(\log \tensorrank_i(b_j))_{j \in [r]}$ lies in $\log \Delta(b_1, \ldots, b_r)$. That is, $\log \Delta(b_1, \ldots, b_r)$ is star-convex with respect to each of the points $(\log \tensorrank_i(b_j))_{j \in [r]}$ with $i \in [3]$. Since its kernel is convex, this proves the claim.
\end{proof}

The tensors $b_1 = \langle 2,1,1 \rangle$, $b_2 = \langle 1,2,1 \rangle$ and $b_3 = \langle 1,1,2 \rangle$ of \cref{sec:mm-complex} are of the form~$\pi\polymult_{1,2}$ with $\pi \in \mathfrak{S}_3$, so \cref{th:star-convex} is the case $r = 3$ of \cref{th:poly-star-convex} with these tensors.

\section{Log-star-convexity for matrix multiplication and graph tensors over any field}
\label{sec:star-convex-any-field}
In \cref{sec:mm-complex} we considered the asymptotic spectrum $\Delta(b_1,b_2,b_3)$ of rectangular matrix multiplication over~$\CC$, where $b_1 = \langle 2,1,1 \rangle$, $b_2 = \langle 1,2,1 \rangle$ and $b_3 = \langle 1,1,2 \rangle$. We proved that $\log \Delta(b_1,b_2,b_3)$ contains $\Gamma = \conv\{(1,1,0), (0,1,1), (1,0,1)\}$ and is star-convex with respect to every point of~$\Gamma$ (\cref{th:star-convex}). 

In this section we will prove a version of \cref{th:star-convex} over an arbitrary field, with any nonzero tensors $b_1, \ldots, b_r$ in place of $\langle 2,1,1 \rangle$, $\langle 1,2,1 \rangle$ and $\langle 1,1,2 \rangle$, under an ``irreducibility condition'' on all products of copies of $b_1, \ldots, b_r$.  In \cref{sec:absirred} we will give a sufficient ``absolute irreducibility condition'' on just the generators $b_1, \ldots, b_r$ themselves. 

We will then prove the absolute irreducibility condition for rectangular matrix multiplication in \cref{sec:mm-any-field}, which recovers Strassen's log-star-convexity theorem over every field \cite[Theorem~3.10 and Corollary~6.6]{strassen1988asymptotic}, and for graph tensors in \cref{sec:graph-tensors}, which recovers the log-star-convexity theorem of Wigderson and Zuiddam \cite[Theorem~12.23]{wigderson2022asymptotic}.

Throughout this section~$\FF$ is an arbitrary field, $k \geq 2$, and~$\mathcal{S}_\FF$ and~$\Delta_\FF$ are the semiring of $k$-tensors over~$\FF$ and its asymptotic spectrum, as in \cref{sec:tensors}. For nonzero $b_1, \ldots, b_r \in \mathcal{S}_\FF$ let
\[
\Delta_\FF(b_1, \ldots, b_r) = \{ (g(b_1), \ldots, g(b_r)) : g \in \Delta_\FF \} \subseteq \RR^r.
\] For $u \in \NN^r$ we write $b^u \coloneqq b_1^{\otimes u_1} \otimes \cdots \otimes b_r^{\otimes u_r}$. 

\subsection{Irreducibility condition for log-star-convexity}\label{sec:star-convex-statement}

As in \cref{sec:tensors} (but now over the arbitrary field~$\FF$), for $i \in [k]$, we say a $k$-tensor in~$\mathcal{S}_\FF$ is \emph{$i$-irreducible} if it has a concise representative $a \in V_1 \otimes \cdots \otimes V_k$ for which there are a finite group~$G$ and, for every $j \in [k]$, a representation $\rho_j : G \to \GL(V_j)$, such that $(\rho_1(g) \otimes \cdots \otimes \rho_k(g))a = a$ for every $g \in G$, and such that~$\rho_i$ is irreducible.

\begin{theorem}\label{th:star-convex-general}
Let $b_1, \ldots, b_r \in \mathcal{S}_\FF$ be nonzero and let $I \subseteq [k]$. Suppose that for every $i \in I$ and every $u \in \NN^r$, the tensor~$b^u$ is $i$-irreducible. Then $\log \Delta_\FF(b_1, \ldots, b_r)$ is star-convex with respect to every point of
\[
\conv\bigl\{ (\log \tensorrank_i(b_j))_{j \in [r]} : i \in I \bigr\}.
\]
\end{theorem}

Every nonzero tensor~$s$ with $\tensorrank_i(s) = 1$ is $i$-irreducible, with the trivial group, so the hypothesis of \cref{th:star-convex-general} concerns only the~$b^u$ with $\tensorrank_i(b^u) > 1$. We will prove \cref{th:star-convex-general} in \cref{sec:mm-proof}, by applying \cref{th:interpolation}.

We note that for proving \cref{th:star-convex-general}  we cannot follow directly the route we used for the complex version \cref{th:star-convex}, for two reasons.

The first reason lies in the proof of decomposability. The proof of \cref{lem:tensor-decomp} decomposes every~$V_i^{\otimes n}$ into irreducible $\mathfrak{S}_n$-subrepresentations, and over a field of positive characteristic such a decomposition need not exist. Instead we will decompose only the subsemiring $\NN[b_1, \ldots, b_r] \subseteq \mathcal{S}_\FF$ generated by $b_1, \ldots, b_r$, with the preorder induced from~$\mathcal{S}_\FF$. For each of its elements~$s$, the multinomial expansion of~$s^n$ decomposes it into terms $\langle m \rangle \otimes b^u$ with~$m \geq 1$ and $u \in \NN^r$. By an extension theorem of Strassen, \cref{prop:extension}, every spectral point of this subsemiring is the restriction of an element of~$\Delta_\FF$, so working in the subsemiring suffices.

The second reason lies in the proof of the $\mathcal{A}$-cover property. Over~$\CC$ it rests on reverse monotonicity, \cref{lem:restriction}, whose proof needs for every Kronecker power of a tensor a finite group acting irreducibly on the first leg. Over~$\CC$ the product of the groups for the factors does this, because a tensor product of irreducible complex representations is irreducible for the product group. Over~$\RR$, for example, this can fail. This is why the hypothesis of \cref{th:star-convex-general} concerns every product~$b^u$, and not only $b_1, \ldots, b_r$.

\subsection{\texorpdfstring{$\mathcal{A}$}{A}-decomposability}\label{sec:mm-decomp}

Let $b_1, \ldots, b_r \in \mathcal{S}_\FF$ be nonzero.
To apply \cref{th:interpolation}, we will need a subset $\mathcal{A}\subseteq \NN[b_1, \ldots, b_r]$ so that the semiring $\NN[b_1, \ldots, b_r]$ is $\mathcal{A}$-decomposable. Recall our notation $b^u = b_1^{\otimes u_1} \otimes \cdots \otimes b_r^{\otimes u_r}$ with $u \in \NN^r$. Let
\[
\mathcal{A} \coloneqq \{ \langle m \rangle \otimes b^{u} : m \geq 1,\ u \in \NN^r \}.
\]
Note that every~$b_j$ lies in~$\mathcal{A}$.

\begin{lemma}\label{lem:mm-decomp}
The semiring $\NN[b_1, \ldots, b_r]$ is $\mathcal{A}$-decomposable.
\end{lemma}
\begin{proof}
The set~$\mathcal{A}$ consists of nonzero tensors, since the~$b_j$ are nonzero. It is multiplicatively closed, since $(\langle m \rangle \otimes b^{u})\otimes(\langle \ell \rangle \otimes b^{v}) = \langle m\ell \rangle \otimes b^{u+v}$. It contains every positive integer. Every nonzero $s \in \NN[b_1, \ldots, b_r]$ satisfies $1 \leq s$, so property~(\ref{it:dec-finite}) of \cref{def:decomposable} follows from~(\ref{it:dec-nat}). Let $s \in \NN[b_1, \ldots, b_r] \setminus \{0\}$, say $s = \sum_{h=1}^q c_h b^{u_h}$ with $c_h \geq 1$ and $u_h \in \NN^r$. Expanding, we get
\[
s^n = \bigoplus_{\beta} \Bigl\langle \frac{n!}{\beta_1! \cdots \beta_q!}\, c_1^{\beta_1} \cdots c_q^{\beta_q} \Bigr\rangle \otimes b^{\beta_1 u_1 + \cdots + \beta_q u_q},
\]
where $\beta = (\beta_1, \ldots, \beta_q)$ runs over the $\binom{n+q-1}{q-1}$ elements of~$\NN^q$ with $\beta_1 + \cdots + \beta_q = n$. This is an equality in $\NN[b_1, \ldots, b_r]$ whose terms lie in~$\mathcal{A}$. Every term is at most~$s^n$, being one of its summands. So the terms, one for each~$\beta$, give an $\mathcal{A}$-decomposition of~$s^n$ of polynomial length in~$n$. This shows $\asympdeclength_{\mathcal{A}}(s) = 1$.
\end{proof}

\subsection{Reverse monotonicity}\label{sec:reverse-any-field}

The next lemma extends \cref{lem:restriction} to an arbitrary field. It assumes that all Kronecker powers of the tensor are $1$-irreducible, because its proof will apply \cref{lem:spanning} to them.

\begin{lemma}\label{lem:restriction-any-field}
Let $a \in V_1 \otimes \cdots \otimes V_k$ be concise, and suppose that~$a^{\otimes n}$ is $1$-irreducible for every $n \geq 1$. Let $\pi : V_1 \to W$ be a linear map and let $b \coloneqq (\pi \otimes \Id \otimes \cdots \otimes \Id)a$. Then for every $f \in \Delta_\FF$ we have
\[
\frac{\tensorrank_1(b)}{\tensorrank_1(a)} f(a) \leq f(b).
\]
\end{lemma}
\begin{proof}
Let $n \geq 1$. The tensor $a^{\otimes n} \in V_1^{\otimes n} \otimes \cdots \otimes V_k^{\otimes n}$ is concise, and by assumption it is equivalent to a concise tensor that is fixed by a finite group acting irreducibly on its first leg. Two equivalent concise tensors are related by invertible linear maps on the legs, so~$a^{\otimes n}$ itself is fixed by a finite group acting irreducibly on its first leg $V_1^{\otimes n}$. With these groups, the proof of \cref{lem:restriction} applies over~$\FF$, since it uses only \cref{lem:spanning}, which holds over every field.
\end{proof}

\subsection{\texorpdfstring{$\mathcal{A}$}{A}-cover property}\label{sec:cover-any-field}

In this subsection we will prove the second hypothesis of \cref{th:interpolation}, the $\mathcal{A}$-cover property, with~$\mathcal{A}$ as in \cref{sec:mm-decomp} and under the hypothesis of \cref{th:star-convex-general}. As in the proof of \cref{lem:tensor-cover}, we will derive it from reverse monotonicity, now in the form of \cref{lem:restriction-any-field}. Relabel the legs so that the leg under consideration is the first. To apply \cref{lem:restriction-any-field} to an element $\langle m \rangle \otimes b^v$ of~$\mathcal{A}$ with $\tensorrank_1(b^v) > 1$, we need, for this element and for each of its Kronecker powers, a finite group that fixes it and acts irreducibly on its first leg. These powers are elements of the same kind, since $(\langle m \rangle \otimes b^v)^{\otimes n} = \langle m^n \rangle \otimes b^{nv}$ and $\tensorrank_1(b^{nv}) = \tensorrank_1(b^v)^n$, and the hypothesis of \cref{th:star-convex-general} gives such a group for every~$b^v$ with $\tensorrank_1(b^v) > 1$. The element is a direct sum of~$m$ copies of~$b^v$, and the next lemma constructs such a group from one for~$b^v$. We will handle the elements with $\tensorrank_1(b^v) = 1$ directly.

\begin{lemma}\label{lem:wreath-irred}
Let $\rho : G \to \GL(V)$ be an irreducible representation of a finite group~$G$ with $\dim V > 1$, and let $m \geq 1$. Let the wreath product $G \wr \mathfrak{S}_m = G^{\times m} \rtimes \mathfrak{S}_m$, in which~$\mathfrak{S}_m$ permutes the factors of~$G^{\times m}$, act on $V^{\oplus m}$ as follows: for $(g_1, \ldots, g_m) \in G^{\times m}$, $\sigma \in \mathfrak{S}_m$ and $x_1, \ldots, x_m \in V$,
\begin{align*}
(g_1, \ldots, g_m) \cdot (x_1, \ldots, x_m) &\coloneqq (\rho(g_1) x_1, \ldots, \rho(g_m) x_m),\\
\sigma \cdot (x_1, \ldots, x_m) &\coloneqq (x_{\sigma^{-1}(1)}, \ldots, x_{\sigma^{-1}(m)}).
\end{align*}
Then this representation of $G \wr \mathfrak{S}_m$ is irreducible.
\end{lemma}
\begin{proof}
First, no nonzero vector of~$V$ is fixed by~$\rho(G)$, since its span would be a $\rho(G)$-invariant subspace of dimension $1 < \dim V$.

Now let $W \subseteq V^{\oplus m}$ be a nonzero invariant subspace, let $w = (w_1, \ldots, w_m) \in W$ be nonzero, and let $j \in [m]$ with $w_j \neq 0$. Some $g \in G$ satisfies $\rho(g) w_j \neq w_j$. Let $y = (1, \ldots, 1, g, 1, \ldots, 1) \cdot w \in W$, with~$g$ in position~$j$. Then $y - w$ is a nonzero vector of~$W$ that is zero outside the $j$th copy. So the vectors $x \in V$ for which~$W$ contains the vector with~$x$ in the $j$th copy and~$0$ elsewhere are a nonzero $\rho(G)$-invariant subspace of~$V$, which is~$V$ by irreducibility. Hence~$W$ contains the $j$th copy of~$V$, and by applying permutations, every copy. So $W = V^{\oplus m}$.
\end{proof}

\begin{lemma}\label{lem:mm-cover}
Let $i \in [k]$. Suppose that~$b^u$ is $i$-irreducible for every $u \in \NN^r$. Then for every $f_0 \in \Delta_\FF$ the pair $f_0, \tensorrank_i$ has the $\mathcal{A}$-cover property in~$\mathcal{S}_\FF$ with $\kappa_a(n) = 1$.
\end{lemma}
\begin{proof}
Relabel the legs so that $i = 1$. The proof has three steps. First, we will reduce the $\mathcal{A}$-cover property to one inequality for every restriction $a \leq s_1 \oplus s_2$ with $a \in \mathcal{A}$ nonzero. Second, we will apply a quotient map to the first leg of~$a$ that removes the contribution of~$s_2$, which gives a tensor~$c$ with $c \leq s_1$. Third, writing $a = \langle m \rangle \otimes b^v$, we will bound~$f_0(c)$ from below in terms of~$f_0(a)$, with \cref{lem:restriction-any-field} if $\tensorrank_1(b^v) > 1$, and directly if $\tensorrank_1(b^v) = 1$.

For the first step, we show that it suffices to prove that for every nonzero $a \in \mathcal{A}$ and all $s_1, s_2 \in \mathcal{S}_\FF$ with $a \leq s_1 \oplus s_2$, we have
\begin{equation}\label{eq:mm-cover-claim}
1 \leq \frac{f_0(s_1)}{f_0(a)} + \frac{\tensorrank_1(s_2)}{\tensorrank_1(a)}.
\end{equation}
This inequality gives the $\mathcal{A}$-cover property with $\kappa_a(n) = 1$. Indeed, for every nonzero $a \in \mathcal{A}$ and every $n \in \NN$, the element~$a^{\otimes n}$ is nonzero and lies in~$\mathcal{A}$, and $f_0(a^{\otimes n}) = f_0(a)^n$ and $\tensorrank_1(a^{\otimes n}) = \tensorrank_1(a)^n$. So \eqref{eq:mm-cover-claim} with~$a^{\otimes n}$ in place of~$a$ is the inequality of \cref{def:cover}.

For the second step, write $a = \langle m \rangle \otimes M$ with $m \geq 1$ and $M = b^v$ for some $v \in \NN^r$. Take a concise representative of~$M$, which we also denote by~$M$, and write~$V$ for its first leg. We represent~$a$ by the direct sum $a = M^{\oplus m} \in V_1 \otimes \cdots \otimes V_k$ of~$m$ copies of~$M$, which is concise. So $V_1 = V^{\oplus m}$, with one block for every copy of~$M$, and similarly for $V_2, \ldots, V_k$. Take concise representatives $s_1 \in X_1 \otimes \cdots \otimes X_k$ and $s_2 \in Y_1 \otimes \cdots \otimes Y_k$. Since $a \leq s_1 \oplus s_2$, there are linear maps $A_j : X_j \oplus Y_j \to V_j$ for $j \in [k]$ with $a = (A_1 \otimes \cdots \otimes A_k)(s_1 \oplus s_2)$. Let $U = A_1(Y_1)$ and $u = \dim U$, so that $u \leq \dim Y_1 = \tensorrank_1(s_2)$. Let $\pi : V_1 \to V_1/U$ be the quotient map and let $c = (\pi \otimes \Id \otimes \cdots \otimes \Id)a$. The map $\pi A_1$ vanishes on~$Y_1$, so~$c$ is the image of~$s_1$ under $\pi A_1|_{X_1} \otimes A_2|_{X_2} \otimes \cdots \otimes A_k|_{X_k}$, and hence $c \leq s_1$.

Let $d = \tensorrank_1(a) = \dim V_1$. If $\tensorrank_1(s_2) \geq d$, then the second term in \eqref{eq:mm-cover-claim} is at least~$1$. So let $\tensorrank_1(s_2) < d$, so that $u < d$. Since~$a$ is $1$-concise, $\tensorrank_1(c) = \dim(V_1/U) = d - u$. Since $u \leq \tensorrank_1(s_2)$, it suffices to prove
\begin{equation}\label{eq:mm-cover-goal}
f_0(s_1) \geq \Bigl(1 - \frac{u}{d}\Bigr) f_0(a).
\end{equation}

For the third step, suppose first that $\tensorrank_1(M) > 1$. By assumption, $M$ is equivalent to a concise tensor that is fixed by a finite group acting irreducibly on its first leg. Equivalent concise tensors are related by invertible linear maps on the legs, so~$M$ itself is fixed by a finite group~$G$ whose representation~$\rho$ on~$V$ is irreducible. The group $G \wr \mathfrak{S}_m$ fixes $a = M^{\oplus m}$, where $(g_1, \ldots, g_m) \in G^{\times m}$ acts on the $j$th copy of~$M$ by~$g_j$ and~$\mathfrak{S}_m$ permutes the copies. Its representation on $V_1 = V^{\oplus m}$ is irreducible by \cref{lem:wreath-irred}, since $\dim V = \tensorrank_1(M) > 1$. So~$a$ is $1$-irreducible. This argument uses only that $M = b^v$ for some $v \in \NN^r$ and that $\tensorrank_1(M) > 1$. For $n \geq 1$, the element~$a^{\otimes n}$ equals $\langle m^n \rangle \otimes M^{\otimes n}$, where $M^{\otimes n} = b^{nv}$ and $\tensorrank_1(M^{\otimes n}) = \tensorrank_1(M)^n > 1$. So~$a^{\otimes n}$ is $1$-irreducible for every $n \geq 1$, and \cref{lem:restriction-any-field} applies to~$a$ and~$\pi$. It gives $(\tensorrank_1(c)/\tensorrank_1(a))\, f_0(a) \leq f_0(c)$, that is, $(1 - u/d) f_0(a) \leq f_0(c)$. Since $c \leq s_1$, we have $f_0(c) \leq f_0(s_1)$, which proves \eqref{eq:mm-cover-goal}.

Suppose now that $\tensorrank_1(M) = 1$, so that $d = m$. Identify~$V$ with~$\FF$, so that $M = 1 \otimes N$ for a tensor~$N$ in the tensor product of the other legs of~$M$. Then $V_1 = \FF^m$ and $a = \sum_{j=1}^m e_j \otimes N_j$, where $e_1, \ldots, e_m$ is the standard basis of~$\FF^m$ and~$N_j$ is the copy of~$N$ in the tensor product of the $j$th blocks of $V_2, \ldots, V_k$. The images $\pi(e_1), \ldots, \pi(e_m)$ span $V_1/U$, which has dimension $m - u$, so there is $J \subseteq [m]$ with $|J| = m - u$ such that the~$\pi(e_j)$ with $j \in J$ are a basis of $V_1/U$. Mapping this basis to the standard basis of~$\FF^J$ on the first leg, and projecting $V_2, \ldots, V_k$ onto their blocks with index in~$J$, maps~$c$ to $\sum_{j \in J} e_j \otimes N_j$, which is equivalent to $\langle m-u \rangle \otimes M$. So $\langle m-u \rangle \otimes M \leq c \leq s_1$, and $f_0(s_1) \geq (m-u) f_0(M) = (1 - u/d) f_0(a)$, which is \eqref{eq:mm-cover-goal}.
\end{proof}

\subsection{Log-star-convexity}\label{sec:mm-proof}

In this subsection we will prove \cref{th:star-convex-general}. For a subsemiring~$\mathcal{T}$ of~$\mathcal{S}_\FF$, with the preorder induced from~$\mathcal{S}_\FF$, we write $\Delta(\mathcal{T})$ for the asymptotic spectrum of~$\mathcal{T}$.

The proof of \cref{th:star-convex-general} will use \cref{prop:extension} as follows. We will apply \cref{th:interpolation} to the subsemiring $\mathcal{T} = \NN[b_1, \ldots, b_r]$ of~$\mathcal{S}_\FF$, instead of to~$\mathcal{S}_\FF$ itself. The interpolants given by \cref{th:interpolation} lie in $\Delta(\mathcal{T})$, while \cref{th:star-convex-general} is a statement about~$\Delta_\FF$. By \cref{prop:extension}, every interpolant is the restriction of an element of~$\Delta_\FF$.

\begin{proof}[Proof of \cref{th:star-convex-general}]
Let $i \in I$ and $f_0 \in \Delta_\FF$. By \cref{lem:mm-cover}, the pair $f_0, \tensorrank_i$ has the $\mathcal{A}$-cover property in~$\mathcal{S}_\FF$. Taking the elements $s_1, s_2$ of \cref{def:cover} in $\NN[b_1, \ldots, b_r]$ shows that the restrictions of~$f_0$ and~$\tensorrank_i$ to $\NN[b_1, \ldots, b_r]$ have the $\mathcal{A}$-cover property in $\NN[b_1, \ldots, b_r]$. With \cref{lem:mm-decomp}, this gives the hypotheses of parts~(\ref{it:int-spectral}) and~(\ref{it:int-family}) of \cref{th:interpolation} on $\NN[b_1, \ldots, b_r]$, for the restriction of~$f_0$ and with~$f_1$ the restriction of~$\tensorrank_i$.

Let~$F_t$ be the interpolants obtained from the restriction of~$f_0$. Since $b_1, \ldots, b_r$ lie in~$\mathcal{A}$, \cref{th:interpolation}~(\ref{it:int-family}) gives $F_t(b_j) = f_0(b_j)^{1-t} \tensorrank_i(b_j)^t$ for every $j \in [r]$. Every~$F_t$ lies in $\Delta(\NN[b_1, \ldots, b_r])$, so by \cref{prop:extension} it is the restriction of an element of~$\Delta_\FF$, and therefore $(\log F_t(b_j))_{j \in [r]} \in \log \Delta_\FF(b_1, \ldots, b_r)$. So for every $f_0 \in \Delta_\FF$ and every $i \in I$, the segment from $(\log f_0(b_j))_{j \in [r]}$ to $(\log \tensorrank_i(b_j))_{j \in [r]}$ lies in $\log \Delta_\FF(b_1, \ldots, b_r)$. That is, $\log \Delta_\FF(b_1, \ldots, b_r)$ is star-convex with respect to each of the points $(\log \tensorrank_i(b_j))_{j \in [r]}$ with $i \in I$. Since its kernel is convex, this proves the claim.
\end{proof}

\subsection{Absolute irreducibility condition on generators for log-star-convexity}\label{sec:absirred}

In this subsection we will give an absolute irreducibility condition on the generators $b_1, \ldots, b_r$ that implies the irreducibility condition of \cref{th:star-convex-general}. Let $i \in [k]$. Asking that every~$b_j$ is $i$-irreducible is not enough for our argument. For $u \in \NN^r$, finite groups that fix $b_1, \ldots, b_r$ give a finite group that fixes~$b^u$, and its representation on the $i$th leg of~$b^u$ is the tensor product of the representations on the $i$th legs of the factors, and this tensor product can be reducible. For example, the representation of the cyclic group~$C_4$ on~$\RR^2$ in which a generator acts by the rotation by $\pi/2$ is irreducible, but the corresponding representation of $C_4 \times C_4$ on $\RR^2 \otimes \RR^2$ is not. So we use the following stronger notion.

A representation $\rho : G \to \GL(V)$ of a finite group~$G$ on an $\FF$-vector space~$V$ is \emph{absolutely irreducible} if $V \neq 0$ and the $\FF$-span of~$\rho(G)$ equals~$\End(V)$. For $i \in [k]$, a $k$-tensor in~$\mathcal{S}_\FF$ is \emph{absolutely $i$-irreducible} if it satisfies the definition of $i$-irreducibility with~$\rho_i$ absolutely irreducible. The \emph{dual} of a representation $\rho : G \to \GL(V)$ is the representation of~$G$ on~$V^*$ that maps~$g$ to $\rho(g^{-1})^{\mathsf{T}}$, and the \emph{outer tensor product} of~$\rho$ and a representation $\sigma : H \to \GL(W)$ is the representation of $G \times H$ on $V \otimes W$ that maps $(g,h)$ to $\rho(g) \otimes \sigma(h)$. We use the following three properties of absolute irreducibility.

\begin{lemma}\label{lem:absirred}
Let $\rho : G \to \GL(V)$ be an absolutely irreducible representation of a finite group~$G$. Then
\begin{enumerate}[\upshape(i)]
\item\label{it:absirred-irred} $\rho$ is irreducible.
\item\label{it:absirred-dual} The dual of~$\rho$ is absolutely irreducible.
\item\label{it:absirred-tensor} For every absolutely irreducible representation $\sigma : H \to \GL(W)$ of a finite group~$H$, the outer tensor product of~$\rho$ and~$\sigma$ is absolutely irreducible.
\end{enumerate}
\end{lemma}
\begin{proof}
(i) We have $V \neq 0$, and every $\rho(G)$-invariant subspace of~$V$ is invariant under the $\FF$-span of~$\rho(G)$, which equals~$\End(V)$, so it is~$0$ or~$V$.

(ii) Transposition is a linear bijection from~$\End(V)$ to~$\End(V^*)$, and $\rho(g^{-1})$ runs over~$\rho(G)$ as~$g$ runs over~$G$. So the $\FF$-span of the maps $\rho(g^{-1})^{\mathsf{T}}$ equals~$\End(V^*)$.

(iii) The $\FF$-span of the maps $\rho(g) \otimes \sigma(h)$ is $\End(V) \otimes \End(W) = \End(V \otimes W)$.
\end{proof}

\begin{corollary}\label{cor:absirred-generators}
Let $b_1, \ldots, b_r \in \mathcal{S}_\FF$ be nonzero and let $I \subseteq [k]$. Suppose that for every $i \in I$ and every $j \in [r]$, the tensor~$b_j$ is absolutely $i$-irreducible. Then $\log \Delta_\FF(b_1, \ldots, b_r)$ is star-convex with respect to every point of
\[
\conv\bigl\{ (\log \tensorrank_i(b_j))_{j \in [r]} : i \in I \bigr\}.
\]
\end{corollary}
\begin{proof}
Let $i \in I$ and let $u \in \NN^r$ be nonzero, which suffices since $b^0 = \langle 1 \rangle$ is $i$-irreducible with the trivial group. For every $j \in [r]$ take a concise representative of~$b_j$ and a finite group~$G_j$ that fixes it and whose representation on the $i$th leg is absolutely irreducible. The group $G_1^{\times u_1} \times \cdots \times G_r^{\times u_r}$, with one copy acting on each factor, fixes the Kronecker product of these representatives, which is a concise representative of~$b^u$. Its representation on the $i$th leg is an outer tensor product of absolutely irreducible representations, so it is absolutely irreducible by \cref{lem:absirred}~(\ref{it:absirred-tensor}), and hence irreducible by \cref{lem:absirred}~(\ref{it:absirred-irred}). So~$b^u$ is $i$-irreducible, and \cref{th:star-convex-general} applies.
\end{proof}

Over~$\CC$ every irreducible representation of a finite group is absolutely irreducible, by Burnside's theorem, so \cref{cor:absirred-generators} holds with ``$i$-irreducible'' in place of ``absolutely $i$-irreducible''.

\subsection{Rectangular matrix multiplication}\label{sec:mm-any-field}

In this subsection $k = 3$. Matrix multiplication tensors are defined over~$\FF$ by the same formula \eqref{eq:mm-tensor} as over~$\CC$, and we let $b_1 = \langle 2,1,1 \rangle$, $b_2 = \langle 1,2,1 \rangle$ and $b_3 = \langle 1,1,2 \rangle$ over~$\FF$. The set $\Delta_\FF(b_1,b_2,b_3)$ is the asymptotic spectrum of rectangular matrix multiplication over~$\FF$. As in \cref{sec:mm-complex}, let~$\Gamma \subseteq \RR^3$ be the convex hull of the points $(\log \tensorrank_i(b_j))_{j \in [3]}$ with $i \in [3]$, which are $(1,1,0)$, $(0,1,1)$ and $(1,0,1)$. Strassen proved the following.

\begin{theorem}[{\cite[Theorem~3.10 and Corollary~6.6]{strassen1988asymptotic}}]\label{th:star-convex-any-field}
Let~$\FF$ be an arbitrary field. Then the set $\log \Delta_\FF(b_1,b_2,b_3)$ is star-convex with respect to every point of~$\Gamma$.
\end{theorem}

We will deduce \cref{th:star-convex-any-field} from \cref{cor:absirred-generators}, by constructing finite groups that fix matrix multiplication tensors and act absolutely irreducibly on their legs. For this we will use finite groups of matrices that span all matrices.

\begin{lemma}\label{lem:spanning-group}
For every field~$\FF$ and every $n \geq 1$, there is a finite subgroup~$H$ of $\GL(\FF^n)$ whose $\FF$-span is $\End(\FF^n)$.
\end{lemma}
\begin{proof}
We identify $\End(\FF^n)$ with the $n \times n$ matrices over~$\FF$ via the standard basis, and write~$E_{ij}$ for the matrix with entry~$1$ in position $(i,j)$ and~$0$ elsewhere and~$I$ for the identity matrix. We give two constructions. The first works if the characteristic of~$\FF$ is different from~$2$, and the second works if it is different from~$0$, so together they cover every field.

If~$\FF$ has characteristic different from~$2$, let~$H$ be the group of signed permutation matrices. For $i \in [n]$ let~$D_i$ be the diagonal matrix with $-1$ in position~$i$ and~$1$ elsewhere, so that $E_{ii} = (I - D_i)/2$. For $j \in [n]$ let~$P$ be a permutation matrix with $P_{ij} = 1$, so that $E_{ij} = E_{ii} P = (P - D_i P)/2$. Each of~$I$, $D_i$, $P$ and $D_i P$ is a signed permutation matrix, so the $\FF$-span of~$H$ contains every~$E_{ij}$.

If~$\FF$ has characteristic $p > 0$, then~$\FF$ contains the prime field~$\FF_p$, and we let~$H$ be the finite group of invertible $n \times n$ matrices with entries in~$\FF_p$. The $\FF_p$-span of~$H$ contains~$I$ and $I + E_{ij}$, and hence~$E_{ij}$, for all $i \neq j$. The span of a group of matrices is closed under multiplication, so for $n \geq 2$ it also contains $E_{ii} = E_{ij} E_{ji}$. For $n = 1$ we have $E_{11} = I$. So the $\FF_p$-span of~$H$ consists of all $n \times n$ matrices with entries in~$\FF_p$, and hence its $\FF$-span is $\End(\FF^n)$.
\end{proof}

\begin{lemma}\label{lem:mm-absirred}
For every field~$\FF$ and $n_1,n_2,n_3 \geq1$, there are a finite group~$G$ and representations $\rho_1 : G \to \GL(\FF^{n_1 n_2})$, $\rho_2 : G \to \GL(\FF^{n_2 n_3})$ and $\rho_3 : G \to \GL(\FF^{n_3 n_1})$ such that $(\rho_1(g) \otimes \rho_2(g) \otimes \rho_3(g)) \langle n_1,n_2,n_3 \rangle = \langle n_1,n_2,n_3 \rangle$ for every $g \in G$, and such that~$\rho_i$ is absolutely irreducible for every $i \in [3]$.
\end{lemma}
\begin{proof}
Identify the three legs~$\FF^{n_1 n_2}$, $\FF^{n_2 n_3}$ and~$\FF^{n_3 n_1}$ of $\langle n_1, n_2, n_3 \rangle$ with the spaces of $n_1 \times n_2$, $n_2 \times n_3$ and $n_3 \times n_1$ matrices over~$\FF$, and for every $j \in [3]$ let~$G_j$ be a finite subgroup of $\GL(\FF^{n_j})$ whose $\FF$-span is $\End(\FF^{n_j})$, which exists by \cref{lem:spanning-group}. So the representation of~$G_j$ on~$\FF^{n_j}$ by matrix multiplication is absolutely irreducible. Let $G = G_1 \times G_2 \times G_3$, and let $(g_1, g_2, g_3) \in G$ act on the three legs by $X \mapsto g_1 X g_2^{-1}$, $Y \mapsto g_2 Y g_3^{-1}$ and $Z \mapsto g_3 Z g_1^{-1}$. This action fixes $\langle n_1,n_2,n_3\rangle$. Indeed, for $j \in [3]$ let $z_j = \sum_{x=1}^{n_j} e_x \otimes e_x \in \FF^{n_j} \otimes \FF^{n_j}$. Regrouping the six factors of the three legs by the summation indices of \eqref{eq:mm-tensor} writes $\langle n_1,n_2,n_3\rangle$ as $z_1 \otimes z_2 \otimes z_3$. The element $(g_1, g_2, g_3)$ acts on~$z_j$ by $g_j \otimes (g_j^{-1})^{\mathsf{T}}$, which, writing $u \otimes v$ as the matrix $uv^{\mathsf{T}}$, maps~$X$ to $g_j X g_j^{\smash[t]{-1}}$ and so fixes the identity matrix~$z_j$.

The first leg is $\FF^{n_1} \otimes (\FF^{n_2})^*$ as a representation of $G_1 \times G_2$, and~$G_3$ acts trivially on it. So~$\rho_1$ is absolutely irreducible by \cref{lem:absirred}~(\ref{it:absirred-dual}) and (\ref{it:absirred-tensor}). The other legs are treated in the same way.
\end{proof}

\begin{proof}[Proof of \cref{th:star-convex-any-field}]
The matrix multiplication tensors are concise, so by \cref{lem:mm-absirred} the tensors~$b_1$, $b_2$ and~$b_3$ are absolutely $i$-irreducible for every $i \in [3]$. So \cref{cor:absirred-generators} with $I = [3]$ gives the claim.
\end{proof}

\subsection{Graph tensors}\label{sec:graph-tensors}

In this subsection we will apply \cref{cor:absirred-generators} to graph tensors \cite{MR3924627}, also called tensor networks \cite[Definition~12.1]{wigderson2022asymptotic}, which include the matrix multiplication tensors.
\begin{definition}\label{def:graph-tensor}
Let~$\mathsf{H}$ be a simple graph with vertex set $[k]$ and edge set~$\mathsf{E}$, and let $n \in \NN^\mathsf{E}$ with $n_{\mathsf{e}} \geq 1$ for every $\mathsf{e} \in \mathsf{E}$. For $\mathsf{e} \in \mathsf{E}$ let $z_{\mathsf{e}} = \sum_{x=1}^{n_{\mathsf{e}}} e_x \otimes e_x \in \FF^{n_{\mathsf{e}}} \otimes \FF^{n_{\mathsf{e}}}$. For $i \in [k]$ let $V_i = \bigotimes_{\mathsf{e} \ni i} \FF^{n_{\mathsf{e}}}$, with the factors indexed by the edges at~$i$, so that $V_i = \FF$ if no edge contains~$i$. Collecting the two factors of each edge gives an isomorphism from $V_1 \otimes \cdots \otimes V_k$ to $\bigotimes_{\mathsf{e} \in \mathsf{E}} (\FF^{n_{\mathsf{e}}} \otimes \FF^{n_{\mathsf{e}}})$. The \emph{graph tensor} $\langle \mathsf{H}; n \rangle$ is the $k$-tensor in $V_1 \otimes \cdots \otimes V_k$ that corresponds to $\bigotimes_{\mathsf{e} \in \mathsf{E}} z_{\mathsf{e}}$ under this isomorphism.
\end{definition}
The graph tensor $\langle \mathsf{H}; n \rangle$ has one leg for every vertex of~$\mathsf{H}$, and every edge $\mathsf{e} = \{i,j\}$ places~$z_{\mathsf{e}}$ on the factors~$\FF^{n_{\mathsf{e}}}$ of~$V_i$ and~$V_j$.

If $k = 3$ and~$\mathsf{H}$ is the $3$-cycle, with edges $\{3,1\}$, $\{1,2\}$ and $\{2,3\}$, and~$n$ takes the values~$n_1$, $n_2$ and~$n_3$ on these edges, then $\langle \mathsf{H}; n \rangle$ is equivalent to $\langle n_1, n_2, n_3 \rangle$.

For $\mathsf{e} \in \mathsf{E}$ let $b_{\mathsf{e}} = \langle \mathsf{H}; n \rangle$, where~$n$ takes the value~$2$ on~$\mathsf{e}$ and~$1$ on every other edge. For $w \in [k]$ let $\chi_w \in \{0,1\}^\mathsf{E}$ be the indicator vector of the set of edges containing~$w$. Then $\log \tensorrank_w(b_{\mathsf{e}}) = \chi_w(\mathsf{e})$ for every $\mathsf{e} \in \mathsf{E}$, and we write $\Delta_\FF((b_{\mathsf{e}})_{\mathsf{e} \in \mathsf{E}}) = \{ (g(b_{\mathsf{e}}))_{\mathsf{e} \in \mathsf{E}} : g \in \Delta_\FF \} \subseteq \RR^\mathsf{E}$.

\begin{theorem}[{\cite[Theorem~12.23 and Lemma~8.8]{wigderson2022asymptotic}}]\label{th:graph-tensors}
Let~$\FF$ be a field, let $k \geq 2$, and let~$\mathsf{H}$ be a graph with vertex set $[k]$ and edge set~$\mathsf{E}$. Then $\log \Delta_\FF((b_{\mathsf{e}})_{\mathsf{e} \in \mathsf{E}})$ is star-convex with respect to every point of $\conv\{\chi_w : w \in [k]\}$.
\end{theorem}

Wigderson and Zuiddam proved \cref{th:graph-tensors} for the points of the convex hull of the~$\chi_w$ with~$w$ contained in an edge. The vertices in no edge follow by applying their theorem to the complete graph on $[k]$, whose generators~$b_{\mathsf{e}}$ with $\mathsf{e} \in \mathsf{E}$ are the same tensors, and forgetting the coordinates outside~$\mathsf{E}$.

Let $k = 3$ and let~$\mathsf{H}$ be the $3$-cycle. Then $b_{\{3,1\}}$, $b_{\{1,2\}}$ and $b_{\{2,3\}}$ are equivalent to the tensors $b_1 = \langle 2,1,1 \rangle$, $b_2 = \langle 1,2,1 \rangle$ and $b_3 = \langle 1,1,2 \rangle$ of \cref{sec:mm-any-field}. With the edges in this order, $\chi_1 = (1,1,0)$, $\chi_2 = (0,1,1)$ and $\chi_3 = (1,0,1)$, so $\conv\{\chi_1, \chi_2, \chi_3\} = \Gamma$, and \cref{th:star-convex-any-field} is the special case of \cref{th:graph-tensors} for the $3$-cycle.

We will deduce \cref{th:graph-tensors} from \cref{cor:absirred-generators}, using finite groups that fix graph tensors and act absolutely irreducibly on their legs.

\begin{lemma}\label{lem:graph-absirred}
Let~$\mathsf{H}$ be a graph with vertex set $[k]$ and edge set~$\mathsf{E}$, and let $n \in \NN^\mathsf{E}$ with $n_{\mathsf{e}} \geq 1$ for every $\mathsf{e} \in \mathsf{E}$. Then there are a finite group~$G$ and representations $\rho_i : G \to \GL(V_i)$ for $i \in [k]$ such that $(\rho_1(g) \otimes \cdots \otimes \rho_k(g)) \langle \mathsf{H}; n \rangle = \langle \mathsf{H}; n \rangle$ for every $g \in G$, and such that~$\rho_i$ is absolutely irreducible for every $i \in [k]$.
\end{lemma}
\begin{proof}
Orient every edge of~$\mathsf{H}$. By \cref{lem:spanning-group}, for every $\mathsf{e} \in \mathsf{E}$ there is a finite subgroup $G_{\mathsf{e}} \subseteq \GL(\FF^{n_{\mathsf{e}}})$ whose $\FF$-span is $\End(\FF^{n_{\mathsf{e}}})$. Let $G = \prod_{\mathsf{e} \in \mathsf{E}} G_{\mathsf{e}}$, and let $g \in G$ act on the factor~$\FF^{n_{\mathsf{e}}}$ of~$V_i$ by~$g_{\mathsf{e}}$ if~$\mathsf{e}$ starts at~$i$, and by $(g_{\mathsf{e}}^{-1})^{\mathsf{T}}$ if~$\mathsf{e}$ ends at~$i$. Under the identification of $\FF^{n_{\mathsf{e}}} \otimes \FF^{n_{\mathsf{e}}}$ with the $n_{\mathsf{e}} \times n_{\mathsf{e}}$ matrices, $z_{\mathsf{e}}$ is the identity matrix and $g_{\mathsf{e}} \otimes (g_{\mathsf{e}}^{-1})^{\mathsf{T}}$ acts by $X \mapsto g_{\mathsf{e}} X g_{\mathsf{e}}^{-1}$, so~$G$ fixes $\langle \mathsf{H}; n \rangle$. The representation of~$G_{\mathsf{e}}$ on~$\FF^{n_{\mathsf{e}}}$ by matrix multiplication is absolutely irreducible, and hence so is the one that maps~$g$ to $(g^{-1})^{\mathsf{T}}$, by \cref{lem:absirred}~(\ref{it:absirred-dual}). So~$\rho_i$ is an outer tensor product of absolutely irreducible representations of the factors~$G_{\mathsf{e}}$ with $\mathsf{e} \ni i$, and hence absolutely irreducible by \cref{lem:absirred}~(\ref{it:absirred-tensor}). If no edge contains~$i$, then~$\rho_i$ is the trivial representation on $V_i = \FF$, which is absolutely irreducible.
\end{proof}

\begin{proof}[Proof of \cref{th:graph-tensors}]
Every graph tensor is concise: its flattening at~$w \in [k]$ is the tensor product of the full-rank flattenings of the~$z_{\mathsf{e}}$ with $\mathsf{e} \ni w$ and the nonzero tensor $\bigotimes_{\mathsf{e} \not\ni w} z_{\mathsf{e}}$, so it has rank $\prod_{\mathsf{e} \ni w} n_{\mathsf{e}} = \dim V_w$. So by \cref{lem:graph-absirred}, every~$b_{\mathsf{e}}$ is absolutely $w$-irreducible for every $w \in [k]$. Since $(\log \tensorrank_w(b_{\mathsf{e}}))_{\mathsf{e} \in \mathsf{E}} = \chi_w$ for every $w \in [k]$, \cref{cor:absirred-generators}, with the generators~$b_{\mathsf{e}}$ for $\mathsf{e} \in \mathsf{E}$ and with $I = [k]$, gives the claim.
\end{proof}

\Cref{th:graph-tensors} extends to the following hypergraph tensors.

\begin{definition}\label{def:hypergraph-tensor}
A \emph{hypergraph}~$\mathsf{H}$ with vertex set~$[k]$ is a finite family~$\mathsf{E}$ of subsets of~$[k]$ with at least two elements each, called its \emph{edges}; the same subset may occur as several edges. For a subset $\mathsf{e} \subseteq [k]$ with at least two elements and an integer $n \geq 1$, let~$\langle n \rangle_{\mathsf{e}}$ be the $k$-tensor $\sum_{x=1}^n v_{1,x} \otimes \cdots \otimes v_{k,x}$ with $v_{i,x} = e_x \in \FF^n$ for $i \in \mathsf{e}$ and $v_{i,x} = 1 \in \FF$ for $i \notin \mathsf{e}$. For $n \in \NN^{\smash{\mathsf{E}}}$ with positive entries, the \emph{hypergraph tensor} $\langle \mathsf{H}; n \rangle$ is the Kronecker product of the tensors~$\langle n_{\mathsf{e}} \rangle_{\mathsf{e}}$ with $\mathsf{e} \in \mathsf{E}$.
\end{definition}

Thus~$\langle n \rangle_{\mathsf{e}}$ is the unit tensor~$\langle n \rangle$ on the legs in~$\mathsf{e}$, and its other legs are one-dimensional.

\begin{corollary}\label{cor:hypergraph-tensors-star}
Let~$\FF$ be a field, let $k \geq 2$, and let~$\mathsf{E}$ be a nonempty set of subsets of~$[k]$ with at least two elements each. For $\mathsf{e} \in \mathsf{E}$ let $b_{\mathsf{e}} = \langle 2 \rangle_{\mathsf{e}}$, and for $w \in [k]$ let $\chi_w \in \{0,1\}^{\mathsf{E}}$ be the indicator vector of the set of $\mathsf{e} \in \mathsf{E}$ that contain~$w$. Then $\log \Delta_\FF((b_{\mathsf{e}})_{\mathsf{e} \in \mathsf{E}})$ is star-convex with respect to every point of $\conv\{\chi_w : w \in [k]\}$.
\end{corollary}
\begin{proof}
By Strassen \cite[Theorem~3.10]{strassen1988asymptotic}, extending~$\FF$ does not change the joint asymptotic spectrum~$\Delta_\FF((b_{\mathsf{e}})_{\mathsf{e} \in \mathsf{E}})$. So we may assume that~$\FF$ contains a root of unity $\zeta \neq 1$: we take $\zeta = -1$ if the characteristic of~$\FF$ is not~$2$, and otherwise we adjoin a primitive third root of unity.

Let $\mathsf{e} \in \mathsf{E}$, let $D = \mathrm{diag}(\zeta, 1)$, and let~$P$ be the permutation matrix that swaps the two standard basis vectors of~$\FF^2$. Let $G \subseteq \GL(\FF^2)^{\mathsf{e}}$ be the group generated by the tuple with~$P$ in every entry and by the tuples with~$D$ in one entry, $D^{-1}$ in another entry and~$I$ elsewhere. Every entry of an element of~$G$ is a permutation matrix times a diagonal matrix whose diagonal entries are powers of~$\zeta$, so~$G$ is finite. Let $g \in G$ act on the leg $j \in \mathsf{e}$ of~$b_{\mathsf{e}}$ by its $j$th entry and trivially on the other legs. The tensor~$b_{\mathsf{e}}$ is the sum of two product tensors, whose factors on the legs in~$\mathsf{e}$ are all~$e_1$, respectively all~$e_2$. Every generator fixes~$b_{\mathsf{e}}$: the first swaps the two summands, and for the others the factors~$\zeta$ and~$\zeta^{-1}$ on the first summand cancel.

Let $j \in \mathsf{e}$. Since~$\mathsf{e}$ has at least two elements, the $j$th entries of the elements of~$G$ include $I$, $D$, $P$ and $DP$. Since $\zeta \neq 1$, these four matrices span $\End(\FF^2)$, so the representation of~$G$ on the leg~$j$ is absolutely irreducible. The other legs are one-dimensional, and the trivial representation on~$\FF$ is absolutely irreducible. The tensor~$b_{\mathsf{e}}$ is concise, and $\log \tensorrank_w(b_{\mathsf{e}}) = \chi_w(\mathsf{e})$ for every $w \in [k]$. So~$b_{\mathsf{e}}$ is absolutely $w$-irreducible for every $w \in [k]$, and \cref{cor:absirred-generators}, with the generators~$b_{\mathsf{e}}$ for $\mathsf{e} \in \mathsf{E}$ and with $I = [k]$, gives the claim.
\end{proof}

\section{Singleton upper quantum functionals}\label{sec:qf}
In this section we will show that repeated interpolation (\cref{th:interpolation}) between the flattening ranks $\tensorrank_1, \ldots, \tensorrank_k$ of complex tensors produces the singleton quantum functionals of Christandl, Vrana and Zuiddam \cite{MR4495838}. This gives a new proof that these functionals are spectral points \cite[Corollary~3.31]{MR4495838}.

Here $\mathcal{S} = \mathcal{S}_\CC$ is the semiring of complex $k$-tensors of \cref{sec:tensors}, with asymptotic spectrum $\Delta = \Delta_\CC$, and the sets $\mathcal{A}_1, \ldots, \mathcal{A}_k$ and~$\mathcal{B}$, and the flattening ranks $\tensorrank_1, \ldots, \tensorrank_k$, are as defined there. We work with~$\mathcal{B}$ because it lies in every~$\mathcal{A}_i$: when we interpolate towards $\tensorrank_1, \ldots, \tensorrank_k$ in turn, \cref{th:interpolation}~(\ref{it:int-family}) will give the interpolant at every step as a geometric average on~$\mathcal{B}$. Moreover~$\mathcal{S}$ is $\mathcal{B}$-decomposable by \cref{lem:tensor-decomp}, so every spectral point is determined by its values on~$\mathcal{B}$ (\cref{th:w-good-spec}).

\subsection{\texorpdfstring{$\mathcal{B}$}{B}-good geometric averages of a spectral point and the flattening ranks}\label{sec:qf-averages}
In this subsection we will show that for every geometric average~$w$ of a spectral point $f \in \Delta$ and the flattening ranks $\tensorrank_1, \ldots, \tensorrank_k$, the function
\[
F_w(s) = \sup \{ w(a)^{1/n} : n \geq 1,\ a \in \mathcal{B},\ a \leq s^n \}
\]
of \cref{def:Fw}, with~$\mathcal{B}$ in place of~$\mathcal{A}$, is the unique spectral point that agrees with~$w$ on~$\mathcal{B}$. The proof will first construct a spectral point that agrees with~$w$ on~$\mathcal{B}$, by repeated interpolation.

\begin{proposition}\label{prop:qf}
Let $f \in \Delta$. Let $\theta \in \mathcal{P}(\{0, 1, \ldots, k\})$. Let 
\[
w(s) \coloneqq f(s)^{\theta_0} \prod_{i=1}^k \tensorrank_i(s)^{\theta_i}
\]
for $s \in \mathcal{S}$. Then~$w$ is $\mathcal{B}$-good, and~$F_w$ is the unique element of~$\Delta$ agreeing with~$w$ on~$\mathcal{B}$.
\end{proposition}
\begin{proof}
We first construct $g \in \Delta$ such that $g = w$ on~$\mathcal{B}$, by induction on the number of $i \in [k]$ for which $\theta_i > 0$.

If $\theta_i = 0$ for every $i \in [k]$ then $\theta_0 = 1$ and $w = f$, so we take $g = f$. If $\theta_i = 1$ for some $i \in [k]$ then $w = \tensorrank_i$, so we take $g = \tensorrank_i$.

Otherwise choose $i \in [k]$ with $0 < \theta_i < 1$, let $\eta_i = 0$ and $\eta_j = \theta_j/(1 - \theta_i)$ for $j \in \{0, 1, \ldots, k\} \setminus \{i\}$, and let $v(s) = f(s)^{\eta_0} \prod_{j=1}^{\smash[t]{k}} \tensorrank_j(s)^{\eta_j}$ for nonzero $s \in \mathcal{S}$ and $v(0) = 0$, so that $w = v^{\smash{1-\theta_i}} \tensorrank_i^{\smash[t]{\theta_i}}$. The~$\eta_j$ are nonnegative and sum to~$1$, and fewer of $\eta_1, \ldots, \eta_k$ are positive, so the induction hypothesis gives $h \in \Delta$ with $h = v$ on~$\mathcal{B}$. By \cref{lem:tensor-decomp} and \cref{lem:tensor-cover}, after relabelling the legs, the hypotheses of \cref{th:interpolation} hold with~$\mathcal{A}_i$ and $f_1 = \tensorrank_i$. Apply it with $f_0 = h$, and let $g = F_t$ for $t = \theta_i$. Then $g \in \Delta$. For $a \in \mathcal{B}$ we have $a \in \mathcal{A}_i$, so \cref{th:interpolation}~(\ref{it:int-family}) gives $g(a) = h(a)^{1-\theta_i} \tensorrank_i(a)^{\theta_i} = v(a)^{1-\theta_i} \tensorrank_i(a)^{\theta_i} = w(a)$.

Next we show that~$w$ is $\mathcal{B}$-good. It is multiplicative because~$f$ and the~$\tensorrank_i$ are, and $w(m) = m^{\theta_0 + \theta_1 + \cdots + \theta_k} = m$ for every $m \in \NN$. If $a, a_1, \ldots, a_q \in \mathcal{B}$ and $a \leq a_1 + \cdots + a_q$, then $w(a) = g(a) \leq \sum_{j=1}^q g(a_j) = \sum_{j=1}^q w(a_j)$, because~$g$ is monotone and additive and agrees with~$w$ on~$\mathcal{B}$. Since~$\mathcal{S}$ is $\mathcal{B}$-decomposable by \cref{lem:tensor-decomp}, \cref{th:w-good-spec}~(\ref{it:Fw-unique}) now gives the claim.
\end{proof}

For $\theta \in \mathcal{P}([k])$ we now let~$w_\theta$ be the $\theta$-weighted geometric average of $\tensorrank_1, \ldots, \tensorrank_k$, so that we have $w_\theta(s) = \prod_{i=1}^k \tensorrank_i(s)^{\theta_i}$ for $s \in \mathcal{S}$.

\begin{corollary}\label{cor:qf}
For every $\theta \in \mathcal{P}([k])$, the function~$F_{w_\theta}$ is the unique element of~$\Delta$ agreeing with~$w_\theta$ on~$\mathcal{B}$.
\end{corollary}
\begin{proof}
Take $\theta_0 = 0$ in \cref{prop:qf}.
\end{proof}

The spectral points of \cref{cor:qf} include the flattening ranks and give continuous paths between them: If~$\theta$ is the $i$th standard basis vector, then $w_\theta = \tensorrank_i$, so $F_{w_\theta} = \tensorrank_i$. Now let $i, j \in [k]$ be distinct, and for $t \in [0,1]$ let $\theta(t) \in \mathcal{P}([k])$ have $i$th entry $1 - t$, $j$th entry~$t$ and all other entries~$0$. \Cref{th:interpolation} with~$\mathcal{A}_j$, with $f_1 = \tensorrank_j$ and with $f_0 = \tensorrank_i$ gives spectral points~$F_t$ with $F_t(a) = \tensorrank_i(a)^{1-t} \tensorrank_j(a)^t = w_{\theta(t)}(a)$ for every $a \in \mathcal{B}$, since $\mathcal{B} \subseteq \mathcal{A}_j$. So $F_t = F_{w_{\theta(t)}}$ by \cref{cor:qf}, and by \cref{th:interpolation}~(\ref{it:int-cont}) the map $t \mapsto F_{w_{\theta(t)}}$ is a continuous path in~$\Delta$ from~$\tensorrank_i$ to~$\tensorrank_j$.

\begin{corollary}\label{cor:qf-star-convex}
Give~$\RR^{\mathcal{B}}$ the product topology, and let
\[
\Delta(\mathcal{B}) \coloneqq \{ (f(b))_{b \in \mathcal{B}} : f \in \Delta \} \subseteq \RR^{\mathcal{B}}.
\]
Then $f \mapsto (f(b))_{b \in \mathcal{B}}$ is a homeomorphism from~$\Delta$ onto $\Delta(\mathcal{B})$, and $\log \Delta(\mathcal{B})$ is star-convex with respect to every point of $\conv\{ (\log \tensorrank_i(b))_{b \in \mathcal{B}} : i \in [k] \}$.
\end{corollary}
\begin{proof}
As in the proof of \cref{cor:convex}, the map $f \mapsto (f(b))_{b \in \mathcal{B}}$ is a homeomorphism onto $\Delta(\mathcal{B})$, since~$\mathcal{S}$ is $\mathcal{B}$-decomposable by \cref{lem:tensor-decomp}. Let $f \in \Delta$, $\varphi \in \mathcal{P}([k])$ and $t \in [0,1]$, and let $\theta \in \mathcal{P}(\{0, 1, \ldots, k\})$ have $\theta_0 = 1 - t$ and $\theta_i = t \varphi_i$ for $i \in [k]$. By \cref{prop:qf}, the spectral point~$F_w$ agrees with~$w$ on~$\mathcal{B}$, so every $b \in \mathcal{B}$ satisfies
\[
\log F_w(b) = (1-t) \log f(b) + t \sum_{i=1}^k \varphi_i \log \tensorrank_i(b).
\]
So $\log \Delta(\mathcal{B})$ contains the segment from $(\log f(b))_{b \in \mathcal{B}}$ to $\sum_{i=1}^k \varphi_i\, (\log \tensorrank_i(b))_{b \in \mathcal{B}}$.
\end{proof}

\subsection{Identification with the singleton quantum functionals}\label{sec:qf-identify}

In this subsection we will identify the~$F_{w_\theta}$ with the singleton quantum functionals of \cite{MR4495838}. 

The singleton quantum functionals are defined as follows. For $n \geq 1$ we identify the space $(V_1 \otimes \cdots \otimes V_k)^{\otimes n}$ with $V_1^{\otimes n} \otimes \cdots \otimes V_k^{\otimes n}$ by regrouping the factors, so that the $n$th tensor power of~$t$ becomes its Kronecker power~$t^{\otimes n}$. A \emph{partition} of~$n$ is a sequence $\lambda = (\lambda_1, \ldots, \lambda_\ell)$ of positive integers, its \emph{parts}, with $\lambda_1 \geq \cdots \geq \lambda_\ell$ and $\lambda_1 + \cdots + \lambda_\ell = n$, and we write $\lambda \vdash n$. We set $\lambda_i = 0$ for $i > \ell$. For $\lambda \vdash n$ let $[\lambda]$ be the irreducible complex representation of~$\mathfrak{S}_n$ labelled by~$\lambda$. For $i \in [k]$ and $\lambda \vdash n$, let~$P^{V_i}_\lambda$ be the projection of $V_1^{\otimes n} \otimes \cdots \otimes V_k^{\otimes n}$ onto the isotypic component of $[\lambda]$ along the other isotypic components, for the action of~$\mathfrak{S}_n$ on $V_i^{\otimes n}$ by permuting the factors. For $\theta \in \mathcal{P}([k])$ and nonzero $t \in V_1 \otimes \cdots \otimes V_k$, let
\[
E^\theta(t) \coloneqq \sup \Bigl\{ \sum_{i=1}^k \theta_i H(\lambda^{(i)}/n) : n \geq 1,\ \lambda^{(1)}, \ldots, \lambda^{(k)} \vdash n,\ \Bigl(\prod_{i=1}^k P^{V_i}_{\lambda^{(i)}}\Bigr) t^{\otimes n} \neq 0 \Bigr\},
\]
where $\lambda/n$ is the probability distribution $(\lambda_1/n, \lambda_2/n, \ldots)$ on the parts of a partition $\lambda \vdash n$, and $H(p) = -\sum_{x \in X} p(x) \log p(x)$ is the Shannon entropy of a probability distribution~$p$ on a finite set~$X$, with $0 \log 0 = 0$. Let $F^\theta(t) = 2^{\smash{E^\theta(t)}}$ and $F^\theta(0) = 0$. For $k \geq 3$, in the terminology of \cite{MR4495838}, $F^\theta$ is the upper quantum functional \cite[Definition~3.14]{MR4495838} for the probability distribution that assigns weight~$\theta_i$ to the bipartition $\{\{i\}, [k] \setminus \{i\}\}$ for each $i \in [k]$. 

\begin{theorem}\label{th:cvz}
For every $\theta \in \mathcal{P}([k])$ we have $F_{w_\theta} = F^\theta$. In particular, $F^\theta \in \Delta$.
\end{theorem}

\Cref{th:cvz} gives a new proof that $F^\theta \in \Delta$, which is part of \cite[Corollary~3.31]{MR4495838}. We prove \cref{th:cvz} at the end of this subsection, without using that~$F^\theta$ is a spectral point. The proof will combine the decomposition of \cref{lem:qf-explicit} and the bounds of \cref{th:w-good-spec}~(\ref{it:Fw-limit}) with the definition of~$F^\theta$, Schur--Weyl duality in the following form, and the two lemmas after it. For a partition~$\lambda$ and $N \geq 1$ we write $N\lambda$ for the partition with parts $N\lambda_j$.

\begin{theorem}[{\cite[Theorem~6.3 and Exercise~6.30]{MR1153249}}]\label{th:schur-weyl}
Let~$V$ be a nonzero finite-dimensional complex vector space and let $n \geq 1$. For every $\lambda \vdash n$ there is a representation~$\mathbb{S}_\lambda(V)$ of~$\GL(V)$, which is zero if~$\lambda$ has more than $\dim V$ parts and irreducible otherwise, such that the nonzero~$\mathbb{S}_\lambda(V)$ are pairwise non-isomorphic and
\[
V^{\otimes n} \cong \bigoplus_{\lambda \vdash n} \mathbb{S}_\lambda(V) \otimes [\lambda]
\]
as representations of $\GL(V) \times \mathfrak{S}_n$, with~$\mathfrak{S}_n$ permuting the factors. In particular, the projection of~$V^{\otimes n}$ onto the isotypic component of~$[\lambda]$ for~$\mathfrak{S}_n$ is nonzero if and only if~$\lambda$ has at most $\dim V$ parts, in which case it is the projection onto the isotypic component of~$\mathbb{S}_\lambda(V)$ for~$\GL(V)$, and the multiplicity of~$[\lambda]$ in~$V^{\otimes n}$ is~$\dim \mathbb{S}_\lambda(V)$.
\end{theorem}

\begin{lemma}\label{lem:scaling}
Let $t \in V_1 \otimes \cdots \otimes V_k$, let $n \geq 1$ and let $\lambda^{\smash{(1)}}, \ldots, \lambda^{\smash{(k)}} \vdash n$ with $\bigl(\prod_{i=1}^k P^{V_i}_{\lambda^{\smash{(i)}}}\bigr) t^{\otimes n} \neq 0$. Then $\bigl(\prod_{i=1}^{\smash[t]{k}} P^{\smash[t]{V_i}}_{N\lambda^{\smash{(i)}}}\bigr) t^{\otimes Nn} \neq 0$ for every $N \geq 1$.
\end{lemma}
\begin{proof}
Let $V = V_1 \otimes \cdots \otimes V_k$ and $G = \GL(V_1) \times \cdots \times \GL(V_k)$. By \cref{th:schur-weyl}, $P^{\smash[t]{V_i}}_{\lambda}$ is also the projection onto the isotypic component of~$\mathbb{S}_\lambda(V_i)$, and $\prod_{i=1}^{\smash[t]{k}} P^{\smash[t]{V_i}}_{\lambda^{\smash{(i)}}}$ is the projection of $V^{\otimes n}$ onto the isotypic component of the irreducible $G$-representation $W_\lambda = \mathbb{S}_{\lambda^{\smash{(1)}}}(V_1) \otimes \cdots \otimes \mathbb{S}_{\lambda^{\smash{(k)}}}(V_k)$.

Let~$G$ act on the space $\CC[V]_n$ of homogeneous polynomials of degree~$n$ on~$V$ by $(g f)(x) = f(g^{-1} x)$. The pairing $\langle f, x^{\otimes n} \rangle = f(x)$ gives a $G$-equivariant isomorphism $\CC[V]_n \cong \Sym^n(V)^*$. It pairs the $W_\lambda$-isotypic component of~$\Sym^n(V)$ with the $W_\lambda^*$-isotypic component of $\CC[V]_n$, and it is zero between isotypic components whose types are not dual to each other. The vector $t^{\otimes n}$ lies in~$\Sym^n(V)$, and by the hypothesis $\bigl(\prod_{i=1}^k P^{V_i}_{\lambda^{\smash{(i)}}}\bigr) t^{\otimes n} \neq 0$ its component in the $W_\lambda$-isotypic component is nonzero. So the pairing gives an~$f$ in the $W_\lambda^*$-isotypic component of $\CC[V]_n$ with $f(t) \neq 0$.

Choose a basis of every~$V_i$, and take highest weight vectors with respect to the subgroup of~$G$ of tuples of matrices that are upper triangular in these bases. The $W_\lambda^*$-isotypic component is spanned by the $G$-translates of its highest weight vectors, so there are a highest weight vector~$h$ in it and $g \in G$ with $h(g^{-1} t) \neq 0$. The weight of~$h$ is the highest weight~$\mu$ of $W_\lambda^*$, and~$N\mu$ is the highest weight of $W_{N\lambda}^*$, since the highest weight of the dual of $\mathbb{S}_\lambda(\CC^d)$ is $(-\lambda_d, \ldots, -\lambda_1)$. The polynomial~$h^N$ is a highest weight vector of weight~$N\mu$ in $\CC[V]_{Nn}$, and $h^N(g^{-1} t) = h(g^{-1} t)^N \neq 0$. So $g h^N$ lies in the $W_{N\lambda}^*$-isotypic component of $\CC[V]_{Nn}$ and $(g h^N)(t) \neq 0$. By the same pairing in degree $Nn$, $\bigl(\prod_{i=1}^k P^{V_i}_{N\lambda^{\smash{(i)}}}\bigr) t^{\otimes Nn} \neq 0$.
\end{proof}

For integers $d, n \geq 1$ let $q_d(n) = (n + d)^{d(d+1)/2}$. We will use the following bounds of Hayashi and Harrow.

\begin{lemma}[{\cite[Lemma~10, (B1)]{hayashi2002exponentsquantumfixedlengthpure} and \cite[Section~6.2, (6.16)]{harrow2005applicationscoherentclassicalcommunication}}]\label{lem:dim-irrep}
Let $d, n \geq 1$ and let $\lambda \vdash n$ have at most~$d$ parts. Then
\[
q_d(n)^{-1}\, 2^{nH(\lambda/n)} \leq \dim [\lambda] \leq 2^{nH(\lambda/n)}.
\]
\end{lemma}
\begin{proof}
Let $\ell_i = \lambda_i + d - i$ for $i \in [d]$. Since~$\lambda$ has at most~$d$ parts, the Frobenius formula \cite[Section~4.1, (4.11)]{MR1153249} gives
\[
\dim[\lambda] = \frac{n!}{\ell_1! \cdots \ell_d!} \prod_{1 \leq i < j \leq d} (\ell_i - \ell_j).
\]
For $i < j$ the factor $\ell_i - \ell_j = \lambda_i - \lambda_j + j - i$ is an integer with $1 \leq \ell_i - \ell_j \leq \lambda_i + j - i$. Multiplying these bounds over $j = i + 1, \ldots, d$ gives, for every $i \in [d]$,
\[
1 \leq \prod_{j=i+1}^{d} (\ell_i - \ell_j) \leq (\lambda_i + 1) \cdots (\lambda_i + d - i) = \frac{\ell_i!}{\lambda_i!} \leq (n+d)^{d-i}.
\]
With the multinomial coefficient $C = n!/(\lambda_1! \cdots \lambda_d!)$, the second inequality gives $\dim[\lambda] \leq C$, and the first and the last give $\dim[\lambda] \geq n!/(\ell_1! \cdots \ell_d!) \geq (n+d)^{-d(d-1)/2}\, C$. Moreover, $C$ is the number of sequences in~$[d]^n$ in which every $i \in [d]$ occurs~$\lambda_i$ times, so $(n+1)^{-d}\, 2^{nH(\lambda/n)} \leq C \leq 2^{nH(\lambda/n)}$ by Csisz\'ar and K\"orner \cite[Lemma~2.3]{MR2839250}. The lemma follows, since $n + 1 \leq n + d$ and $d(d-1)/2 + d = d(d+1)/2$.
\end{proof}

\begin{proof}[Proof of \cref{th:cvz}]
Both sides vanish at~$0$. Let $t \in V_1 \otimes \cdots \otimes V_k$ be nonzero. The proof has three steps. First, we write~$F_{w_\theta}(t)$ as the limit of~$L_n^{\smash[t]{1/n}}$, where~$L_n$ is the largest value of~$w_\theta$ on the tensors of the decomposition of \cref{lem:qf-explicit}. Second, we write~$L_n$ as a maximum over tuples of partitions. Third, we compare this maximum with~$E^\theta(t)$, using \cref{lem:dim-irrep} and \cref{lem:scaling}.

For the first step, for $n \geq 1$ let the~$U^{\smash{(i)}}_{\!j}$, the~$p^{\smash{(i)}}_{\!j}$ and the $a_{j_1, \ldots, j_k}$ be as in \cref{lem:qf-explicit}. Let
\[
L_n \coloneqq \max\bigl\{ w_\theta(a_{j_1, \ldots, j_k}) : a_{j_1, \ldots, j_k} \neq 0 \bigr\}.
\]
By \cref{lem:qf-explicit}, the nonzero $a_{j_1, \ldots, j_k}$ give a $\mathcal{B}$-decomposition of~$t^{\otimes n}$ of length~$c_n$, where~$c_n$ is at most a polynomial in~$n$. So \cref{th:w-good-spec}~(\ref{it:Fw-limit}) gives $L_n \leq F_{w_\theta}(t)^n \leq c_n L_n$, and hence $F_{w_\theta}(t) = \lim_{n \to \infty} L_n^{\smash[t]{1/n}}$.

For the second step, note that every~$U^{\smash{(i)}}_{\!\smash[b]{j}}$ is isomorphic to $[\lambda]$ for some $\lambda \vdash n$, and~$P^{V_i}_\lambda$ is the sum of the~$p^{\smash{(i)}}_{\!j}$ with $U^{\smash{(i)}}_{\!j} \cong [\lambda]$. So for $\lambda^{\smash{(1)}}, \ldots, \lambda^{\smash{(k)}} \vdash n$, the vector $\bigl(\prod_{i=1}^{\smash[t]{k}} P^{\smash[t]{V_i}}_{\lambda^{\smash{(i)}}}\bigr) t^{\otimes n}$ is the sum of the $a_{j_1, \ldots, j_k}$ with $U^{\smash{(i)}}_{\!j_i} \cong [\lambda^{\smash{(i)}}]$ for every $i \in [k]$. These lie in the subspaces $U^{\smash{(1)}}_{\!j_1} \otimes \cdots \otimes U^{\smash{(k)}}_{\!j_k}$, whose sum is direct, so the vector is nonzero if and only if one of them is. A nonzero $a_{j_1, \ldots, j_k}$ is concise in $U^{\smash{(1)}}_{\!j_1} \otimes \cdots \otimes U^{\smash{(k)}}_{\!j_k}$, so $w_\theta(a_{j_1, \ldots, j_k}) = \prod_{i=1}^{\smash[t]{k}} (\dim [\lambda^{\smash{(i)}}])^{\smash{\theta_i}}$. Hence
\[
L_n = \max \Bigl\{ \prod_{i=1}^k (\dim [\lambda^{(i)}])^{\theta_i} : \lambda^{(1)}, \ldots, \lambda^{(k)} \vdash n,\ \Bigl(\prod_{i=1}^k P^{V_i}_{\lambda^{(i)}}\Bigr) t^{\otimes n} \neq 0 \Bigr\}.
\]

For the third step, let $d = \max_i \dim V_i$. Every partition~$\lambda^{\smash{(i)}}$ in the maximum defining~$L_n$ has at most~$d$ parts, since $P^{V_i}_\lambda = 0$ if~$\lambda$ has more than $\dim V_i$ parts. So \cref{lem:dim-irrep} applies to these partitions.

The upper bound in \cref{lem:dim-irrep} gives $L_n \leq 2^{n E^\theta(t)}$, so $F_{w_\theta}(t) \leq F^\theta(t)$. For the reverse inequality, let $\lambda^{\smash{(1)}}, \ldots, \lambda^{\smash{(k)}} \vdash n$ with $\bigl(\prod_{i=1}^{\smash[t]{k}} P^{V_i}_{\lambda^{\smash{(i)}}}\bigr) t^{\otimes n} \neq 0$. By \cref{lem:scaling}, for every $N \geq 1$ the partitions $N\lambda^{\smash{(1)}}, \ldots, N\lambda^{\smash{(k)}} \vdash Nn$ satisfy the condition in the maximum defining~$L_{Nn}$. Since $\sum_i \theta_i = 1$, the lower bound in \cref{lem:dim-irrep} gives
\[
L_{Nn} \geq q_d(Nn)^{-1}\, 2^{Nn \sum_i \theta_i H(\lambda^{(i)}/n)}.
\]
Taking $Nn$th roots and letting $N \to \infty$, the polynomial factor tends to~$1$, and the first step gives $F_{w_\theta}(t) \geq 2^{\sum_i \theta_i H(\lambda^{\smash{(i)}}/n)}$. Taking the supremum over~$n$ and the~$\lambda^{\smash{(i)}}$ gives $F_{w_\theta}(t) \geq F^\theta(t)$. Finally, $F_{w_\theta} \in \Delta$ by \cref{cor:qf}.
\end{proof}

As noted after \cref{cor:interval}, an interpolation can give a better upper bound on asymptotic subrank than its end points. For example, let $k = 3$ and let $W = e_1 \otimes e_1 \otimes e_2 + e_1 \otimes e_2 \otimes e_1 + e_2 \otimes e_1 \otimes e_1$, so that $\tensorrank_i(W) = 2$ for every $i \in [3]$. The tensor~$W$ is equivalent to the multiplication tensor of~$\CC[T]/(T^2)$. Strassen \cite[Theorem~6.7]{strassen1991degeneration} determined the interval for this tensor. With the duality of \cref{sec:asymptotic-rank}, his theorem gives
\[
\Delta(W) = [\asympsubrank(W), 2] = [3 \cdot 2^{-2/3}, 2].
\]
Christandl, Vrana and Zuiddam \cite[Example~4.8]{MR4495838} showed that $F^\theta(W) = 3 \cdot 2^{-2/3} \approx 1.89$ for $\theta = (1/3, 1/3, 1/3)$. So by \cref{th:cvz}, the interpolation~$F_{w_\theta}$ between the flattening ranks has value~$\asympsubrank(W)$ at~$W$, below the value~$2$ of every flattening rank.

\section{Non-crossing upper quantum functionals}\label{sec:upper}

In \cref{sec:qf} we obtained the singleton upper quantum functionals by repeated interpolation between the flattening ranks of the single legs. Christandl, Vrana and Zuiddam defined upper quantum functionals more generally for weights on the bipartitions of the legs \cite[Definition~3.14]{MR4495838}, and the singleton ones are those with weights on the bipartitions that separate one leg from the others. 

In this section we will prove that the non-crossing upper quantum functionals, those whose weights are supported on a non-crossing (to be defined in a moment) family of bipartitions, are spectral points (\cref{th:upper-spectral}). Previously, this was proven for a subsemiring of all $k$-tensors, which is strict for $k \geq 4$, by work of Botero, Christandl, Fraser, Leigh and Nieuwboer~\cite[Proposition~4.2 and Theorem~3.1]{botero2026quantumfunctionalshigherordertensors} (namely the subsemiring of tensors at which the upper quantum functionals agree with the so-called lower quantum functionals \cite[Definition~3.3]{MR4495838} for all weights supported on non-crossing families). 

As in \cref{sec:qf}, we will construct spectral points by repeated interpolation, now between the flattening ranks for any non-crossing bipartition family, and then identify them with the non-crossing upper quantum functionals. We will define the non-crossing upper quantum functionals and state \cref{th:upper-spectral} in \cref{sec:upper-def}, and prove it in \cref{sec:upper-identify}.

Here $\mathcal{S} = \mathcal{S}_\CC$ is again the semiring of complex $k$-tensors, with asymptotic spectrum $\Delta = \Delta_\CC$, and~$\mathcal{B}$ the set of jointly irreducible tensors, as in \cref{sec:tensors}.

\subsection{Non-crossing families of bipartitions}\label{sec:upper-noncrossing}

In this subsection we will define non-crossing families of bipartitions and show how they correspond to trees.

A \emph{bipartition} of $[k]$ is a set $\{X_1, X_2\}$ of nonempty disjoint subsets of $[k]$ with $X_1 \cup X_2 = [k]$, and $\bipart([k])$ denotes the set of bipartitions of~$[k]$. The bipartitions $\{\{i\}, [k] \setminus \{i\}\}$ with $i \in [k]$ we call \emph{singleton bipartitions}. We say that $\{X_1, X_2\}, \{Y_1, Y_2\} \in \bipart([k])$ are \emph{non-crossing} if $X_i \subseteq Y_j$ for some $i, j \in \{1, 2\}$, and we say they are  \emph{crossing} otherwise. A family $B \subseteq \bipart([k])$ is \emph{non-crossing} (also called \emph{laminar}) if it is pairwise non-crossing.

A bipartition is determined by either of its parts: if~$S$ is a part of $b \in \bipart([k])$, then $b = \{S, [k] \setminus S\}$. It will be useful to choose the parts of all bipartitions in a uniform way: for every $b \in \bipart([k])$ we let~$S_b$ be the part of~$b$ not containing~$k$. If $B \subseteq \bipart([k])$ is non-crossing, then any two of the sets~$S_b$ with $b \in B$ are nested or disjoint, by the definition of non-crossing and since no~$S_b$ contains~$k$. So, ordered by inclusion, these sets and~$[k]$ form a tree with root~$[k]$. Such a correspondence between bipartitions and trees appears in phylogenetics \cite[Lemmas~3 and~5]{buneman1971recovery}, \cite[Theorem~3.1.4, Proposition~3.5.4 and Theorem~3.5.2]{MR2060009}.

We give three examples of non-crossing $B\subseteq \bipart([k])$, for $k=3,5,6$ respectively, and the corresponding tree of parts $S_b$ for $b \in B$ under inclusion with $[k]$ at the root: \begin{center}
\begin{tabular}{c@{\hspace{6em}}c@{\hspace{6em}}c}
\addlinespace[6pt]
\begin{tabular}[t]{@{}c@{}} $\{\{1\},\{2,3\}\}$ \\ $\{\{2\},\{1,3\}\}$ \\ $\{\{1,2\},\{3\}\}$ \end{tabular}
& \begin{tabular}[t]{@{}c@{}} $\{\{1,2\},\{3,4,5\}\}$ \\ $\{\{3,4\},\{1,2,5\}\}$ \\ $\{\{3\},\{1,2,4,5\}\}$ \end{tabular}
& \begin{tabular}[t]{@{}c@{}} $\{\{1,2\},\{3,4,5,6\}\}$ \\ $\{\{3\},\{1,2,4,5,6\}\}$ \\ $\{\{4,5\},\{1,2,3,6\}\}$ \\ $\{\{4\},\{1,2,3,5,6\}\}$ \\ $\{\{5\},\{1,2,3,4,6\}\}$ \end{tabular} \\
\addlinespace[10pt]
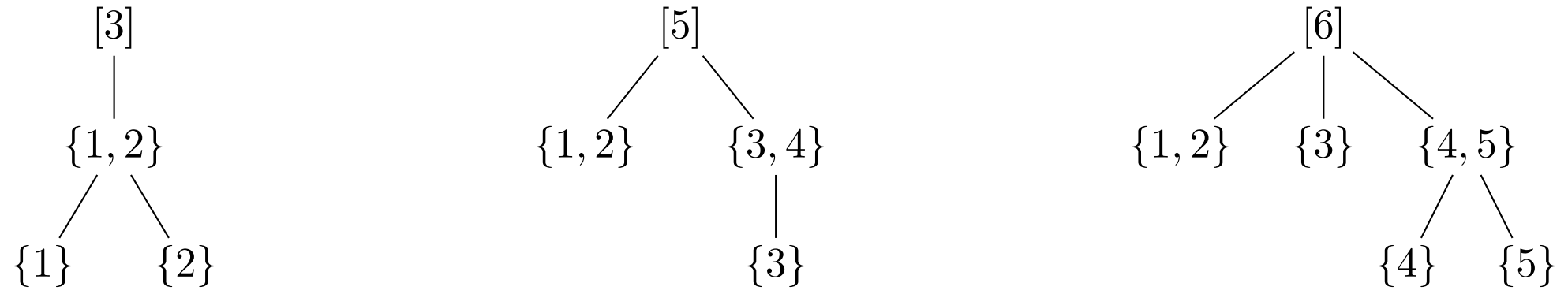
\begin{tikzpicture}[baseline=(r.base), y=1.25cm, s/.style={inner sep=1.5pt, font=\small}]
\node[s] (r) at (0.6,1.6) {$[3]$};
\node[s] (c) at (0.6,0.8) {$\{1,2\}$};
\node[s] (a) at (0,0) {$\{1\}$};
\node[s] (b) at (1.2,0) {$\{2\}$};
\draw (r) -- (c);
\draw (a) -- (c) -- (b);
\end{tikzpicture}
& \begin{tikzpicture}[baseline=(r.base), y=1.25cm, s/.style={inner sep=1.5pt, font=\small}]
\node[s] (r) at (0.8,1.6) {$[5]$};
\node[s] (a) at (0,0.8) {$\{1,2\}$};
\node[s] (b) at (1.6,0.8) {$\{3,4\}$};
\node[s] (c) at (1.6,0) {$\{3\}$};
\draw (a) -- (r) -- (b) -- (c);
\end{tikzpicture}
& \begin{tikzpicture}[baseline=(r.base), y=1.25cm, s/.style={inner sep=1.5pt, font=\small}]
\node[s] (r) at (1.2,1.6) {$[6]$};
\node[s] (a) at (0,0.8) {$\{1,2\}$};
\node[s] (b) at (1.2,0.8) {$\{3\}$};
\node[s] (c) at (2.4,0.8) {$\{4,5\}$};
\node[s] (d) at (1.9,0) {$\{4\}$};
\node[s] (e) at (2.9,0) {$\{5\}$};
\draw (a) -- (r) -- (b);
\draw (r) -- (c) -- (d);
\draw (c) -- (e);
\end{tikzpicture}
\end{tabular}
\end{center}

\subsection{Bipartition flattening ranks; geometric averages need not be \texorpdfstring{$\mathcal{B}$}{B}-good}\label{sec:upper-main}

In this subsection we will explain why, unlike for the singleton quantum functionals in \cref{sec:qf}, we cannot obtain the spectral points of this section directly from \cref{th:w-good-spec}: weighted geometric averages of bipartition flattening ranks need not be $\mathcal{B}$-good.

For $b = \{S, [k] \setminus S\} \in \bipart([k])$, let $\tensorrank_b(t)$ be the rank of the flattening of~$t$ that groups the legs in~$S$ into one leg and the legs in $[k] \setminus S$ into another. Then $\tensorrank_b \in \Delta$, and $\tensorrank_b = \tensorrank_i$ for the singleton bipartition $b = \{\{i\}, [k] \setminus \{i\}\}$. 

For a non-crossing family $B \subseteq \bipart([k])$ and $\theta \in \mathcal{P}(B)$, let $w_\theta = \prod_{b \in B} \tensorrank_b^{\smash[t]{\theta(b)}}$.
In \cref{sec:qf} we obtained the singleton quantum functionals as the spectral points~$F_{w_\theta}$ of \cref{th:w-good-spec}, using that, for the family of singleton bipartitions, $w_\theta = \prod_{i \in [k]} \tensorrank_i^{\smash[t]{\theta_i}}$ is $\mathcal{B}$-good (\cref{cor:qf} and \cref{th:cvz}). 
For $k \leq 4$ this extends to every non-crossing family.

\begin{proposition}\label{prop:upper-small-k}
Let $k \leq 4$, let $B \subseteq \bipart([k])$ be non-crossing, let $\theta \in \mathcal{P}(B)$, and let $w_\theta = \prod_{b \in B} \tensorrank_b^{\smash[t]{\theta(b)}}$. Then~$w_\theta$ is $\mathcal{B}$-good.
\end{proposition}
\begin{proof}
For $k \leq 4$, any two bipartitions without a singleton part are crossing, so at most one $b \in B$ has no singleton part. Hence \cref{prop:qf} applies, with $f = \tensorrank_b$ for this~$b$, or with $\theta_0 = 0$ if there is none.
\end{proof}

For $k \leq 4$ we can therefore apply \cref{th:w-good-spec} to~$w_\theta$, as in \cref{sec:qf}: it gives the spectral point~$F_{w_\theta}$, the unique element of~$\Delta$ that agrees with~$w_\theta$ on~$\mathcal{B}$. For $k \geq 5$, however, $w_\theta$ need not be $\mathcal{B}$-good, as the following example shows. 

\begin{example}\label{ex:not-B-good}
Let $k = 5$, and let~$B$ consist of $\{\{1,2\},\{3,4,5\}\}$ and $\{\{3,4\},\{1,2,5\}\}$, and let $\theta$ assign weight~$\tfrac12$ to each. Let~$e_1, e_2$ be a basis of~$\CC^2$, and let $u = e_1 \otimes e_2 + e_2 \otimes e_1$ and $v = e_1 \otimes e_1 \otimes e_1 + e_2 \otimes e_2 \otimes e_2$. Let $a_1 = u_{12} \otimes v_{345}$ and $a_2 = v_{125} \otimes u_{34}$, where the subscripts indicate the legs on which we place the factors, and let $a = a_1 + a_2$. Let~$\omega$ be a primitive cube root of unity, and let~$\mathfrak{S}_3$ act on~$\CC^2$ by letting $(123)$ map~$e_1$ to~$\omega e_1$ and~$e_2$ to~$\omega^2 e_2$, and letting $(12)$ exchange~$e_1$ and~$e_2$. This representation is irreducible, and the simultaneous action of~$\mathfrak{S}_3$ on all five legs fixes~$a_1$, $a_2$ and~$a$. These three tensors are concise, so they lie in~$\mathcal{B}$. We have $a \leq a_1 \oplus a_2$. The flattening ranks for the two bipartitions are $3$ and~$3$ for~$a$, $1$ and~$2$ for~$a_1$, and $2$ and~$1$ for~$a_2$. Thus $w_\theta(a) = 3 > 2\sqrt{2} = w_\theta(a_1) + w_\theta(a_2)$, so~$w_\theta$ is not $\mathcal{B}$-good.
\end{example}

Because of \cref{ex:not-B-good}, we cannot directly apply \cref{th:w-good-spec} to obtain the spectral points of this section. In that example, no spectral point agrees with~$w_\theta$ on~$\mathcal{B}$, since the restriction of every spectral point to~$\mathcal{B}$ is $\mathcal{B}$-good (see after \cref{th:w-good-spec}). Instead, we will use an iterative procedure: we will merge the legs in each~$S_b$ into one leg, one~$S_b$ at a time, and use repeated interpolation. We will prepare for this procedure by discussing the merging operation and its properties in \cref{sec:upper-merge}, carry it out for one example in \cref{sec:upper-example}, and for every non-crossing family in \cref{sec:upper-interp}. 

\subsection{Merging legs along \texorpdfstring{set}{set} partitions: spectral points,  irreducibility}\label{sec:upper-merge}

In this subsection we will set up the merging operation, which merges groups of legs into single legs, and its properties. We will need them for the iterative construction of spectral points in \cref{sec:upper-example} and \cref{sec:upper-interp}. We will show that composing a spectral point with a merging map gives a spectral point, extend the sets~$\mathcal{A}_i$ of \cref{sec:tensors} from single legs to groups of legs, and show that merging maps these sets to the corresponding sets of merged tensors.

Let $\parts([k])$ be the set of set partitions of~$[k]$, so that $\bipart([k]) \subseteq \parts([k])$. For $p \in \parts([k])$, a \emph{$p$-tensor} is a $|p|$-tensor whose legs are labelled by the parts of~$p$, and for a $k$-tensor~$t$, $\merge_{p}(t)$ denotes the $p$-tensor obtained from~$t$ by merging the legs in each part of~$p$ into one leg. The tensor~$\merge_p(t)$ is also called a \emph{partial flattening} of~$t$. We say that $q \in \parts([k])$ \emph{refines}~$p$ if every part of~$p$ is a union of parts of~$q$. Then~$\merge^{q}_{p}$ denotes the map from $q$-tensors to $p$-tensors that merges the legs that lie in the same part of~$p$. 
\begin{lemma}\label{lem:merge-compose}
Let $p, q \in \parts([k])$ such that~$q$ refines~$p$. Then $\merge^{q}_{p} \circ \merge_{q} = \merge_p$.
\end{lemma}
\begin{proof}
Both sides merge the legs in each part of~$p$ into one leg. The left-hand side first merges the legs in each part of~$q$, and then the legs of the $q$-tensor that lie in the same part of~$p$.
\end{proof}

\begin{lemma}\label{lem:merge-spectral}
Let $p \in \parts([k])$ with $|p| \geq 2$, and let $g \in \Delta_{\CC,|p|}$. Then $g \circ \merge_{p} \in \Delta_{\CC,k}$. If $q \in \parts([k])$ refines~$p$, then also $g \circ \merge^{q}_{p} \in \Delta_{\CC,|q|}$.
\end{lemma}
\begin{proof}
If $s \leq t$ then $\merge_{p}(s) \leq \merge_{p}(t)$, and $\merge_{p}(s \otimes t) = \merge_{p}(s) \otimes \merge_{p}(t)$. Now let $s \in W_1 \otimes \cdots \otimes W_k$ and $t \in V_1 \otimes \cdots \otimes V_k$, and for a part~$S$ of~$p$ write $W_S = \bigotimes_{i \in S} W_i$ and $V_S = \bigotimes_{i \in S} V_i$. The merged leg~$S$ of $s \oplus t$ is $\bigotimes_{i \in S} (W_i \oplus V_i)$, which contains $W_S \oplus V_S$ as a direct summand. The tensor $\merge_{p}(s \oplus t)$ lies in $\bigotimes_{S \in p} (W_S \oplus V_S)$, where it equals $\merge_{p}(s) \oplus \merge_{p}(t)$. Projecting every merged leg onto $W_S \oplus V_S$, and including it back, shows that $\merge_{p}(s \oplus t)$ and $\merge_{p}(s) \oplus \merge_{p}(t)$ are equivalent. Finally, merging the legs of $\langle m \rangle = \sum_{x=1}^m e_x \otimes \cdots \otimes e_x$ gives $\sum_{x=1}^m f_x \otimes \cdots \otimes f_x$, where on every merged leg the vectors $f_x = e_x \otimes \cdots \otimes e_x$ are linearly independent. So $\merge_{p}(\langle m \rangle)$ is equivalent to the unit tensor~$\langle m \rangle$ of order~$|p|$. Since~$g$ is monotone, multiplicative, additive and normalised, so is $g \circ \merge_{p}$. The second assertion is the first one for $|q|$-tensors, whose legs are the parts of~$q$, and the set partition of these parts given by~$p$.
\end{proof}

Let $b \in \bipart([k])$. If $p \in \parts([k])$ refines~$b$, then $\merge^p_b$ maps $p$-tensors to matrices, and we let $\tensorrank^p_b = \tensorrank \circ \merge^p_b$, where~$\tensorrank$ is the matrix rank. By \cref{lem:merge-compose}, $\tensorrank^p_b \circ \merge_p = \tensorrank_b$.

Let $S \subseteq [k]$ be nonempty with $S \neq [k]$, and let $t \in V_1 \otimes \cdots \otimes V_k$. Identify $V_1 \otimes \cdots \otimes V_k$ with $\bigl( \bigotimes_{i \in S} V_i \bigr) \otimes \bigl( \bigotimes_{i \notin S} V_i \bigr)$. The \emph{$S$-support} $\supp_S(t)$ of~$t$ is the smallest subspace $U \subseteq \bigotimes_{i \in S} V_i$ with $t \in U \otimes \bigotimes_{i \notin S} V_i$. Thus~$\supp_{\{i\}}(t)$ is the $i$-support of~$t$, and $\dim \supp_S(t) = \tensorrank_b(t)$ for the bipartition $b = \{S, [k] \setminus S\}$. 

Extending the definition of $i$-irreducibility in \cref{sec:tensors}, we say that an element of~$\mathcal{S}$ is \emph{$S$-irreducible} if it has a concise representative $t \in V_1 \otimes \cdots \otimes V_k$ for which there are a finite group~$G$ and, for every $i \in [k]$, a representation $\rho_i : G \to \GL(V_i)$, such that $(\rho_1(g) \otimes \cdots \otimes \rho_k(g)) t = t$ for every $g \in G$, and such that the restriction of $\bigotimes_{i \in S} \rho_i$ to~$\supp_S(t)$ is irreducible. Let $\mathcal{A}_S \subseteq \mathcal{S}$ be the set of $S$-irreducible tensors. For a concise~$t$, the $\{i\}$-support is~$V_i$, so $\mathcal{A}_{\{i\}} = \mathcal{A}_i$ for every $i \in [k]$. The representative need not be concise: if a representative~$t$, not necessarily concise, has a group~$G$ and representations~$\rho_i$ as above, then the element that~$t$ represents is $S$-irreducible. Indeed, for $j \in [k]$, the $j$-support of~$t$ is invariant under~$\rho_j$, since~$t$ is fixed. Replacing every~$V_j$ by the $j$-support of~$t$ and every~$\rho_j$ by its restriction gives a concise representative, with the same $S$-support and the same action of~$G$ on it. We will use this in the proofs of \cref{lem:merge-family} and \cref{lem:atleast}. For $p \in \parts([k])$ and a nonempty union~$S \neq [k]$ of parts of~$p$, we define the set $\mathcal{A}^p_S \subseteq \mathcal{S}_{\CC,|p|}$ of $p$-tensors in the same way, with the legs of $p$-tensors in place of $1, \ldots, k$. For the set partition~$p$ of~$[k]$ into singletons, $\mathcal{A}^p_S = \mathcal{A}_S$.

\begin{lemma}\label{lem:merge-family}
Let $p \in \parts([k])$ with $|p| \geq 2$, let $S \in p$ and let $t \in \mathcal{A}_S$. Then $\merge_{p}(t) \in \mathcal{A}^p_S$.
\end{lemma}
\begin{proof}
Let $t \in V_1 \otimes \cdots \otimes V_k$ be a representative and let~$G$ and the~$\rho_i$ be as in the definition of~$\mathcal{A}_S$. Merging the legs in each part of~$p$ gives the $p$-tensor $\merge_{p}(t)$, whose leg for a part of~$p$ is the tensor product of the~$V_i$ over that part. On this leg, $G$ acts by the tensor product of the corresponding~$\rho_i$. The simultaneous action of~$G$ on these legs is its action on $V_1 \otimes \cdots \otimes V_k$, so it fixes~$\merge_{p}(t)$. The $S$-support of~$\merge_{p}(t)$ is~$\supp_S(t)$. By the choice of~$G$ and the~$\rho_i$, the action $\bigotimes_{i \in S} \rho_i$ of~$G$ on~$\supp_S(t)$ is irreducible. Since the representative need not be concise, $\merge_{p}(t) \in \mathcal{A}^p_S$.
\end{proof}

\subsection{Example interpolation among non-crossing bipartition flattening ranks}\label{sec:upper-example}

In this subsection we will illustrate with an example how to construct an interpolation among flattening ranks corresponding to a family of non-crossing bipartitions. We will give the general construction in \cref{sec:upper-interp}, and the identification with the non-crossing upper quantum functionals in \cref{sec:upper-identify}. Recall that $\parts([k])$ denotes the set partitions of $[k]$ and $\bipart([k])$ the bipartitions.

The short idea is this.
\Cref{th:interpolation} lets us interpolate, for any $k\geq3$, between any element in $\Delta_{\CC,k}$ and any singleton flattening rank $\tensorrank_i \in \Delta_{\CC,k}$ for $i \in [k]$. To interpolate towards~$\tensorrank_b$ for $b = \{S, [k] \setminus S\} \in \bipart([k])$, we will merge the legs in~$S$ into one leg. The flattening rank of a merged tensor at this single leg equals~$\tensorrank_b$ of the original tensor, so \cref{th:interpolation} applies to the merged tensors. Composing with the merging map then gives a spectral point of the original $k$-tensors. We will repeat this with finer and finer set partitions.

We give the example. Let $k = 5$. We write $12|345$ for the set partition $\{\{1,2\},\{3,4,5\}\}$ of~$[5]$, and similarly for other set partitions. Let~$B$ consist of the bipartitions $12|345$, $34|125$ and $3|1245$, and let~$\theta$ assign weight~$1/3$ to each. We will successively interpolate between the bipartition flattening ranks 
\[
\tensorrank^{1|2|3|4|5}_{12|345}\!,\,\,
\tensorrank^{1|2|3|4|5}_{34|125}\!,\,\,\textnormal{and}\, \tensorrank^{1|2|3|4|5}_{3|1245}\!.
\]
\begin{enumerate}[(1)]
\item 
Let $g_1 = \tensorrank \in \Delta_{\CC,2}$ be the matrix rank, so that $g_1 \circ \merge^{1|2|3|4|5}_{12|345} = \tensorrank^{1|2|3|4|5}_{12|345} \in \Delta_{\CC,5}$.
\item To interpolate between $\tensorrank^{1|2|3|4|5}_{12|345}$ and~$\tensorrank^{1|2|3|4|5}_{34|125}$, we unmerge $g_1$ from $12|345$ to $12|34|5$: By \cref{lem:merge-spectral} 
\[h_2 \coloneqq g_1 \circ \merge^{12|34|5}_{12|345} \in \Delta_{\CC,3}.
\] 
Apply \cref{th:interpolation} with $\mathcal{A}^{12|34|5}_{\{3,4\}} \subseteq \mathcal{S}_{\CC,3}$ and $w_2 = h_2^{1/2} (\tensorrank^{12|34|5}_{34|125})^{1/2} \colon \mathcal{S}_{\CC,3} \to \RR_{\geq 0}$ to get
\[
g_2 \coloneqq F_{w_2} \in \Delta_{\CC,3}.
\]
By \cref{th:w-good-spec}~(\ref{it:Fw-below}),
\[
g_2 \circ \merge^{1|2|3|4|5}_{12|34|5} \leq w_2 \circ \merge^{1|2|3|4|5}_{12|34|5} = (\tensorrank^{1|2|3|4|5}_{12|345} \tensorrank^{1|2|3|4|5}_{34|125})^{1/2}.
\]
\item To interpolate towards~$\tensorrank^{1|2|3|4|5}_{3|1245}$, we unmerge $g_2$ from $12|34|5$ to $12|3|4|5$: By \cref{lem:merge-spectral}
\[
h_3 \coloneqq g_2 \circ \merge^{12|3|4|5}_{12|34|5} \in \Delta_{\CC,4}.
\]
Apply \cref{th:interpolation} with $\mathcal{A}^{12|3|4|5}_{\{3\}} \subseteq \mathcal{S}_{\CC,4}$ and $w_3 = h_3^{2/3} (\tensorrank^{12|3|4|5}_{3|1245})^{1/3} \colon \mathcal{S}_{\CC,4} \to \RR_{\geq 0}$ to get
\[
g_3 \coloneqq F_{w_3} \in \Delta_{\CC,4}.
\]
The exponent~$2/3$ is the total weight of $12|345$ and $34|125$, the bipartitions used in step~(2).
\item Finally, we unmerge $g_3$ from $12|3|4|5$ to $1|2|3|4|5$: By \cref{lem:merge-spectral}
\[
f \coloneqq g_3 \circ \merge^{1|2|3|4|5}_{12|3|4|5} \in \Delta_{\CC,5}.
\]
By \cref{th:w-good-spec}~(\ref{it:Fw-below}) again,
\begin{align*}
f &\leq (g_2 \circ \merge^{1|2|3|4|5}_{12|34|5})^{2/3} \, (\tensorrank^{1|2|3|4|5}_{3|1245})^{1/3} \\
&\leq \bigl( (\tensorrank^{1|2|3|4|5}_{12|345} \tensorrank^{1|2|3|4|5}_{34|125})^{1/2} \bigr)^{2/3} (\tensorrank^{1|2|3|4|5}_{3|1245})^{1/3} = (\tensorrank^{1|2|3|4|5}_{12|345} \tensorrank^{1|2|3|4|5}_{34|125} \tensorrank^{1|2|3|4|5}_{3|1245})^{1/3},
\end{align*}
and on the tensors in $\mathcal{A}^{1|2|3|4|5}_{\smash[b]{\{1,2\}}} \cap \mathcal{A}^{1|2|3|4|5}_{\smash[b]{\{3,4\}}} \cap \mathcal{A}^{1|2|3|4|5}_{\smash[b]{\{3\}}}$ both inequalities are equalities, by \cref{lem:merge-family} and \cref{th:interpolation}~(\ref{it:int-family}).
\end{enumerate}
In \cref{sec:upper-identify} we will show that~$f$ is the largest spectral point dominated by
\[
\bigl( \tensorrank^{1|2|3|4|5}_{12|345} \tensorrank^{1|2|3|4|5}_{34|125} \tensorrank^{1|2|3|4|5}_{3|1245} \bigr)^{1/3},
\]
and that for every $t \in V_1 \otimes \cdots \otimes V_5$
\[
f(t) = \lim_{n \to \infty} \Bigl( \max_{\lambda} \bigl( \tensorrank^{1|2|3|4|5}_{12|345}(y_\lambda) \tensorrank^{1|2|3|4|5}_{34|125}(y_\lambda) \tensorrank^{1|2|3|4|5}_{3|1245}(y_\lambda) \bigr)^{1/3} \Bigr)^{1/n},
\]
where~$\lambda$ runs over the triples $(\lambda_1, \lambda_2, \lambda_3)$ of partitions of~$n$, and $y_\lambda = P^{V_1 \otimes V_2}_{\lambda_1} P^{V_3 \otimes V_4}_{\lambda_2} P^{V_3}_{\lambda_3}\, t^{\otimes n}$, with the isotypic projections of \cref{sec:qf} for the merged legs $V_1 \otimes V_2$ and $V_3 \otimes V_4$ and for the leg~$V_3$. These three projections commute, since any two of the sets $\{1,2\}$, $\{3,4\}$ and~$\{3\}$ are nested or disjoint (\cref{lem:side}~(\ref{it:side-commute})). We will also identify~$f$ with the non-crossing upper quantum functional~$F^\theta$, which we will define in \cref{sec:upper-def}.

\subsection{Spectral points by repeated interpolation}\label{sec:upper-interp}
In this subsection we will do for every non-crossing family of bipartitions what \cref{sec:upper-example} did for one example. The result is a spectral point that is at most the weighted geometric average of the flattening ranks of the family. It equals this average on a set of tensors that we will show includes the graph tensors (\cref{prop:graph-in-AC}).

Let $B \subseteq \bipart([k])$ be non-crossing, and let $\theta \in \mathcal{P}(B)$ with $\theta(b) > 0$ for every $b \in B$. Enumerate $B = \{b_1, \ldots, b_N\}$, and let $C_j = S_{b_j}$. We choose the enumeration so that if $C_j \supseteq C_\ell$, then $j \leq \ell$. We can for example do this by enumerating by decreasing size of~$C_j$, since distinct bipartitions have distinct sets~$S_b$. Write $\theta_j = \theta(b_j)$ and $\theta_{\leq j} = \theta_1 + \cdots + \theta_j$, so $\theta_{\leq N} = 1$. For $j \in [N]$, let $p_j \in \parts([k])$ be the coarsest common refinement of $b_1, \ldots, b_j$, so that two legs lie in the same part of~$p_j$ if and only if none of $b_1, \ldots, b_j$ separates them. For $\ell < j$, the set~$C_\ell$ contains~$C_j$ or is disjoint from it, since the sets~$S_b$ are nested or disjoint and the enumeration excludes $C_\ell \subseteq C_j$. Hence~$C_j$ is a part of~$p_j$, and~$p_j$ refines~$p_{j-1}$ for $j \geq 2$. So~$p_j$ refines~$b_j$, and $\tensorrank^{p_j}_{\smash[b]{b_j}}$ is the flattening rank at the single leg~$C_j$ of $p_j$-tensors. 

In terms of the tree of \cref{sec:upper-noncrossing}, $C_1, \ldots, C_N$ are the vertices other than the root, each listed before the vertices below it, and~$p_j$ merges two legs if every set among $C_1, \ldots, C_j$ contains both or neither of them. In the example of \cref{sec:upper-example}, $p_1$, $p_2$, $p_3$ are $12|345$, $12|34|5$ and $12|3|4|5$.

We construct spectral points $g_1, \ldots, g_N$ with $g_j \in \Delta_{\CC,|p_j|}$, as in the example of \cref{sec:upper-example}.
\begin{enumerate}[(1)]
\item Let $g_1 = \tensorrank \in \Delta_{\CC,2}$ be the matrix rank, so that $g_1 \circ \merge_{p_1} = \tensorrank_{b_1}$.
\item For $j = 2, \ldots, N$ in turn, to interpolate towards~$\tensorrank_{b_j}$, we unmerge $g_{j-1}$ from $p_{j-1}$ to $p_j$: By \cref{lem:merge-spectral}
\[
h_j \coloneqq g_{j-1} \circ \merge^{p_j}_{p_{j-1}} \in \Delta_{\CC,|p_j|},
\]
and $h_j \circ \merge_{\smash{p_j}} = g_{j-1} \circ \merge_{\smash{p_{j-1}}}$ by \cref{lem:merge-compose}. Let $\tau_j = \theta_j / \theta_{\leq j}$. Apply \cref{th:interpolation} with $\mathcal{A}^{p_j}_{C_j} \subseteq \mathcal{S}_{\CC,|p_j|}$ and $w_j = h_j^{\smash{1-\tau_j}} (\tensorrank^{p_j}_{b_j})^{\tau_j} \colon \mathcal{S}_{\CC,|p_j|} \to \RR_{\geq 0}$ to get
\[
g_j \coloneqq F_{w_j} \in \Delta_{\CC,|p_j|}.
\]
The hypotheses of \cref{th:interpolation} hold by \cref{lem:tensor-decomp} and \cref{lem:tensor-cover}, after relabelling the legs.
\item Finally, we unmerge $g_N$ from $p_N$ to the set partition of~$[k]$ into singletons: By \cref{lem:merge-spectral}
\[
f_\theta \coloneqq g_N \circ \merge_{p_N} \in \Delta_{\CC,k}.
\]
We call~$f_\theta$ the \emph{repeated interpolant} for~$B$ and~$\theta$.
\end{enumerate}
The notation $f_\theta$ does not show the enumeration of $B$, and \cref{th:upper-spectral} will show that~$f_\theta$ does not depend on it. If $\theta(b) = 0$ for some $b \in B$, we let~$f_\theta$ be the repeated interpolant for the restriction of~$\theta$ to $\{b \in B : \theta(b) > 0\}$. 

The parameter~$\tau_j$ in the construction is the share of the new weight~$\theta_j$ in the accumulated weight~$\theta_{\leq j}$, and $1 - \tau_j = \theta_{\leq j-1}/\theta_{\leq j}$ is the share of the earlier weights. In the example of \cref{sec:upper-example}, $\tau_2 = 1/2$ and $\tau_3 = 1/3$.
\begin{theorem}\label{th:refine}
Let $B \subseteq \bipart([k])$ be non-crossing, let $\theta \in \mathcal{P}(B)$, and for $b \in B$ let~$S_b$ be the part of~$b$ not containing~$k$. Let $w_\theta(t) = \prod_{b \in B} \tensorrank_b(t)^{\theta(b)}$. Then the repeated interpolant~$f_\theta$ for~$B$ and~$\theta$ is a spectral point with $f_\theta \leq w_\theta$, and $f_\theta = w_\theta$ on $\bigcap_{b \in B} \mathcal{A}_{S_b}$.
\end{theorem}
\begin{proof}
Discarding the~$b$ with $\theta(b) = 0$ changes neither~$w_\theta$ nor~$f_\theta$, and it can only enlarge the intersection $\bigcap_{b \in B} \mathcal{A}_{S_b}$, so we may assume $\theta(b) > 0$ for every $b \in B$. We use the notation of the construction of~$f_\theta$ and the facts established there. Since $g_N \in \Delta_{\CC,|p_N|}$, \cref{lem:merge-spectral} gives $f_\theta \in \Delta_{\CC,k}$.

We will now show that $f_\theta \leq w_\theta$. The function~$w_j$ is a product of powers of spectral points, so it is monotone and multiplicative, and \cref{th:w-good-spec}~(\ref{it:Fw-below}) gives $g_j \leq w_j$. So 
\[
g_j \circ \merge_{p_j} \leq w_j \circ \merge_{p_j} = (g_{j-1} \circ \merge_{p_{j-1}})^{1-\tau_j} \tensorrank_{b_j}^{\tau_j}
\]
on~$\mathcal{S}$, using that $h_j \circ \merge_{p_j} = g_{j-1} \circ \merge_{p_{j-1}}$ and that $\tensorrank^{p_j}_{\smash[b]{b_j}} \circ \merge_{p_j} = \tensorrank_{b_j}$. Since $g_1 \circ \merge_{p_1} = \tensorrank_{b_1}$ and $1 - \tau_j = \theta_{\leq j-1}/\theta_{\leq j}$, induction on~$j$ gives
\[
g_j \circ \merge_{p_j} \leq \prod_{\mathclap{\ell \leq j}} \tensorrank_{b_\ell}^{\theta_\ell/\theta_{\leq j}},
\]
and $j = N$ gives $f_\theta \leq w_\theta$.

It remains to show that $f_\theta = w_\theta$ on $\bigcap_{b \in B} \mathcal{A}_{S_b}$. We will show by induction on~$j$ that for every $t \in \bigcap_{\ell \leq j} \mathcal{A}_{C_\ell}$ we have
\[
g_j(\merge_{p_j}(t)) = \prod_{\ell \leq j} \tensorrank_{b_\ell}(t)^{\theta_\ell / \theta_{\leq j}}.
\]
For the base case $j = 1$ this is the definition of~$g_1$. For $j \geq 2$, let~$t$ lie in the intersection. Then $\merge_{p_j}(t) \in \mathcal{A}^{p_j}_{C_j}$ by \cref{lem:merge-family}, so \cref{th:interpolation}~(\ref{it:int-family}) gives
\[
g_j(\merge_{p_j}(t)) = g_{j-1}(\merge_{p_{j-1}}(t))^{1 - \tau_j} \tensorrank_{b_j}(t)^{\tau_j},
\]
using that $h_j(\merge_{p_j}(t)) = g_{j-1}(\merge_{p_{j-1}}(t))$, and that $\tensorrank^{p_j}_{\smash[b]{b_j}}(\merge_{p_j}(t)) = \tensorrank_{b_j}(t)$. Now we have $1 - \tau_j = \theta_{\leq j-1}/\theta_{\leq j}$, so the induction hypothesis turns the right-hand side into $\prod_{\ell \leq j} \tensorrank_{b_\ell}(t)^{\theta_\ell / \theta_{\leq j}}$. For $j = N$ this gives $f_\theta = w_\theta$ on $\bigcap_{b \in B} \mathcal{A}_{S_b}$, since $\theta_{\leq N} = 1$.
\end{proof}

\subsection{Isotypic projections on groups of legs}\label{sec:upper-projections}

To define the non-crossing upper quantum functionals in \cref{sec:upper-def}, we need isotypic projections on groups of legs. In this subsection we will treat a nonempty set~$S\subseteq[k]$ of legs as one leg and use the isotypic projections of \cref{sec:qf} for this leg. We also need two facts that \cref{sec:qf} did not need: on invariant tensors the projections for~$S$ and for $[k] \setminus S$ agree, and the projections for nested or disjoint sets commute.

Let $t \in V_1 \otimes \cdots \otimes V_k$ and $n \geq 1$. As in \cref{sec:qf}, the projections that we will define act on $V_1^{\otimes n} \otimes \cdots \otimes V_k^{\otimes n}$, which contains~$t^{\otimes n}$. For a nonempty $S \subseteq [k]$ write $V_S = \bigotimes_{i \in S} V_i$, so that $V_S^{\otimes n} = \bigotimes_{i \in S} V_i^{\otimes n}$ after regrouping. 

For $\sigma \in \mathfrak{S}_n$ let~$\pi_S(\sigma)$ be the linear map on $V_1^{\otimes n} \otimes \cdots \otimes V_k^{\otimes n}$ that permutes the~$n$ factors of~$V_i^{\otimes n}$ according to~$\sigma$ for every $i \in S$, and is the identity on~$V_i^{\otimes n}$ for $i \notin S$. For $\lambda \vdash n$, let~$P^{\smash[t]{V_S}}_\lambda$ be the projection onto the isotypic component of~$[\lambda]$ for this action~$\pi_S$ of~$\mathfrak{S}_n$, along the other isotypic components. For $S = \{i\}$ this is the projection~$P^{\smash[t]{V_i}}_\lambda$ of \cref{sec:qf}. With~$\chi_\lambda$ the character of~$[\lambda]$, the projection formula for isotypic components \cite[Section~2.4]{MR1153249} gives, since $\chi_\lambda(\sigma^{-1})$ is the complex conjugate of~$\chi_\lambda(\sigma)$,
\[
P^{V_S}_\lambda = \frac{\dim[\lambda]}{n!} \sum_{\sigma \in \mathfrak{S}_n} \chi_\lambda(\sigma^{-1})\, \pi_S(\sigma).
\]
For $S = [k]$, the action~$\pi_S$ permutes the factors of all legs simultaneously. We call the elements of the $\mathfrak{S}_n$-invariant subspace for this action \emph{invariant}. For example, $t^{\otimes n}$ is invariant.

The following lemma is due to Christandl, Vrana and Zuiddam \cite[Lemmas~3.12 and~3.13]{MR4495838}, who proved commutation for non-crossing bipartitions on the invariant subspace. We also need commutation on the full tensor space for nested or disjoint sets, and include the short proof.
\begin{lemma}\label{lem:side}
Let $n \geq 1$ and let $\lambda, \mu \vdash n$.
\begin{enumerate}[\upshape(i)]
\item\label{it:side-invariant} Let $\emptyset \neq S \subsetneq [k]$, and let $y \in V_1^{\otimes n} \otimes \cdots \otimes V_k^{\otimes n}$ be invariant. Then $P^{V_S}_\lambda y = P^{V_{[k] \setminus S}}_\lambda y$, and this vector is again invariant.
\item\label{it:side-commute} Let $S, T \subseteq [k]$ be nonempty, and nested or disjoint. Then~$P^{V_S}_\lambda$ and~$P^{V_T}_\mu$ commute.
\end{enumerate}
\end{lemma}
\begin{proof}
(\ref{it:side-invariant}) We have $\pi_{[k]}(\sigma) = \pi_S(\sigma) \pi_{[k] \setminus S}(\sigma)$, and the two factors commute, so for invariant~$y$ we have $\pi_S(\sigma) y = \pi_{[k] \setminus S}(\sigma^{-1}) y$ for every $\sigma \in \mathfrak{S}_n$. Substituting~$\sigma^{-1}$ for~$\sigma$ in the formula for~$P^{\smash[t]{V_S}}_\lambda$ and using $\chi_\lambda(\sigma) = \chi_\lambda(\sigma^{-1})$, which holds because the characters of~$\mathfrak{S}_n$ are real, gives $P^{\smash[t]{V_S}}_\lambda y = P^{\smash[t]{V_{[k] \setminus S}}}_\lambda y$. The projection~$P^{\smash[t]{V_S}}_\lambda$ commutes with every~$\pi_{[k]}(\sigma)$, so this vector is invariant.

(\ref{it:side-commute}) If~$S$ and~$T$ are disjoint, then~$\pi_S$ and~$\pi_T$ act on different tensor factors, so the two projections commute. Let $S \subseteq T$; the case $T \subseteq S$ is symmetric. For every $\sigma \in \mathfrak{S}_n$,
\[
\pi_S(\sigma)\, P^{V_S}_\lambda\, \pi_S(\sigma)^{-1} = \frac{\dim[\lambda]}{n!} \sum_{\tau \in \mathfrak{S}_n} \chi_\lambda(\tau^{-1})\, \pi_S(\sigma \tau \sigma^{-1}) = P^{V_S}_\lambda,
\]
where the second equality substitutes~$\sigma^{-1} \tau \sigma$ for~$\tau$ and uses that~$\chi_\lambda$ is constant on conjugacy classes. So~$P^{\smash[t]{V_S}}_\lambda$ commutes with~$\pi_S(\sigma)$. If $S \neq T$, it also commutes with~$\pi_{T \setminus S}(\sigma)$, which acts on other tensor factors, and hence with $\pi_T(\sigma) = \pi_S(\sigma) \pi_{T \setminus S}(\sigma)$. So in both cases~$P^{\smash[t]{V_S}}_\lambda$ commutes with every~$\pi_T(\sigma)$. Since~$P^{\smash[t]{V_T}}_\mu$ is a linear combination of the~$\pi_T(\sigma)$, the projections~$P^{\smash[t]{V_S}}_\lambda$ and~$P^{\smash[t]{V_T}}_\mu$ commute.
\end{proof}

\subsection{Non-crossing upper quantum functionals}\label{sec:upper-def}

In this subsection we will state the definition of the non-crossing upper quantum functionals.

Fix a non-crossing family $B \subseteq \bipart([k])$ and $\theta \in \mathcal{P}(B)$. Recall that~$S_b$ is the part of~$b$ not containing~$k$, and for $b \in B$ and $\lambda \vdash n$ we let~$P^b_\lambda$ be the projection~$P^{V_S}_\lambda$ of \cref{sec:upper-projections} for $S = S_b$. Any two of the sets~$S_b$ are nested or disjoint, so the projections~$P^b_\lambda$ commute by \cref{lem:side}~(\ref{it:side-commute}).
The \emph{non-crossing upper quantum functional}~$F^\theta$ is defined by $F^\theta(0) = 0$ and, for nonzero $t \in V_1 \otimes \cdots \otimes V_k$, by
\[
\log F^\theta(t) \coloneqq \sup \Bigl\{ \sum_{b \in B} \theta(b) H(\lambda_b/n) : n \geq 1,\ \lambda_b \vdash n \text{ for } b \in B,\ \Bigl( \prod_{b \in B} P^b_{\lambda_b} \Bigr) t^{\otimes n} \neq 0 \Bigr\}.
\]
Replacing any of the sets~$S_b$ by their complements does not change the vectors $\bigl( \prod_{b \in B} P^b_{\lambda_b} \bigr) t^{\otimes n}$, by \cref{lem:side}~(\ref{it:side-invariant}) applied to~$t^{\smash{\otimes n}}$ and to the vectors obtained from it by applying the projections~$P^b_{\smash{\lambda_b}}$ one at a time, which are all invariant under the simultaneous action of~$\mathfrak{S}_n$. So~$F^\theta$ does not depend on which part of each $b \in B$ is used.

For $k \geq 3$, if~$B = \{\{\{i\}, [k] \setminus \{i\}\} : i \in [k]\}$, all the singleton bipartitions,  then~$F^\theta$ is the singleton quantum functional of \cref{sec:qf} for the weights $\theta_i = \theta(\{\{i\}, [k] \setminus \{i\}\})$ with $i \in [k]$.

The main result of this section is the following theorem. We will prove it in \cref{sec:upper-identify}.

\begin{theorem}\label{th:upper-spectral}
Let $B \subseteq \bipart([k])$ be non-crossing, let $\theta \in \mathcal{P}(B)$, and let $w_\theta = \prod_{b \in B} \tensorrank_b^{\theta(b)}$. Then the non-crossing upper quantum functional~$F^\theta$ is a spectral point, it is the largest element of~$\Delta$ dominated by~$w_\theta$, and it equals the repeated interpolant~$f_\theta$ for~$B$ and~$\theta$, for every enumeration allowed in the construction of~$f_\theta$.
\end{theorem}

\subsection{Adding partitions up to dominance}\label{sec:upper-dominance}
In this subsection we will show that the tuples of partitions from the definition of the non-crossing upper quantum functional~$F^\theta$ can be added ``up to dominance''. We will use this fact in the proof of \cref{th:upper-spectral}, to pass from partitions of~$n$ to partitions of multiples of~$n$.

For partitions~$\alpha = (\alpha_1, \ldots, \alpha_d)$ and~$\beta = (\beta_1, \ldots, \beta_d)$, which we may assume have the same length $d$ after padding by zeroes, let $\alpha + \beta = (\alpha_1 + \beta_1, \ldots, \alpha_d + \beta_d)$. For partitions~$\alpha$ and~$\beta$ of the same integer, write $\alpha \unlhd \beta$ if $\alpha_1 + \cdots + \alpha_\ell \leq \beta_1 + \cdots + \beta_\ell$ for every $\ell \geq 1$. This is the dominance order. Vrana \cite[Proposition~5.10]{vrana2025errorexponentstripartitetobipartiteentanglement} proved an analogue of the following lemma for the isotypic projections of the single legs of $3$-tensors. We will follow his argument.
\begin{lemma}[addition up to dominance]\label{lem:dominance}
Let $B \subseteq \bipart([k])$ be non-crossing, let $t \in \mathcal{S} \setminus \{0\}$, let $n, m \geq 1$, and let $\lambda = (\lambda_b)_{b \in B}$ and $\mu = (\mu_b)_{b \in B}$ be tuples of partitions of~$n$ and of~$m$, respectively, with $\bigl( \prod_b P^b_{\smash{\lambda_b}} \bigr) t^{\otimes n} \neq 0$ and $\bigl( \prod_b P^b_{\smash{\mu_b}} \bigr) t^{\otimes m} \neq 0$. Then there is a tuple $\nu = (\nu_b)_{b \in B}$ of partitions of $n+m$ with $\bigl( \prod_b P^b_{\nu_b} \bigr) t^{\smash{\otimes (n+m)}} \neq 0$ and $\nu_b \unlhd \lambda_b + \mu_b$ for every $b \in B$.
\end{lemma}
\begin{proof}
Let $x = \bigl( \prod_b P^b_{\smash{\lambda_{b}}} \bigr) t^{\otimes n} \otimes \bigl( \prod_b P^b_{\smash{\mu_{b}}} \bigr) t^{\otimes m}$, a nonzero vector of $V^{\otimes n} \otimes V^{\otimes m} = V^{\otimes (n+m)}$, where $V = V_1 \otimes \cdots \otimes V_k$, and let $Q = \prod_{b \in B} Q_b$ with $Q_b = P^b_{\smash{\lambda_{b}}} \otimes P^b_{\smash{\mu_{b}}}$, so that $x = Q\, t^{\smash{\otimes (n+m)}}$.

For every~$b \in B$, the projections~$P^b_\xi$ on~$V^{\smash{\otimes (n+m)}}$ with $\xi \vdash n+m$ sum to the identity, so $t^{\otimes (n+m)} = \sum_{\nu} y_{\nu}$ with $y_{\nu} = \bigl( \prod_b P^b_{\nu_{b}} \bigr) t^{\otimes (n+m)}$, the sum being over all tuples~$\nu$ of partitions of $n+m$. Hence $0 \neq x = \sum_{\nu} Q\, y_{\nu}$, so $Q\, y_{\nu} \neq 0$ for some~$\nu$. In particular $y_{\nu} \neq 0$.

The~$Q_b$ commute with one another: for $b, c \in B$, the sets~$S_b$ and~$S_c$ are nested or disjoint, so by \cref{lem:side}~(\ref{it:side-commute}) the projections~$P^b_{\lambda_b}$ and~$P^c_{\lambda_c}$ on~$V^{\otimes n}$ commute, and so do~$P^b_{\mu_b}$ and~$P^c_{\smash{\mu_c}}$ on~$V^{\otimes m}$. Now fix $b \in B$ and write $S = S_b$. Since the~$Q_c$ with $c \in B$ commute, $Q\, y_{\nu} \neq 0$ forces $Q_b\, y_{\nu} \neq 0$. Since the~$P^c_{\nu_c}$ with $c \in B$ commute, $y_{\nu} = P^b_{\nu_b}\, y_{\nu}$, so~$y_{\nu}$ lies in the $\nu_{b}$-isotypic component for the action of $\mathfrak{S}_{n+m}$ on $V_S^{\smash{\otimes (n+m)}}$, and on that component~$Q_b$ acts as the $(\lambda_{b}, \mu_{b})$-isotypic projection of $\CC[\mathfrak{S}_n \times \mathfrak{S}_m]$. The restriction of $[\xi]$ to $\mathfrak{S}_n \times \mathfrak{S}_m$ is $\bigoplus_{\alpha, \beta} c^{\smash[t]{\xi}}_{\alpha \beta} [\alpha] \otimes [\beta]$ for $\xi \vdash n+m$, where the~$c^{\xi}_{\alpha \beta}$ are the Littlewood--Richardson coefficients \cite[Exercise~4.43]{MR1153249}, so $Q_b\, y_{\nu} \neq 0$ forces $c^{\smash[t]{\nu_b}}_{\lambda_b \mu_b} \neq 0$. Every partition~$\xi$ with $c^{\smash[t]{\xi}}_{\alpha\beta} \neq 0$ satisfies $\xi \unlhd \alpha + \beta$, since the product of Schur polynomials is $s_\alpha s_\beta = \sum_\tau c^\tau_{\smash[b]{\alpha\beta}} s_\tau$ with nonnegative coefficients, and~$s_\xi$ contains the monomial~$x^\xi$. So $c^{\smash[t]{\xi}}_{\alpha\beta} \neq 0$ implies that~$x^\xi$ occurs in $s_\alpha s_\beta$, that is, $\xi = u + v$ for a weight~$u$ of~$\mathbb{S}_\alpha$ and a weight~$v$ of~$\mathbb{S}_\beta$. Every weight~$u$ of~$\mathbb{S}_\alpha$ satisfies $\sum_{i \leq \ell} u_i \leq \sum_{i \leq \ell} \alpha_i$ for every $\ell \geq 1$, because the columns of a semistandard tableau are strictly increasing, and similarly for~$v$ and~$\beta$.
\end{proof}

If the sets~$S_b$ with $b \in B$ are pairwise disjoint, one can take $\nu = \lambda + \mu$ in \cref{lem:dominance}. Indeed, by Schur--Weyl duality (\cref{th:schur-weyl}), $\bigl( \prod_b P^b_{\lambda_b} \bigr) t^{\otimes n} \neq 0$ means that the closure of the orbit of~$t$ under $\prod_{b \in B} \GL(V_{S_b})$, acting trivially on the other legs, has a nonzero covariant of degree~$n$ with values in the simple module of highest weight $(\lambda_b)_{b \in B}$, that is, an equivariant map to this module given by homogeneous polynomials of degree~$n$. The pairs of degree and highest weight of such covariants are closed under addition \cite[proof of Proposition~2.1]{MR932055}. For nested sets~$S_b$ the projections do not come from one group of this form.

\subsection{Identification with the non-crossing upper quantum functionals}\label{sec:upper-identify}

In this subsection we will show that the repeated interpolant~$f_\theta$ of \cref{sec:upper-interp} is the non-crossing upper quantum functional~$F^\theta$ of \cref{sec:upper-def}, and prove \cref{th:upper-spectral}.

Assume $\theta(b) > 0$ for every $b \in B$. Let $w_\theta(t) = \prod_{b \in B} \tensorrank_b(t)^{\theta(b)}$, and let the enumeration $b_1, \ldots, b_N$ of~$B$, the sets $C_j = S_{b_j}$, the set partitions $p_1, \ldots, p_N$, the numbers $\theta_j = \theta(b_j)$, their partial sums $\theta_{\leq j} = \theta_1 + \cdots + \theta_j$, the spectral points $g_j \in \Delta_{\CC,|p_j|}$ and the repeated interpolant $f_\theta = g_N \circ \merge_{p_N}$ be as in \cref{sec:upper-interp}.

\begin{lemma}\label{lem:dominated}
Let $B \subseteq \bipart([k])$ be non-crossing, let $\theta \in \mathcal{P}(B)$, and let $w_\theta = \prod_{b \in B} \tensorrank_b^{\theta(b)}$. Let $t \in V_1 \otimes \cdots \otimes V_k$ be nonzero, and for $n \geq 1$ and a tuple $\lambda = (\lambda_b)_{b \in B}$ of partitions of~$n$ let $y_\lambda = \bigl( \prod_{b \in B} P^b_{\smash{\lambda_b}} \bigr) t^{\otimes n}$. There is a polynomial~$P$ such that for every $n \geq 1$, the~$y_\lambda$ sum to~$t^{\otimes n}$, at most~$P(n)$ of them are nonzero, and $w_\theta(y_\lambda) \leq P(n)\, F^\theta(t)^n$ for every~$\lambda$. Moreover, every $h \in \Delta$ with $h \leq w_\theta$ satisfies $h \leq F^\theta$.
\end{lemma}
\begin{proof}
Let $n \geq 1$. For every~$b$ the projections~$P^b_{\lambda}$ with $\lambda \vdash n$ sum to the identity, and the projections for different~$b$ commute, so $t^{\otimes n} = \sum_\lambda y_\lambda$.

Let~$\lambda$ have $y_\lambda \neq 0$ and let $b \in B$. The $S_b$-support of~$y_\lambda$ lies in the $\lambda_b$-part of $V_{S_b}^{\otimes n}$, which by \cref{th:schur-weyl} has dimension $\dim[\lambda_b] \dim \mathbb{S}_{\lambda_b}(V_{S_b})$, so $\tensorrank_b(y_\lambda) \leq \dim[\lambda_b] \dim \mathbb{S}_{\lambda_b}(V_{S_b})$. With $D_b = \dim V_{S_b}$, the dimension of $\mathbb{S}_{\lambda_b}(V_{S_b})$ is the multiplicity of~$[\lambda_b]$ in $V_{S_b}^{\otimes n}$, so \cref{lem:irrep-count} gives
\[
\tensorrank_b(y_\lambda) \leq (n+1)^{D_b^2} \dim[\lambda_b].
\] Since the weights sum to~$1$, this gives the first inequality in
\[
w_\theta(y_\lambda) \leq \Bigl( \max_{b \in B} (n+1)^{D_b^2} \Bigr) \prod_{b \in B} \dim[\lambda_b]^{\theta(b)} \leq \Bigl( \max_{b \in B} (n+1)^{D_b^2} \Bigr) F^\theta(t)^n,
\]
and the second follows from the upper bound in \cref{lem:dim-irrep}, since $\sum_b \theta(b) H(\lambda_b/n) \leq \log F^\theta(t)$ by the definition of~$F^\theta$, as $y_\lambda \neq 0$. There are at most $\prod_{b \in B} (n+1)^{D_b}$ tuples with $y_\lambda \neq 0$. So $P(n) = \prod_{b \in B} (n+1)^{\smash{D_b + D_b^2}}$ satisfies the first assertion, since $w_\theta(0) = 0$.

For the second assertion it suffices to show $h(t) \leq F^\theta(t)$ for the nonzero~$t$ fixed above, since $h(0) = 0 = F^\theta(0)$. A sum of tensors in one space is a restriction of their direct sum, so $t^{\otimes n} \leq \bigoplus_\lambda y_\lambda$, the direct sum over those~$\lambda$ with $y_\lambda \neq 0$. Since~$h$ is multiplicative, monotone and additive and $h \leq w_\theta$,
\[
h(t)^n = h(t^{\otimes n}) \leq \sum_\lambda h(y_\lambda) \leq \sum_\lambda w_\theta(y_\lambda) \leq P(n)^2\, F^\theta(t)^n.
\]
Taking~$n$th roots and letting $n \to \infty$ gives $h(t) \leq F^\theta(t)$.
\end{proof}

\begin{lemma}\label{lem:atleast}
Let $B \subseteq \bipart([k])$ be non-crossing, and let $\theta \in \mathcal{P}(B)$ with $\theta(b) > 0$ for every $b \in B$. Let $n \geq 1$, let $y \in V_1^{\otimes n} \otimes \cdots \otimes V_k^{\otimes n}$ be invariant under the action of~$\mathfrak{S}_n$ permuting the~$n$ factors of every leg simultaneously, let $j \in [N]$, and let $\lambda_1, \ldots, \lambda_j$ be partitions of~$n$ with $\bigl( \prod_{\ell \leq j} P^{\smash[t]{b_\ell}}_{\lambda_\ell} \bigr) y \neq 0$. Then
\[
g_j\bigl( \merge_{p_j}(y) \bigr) \geq \prod_{\ell \leq j} \dim[\lambda_\ell]^{\theta_\ell / \theta_{\leq j}}.
\]
\end{lemma}
\begin{proof}
Induction on~$j$. Write $Q_j = \prod_{\ell \leq j} P^{b_\ell}_{\lambda_\ell}$.

Let $j = 1$. The $C_1$-support of~$y$ is invariant under the action of~$\mathfrak{S}_n$ on $V_{C_1}^{\smash[t]{\otimes n}}$, because~$y$ is invariant, and $P^{\smash[t]{b_1}}_{\lambda_1} y \neq 0$, so that support has a nonzero $\lambda_1$-part and therefore contains a subrepresentation isomorphic to $[\lambda_1]$. Since $g_1(\merge_{p_1}(y)) = \tensorrank_{b_1}(y)$, it is the dimension of that support, which is at least $\dim[\lambda_1]$. This is the claim, since $\theta_{\leq 1} = \theta_1$.

Let $j \geq 2$. Decompose $V_{\smash[b]{C_j}}^{\otimes n}$ into irreducible $\mathfrak{S}_n$-subrepresentations, refining the decomposition into isotypic components. The tensor $Q_j y$ lies in the $\lambda_j$-part of $V_{\smash[b]{C_j}}^{\otimes n}$, so there is an equivariant projection~$P_U$ onto a summand $U \cong [\lambda_j]$, acting on the legs in~$C_j$ only, with $P_U Q_j y \neq 0$. Because the decomposition refines the isotypic one, $P_U P^{b_j}_{\smash{\lambda_j}} = P_U$.

For $\ell < j$ the projections~$P_U$ and~$P^{\smash[t]{b_\ell}}_{\lambda_\ell}$ commute. Indeed, $C_j \subseteq C_\ell$ or $C_j \cap C_\ell = \emptyset$ by the enumeration. In the second case the two act on disjoint sets of legs. In the first case $V_{C_\ell}^{\otimes n} = V_{C_j}^{\otimes n} \otimes V_{C_\ell \setminus C_j}^{\otimes n}$, and every $\sigma \in \mathfrak{S}_n$ acts on the two factors separately, so~$P_U$, which is $\sigma$-equivariant on the first factor and the identity on the second, commutes with the action of~$\mathfrak{S}_n$ on $V_{C_\ell}^{\otimes n}$ and hence with~$P^{b_\ell}_{\lambda_\ell}$, which is a linear combination of the operators $\pi_{C_\ell}(\sigma)$.

Let $a = P_U y$. Then $Q_{j-1} a = P_U \, Q_{j-1} y = P_U \, Q_j y \neq 0$, and~$a$ is invariant under the simultaneous action of~$\mathfrak{S}_n$ because~$P_U$ is equivariant. The $C_j$-support of~$a$ is contained in~$U$ and is a nonzero invariant subspace of it, so it equals~$U$. Hence $\merge_{p_j}(a) \in \mathcal{A}^{p_j}_{\smash[b]{C_j}}$: the finite group is~$\mathfrak{S}_n$, acting on each leg of~$p_j$ by permuting its $n$ tensor factors, the simultaneous action fixes~$\merge_{p_j}(a)$, and the action on the $C_j$-support is irreducible. The flattening rank of~$\merge_{p_j}(a)$ at the leg~$C_j$ is $\dim[\lambda_j]$.

Since~$C_j$ is one leg of~$p_j$ and~$P_U$ acts on that leg alone, $\merge_{p_j}(a) \leq \merge_{p_j}(y)$. So~$g_j$ being monotone, \cref{th:interpolation}~(\ref{it:int-family}) at order $|p_j|$ applied to~$\merge_{p_j}(a)$, and then the induction hypothesis applied to~$a$, give
\begin{align*}
g_j\bigl( \merge_{p_j}(y) \bigr)
&\geq g_j\bigl( \merge_{p_j}(a) \bigr)\\
&= g_{j-1}\bigl( \merge_{p_{j-1}}(a) \bigr)^{1 - \tau_j} \dim[\lambda_j]^{\tau_j}\\
&\geq \Bigl( \prod_{\ell < j} \dim[\lambda_\ell]^{\theta_\ell / \theta_{\leq j-1}} \Bigr)^{\theta_{\leq j-1}/\theta_{\leq j}} \dim[\lambda_j]^{\theta_j / \theta_{\leq j}},
\end{align*}
using $\tau_j = \theta_j/\theta_{\leq j}$ and $1 - \tau_j = \theta_{\leq j-1}/\theta_{\leq j}$. The right-hand side is $\prod_{\ell \leq j} \dim[\lambda_\ell]^{\theta_\ell/\theta_{\leq j}}$.
\end{proof}

\begin{proof}[Proof of \cref{th:upper-spectral}]
The bipartitions~$b$ with $\theta(b) = 0$ can be removed from~$B$ without changing~$w_\theta$ or~$F^\theta$, since their projections commute with the others and sum to the identity. So assume $\theta(b) > 0$ for every $b \in B$.

Let $t \neq 0$ and $n \geq 1$, and let~$\lambda$ be a tuple of partitions of~$n$ with $\bigl( \prod_{b \in B} P^b_{\lambda_b} \bigr) t^{\otimes n} \neq 0$. Let $m \geq 1$. Applying \cref{lem:dominance} $m - 1$ times, each time adding~$\lambda$, gives a tuple~$\nu$ of partitions of~$mn$ with $\bigl( \prod_{b \in B} P^b_{\nu_b} \bigr) t^{\otimes mn} \neq 0$ and $\nu_b \unlhd m\lambda_b$ for every $b \in B$, since $\alpha \unlhd \beta$ implies $\alpha + \gamma \unlhd \beta + \gamma$. The tensor~$t^{\otimes mn}$ is invariant under the simultaneous action of~$\mathfrak{S}_{mn}$, so \cref{lem:atleast} with $j = N$ and $y = t^{\otimes mn}$ gives, using $\theta_{\leq N} = 1$ and multiplicativity of~$f_\theta$,
\[
f_\theta(t)^{mn} = f_\theta(t^{\otimes mn}) = g_N\bigl( \merge_{p_N}(t^{\otimes mn}) \bigr) \geq \prod_{b \in B} \dim[\nu_b]^{\theta(b)}.
\]
By \cref{th:schur-weyl}, every~$\nu_b$ has at most $\dim V_{S_b}$ parts, since $P^b_{\nu_b} \neq 0$. Let $d = \max_{b \in B} \dim V_{S_b}$. Since $\nu_b \unlhd m\lambda_b$, the distribution $\nu_b/(mn)$ is majorised by~$\lambda_b/n$, and~$H$ is Schur-concave, so $H(\nu_b/(mn)) \geq H(\lambda_b/n)$. The lower bound in \cref{lem:dim-irrep} for every $b \in B$, together with $\sum_b \theta(b) = 1$, therefore gives
\[
f_\theta(t)^{mn} \geq q_d(mn)^{-1}\, 2^{mn \sum_b \theta(b) H(\lambda_b/n)}.
\]
Taking~$mn$th roots and letting $m \to \infty$ gives $f_\theta(t) \geq 2^{\sum_b \theta(b) H(\lambda_b/n)}$. Taking the supremum over~$n$ and~$\lambda$ gives $f_\theta(t) \geq F^\theta(t)$.

By \cref{th:refine} we have $f_\theta \leq w_\theta$, so \cref{lem:dominated} applied to $h = f_\theta$ gives $f_\theta \leq F^\theta$. Hence $F^\theta = f_\theta \in \Delta$, and $F^\theta \leq w_\theta$. Every $h \in \Delta$ with $h \leq w_\theta$ satisfies $h \leq F^\theta$ by \cref{lem:dominated}, so~$F^\theta$ is the largest such element.
\end{proof}

From \cref{th:upper-spectral} and \cref{lem:dominated} we also get that the non-crossing upper quantum functionals are regularised costs of decompositions weighted by~$w_\theta$.

\begin{corollary}\label{cor:upper-cost}
Let $B \subseteq \bipart([k])$ be non-crossing, let $\theta \in \mathcal{P}(B)$, and let $w_\theta(t) = \prod_{b \in B} \tensorrank_b(t)^{\theta(b)}$. For a tensor~$t$ let~$C(t)$ be the infimum of $\sum_{j=1}^{\ell} w_\theta(t_j)$ over all $\ell \geq 1$ and all $t_1, \ldots, t_\ell$ in the space of~$t$ with $t = t_1 + \cdots + t_\ell$. For $n \geq 1$ and a tuple~$\lambda$ of partitions of~$n$ indexed by the $b \in B$ with $\theta(b) > 0$, let $y_\lambda = \bigl( \prod_b P^b_{\lambda_b} \bigr) t^{\otimes n}$, the product over these~$b$. Then for every tensor~$t$,
\[
F^\theta(t) = \lim_{n \to \infty} C(t^{\otimes n})^{1/n} = \lim_{n \to \infty} \bigl( \max_\lambda w_\theta(y_\lambda) \bigr)^{1/n}.
\]
\end{corollary}
\begin{proof}
Discarding the~$b$ with $\theta(b) = 0$ changes neither~$w_\theta$ nor~$C$, nor~$F^\theta$, which by \cref{th:upper-spectral} is the largest element of~$\Delta$ dominated by~$w_\theta$. So we may assume $\theta(b) > 0$ for every $b \in B$. For $t = 0$ all three expressions are~$0$, so let $t \neq 0$ and $n \geq 1$. By \cref{th:upper-spectral}, $F^\theta \in \Delta$ and $F^\theta \leq w_\theta$. If $t^{\otimes n} = t_1 + \cdots + t_\ell$, then $t^{\otimes n} \leq t_1 \oplus \cdots \oplus t_\ell$, so $F^\theta(t)^n \leq \sum_{\smash{j}} F^\theta(t_j) \leq \sum_{\smash{j}} w_\theta(t_j)$. Hence $F^\theta(t)^n \leq C(t^{\otimes n})$. Conversely, let~$P$ be the polynomial of \cref{lem:dominated} and let $M_n = \max_\lambda w_\theta(y_\lambda)$. By \cref{lem:dominated}, $M_n \leq P(n) F^\theta(t)^n$, and the nonzero~$y_\lambda$ are at most~$P(n)$ tensors with sum~$t^{\otimes n}$, so $F^\theta(t)^n \leq C(t^{\otimes n}) \leq P(n) M_n \leq P(n)^2 F^\theta(t)^n$. Taking~$n$th roots and letting $n \to \infty$ proves the claim.
\end{proof}

For~$B$ consisting of the singleton bipartitions and $\theta(b) > 0$ for every $b \in B$, the right-hand side of \cref{cor:upper-cost}, with~$\limsup$ in place of~$\lim$, is the isotypic interpolation of Van den Berg \cite[Definition~5.1]{vandenberg2023asymptotic}, and its equality with~$F^\theta$ follows from \cite[Theorems~4.36 and~5.4]{vandenberg2023asymptotic}.

\subsection{Graph tensors are \texorpdfstring{$S$}{S}-irreducible for every \texorpdfstring{$S$}{S}}\label{sec:upper-graph-tensors}
In this subsection we will show that the graph tensors of \cref{def:graph-tensor} are $S$-irreducible, as defined in \cref{sec:upper-merge}, for every nonempty $S \subsetneq [k]$, and compute the non-crossing upper quantum functionals on them. 

For a graph~$\mathsf{H}$ with vertex set~$[k]$ and edge set $\mathsf{E}$, and any set $S \subseteq [k]$, let~$\partial S \subseteq \mathsf{E}$ be the set of edges of~$\mathsf{H}$ with exactly one end in~$S$.

\begin{lemma}\label{lem:graph-flattening}
Let~$\mathsf{H}$ be a graph with vertex set~$[k]$ and edge set~$\mathsf{E}$, and let $n \in \NN^{\mathsf{E}}$ with positive entries. Then $\tensorrank_b(\langle \mathsf{H}; n \rangle) = \prod_{\mathsf{e} \in \partial S} n_{\mathsf{e}}$ for every $b = \{S, [k] \setminus S\} \in \bipart([k])$.
\end{lemma}
\begin{proof}
By \cref{def:graph-tensor}, collecting the two factors of each edge identifies $\langle \mathsf{H}; n \rangle$ with $\bigotimes_{\mathsf{e} \in \mathsf{E}} z_{\mathsf{e}}$, where $z_{\mathsf{e}} = \sum_{x=1}^{n_{\mathsf{e}}} e_x \otimes e_x \in \CC^{n_{\mathsf{e}}} \otimes \CC^{n_{\mathsf{e}}}$. The flattening for~$b$ is then the tensor product over $\mathsf{e} \in \mathsf{E}$ of the flattenings of the~$z_{\mathsf{e}}$. For $\mathsf{e} \in \partial S$ this is the $n_{\mathsf{e}} \times n_{\mathsf{e}}$ identity matrix. For the other edges both factors of~$z_{\mathsf{e}}$ lie on the same side, so the flattening of~$z_{\mathsf{e}}$ has rank~$1$. Ranks multiply under tensor products of matrices.
\end{proof}

\begin{proposition}\label{prop:graph-in-AC}
Let~$\mathsf{H}$ be a graph with vertex set $[k]$ and edge set~$\mathsf{E}$, let $n \in \NN^{\mathsf{E}}$ with positive entries, and let $\langle \mathsf{H}; n \rangle$ be its complex graph tensor. Then $\langle \mathsf{H}; n \rangle \in \mathcal{A}_S$ for every nonempty $S \subsetneq [k]$, and $\langle \mathsf{H}; n \rangle \in \mathcal{B}$.
\end{proposition}

Membership in~$\mathcal{B}$ gives membership in~$\mathcal{A}_S$ if~$S$ or $[k] \setminus S$ is a singleton, since the supports on complementary sets are dual representations, but not for every~$S$. For example, with~$B$ and~$\theta$ as in \cref{ex:not-B-good}, the tensor $a \in \mathcal{B}$ there does not lie in both $\mathcal{A}_{\{1,2\}}$ and~$\mathcal{A}_{\{3,4\}}$: otherwise \cref{th:refine} would give $f_\theta(a) = w_\theta(a)$, whereas $f_\theta(a) \leq f_\theta(a_1) + f_\theta(a_2) \leq w_\theta(a_1) + w_\theta(a_2) < w_\theta(a)$. In the proof below, one group serves every~$S$, and the case $S = \{i\}$ gives membership in~$\mathcal{B}$.
\begin{proof}
By \cref{def:graph-tensor}, $V_i = \bigotimes_{\mathsf{e} \ni i} \CC^{n_{\mathsf{e}}}$, and collecting the two factors of each edge gives an isomorphism from $V_1 \otimes \cdots \otimes V_k$ to $\bigotimes_{\mathsf{e} \in \mathsf{E}} (\CC^{n_{\mathsf{e}}} \otimes \CC^{n_{\mathsf{e}}})$, under which $\langle \mathsf{H}; n \rangle = \bigotimes_{\mathsf{e} \in \mathsf{E}} z_{\mathsf{e}}$ with $z_{\mathsf{e}} = \sum_{x=1}^{n_{\mathsf{e}}} e_x \otimes e_x$.

Let $S \subseteq [k]$ be nonempty with $S \neq [k]$, and let~$\mathsf{E}_S$ be the set of edges with both ends in~$S$. For $\mathsf{e} \in \mathsf{E}_S$ both factors of $\CC^{n_{\mathsf{e}}} \otimes \CC^{n_{\mathsf{e}}}$ lie on legs in~$S$, for $\mathsf{e} \in \partial S$ exactly one of them does, and for the remaining edges neither does, so
\[
\supp_S(\langle \mathsf{H}; n \rangle) = \Bigl( \bigotimes_{\mathsf{e} \in \mathsf{E}_S} \CC z_{\mathsf{e}} \Bigr) \otimes \bigotimes_{\mathsf{e} \in \partial S} \CC^{n_{\mathsf{e}}}.
\]
Let~$G$ be the group of tuples $g = (g_{\mathsf{e}})_{\mathsf{e} \in \mathsf{E}}$ in which every~$g_{\mathsf{e}}$ is an $n_{\mathsf{e}} \times n_{\mathsf{e}}$ signed permutation matrix, and let $\rho_i : G \to \GL(V_i)$ map~$g$ to $\bigotimes_{\mathsf{e} \ni i} g_{\mathsf{e}}$. Every $\mathsf{e} \in \mathsf{E}$ has two ends, so the simultaneous action of the~$\rho_i$ maps $\langle \mathsf{H}; n \rangle$ to $\bigotimes_{\mathsf{e} \in \mathsf{E}} (g_{\mathsf{e}} \otimes g_{\mathsf{e}}) z_{\mathsf{e}} = \langle \mathsf{H}; n \rangle$. On $\supp_S(\langle \mathsf{H}; n \rangle)$ the factors of~$G$ indexed by edges outside $\partial S$ act trivially and the factor indexed by $\mathsf{e} \in \partial S$ acts on its~$\CC^{n_{\mathsf{e}}}$ by the signed permutation matrices, which act irreducibly by \cref{lem:monomial-irred}. A tensor product of irreducible complex representations of distinct direct factors is irreducible, so the restriction of $\bigotimes_{i \in S} \rho_i$ to $\supp_S(\langle \mathsf{H}; n \rangle)$ is irreducible. Hence $\langle \mathsf{H}; n \rangle \in \mathcal{A}_S$ for every nonempty $S \neq [k]$. Taking $S = \{i\}$ in the formula for $\supp_S(\langle \mathsf{H}; n \rangle)$ gives $\supp_{\{i\}}(\langle \mathsf{H}; n \rangle) = V_i$ for every $i \in [k]$. So $\langle \mathsf{H}; n \rangle$ is concise. The same group works for all~$S$ at once, so in particular $\langle \mathsf{H}; n \rangle \in \mathcal{B}$.
\end{proof}

\begin{corollary}\label{cor:upper-graph-values}
Let~$\mathsf{H}$ be a graph with vertex set~$[k]$ and edge set~$\mathsf{E}$, let $n \in \NN^{\mathsf{E}}$ with positive entries, and let $\langle \mathsf{H}; n \rangle$ be its complex graph tensor. Let $B \subseteq \bipart([k])$ be non-crossing, let $\theta \in \mathcal{P}(B)$, and for $b \in B$ let~$S_b$ be the part of~$b$ not containing~$k$. Then the non-crossing upper quantum functional~$F^\theta$ and the repeated interpolant~$f_\theta$ for~$B$ and~$\theta$ satisfy
\[
F^\theta(\langle \mathsf{H}; n \rangle) = f_\theta(\langle \mathsf{H}; n \rangle) = \prod_{b \in B} \Bigl( \prod_{\mathsf{e} \in \partial S_b} n_{\mathsf{e}} \Bigr)^{\theta(b)}.
\]
\end{corollary}
\begin{proof}
Let $w_\theta = \prod_{b \in B} \tensorrank_b^{\theta(b)}$. By \cref{lem:graph-flattening}, $w_\theta(\langle \mathsf{H}; n \rangle) = \prod_{b \in B} \bigl( \prod_{\mathsf{e} \in \partial S_b} n_{\mathsf{e}} \bigr)^{\theta(b)}$. By \cref{prop:graph-in-AC}, $\langle \mathsf{H}; n \rangle \in \mathcal{A}_{S_b}$ for every $b \in B$, so \cref{th:refine} gives $f_\theta(\langle \mathsf{H}; n \rangle) = w_\theta(\langle \mathsf{H}; n \rangle)$. Finally, $F^\theta = f_\theta$ by \cref{th:upper-spectral}.
\end{proof}

\section{Graphs, directed graphs and hypergraphs}

So far we have applied the method of \cref{sec:dec} to semirings of tensors. To illustrate it in other settings, we now apply it to graphs, to directed graphs and to uniform hypergraphs. In each subsection, $\mathcal{S}$ is the semiring defined there and~$\Delta$ is its asymptotic spectrum.

\subsection{Graphs}\label{sec:graphs}

In this subsection we will prove that the asymptotic spectrum of graphs is path-connected and contractible, and show that our interpolation coincides with that of Vrana \cite{MR4357434}.

A \emph{graph} $s$ consists of a finite set~$\mathsf{V}(s)$ and a set~$\mathsf{E}(s)$ of $2$-element subsets of~$\mathsf{V}(s)$, the edges of~$s$, and two vertices $u, v$ are adjacent if $\{u, v\}$ is an edge. For $U \subseteq \mathsf{V}(s)$, the induced subgraph $s[U]$ has vertex set~$U$ and the edges of~$s$ contained in~$U$. An automorphism of~$s$ is a permutation of~$\mathsf{V}(s)$ that maps edges to edges, and~$s$ is \emph{vertex-transitive} if its automorphisms act transitively on~$\mathsf{V}(s)$. 

Let~$\mathcal{S}$ be the semiring of finite graphs up to isomorphism, with disjoint union~$\sqcup$ and the following strong product~$\boxtimes$. The vertex set of $s \boxtimes t$ is $\mathsf{V}(s) \times \mathsf{V}(t)$, and distinct $(u_1,v_1), (u_2,v_2) \in \mathsf{V}(s) \times \mathsf{V}(t)$ are adjacent if and only if $(u_1 = u_2 \text{ or } \{u_1,u_2\} \in \mathsf{E}(s))$ and $(v_1 = v_2 \text{ or } \{v_1,v_2\} \in \mathsf{E}(t))$. For graphs~$s$ and~$t$, a \emph{cohomomorphism} from~$s$ to~$t$ is a map $\varphi : \mathsf{V}(s) \to \mathsf{V}(t)$ such that for all $u, v \in \mathsf{V}(s)$, if $u \neq v$ and $u, v$ are non-adjacent, then $\varphi(u) \neq \varphi(v)$ and $\varphi(u), \varphi(v)$ are non-adjacent. The \emph{cohomomorphism preorder} on~$\mathcal{S}$ is given by $s \leq t$ if there is a cohomomorphism from~$s$ to~$t$. The integer $m \in \mathcal{S}$ is the edgeless graph on~$m$ vertices. This is a Strassen semiring \cite{MR4039606}. Its zero is the empty graph, which has no vertices, and we call a graph nonempty if it has at least one vertex. For a graph~$s$, the subrank~$\subrank(s)$ is the independence number of~$s$, the largest number of pairwise non-adjacent vertices. The rank~$\tensorrank(s)$ is the clique cover number of~$s$, the smallest number of cliques that cover~$\mathsf{V}(s)$. So the Shannon capacity of~$s$ \cite{MR0089131} is
\[
\asympsubrank(s) = \lim_{n \to \infty} \subrank(s^{\boxtimes n})^{1/n}.
\]
McEliece and Posner \cite[Theorem~1]{MR403804} (see also \cite[Theorem~67.17]{MR1956925}) proved that the fractional clique cover number of~$s$ is
\[
\asymprank(s) = \lim_{n \to \infty} \tensorrank(s^{\boxtimes n})^{1/n}.
\] 

Let $\mathcal{A} \subseteq \mathcal{S}$ consist of the nonempty vertex-transitive graphs.

Following Csisz\'ar and K\"orner \cite[Definition~2.1]{MR2839250}, for a nonempty graph~$s$ and $n \geq 1$, the \emph{type classes} of $\mathsf{V}(s)^n$ are the sets of sequences in which every vertex of~$s$ occurs a prescribed number of times. They partition $\mathsf{V}(s)^n$ and are indexed by the vectors $(c_v)_{v \in \mathsf{V}(s)}$ of nonnegative integers with $\sum_{v \in \mathsf{V}(s)} c_v = n$, so the number of type classes is at most $(n+1)^{\smash{|\mathsf{V}(s)|}}$ \cite[Lemma~2.2]{MR2839250}.

\begin{lemma}\label{lem:graph-types}
For every nonempty graph~$s$ and every $n \geq 1$, the induced subgraphs $s^{\boxtimes n}[U]$, one for each type class~$U$ of $\mathsf{V}(s)^n$, are vertex-transitive and give an $\mathcal{A}$-decomposition of~$s^{\boxtimes n}$ of length at most $(n+1)^{|\mathsf{V}(s)|}$.
\end{lemma}
\begin{proof}
Coordinate permutations preserve adjacency and act transitively on each type class, so the induced subgraph $s^{\boxtimes n}[U]$ is vertex-transitive for every type class~$U$, and $s^{\boxtimes n}[U] \leq s^{\boxtimes n}$. Mapping each vertex of~$s^{\boxtimes n}$ to its copy in its own type class is a cohomomorphism, so $s^{\boxtimes n} \leq \bigsqcup_U s^{\boxtimes n}[U]$. There are at most $(n+1)^{|\mathsf{V}(s)|}$ type classes.
\end{proof}

\begin{lemma}\label{lem:graph-decomp}
The semiring~$\mathcal{S}$ is $\mathcal{A}$-decomposable.
\end{lemma}
\begin{proof}
A strong product of nonempty vertex-transitive graphs is nonempty and vertex-transitive, and the nonempty edgeless graphs are vertex-transitive, so~$\mathcal{A}$ is multiplicatively closed and contains every positive integer. Every nonempty graph~$s$ satisfies $1 \leq s$, so property~(\ref{it:dec-finite}) of \cref{def:decomposable} follows. Let $s \in \mathcal{S} \setminus \{0\}$. By \cref{lem:graph-types}, $1 \leq \declength_{\mathcal{A}}(s^{\boxtimes n}) \leq (n+1)^{|\mathsf{V}(s)|}$ for every $n \geq 1$. Taking~$n$th roots and letting $n \to \infty$ gives $\asympdeclength_{\mathcal{A}}(s) = 1$.
\end{proof}

The proof of the next reverse inequality, and its adaptations in \cref{sec:digraphs} and \cref{sec:hypergraphs}, use the following covering lemma.

\begin{lemma}\label{lem:covering-automorphisms}
Let a finite group~$\Gamma$ act transitively on a finite set~$X$, let $U \subseteq X$ be nonempty, and let $N = \lfloor |X|\, |U|^{-1} \ln |X| \rfloor + 1$. Then there are $\sigma_1, \ldots, \sigma_N \in \Gamma$ with $X = \sigma_1 U \cup \cdots \cup \sigma_N U$. Moreover, for such $\sigma_1, \ldots, \sigma_N$ and every map $j : X \to [N]$ with $v \in \sigma_{j(v)} U$ for every $v \in X$, the map $\varphi : X \to [N] \times U$ given by $\varphi(v) = (j(v), \sigma_{j(v)}^{\smash[t]{-1}} v)$ is injective.
\end{lemma}
\begin{proof}
Choose $\sigma_1, \ldots, \sigma_N$ independently and uniformly at random from~$\Gamma$. Since~$\Gamma$ acts transitively on~$X$, for every $v \in X$ and $j \in [N]$ the element~$\sigma_j^{-1} v$ is uniformly distributed on~$X$, so $v \notin \sigma_j U$ with probability $1 - |U|/|X|$. So the probability that some element of~$X$ lies in none of the~$\sigma_j U$ is at most $|X| (1 - |U|/|X|)^N < |X|\, e^{-N |U|/|X|} \leq 1$, by the choice of~$N$. If $\varphi(u) = \varphi(v)$, then $j(u) = j(v) = j$ for some~$j$ and $\sigma_j^{-1} u = \sigma_j^{-1} v$, so $u = v$.
\end{proof}

For every graph~$a$ and every $U \subseteq \mathsf{V}(a)$, the inclusion of~$U$ into~$\mathsf{V}(a)$ is a cohomomorphism from $a[U]$ to~$a$, so every $f \in \Delta$ satisfies 
\[
f(a[U]) \leq f(a).
\]
For vertex-transitive~$a$, we have the following reverse inequality.

\begin{lemma}[{\cite[Lemma~2.15]{deboer2024asymptoticspectrumdistancegraph}}]\label{lem:graph-reverse}
Let~$a$ be a nonempty vertex-transitive graph and let $U \subseteq \mathsf{V}(a)$. Then for every $f \in \Delta$ we have
\[
\frac{|U|}{|\mathsf{V}(a)|} f(a) \leq f(a[U]).
\]
\end{lemma}
\begin{proof}
We recall the proof, because \cref{sec:digraphs} and \cref{sec:hypergraphs} adapt it. The case $U = \emptyset$ is immediate, so let $U \neq \emptyset$. The automorphism group of~$a$ acts transitively on~$\mathsf{V}(a)$. Let~$N$, the automorphisms $\sigma_1, \ldots, \sigma_N$ of~$a$, a map $j : \mathsf{V}(a) \to [N]$ and the map~$\varphi$ be as in \cref{lem:covering-automorphisms}, and let~$b$ be the disjoint union of~$N$ copies of $a[U]$, with vertex set $[N] \times U$. We first prove the covering relation $a \leq b$ of \cite[Lemma~3.1]{MR4357434}. The map~$\varphi$ is a cohomomorphism from~$a$ to~$b$: distinct non-adjacent vertices $u, v$ with $j(u) = j(v) = j$ are mapped to the distinct non-adjacent vertices~$\sigma_j^{-1}(u)$ and~$\sigma_j^{-1}(v)$ of the copy~$j$, and if $j(u) \neq j(v)$ they are mapped to different copies. So $f(a) \leq f(b) = N f(a[U])$.

We now apply this bound to strong powers. For every $m \geq 1$ the graph~$a^{\boxtimes m}$ is vertex-transitive and $a^{\boxtimes m}[U^m] = a[U]^{\boxtimes m}$, so the same bound for~$a^{\boxtimes m}$ and~$U^m$ in place of~$a$ and~$U$ gives
\[
f(a)^m \leq \Bigl( \frac{|\mathsf{V}(a)|^m}{|U|^m}\, m \ln |\mathsf{V}(a)| + 1 \Bigr) f(a[U])^m.
\]
The~$m$th root of the factor in parentheses tends to $|\mathsf{V}(a)|/|U|$, so taking~$m$th roots and letting $m \to \infty$ gives the claim.
\end{proof}

\begin{lemma}\label{lem:graph-cover}
Every $g_1, \ldots, g_r \in \Delta$ have the $\mathcal{A}$-cover property with $\kappa_a(n) = 1$.
\end{lemma}
\begin{proof}
Let $a \in \mathcal{A}$ with $a \leq s_1 \sqcup \cdots \sqcup s_r$, and let $\varphi : \mathsf{V}(a) \to \mathsf{V}(s_1) \sqcup \cdots \sqcup \mathsf{V}(s_r)$ be a cohomomorphism, and let $U_i = \varphi^{-1}(\mathsf{V}(s_i))$. Then $a[U_i] \leq s_i$, and the sets $U_1, \ldots, U_r$ partition~$\mathsf{V}(a)$. Let $i \in [r]$. Monotonicity of~$g_i$ gives $g_i(s_i) \geq g_i(a[U_i])$, and \cref{lem:graph-reverse} with $f = g_i$ and $U = U_i$ gives $g_i(a[U_i]) \geq (|U_i|/|\mathsf{V}(a)|)\, g_i(a)$. Dividing by~$g_i(a)$ and summing over~$i$ gives
\[
\sum_{i=1}^r \frac{g_i(s_i)}{g_i(a)} \geq \sum_{i=1}^r \frac{|U_i|}{|\mathsf{V}(a)|} = 1.
\]
Applying this to $a^{\boxtimes n} \in \mathcal{A}$ gives the claim.
\end{proof}

A \emph{probabilistic graph} is a pair~$(s,P)$ of a nonempty graph~$s$ and a probability distribution~$P$ on~$\mathsf{V}(s)$. The elements of~$\Delta$ correspond bijectively to functions on probabilistic graphs satisfying a list of affine conditions \cite[Theorems~1.1 and 1.2]{MR4357434}. The set of such functions is convex, so for all $f_0, f_1 \in \Delta$ the bijection gives a family in~$\Delta$ interpolating between~$f_0$ and~$f_1$, and for every fixed nonempty graph~$s$ the map $f \mapsto \log f(s)$ is convex along that family \cite[Section~4.5]{MR4357434}. Path-connectedness of~$\Delta$ follows from this family together with a continuity check. The following corollary and the paragraph after it instead obtain it from the method of \cref{sec:dec}. 

\begin{corollary}\label{cor:graph-paths}
Let $f_1 \in \Delta$ and $s_1, \ldots, s_q \in \mathcal{S}$. Then every point of $\Delta(s_1, \ldots, s_q)$ is joined by a path to $(f_1(s_1), \ldots, f_1(s_q))$. In particular $\Delta(s_1, \ldots, s_q)$ is path-connected. Moreover, for every nonempty graph~$s$,
\[
\Delta(s) = [\asympsubrank(s), \asymprank(s)].
\]
\end{corollary}
\begin{proof}
By \cref{lem:graph-decomp} and \cref{lem:graph-cover}, the hypotheses of \cref{cor:path-connected} hold for every element $f_1 \in \Delta$. Every nonempty graph~$s$ satisfies $1 \leq s$, which is the additional assumption of \cref{sec:asymptotic-rank}. Let~$s$ be a nonempty graph. If~$s$ has two distinct non-adjacent vertices, then mapping the two vertices of~$2$ to them shows $2 \leq s$. Otherwise, every two distinct vertices of~$s$ are adjacent, so the constant map from~$\mathsf{V}(s)$ to the vertex of~$1$ is a cohomomorphism. Then $s \leq 1 \leq s$, so $f(s) = 1$ for every $f \in \Delta$. In both cases~$s$ is gapped, so \cref{cor:interval} gives the last claim.
\end{proof}

\cref{lem:graph-decomp} and \cref{lem:graph-cover} also give the hypotheses of \cref{cor:contractible} and \cref{cor:convex}. So for every $f_1 \in \Delta$ the map $\Delta \times [0,1] \to \Delta : (f,\lambda) \mapsto F^f_\lambda$ is continuous, equals the identity at $\lambda = 0$, and is constant equal to~$f_1$ at $\lambda = 1$. In particular~$\Delta$ is contractible. Moreover, taking logarithms of the values on the nonempty vertex-transitive graphs maps~$\Delta$ homeomorphically onto a compact convex set.

We now compare~$F^{f_0}_\lambda$ with the interpolation of \cite{MR4357434}, which is defined through the notion of probabilistic refinement. Let~$s$ be a nonempty graph and let~$P$ be a probability distribution on~$\mathsf{V}(s)$. For $n \geq 1$ and $x \in \mathsf{V}(s)^n$, the \emph{empirical distribution}~$Q_x$ of~$x$ is given by $Q_x(v) = |\{i \in [n] : x_i = v\}|/n$ for $v \in \mathsf{V}(s)$. For $\varepsilon > 0$ let
\[
T_n(P,\varepsilon) \coloneqq \Bigl\{ x \in \mathsf{V}(s)^n : \frac{1}{2} \sum_{v \in \mathsf{V}(s)} |Q_x(v) - P(v)| < \varepsilon \Bigr\},
\]
a union of type classes. For $f \in \Delta$ the limit
\[
f(s,P) \coloneqq \lim_{\varepsilon \to 0} \limsup_{n \to \infty} f\bigl( s^{\boxtimes n}[T_n(P,\varepsilon)] \bigr)^{1/n}
\]
exists, and $f(s) = \max_P f(s,P)$ \cite[Theorem~1.1]{MR4357434}. For $f_0, f_1 \in \Delta$ and $\lambda \in [0,1]$ the interpolation~$f_\lambda$ is given by $\log f_\lambda(s,P) = (1-\lambda) \log f_0(s,P) + \lambda \log f_1(s,P)$ and $f_\lambda(s) = \max_P f_\lambda(s,P)$ for nonempty~$s$, with $f_\lambda(0) = 0$ \cite[Section~4.5]{MR4357434}.

For $f_0, f_1 \in \Delta$ and $\lambda \in [0,1]$, let $w_\lambda(s) = f_0(s)^{1-\lambda} f_1(s)^{\lambda}$ for $s \in \mathcal{S}$.

\begin{proposition}\label{prop:graph-vrana}
Let $f_0, f_1 \in \Delta$ and $\lambda \in [0,1]$. Then $f_\lambda = F_{w_\lambda}$.
\end{proposition}
\begin{proof}
Let~$a$ be a nonempty vertex-transitive graph and let~$U$ be the uniform distribution on~$\mathsf{V}(a)$. We first show that $f(a,U) = f(a)$ for every $f \in \Delta$. The map $P \mapsto \log f(a,P)$ is concave \cite[Theorem~1.1]{MR4357434}, and by its definition it is invariant under the automorphisms of~$a$. The automorphisms act transitively on~$\mathsf{V}(a)$, so averaging the distributions $P \circ \sigma$ over the automorphisms~$\sigma$ of~$a$ gives~$U$. Hence $f(a,U) \geq f(a,P)$ for every~$P$, and $f(a,U) = \max_P f(a,P) = f(a)$.

Now $\log f_\lambda(a,U) = (1-\lambda) \log f_0(a) + \lambda \log f_1(a)$, so $f_\lambda(a) \geq w_\lambda(a)$. For every~$P$ we have $f_0(a,P) \leq f_0(a)$ and $f_1(a,P) \leq f_1(a)$, so $f_\lambda(a,P) \leq w_\lambda(a)$. Hence $f_\lambda(a) = w_\lambda(a)$, so $f_\lambda = w_\lambda$ on~$\mathcal{A}$. By \cref{lem:graph-cover} and \cref{lem:A-cover} the function~$w_\lambda$ is $\mathcal{A}$-good. By \cite[Theorem~1.2]{MR4357434} we have $f_\lambda \in \Delta$. By \cref{th:w-good-spec}~(\ref{it:Fw-unique}), $F_{w_\lambda}$ is the unique element of~$\Delta$ agreeing with~$w_\lambda$ on~$\mathcal{A}$, so $f_\lambda = F_{w_\lambda}$.
\end{proof}

By \cref{prop:graph-vrana} and \cref{th:interpolation}~(\ref{it:int-family}), $\log f_\lambda(a) = (1-\lambda)\log f_0(a) + \lambda\log f_1(a)$ for every nonempty vertex-transitive graph~$a$. So the convex structure on~$\Delta$ given by \cref{cor:convex} is that of the interpolation~$f_\lambda$.

Since $f_\lambda = F_{w_\lambda}$ by \cref{prop:graph-vrana}, and~$w_\lambda$ is monotone and multiplicative, \cref{th:w-good-spec}~(\ref{it:Fw-largest}) and (\ref{it:Fw-below}) show that~$f_\lambda$ is the largest element of~$\Delta$ dominated by~$w_\lambda$. Let~$s$ be a nonempty graph, and for $n \geq 1$ consider the $\mathcal{A}$-decomposition of~$s^{\boxtimes n}$ of \cref{lem:graph-types}, which consists of the graphs $s^{\boxtimes n}[U]$, one for each type class~$U$. Its length~$c_n$ is at most $(n+1)^{|\mathsf{V}(s)|}$, so \cref{th:w-good-spec}~(\ref{it:Fw-limit}) gives $\max_U w_\lambda(s^{\boxtimes n}[U]) \leq f_\lambda(s)^n \leq c_n \max_U w_\lambda(s^{\boxtimes n}[U])$. Hence
\[
f_\lambda(s) = \lim_{n \to \infty} \Bigl( \max_U w_\lambda(s^{\boxtimes n}[U]) \Bigr)^{1/n}.
\]

\subsection{Directed graphs}\label{sec:digraphs}

In this subsection we will adapt \cref{sec:graphs} to directed graphs, and prove that their asymptotic spectrum is path-connected and contractible. 

A \emph{directed graph} $s$ consists of a finite set~$\mathsf{V}(s)$ and a set~$\mathsf{E}(s)$ of pairs $(u,v) \in \mathsf{V}(s) \times \mathsf{V}(s)$ with $u \neq v$, the arcs of~$s$. The other pairs $(u,v)$ with $u \neq v$ are the non-arcs of~$s$. Induced subgraphs $s[U]$, automorphisms and vertex-transitivity are defined as for graphs. In this subsection, let~$\mathcal{S}$ be the semiring of finite directed graphs up to isomorphism, with disjoint union~$\sqcup$ and the following strong product~$\boxtimes$. The vertex set of $s \boxtimes t$ is $\mathsf{V}(s) \times \mathsf{V}(t)$, and for distinct $(u_1,v_1), (u_2,v_2) \in \mathsf{V}(s) \times \mathsf{V}(t)$ the pair $((u_1,v_1),(u_2,v_2))$ is an arc of $s \boxtimes t$ if and only if $(u_1 = u_2 \text{ or } (u_1,u_2) \in \mathsf{E}(s))$ and $(v_1 = v_2 \text{ or } (v_1,v_2) \in \mathsf{E}(t))$. For $s, t \in \mathcal{S}$ we write $s \leq t$ if there is a map $\varphi : \mathsf{V}(s) \to \mathsf{V}(t)$ that maps every non-arc of~$s$ to a non-arc of~$t$. As for graphs, $\varphi$ need not be injective. The integer $m \in \mathcal{S}$ is the directed graph on~$m$ vertices without arcs, and~$0$ is the empty directed graph. This is the Strassen semiring of \cite[Example~4.17]{wigderson2022asymptotic}. For a directed graph~$s$, the subrank~$\subrank(s)$ is the largest number of vertices of~$s$ no two of which are joined by an arc. The complement of~$s$ is the directed graph on~$\mathsf{V}(s)$ whose arcs are the non-arcs of~$s$. Gargano, K\"orner and Vaccaro \cite{MR1184988} introduced the Sperner capacity of a directed graph. For every $n \geq 1$, a set of vertices of~$s^{\boxtimes n}$ has no two elements joined by an arc exactly when it is a symmetric clique \cite[Definition~11.16]{MR2839250} in the $n$th co-normal power \cite[Definition~11.17]{MR2839250} of the complement of~$s$. So, by \cite[Definition~11.18]{MR2839250}, the Sperner capacity of the complement of~$s$ is
\[
\log \asympsubrank(s) = \lim_{n \to \infty} \frac{1}{n} \log \subrank(s^{\boxtimes n}).
\]
For a directed graph~$s$, let~$\mu(s)$ be the graph on~$\mathsf{V}(s)$ whose edges are the pairs $\{u,v\}$ for which both $(u,v)$ and $(v,u)$ are arcs of~$s$. The rank~$\tensorrank(s)$ is the clique cover number of~$\mu(s)$. For directed graphs~$s$ and~$t$, a pair of distinct vertices of $s \boxtimes t$ is joined by arcs in both directions exactly when in each coordinate the entries are equal or joined by arcs in both directions, so $\mu(s \boxtimes t) = \mu(s) \boxtimes \mu(t)$. Hence $\tensorrank(s^{\boxtimes n})$ is the clique cover number of~$\mu(s)^{\boxtimes n}$ for every $n \geq 1$. So, by the theorem of McEliece and Posner recalled in \cref{sec:graphs}, the fractional clique cover number of~$\mu(s)$ is
\[
\asymprank(s) = \lim_{n \to \infty} \tensorrank(s^{\boxtimes n})^{1/n}.
\]

Let $\mathcal{A} \subseteq \mathcal{S}$ consist of the nonempty vertex-transitive directed graphs.

\begin{lemma}\label{lem:digraph}
The semiring~$\mathcal{S}$ is $\mathcal{A}$-decomposable, and every $g_1, \ldots, g_r \in \Delta$ have the $\mathcal{A}$-cover property with $\kappa_a(n) = 1$.
\end{lemma}
\begin{proof}
The proofs of \cref{lem:graph-types}, \cref{lem:graph-decomp}, \cref{lem:graph-reverse} and \cref{lem:graph-cover} apply, with non-arcs in place of pairs of distinct non-adjacent vertices. For distinct vertices $x, y \in \mathsf{V}(s)^n$, the pair $(x,y)$ is an arc of~$s^{\boxtimes n}$ if and only if for every $i \in [n]$ we have $x_i = y_i$ or $(x_i,y_i) \in \mathsf{E}(s)$. So every permutation of the~$n$ coordinates is an automorphism of~$s^{\boxtimes n}$. The map~$\varphi$ of \cref{lem:covering-automorphisms}, for the automorphism group of~$a$, maps a non-arc $(u,v)$ with $j(u) = j(v) = j$ to the non-arc $(\sigma_j^{\smash[t]{-1}}(u), \sigma_j^{\smash[t]{-1}}(v))$ of $a[U]$ in the copy~$j$, and a non-arc with $j(u) \neq j(v)$ to a pair in different copies.
\end{proof}

\begin{corollary}\label{cor:digraph-paths}
Let $f_1 \in \Delta$ and $s_1, \ldots, s_q \in \mathcal{S}$. Then every point of $\Delta(s_1, \ldots, s_q)$ is joined by a path to $(f_1(s_1), \ldots, f_1(s_q))$, and~$\Delta$ is contractible. Moreover, for every nonempty directed graph~$s$,
\[
\Delta(s) = [\asympsubrank(s), \asymprank(s)].
\]
\end{corollary}
\begin{proof}
By \cref{lem:digraph}, the hypotheses of \cref{cor:path-connected} and \cref{cor:contractible} hold for every element $f_1 \in \Delta$. Every nonempty directed graph~$s$ satisfies $1 \leq s$, which is the additional assumption of \cref{sec:asymptotic-rank}. Let~$s$ be a nonempty directed graph. If~$s$ has a non-arc~$(u,v)$, then $(u,v)$ and $(v,u)$ are distinct vertices of~$s^{\boxtimes 2}$ with no arc between them, so mapping the two vertices of~$2$ to them shows $2 \leq s^{\boxtimes 2}$. Otherwise, every pair of distinct vertices of~$s$ is an arc, so the constant map from~$\mathsf{V}(s)$ to the vertex of~$1$ shows $s \leq 1$. Then $s \leq 1 \leq s$, so $f(s) = 1$ for every $f \in \Delta$. In both cases~$s$ is gapped, so \cref{cor:interval} gives the last claim.
\end{proof}

By \cref{lem:digraph} and \cref{cor:convex}, taking logarithms of the values on the nonempty vertex-transitive directed graphs maps~$\Delta$ homeomorphically onto a compact convex set.

The asymptotic spectrum of graphs is a retract of the asymptotic spectrum of directed graphs, as follows. Let~$\beta$ map a graph~$s$ to the directed graph on~$\mathsf{V}(s)$ with arcs $(u,v)$ and $(v,u)$ for every edge $\{u,v\}$ of~$s$. Then $\mu \circ \beta$ is the identity. The map~$\beta$ is a monotone semiring homomorphism \cite[Example~4.17]{wigderson2022asymptotic}, and so is~$\mu$. Indeed, $\mu$ preserves disjoint unions and strong products. A non-edge of~$\mu(s)$ is a pair $\{u,v\}$ for which $(u,v)$ or $(v,u)$ is a non-arc of~$s$, so a map that sends every non-arc of~$s$ to a non-arc of~$t$ is a cohomomorphism from~$\mu(s)$ to~$\mu(t)$. 

Write~$\Delta_{\mathrm{gr}}$ and~$\Delta_{\mathrm{dir}}$ for the asymptotic spectra of graphs and of directed graphs, and let $\iota : \Delta_{\mathrm{gr}} \to \Delta_{\mathrm{dir}}$ and $r : \Delta_{\mathrm{dir}} \to \Delta_{\mathrm{gr}}$ be the maps $\iota(f) = f \circ \mu$ and $r(g) = g \circ \beta$. Both are continuous in the topology of pointwise convergence, and $r(\iota(f)) = f \circ \mu \circ \beta = f$ for every $f \in \Delta_{\mathrm{gr}}$, so~$\Delta_{\mathrm{gr}}$ is a retract of~$\Delta_{\mathrm{dir}}$. Since a retract of a contractible space is contractible, the contractibility of the asymptotic spectrum of graphs in \cref{sec:graphs} also follows from \cref{cor:digraph-paths}.

\subsection{Uniform hypergraphs}\label{sec:hypergraphs}

In this subsection we will adapt \cref{sec:graphs} to uniform hypergraphs, and prove that their asymptotic spectrum is path-connected and contractible. 

Fix $k \geq 2$. A \emph{$k$-graph} $s$ consists of a finite set~$\mathsf{V}(s)$ and a set~$\mathsf{E}(s)$ of $k$-element subsets of~$\mathsf{V}(s)$, the edges of~$s$. The other $k$-element subsets of~$\mathsf{V}(s)$ are the non-edges of~$s$. Induced subgraphs $s[U]$, automorphisms and vertex-transitivity are defined as for graphs.

Let~$\mathcal{S}$ be the semiring of finite $k$-graphs up to isomorphism, with disjoint union~$\sqcup$ and the following strong product~$\boxtimes$. The vertex set of $s \boxtimes t$ is $\mathsf{V}(s) \times \mathsf{V}(t)$, and a $k$-element set $\{(u_1,v_1), \ldots, (u_k,v_k)\}$ is an edge of $s \boxtimes t$ if and only if $(u_1 = \cdots = u_k \text{ or } \{u_1, \ldots, u_k\} \in \mathsf{E}(s))$ and $(v_1 = \cdots = v_k \text{ or } \{v_1, \ldots, v_k\} \in \mathsf{E}(t))$. For $s, t \in \mathcal{S}$ we write $s \leq t$ if there is an injective map $\varphi : \mathsf{V}(s) \to \mathsf{V}(t)$ that maps every non-edge of~$s$ to a non-edge of~$t$. The integer $m \in \mathcal{S}$ is the edgeless $k$-graph on~$m$ vertices, and~$0$ is the empty $k$-graph, which has no vertices. This is the Strassen semiring of \cite[Example~2.22]{wigderson2022asymptotic}. Its rank is the number of vertices, and its subrank is the largest size of a set of vertices containing no edge.

Unlike for graphs and directed graphs, the maps in the definition of $\leq$ are injective. This makes $\leq$ compatible with~$\boxtimes$, because an injective map~$\varphi$ preserves the number of distinct elements among any $u_1, \ldots, u_k$. For $k = 2$ the semiring is the semiring of graphs of \cref{sec:graphs}, but the injective preorder has fewer relations than the cohomomorphism preorder of \cref{sec:graphs}.

Let $\mathcal{A} \subseteq \mathcal{S}$ consist of the nonempty vertex-transitive $k$-graphs.

\begin{lemma}\label{lem:hypergraph}
The semiring~$\mathcal{S}$ is $\mathcal{A}$-decomposable, and every $g_1, \ldots, g_r \in \Delta$ have the $\mathcal{A}$-cover property with $\kappa_a(n) = 1$.
\end{lemma}
\begin{proof}
With non-edges in place of pairs of distinct non-adjacent vertices, the proofs of \cref{lem:graph-types}, \cref{lem:graph-decomp}, \cref{lem:graph-reverse} and \cref{lem:graph-cover} apply, with three changes.

First, let $s \in \mathcal{S}$ and $n \geq 1$. By the definition of~$\boxtimes$, for all pairwise distinct $x^1, \ldots, x^k \in \mathsf{V}(s)^n$, the set $\{x^1, \ldots, x^k\}$ is an edge of~$s^{\boxtimes n}$ if and only if for every $i \in [n]$ we have $x^1_i = \cdots = x^k_i$ or $\{x^1_i, \ldots, x^k_i\} \in \mathsf{E}(s)$. Hence every permutation of the~$n$ coordinates is an automorphism of~$s^{\boxtimes n}$. So the type classes induce vertex-transitive $k$-graphs, and $a^{\boxtimes m}[U^m] = a[U]^{\boxtimes m}$ for every $k$-graph~$a$, every $U \subseteq \mathsf{V}(a)$ and every $m \geq 1$.

Second, the maps constructed in the adaptations of \cref{lem:graph-types} and \cref{lem:graph-reverse} are injective. The inclusion of a type class, and the map sending every vertex of~$s^{\boxtimes n}$ to its copy in its own type class, are injective. The map~$\varphi$ of \cref{lem:covering-automorphisms}, for the automorphism group of~$a$, is injective by that lemma, and it maps every non-edge~$S$ of~$a$ to a non-edge. Indeed, if~$j$ is constant on~$S$ with value~$j_0$, then the image of~$S$ is the non-edge~$\sigma_{j_0}^{\smash[t]{-1}}(S)$ of $a[U]$ in the copy~$j_0$, and otherwise the image of~$S$ meets two copies.

Third, in the adaptation of \cref{lem:graph-cover}, let $\varphi : \mathsf{V}(a) \to \mathsf{V}(s_1) \sqcup \cdots \sqcup \mathsf{V}(s_r)$ be an injective map showing $a \leq s_1 \sqcup \cdots \sqcup s_r$, and let $U_i = \varphi^{-1}(\mathsf{V}(s_i))$ for $i \in [r]$. Each restriction $\varphi|_{U_i}$ is injective and maps non-edges to non-edges, so $a[U_i] \leq s_i$.
\end{proof}

\begin{corollary}\label{cor:hypergraph-paths}
Let $f_1 \in \Delta$ and $s_1, \ldots, s_q \in \mathcal{S}$. Then every point of $\Delta(s_1, \ldots, s_q)$ is joined by a path to $(f_1(s_1), \ldots, f_1(s_q))$, and~$\Delta$ is contractible.
\end{corollary}
\begin{proof}
By \cref{lem:hypergraph}, the hypotheses of \cref{cor:path-connected} and \cref{cor:contractible} hold for every element $f_1 \in \Delta$.
\end{proof}

By \cref{lem:hypergraph} and \cref{cor:convex}, taking logarithms of the values on the nonempty vertex-transitive $k$-graphs maps~$\Delta$ homeomorphically onto a compact convex set.

\paragraph{Acknowledgements.} JZ was supported by NWO Vidi grant VI.Vidi.243.195, by ERC Starting Grant 101220349 SPECTRA, and by the Simons Institute for the Theory of Computing, UC Berkeley, as part of the 2025 semester programme Complexity and Linear algebra. PV was supported by the Ministry of Culture and Innovation of Hungary from the National Research, Development and Innovation Fund via the research grants FK~146643 and EXCELLENCE~151342. This work used the Dutch national e-infrastructure with the support of the SURF Cooperative using grant no.~EINF-16525. PV and JZ are grateful for the hospitality of the Erdős Center, Rényi Institute, Budapest.

\paragraph{AI methodology.} The research ideas are by the authors. Large language models Claude (Anthropic), ChatGPT (OpenAI) and gpt-6-astra (OpenAI) were used to assist with working out details of proofs and checking proofs. The authors verified the correctness and originality of all content, including references.

\bibliographystyle{alphaurl}

\begin{thebibliography}{vdBCL{\etalchar{+}}25}

\bibitem[AL25]{MR5048041}
Josh Alman and Baitian Li.
\newblock Kronecker powers, orthogonal vectors, and the asymptotic spectrum.
\newblock In {\em 2025 {IEEE} 66th {A}nnual {S}ymposium on {F}oundations of
  {C}omputer {S}cience---{FOCS} 2025}, pages 1411--1441. IEEE Comput. Soc.
  Press, Los Alamitos, CA, 2025.
\newblock \href {https://doi.org/10.1109/FOCS63196.2025.00074}
  {\path{doi:10.1109/FOCS63196.2025.00074}}.

\bibitem[AL26]{alman2026asymptoticrankspeeduptheorems}
Josh Alman and Baitian Li.
\newblock Asymptotic rank speedup theorems, revisited, 2026.
\newblock \href {https://arxiv.org/abs/2605.21738} {\path{arXiv:2605.21738}}.

\bibitem[Alm19]{alman2018limits}
Josh Alman.
\newblock Limits on the universal method for matrix multiplication.
\newblock In {\em Proceedings of the 34th Computational Complexity Conference
  ({CCC} 2019)}, pages 12:1--12:24, 2019.
\newblock \href {https://arxiv.org/abs/1812.08731} {\path{arXiv:1812.08731}},
  \href {https://doi.org/10.4230/LIPIcs.CCC.2019.12}
  {\path{doi:10.4230/LIPIcs.CCC.2019.12}}.

\bibitem[ALP26]{alman2026edgeasymptoticspectrumtensors}
Josh Alman, Baitian Li, and Kevin Pratt.
\newblock The edge of the asymptotic spectrum of tensors, 2026.
\newblock To appear in FOCS 2026.
\newblock \href {https://arxiv.org/abs/2604.01386} {\path{arXiv:2604.01386}}.

\bibitem[AT18]{MR3786301}
Jeffrey Adams and Olivier Ta\"{\i}bi.
\newblock Galois and {C}artan cohomology of real groups.
\newblock {\em Duke Math. J.}, 167(6):1057--1097, 2018.
\newblock \href {https://arxiv.org/abs/1611.05956} {\path{arXiv:1611.05956}},
  \href {https://doi.org/10.1215/00127094-2017-0052}
  {\path{doi:10.1215/00127094-2017-0052}}.

\bibitem[AvdBZ26]{amanov2026multiplyingmatricespolynomialsexposition}
Alimzhan Amanov, Maxim van~den Berg, and Jeroen Zuiddam.
\newblock Multiplying matrices via polynomials: exposition and barriers, 2026.
\newblock \href {https://arxiv.org/abs/2610.30054} {\path{arXiv:2610.30054}}.

\bibitem[AW18]{MR3899623}
Josh Alman and Virginia~Vassilevska Williams.
\newblock Limits on all known (and some unknown) approaches to matrix
  multiplication.
\newblock In {\em 59th {A}nnual {IEEE} {S}ymposium on {F}oundations of
  {C}omputer {S}cience---{FOCS} 2018}, pages 580--591. IEEE Computer Soc., Los
  Alamitos, CA, 2018.
\newblock \href {https://doi.org/10.1109/FOCS.2018.00061}
  {\path{doi:10.1109/FOCS.2018.00061}}.

\bibitem[BCC{\etalchar{+}}17]{MR3631613}
Jonah Blasiak, Thomas Church, Henry Cohn, Joshua~A. Grochow, Eric Naslund,
  William~F. Sawin, and Chris Umans.
\newblock On cap sets and the group-theoretic approach to matrix
  multiplication.
\newblock {\em Discrete Anal.}, 2017.
\newblock \href {https://arxiv.org/abs/1605.06702} {\path{arXiv:1605.06702}},
  \href {https://doi.org/10.19086/da.1245} {\path{doi:10.19086/da.1245}}.

\bibitem[BCF{\etalchar{+}}26]{botero2026quantumfunctionalshigherordertensors}
Alonso Botero, Matthias Christandl, Thomas~C. Fraser, Itai Leigh, and Harold
  Nieuwboer.
\newblock On quantum functionals for higher-order tensors, 2026.
\newblock \href {https://arxiv.org/abs/2604.18283} {\path{arXiv:2604.18283}}.

\bibitem[BCH{\etalchar{+}}25]{björklund2024chromaticnumber19999ntime}
Andreas Björklund, Radu Curticapean, Thore Husfeldt, Petteri Kaski, and Kevin
  Pratt.
\newblock Fast deterministic chromatic number under the asymptotic rank
  conjecture.
\newblock In {\em Proceedings of the 2025 ACM-SIAM Symposium on Discrete
  Algorithms (SODA)}, pages 2804--2818, 2025.
\newblock \href {https://arxiv.org/abs/2404.04987} {\path{arXiv:2404.04987}},
  \href {https://doi.org/10.1137/1.9781611978322.91}
  {\path{doi:10.1137/1.9781611978322.91}}.

\bibitem[BCL{\etalchar{+}}24]{briet_et_al:LIPIcs.ITCS.2024.20}
Jop Bri\"{e}t, Matthias Christandl, Itai Leigh, Amir Shpilka, and Jeroen
  Zuiddam.
\newblock {Discreteness of Asymptotic Tensor Ranks}.
\newblock In {\em 15th Innovations in Theoretical Computer Science Conference
  (ITCS 2024)}, volume 287, pages 20:1--20:14, 2024.
\newblock \href {https://arxiv.org/abs/2306.01718} {\path{arXiv:2306.01718}},
  \href {https://doi.org/10.4230/LIPIcs.ITCS.2024.20}
  {\path{doi:10.4230/LIPIcs.ITCS.2024.20}}.

\bibitem[BCS97]{burgisser1997algebraic}
Peter B\"{u}rgisser, Michael Clausen, and Mohammad~Amin Shokrollahi.
\newblock {\em Algebraic complexity theory}, volume 315 of {\em Grundlehren der
  mathematischen Wissenschaften}.
\newblock Springer-Verlag, Berlin, 1997.
\newblock \href {https://doi.org/10.1007/978-3-662-03338-8}
  {\path{doi:10.1007/978-3-662-03338-8}}.

\bibitem[BDR24]{https://doi.org/10.48550/arxiv.2212.12219}
Andreas Blatter, Jan Draisma, and Filip Rupniewski.
\newblock Countably many asymptotic tensor ranks.
\newblock {\em Linear and Multilinear Algebra}, 0(0):1--6, 2024.
\newblock \href {https://arxiv.org/abs/2212.12219} {\path{arXiv:2212.12219}},
  \href {https://doi.org/10.1080/03081087.2024.2317906}
  {\path{doi:10.1080/03081087.2024.2317906}}.

\bibitem[BDR25]{bdr}
Andreas Blatter, Jan Draisma, and Filip Rupniewski.
\newblock A tensor restriction theorem over finite fields.
\newblock {\em Compos. Math.}, 161(5):1021--1053, 2025.
\newblock \href {https://arxiv.org/abs/2211.12319} {\path{arXiv:2211.12319}},
  \href {https://doi.org/10.1112/S0010437X25007067}
  {\path{doi:10.1112/S0010437X25007067}}.

\bibitem[BFG{\etalchar{+}}18]{MR3899650}
Peter B\"{u}rgisser, Cole Franks, Ankit Garg, Rafael Oliveira, Michael Walter,
  and Avi Wigderson.
\newblock Efficient algorithms for tensor scaling, quantum marginals, and
  moment polytopes.
\newblock In {\em 59th {A}nnual {IEEE} {S}ymposium on {F}oundations of
  {C}omputer {S}cience---{FOCS} 2018}, pages 883--897. IEEE Computer Soc., Los
  Alamitos, CA, 2018.
\newblock \href {https://doi.org/10.1109/FOCS.2018.00088}
  {\path{doi:10.1109/FOCS.2018.00088}}.

\bibitem[BFG{\etalchar{+}}19]{MR4228202}
Peter B\"{u}rgisser, Cole Franks, Ankit Garg, Rafael Oliveira, Michael Walter,
  and Avi Wigderson.
\newblock Towards a theory of non-commutative optimization: geodesic 1st and
  2nd order methods for moment maps and polytopes.
\newblock In {\em 2019 {IEEE} 60th {A}nnual {S}ymposium on {F}oundations of
  {C}omputer {S}cience}, pages 845--861. IEEE Comput. Soc. Press, Los Alamitos,
  CA, 2019.
\newblock \href {https://doi.org/10.1109/FOCS.2019.00055}
  {\path{doi:10.1109/FOCS.2019.00055}}.

\bibitem[BK24]{10.1145/3618260.3649656}
Andreas Bj\"{o}rklund and Petteri Kaski.
\newblock The asymptotic rank conjecture and the set cover conjecture are not
  both true.
\newblock In {\em Proceedings of the 56th Annual ACM Symposium on Theory of
  Computing (STOC 2024)}, page 859–870. Association for Computing Machinery,
  2024.
\newblock \href {https://doi.org/10.1145/3618260.3649656}
  {\path{doi:10.1145/3618260.3649656}}.

\bibitem[Bl{\"a}13]{blaser2013fast}
Markus Bl{\"a}ser.
\newblock {\em Fast Matrix Multiplication}.
\newblock Number~5 in Graduate Surveys. Theory of Computing Library, 2013.
\newblock \href {https://doi.org/10.4086/toc.gs.2013.005}
  {\path{doi:10.4086/toc.gs.2013.005}}.

\bibitem[Bri87]{MR932055}
Michel Brion.
\newblock Sur l'image de l'application moment.
\newblock In {\em S\'{e}minaire d'alg\`ebre {P}aul {D}ubreil et {M}arie-{P}aule
  {M}alliavin ({P}aris, 1986)}, volume 1296 of {\em Lecture Notes in Math.},
  pages 177--192. Springer, Berlin, 1987.
\newblock \href {https://doi.org/10.1007/BFb0078526}
  {\path{doi:10.1007/BFb0078526}}.

\bibitem[Bun71]{buneman1971recovery}
Peter Buneman.
\newblock The recovery of trees from measures of dissimilarity.
\newblock In F.~R. Hodson, D.~G. Kendall, and P.~T{\u{a}}utu, editors, {\em
  Mathematics in the Archaeological and Historical Sciences: Proceedings of the
  Anglo-Romanian Conference, Mamaia, 1970}, pages 387--395. Edinburgh
  University Press, Edinburgh, 1971.

\bibitem[BV21]{MR4286265}
Gergely Bunth and P\'{e}ter Vrana.
\newblock Asymptotic relative submajorization of multiple-state boxes.
\newblock {\em Lett. Math. Phys.}, 111(4):Paper No. 94, 23, 2021.
\newblock \href {https://doi.org/10.1007/s11005-021-01430-0}
  {\path{doi:10.1007/s11005-021-01430-0}}.

\bibitem[CD21]{DBLP:journals/jcta/CostaD21}
Simone Costa and Marco Dalai.
\newblock {A gap in the slice rank of $k$-tensors}.
\newblock {\em J. Comb. Theory, Ser. A}, 177:105335, 2021.
\newblock \href {https://doi.org/10.1016/j.jcta.2020.105335}
  {\path{doi:10.1016/j.jcta.2020.105335}}.

\bibitem[CGL{\etalchar{+}}21]{MR4204580}
Austin Conner, Fulvio Gesmundo, Joseph~M. Landsberg, Emanuele Ventura, and Yao
  Wang.
\newblock Towards a geometric approach to {S}trassen's asymptotic rank
  conjecture.
\newblock {\em Collect. Math.}, 72(1):63--86, 2021.
\newblock \href {https://doi.org/10.1007/s13348-020-00280-8}
  {\path{doi:10.1007/s13348-020-00280-8}}.

\bibitem[CGZ23]{cgz}
Matthias Christandl, Fulvio Gesmundo, and Jeroen Zuiddam.
\newblock A gap in the subrank of tensors.
\newblock {\em {SIAM} J. Appl. Algebra Geom.}, 7(4):742--767, 2023.
\newblock \href {https://arxiv.org/abs/2212.01668} {\path{arXiv:2212.01668}},
  \href {https://doi.org/10.1137/22M1543276} {\path{doi:10.1137/22M1543276}}.

\bibitem[CHN{\etalchar{+}}25]{MR4928468}
Matthias Christandl, Koen Hoeberechts, Harold Nieuwboer, P\'{e}ter Vrana, and
  Jeroen Zuiddam.
\newblock Asymptotic tensor rank is characterized by polynomials.
\newblock In {\em S{TOC}'25---{P}roceedings of the 57th {A}nnual {ACM}
  {S}ymposium on {T}heory of {C}omputing}, pages 750--755. ACM, New York, 2025.
\newblock \href {https://doi.org/10.1145/3717823.3718122}
  {\path{doi:10.1145/3717823.3718122}}.

\bibitem[CK11]{MR2839250}
Imre Csisz\'{a}r and J\'{a}nos K\"{o}rner.
\newblock {\em Information theory}.
\newblock Cambridge University Press, Cambridge, second edition, 2011.
\newblock Coding theorems for discrete memoryless systems.
\newblock \href {https://doi.org/10.1017/CBO9780511921889}
  {\path{doi:10.1017/CBO9780511921889}}.

\bibitem[CKSU05]{cohn2005group}
Henry Cohn, Robert Kleinberg, Bal{\'a}zs Szegedy, and Christopher Umans.
\newblock Group-theoretic algorithms for matrix multiplication.
\newblock In {\em 46th Annual IEEE Symposium on Foundations of Computer Science
  (FOCS 2005)}, pages 379--388. IEEE, 2005.
\newblock \href {https://doi.org/10.1109/SFCS.2005.39}
  {\path{doi:10.1109/SFCS.2005.39}}.

\bibitem[CLGLZ25]{MR4870821}
Matthias Christandl, Fran\c{c}ois Le~Gall, Vladimir Lysikov, and Jeroen
  Zuiddam.
\newblock Barriers for rectangular matrix multiplication.
\newblock {\em Comput. Complexity}, 34(1):Paper No. 4, 38, 2025.
\newblock \href {https://doi.org/10.1007/s00037-025-00264-9}
  {\path{doi:10.1007/s00037-025-00264-9}}.

\bibitem[CU03]{cohn2003group}
Henry Cohn and Christopher Umans.
\newblock A group-theoretic approach to fast matrix multiplication.
\newblock In {\em Proceedings of the 44th Annual IEEE Symposium on Foundations
  of Computer Science ({FOCS} 2003)}, pages 438--449. IEEE, 2003.
\newblock \href {https://doi.org/10.1109/SFCS.2003.1238217}
  {\path{doi:10.1109/SFCS.2003.1238217}}.

\bibitem[CVZ19]{MR3924627}
Matthias Christandl, P\'{e}ter Vrana, and Jeroen Zuiddam.
\newblock Asymptotic tensor rank of graph tensors: beyond matrix
  multiplication.
\newblock {\em Comput. Complexity}, 28(1):57--111, 2019.
\newblock \href {https://doi.org/10.1007/s00037-018-0172-8}
  {\path{doi:10.1007/s00037-018-0172-8}}.

\bibitem[CVZ21]{v017a002}
Matthias Christandl, P\'eter Vrana, and Jeroen Zuiddam.
\newblock Barriers for fast matrix multiplication from irreversibility.
\newblock {\em Theory of Computing}, 17(2):1--32, 2021.
\newblock \href {https://doi.org/10.4086/toc.2021.v017a002}
  {\path{doi:10.4086/toc.2021.v017a002}}.

\bibitem[CVZ23]{MR4495838}
Matthias Christandl, P\'{e}ter Vrana, and Jeroen Zuiddam.
\newblock Universal points in the asymptotic spectrum of tensors.
\newblock {\em J. Amer. Math. Soc.}, 36(1):31--79, 2023.
\newblock \href {https://doi.org/10.1090/jams/996}
  {\path{doi:10.1090/jams/996}}.

\bibitem[dBBZ24]{deboer2024asymptoticspectrumdistancegraph}
David de~Boer, Pjotr Buys, and Jeroen Zuiddam.
\newblock The asymptotic spectrum distance, graph limits, and the {S}hannon
  capacity, 2024.
\newblock \href {https://arxiv.org/abs/2404.16763} {\path{arXiv:2404.16763}}.

\bibitem[DSW26]{doğan2026convexoptimizationmomentpolytopes}
Mahmut~Levent Doğan, Keiya Sakabe, and Michael Walter.
\newblock Convex optimization on moment polytopes: Hadamard mirror descent and
  efficient algorithms for quantum functionals and other tensor parameters,
  2026.
\newblock \href {https://arxiv.org/abs/2609.06633} {\path{arXiv:2609.06633}}.

\bibitem[EG17]{MR3583358}
Jordan~S. Ellenberg and Dion Gijswijt.
\newblock {On large subsets of $F_q^n$ with no three-term arithmetic
  progression}.
\newblock {\em Annals of Mathematics}, page 339–343, 2017.
\newblock \href {https://doi.org/10.4007/annals.2017.185.1.8}
  {\path{doi:10.4007/annals.2017.185.1.8}}.

\bibitem[FH91]{MR1153249}
William Fulton and Joe Harris.
\newblock {\em Representation theory}, volume 129 of {\em Graduate Texts in
  Mathematics}.
\newblock Springer-Verlag, New York, 1991.
\newblock A first course, Readings in Mathematics.
\newblock \href {https://doi.org/10.1007/978-1-4612-0979-9}
  {\path{doi:10.1007/978-1-4612-0979-9}}.

\bibitem[GKV92]{MR1184988}
L.~Gargano, J.~K\"{o}rner, and U.~Vaccaro.
\newblock Qualitative independence and {S}perner problems for directed graphs.
\newblock {\em J. Combin. Theory Ser. A}, 61(2):173--192, 1992.
\newblock \href {https://doi.org/10.1016/0097-3165(92)90016-N}
  {\path{doi:10.1016/0097-3165(92)90016-N}}.

\bibitem[GZ25]{gesmundo2023gap}
Fulvio Gesmundo and Jeroen Zuiddam.
\newblock The next gap in the subrank of 3-tensors.
\newblock {\em Linear Algebra Appl.}, 722:1--11, 2025.
\newblock \href {https://arxiv.org/abs/2307.06115} {\path{arXiv:2307.06115}},
  \href {https://doi.org/10.1016/j.laa.2025.04.027}
  {\path{doi:10.1016/j.laa.2025.04.027}}.

\bibitem[Har05]{harrow2005applicationscoherentclassicalcommunication}
Aram~W. Harrow.
\newblock Applications of coherent classical communication and the {S}chur
  transform to quantum information theory, 2005.
\newblock \href {https://arxiv.org/abs/quant-ph/0512255}
  {\path{arXiv:quant-ph/0512255}}.

\bibitem[Hay02]{hayashi2002exponentsquantumfixedlengthpure}
Masahito Hayashi.
\newblock Exponents of quantum fixed-length pure state source coding, 2002.
\newblock \href {https://arxiv.org/abs/quant-ph/0202002}
  {\path{arXiv:quant-ph/0202002}}, \href
  {https://doi.org/10.1103/PhysRevA.66.032321}
  {\path{doi:10.1103/PhysRevA.66.032321}}.

\bibitem[HS24]{MR4849336}
Hiroshi Hirai and Keiya Sakabe.
\newblock Gradient descent for unbounded convex functions on {H}adamard
  manifolds and its applications to scaling problems.
\newblock In {\em 2024 {IEEE} 65th {A}nnual {S}ymposium on {F}oundations of
  {C}omputer {S}cience---{FOCS} 2024}, pages 2387--2402. IEEE Computer Soc.,
  Los Alamitos, CA, 2024.
\newblock \href {https://doi.org/10.1109/FOCS61266.2024.00139}
  {\path{doi:10.1109/FOCS61266.2024.00139}}.

\bibitem[KM25]{kaski2024universalsequencetensorsasymptotic}
Petteri Kaski and Mateusz Michałek.
\newblock A universal sequence of tensors for the asymptotic rank conjecture.
\newblock In {\em 16th Innovations in Theoretical Computer Science Conference
  (ITCS 2025)}, volume 325 of {\em LIPIcs}, pages 64:1--64:24, 2025.
\newblock \href {https://arxiv.org/abs/2404.06427} {\path{arXiv:2404.06427}},
  \href {https://doi.org/10.4230/LIPIcs.ITCS.2025.64}
  {\path{doi:10.4230/LIPIcs.ITCS.2025.64}}.

\bibitem[LZ21]{MR4231964}
Yinan Li and Jeroen Zuiddam.
\newblock Quantum asymptotic spectra of graphs and non-commutative graphs, and
  quantum {S}hannon capacities.
\newblock {\em IEEE Trans. Inform. Theory}, 67(1):416--432, 2021.
\newblock \href {https://doi.org/10.1109/TIT.2020.3032686}
  {\path{doi:10.1109/TIT.2020.3032686}}.

\bibitem[MP71]{MR403804}
Robert~J. McEliece and Edward~C. Posner.
\newblock Hide and seek, data storage, and entropy.
\newblock {\em Ann. Math. Statist.}, 42:1706--1716, 1971.
\newblock \href {https://doi.org/10.1214/aoms/1177693169}
  {\path{doi:10.1214/aoms/1177693169}}.

\bibitem[NS17]{naslund_sawin_2017}
Eric Naslund and Will Sawin.
\newblock Upper bounds for sunflower-free sets.
\newblock {\em Forum Math. Sigma}, 5:Paper No. e15, 10, 2017.
\newblock \href {https://doi.org/10.1017/fms.2017.12}
  {\path{doi:10.1017/fms.2017.12}}.

\bibitem[{Ope}26]{OAI:Matrix-Multiplication-Nine-Fourths-October-2-2026}
{OpenAI}.
\newblock {An Upper Bound of $9/4$ for the Matrix Multiplication Exponent}.
\newblock OpenAI Math Release preprint
  \href{https://github.com/openai/math/blob/main/preprints/Matrix-Multiplication-Nine-Fourths-October-2-2026/paper.pdf}{OAI:Matrix-Multiplication-Nine-Fourths-October-2-2026},
  2026.

\bibitem[Pra24]{10.1145/3618260.3649620}
Kevin Pratt.
\newblock A stronger connection between the asymptotic rank conjecture and the
  set cover conjecture.
\newblock In {\em Proceedings of the 56th Annual ACM Symposium on Theory of
  Computing (STOC 2024)}, page 871–874, New York, NY, USA, 2024. Association
  for Computing Machinery.
\newblock \href {https://doi.org/10.1145/3618260.3649620}
  {\path{doi:10.1145/3618260.3649620}}.

\bibitem[PVW22]{MR4395416}
Christopher Perry, P\'{e}ter Vrana, and Albert~H. Werner.
\newblock The semiring of dichotomies and asymptotic relative submajorization.
\newblock {\em IEEE Trans. Inform. Theory}, 68(1):311--321, 2022.
\newblock \href {https://doi.org/10.1109/TIT.2021.3117440}
  {\path{doi:10.1109/TIT.2021.3117440}}.

\bibitem[RZ22]{MR4399732}
Robert Robere and Jeroen Zuiddam.
\newblock Amortized circuit complexity, formal complexity measures, and
  catalytic algorithms.
\newblock In {\em 2021 {IEEE} 62nd {A}nnual {S}ymposium on {F}oundations of
  {C}omputer {S}cience---{FOCS} 2021}, pages 759--769. IEEE Computer Soc., Los
  Alamitos, CA, 2022.
\newblock \href {https://doi.org/10.1109/FOCS52979.2021.00079}
  {\path{doi:10.1109/FOCS52979.2021.00079}}.

\bibitem[Sch81]{MR623057}
A.~Sch\"{o}nhage.
\newblock Partial and total matrix multiplication.
\newblock {\em SIAM J. Comput.}, 10(3):434--455, 1981.
\newblock \href {https://doi.org/10.1137/0210032} {\path{doi:10.1137/0210032}}.

\bibitem[Sch03]{MR1956925}
Alexander Schrijver.
\newblock {\em Combinatorial optimization. {P}olyhedra and efficiency. {V}ol.
  {B}}, volume~24 of {\em Algorithms and Combinatorics}.
\newblock Springer-Verlag, Berlin, 2003.
\newblock Matroids, trees, stable sets, Chapters 39--69.

\bibitem[SDW26]{sakabe2026strassenssupportfunctionalscoincide}
Keiya Sakabe, Mahmut~Levent Doğan, and Michael Walter.
\newblock Strassen's support functionals coincide with the quantum functionals,
  2026.
\newblock To appear in FOCS 2026.
\newblock \href {https://arxiv.org/abs/2601.21553} {\path{arXiv:2601.21553}}.

\bibitem[Sha56]{MR0089131}
Claude~E. Shannon.
\newblock The zero error capacity of a noisy channel.
\newblock {\em Institute of Radio Engineers Transactions on Information
  Theory}, IT-2,:8--19, 1956.
\newblock \href {https://doi.org/10.1109/tit.1956.1056798}
  {\path{doi:10.1109/tit.1956.1056798}}.

\bibitem[Shp26]{shpilka2026exposition}
Amir Shpilka.
\newblock Exposition of {OpenAI}'s matrix multiplication in exponent 9/4 (and
  an explicit algorithm).
\newblock ECCC Report TR26-281, Electronic Colloquium on Computational
  Complexity, October 2026.
\newblock URL: \url{https://eccc.weizmann.ac.il/report/2026/281/}.

\bibitem[SS03]{MR2060009}
Charles Semple and Mike Steel.
\newblock {\em Phylogenetics}, volume~24 of {\em Oxford Lecture Series in
  Mathematics and its Applications}.
\newblock Oxford University Press, Oxford, 2003.

\bibitem[ST16]{sawin}
Will Sawin and Terence Tao.
\newblock {Notes on the “slice rank” of tensors}, 2016.
\newblock URL: \url{https://terrytao.wordpress.com/2016/08/24/}.

\bibitem[Str69]{strassen1969gaussian}
Volker Strassen.
\newblock Gaussian elimination is not optimal.
\newblock {\em Numer. Math.}, 13(4):354--356, 1969.
\newblock \href {https://doi.org/10.1007/BF02165411}
  {\path{doi:10.1007/BF02165411}}.

\bibitem[Str86]{DBLP:conf/focs/Strassen86}
Volker Strassen.
\newblock The asymptotic spectrum of tensors and the exponent of matrix
  multiplication.
\newblock In {\em 27th Annual Symposium on Foundations of Computer Science,
  Toronto, Canada, 27-29 October 1986}, pages 49--54. {IEEE} Computer Society,
  1986.
\newblock \href {https://doi.org/10.1109/SFCS.1986.52}
  {\path{doi:10.1109/SFCS.1986.52}}.

\bibitem[Str87]{MR882307}
Volker Strassen.
\newblock Relative bilinear complexity and matrix multiplication.
\newblock {\em J. Reine Angew. Math.}, 375/376:406--443, 1987.
\newblock \href {https://doi.org/10.1515/crll.1987.375-376.406}
  {\path{doi:10.1515/crll.1987.375-376.406}}.

\bibitem[Str88]{strassen1988asymptotic}
Volker Strassen.
\newblock The asymptotic spectrum of tensors.
\newblock {\em J. Reine Angew. Math.}, 384:102--152, 1988.
\newblock \href {https://doi.org/10.1515/crll.1988.384.102}
  {\path{doi:10.1515/crll.1988.384.102}}.

\bibitem[Str91]{strassen1991degeneration}
Volker Strassen.
\newblock {Degeneration and complexity of bilinear maps: some asymptotic
  spectra}.
\newblock {\em J. Reine Angew. Math.}, 413:127–180, 1991.
\newblock \href {https://doi.org/10.1515/crll.1991.413.127}
  {\path{doi:10.1515/crll.1991.413.127}}.

\bibitem[Str94]{MR1341854}
V.~Strassen.
\newblock Algebra and complexity.
\newblock In {\em First {E}uropean {C}ongress of {M}athematics, {V}ol. {II}
  ({P}aris, 1992)}, volume 120 of {\em Progr. Math.}, pages 429--446.
  Birkh\"{a}user, Basel, 1994.

\bibitem[Str12]{DBLP:conf/issac/Strassen12}
Volker Strassen.
\newblock Asymptotic spectrum and matrix multiplication.
\newblock In {\em International Symposium on Symbolic and Algebraic Computation
  (ISSAC 2012)}, pages 6--7. {ACM}, 2012.
\newblock \href {https://doi.org/10.1145/2442829.2442832}
  {\path{doi:10.1145/2442829.2442832}}.

\bibitem[Tao16]{tao}
Terence Tao.
\newblock {A symmetric formulation of the Croot-Lev-Pach-Ellenberg-Gijswijt
  capset bound}, 2016.
\newblock URL: \url{https://terrytao.wordpress.com/2016/05/18/}.

\bibitem[VC15]{vrana2015transformation}
Péter Vrana and Matthias Christandl.
\newblock {Asymptotic entanglement transformation between W and GHZ states}.
\newblock {\em Journal of Mathematical Physics}, 56(2):022204, 02 2015.
\newblock \href {https://arxiv.org/abs/1310.3244} {\path{arXiv:1310.3244}},
  \href {https://doi.org/10.1063/1.4908106} {\path{doi:10.1063/1.4908106}}.

\bibitem[vdB23]{vandenberg2023asymptotic}
Maxim van~den Berg.
\newblock The asymptotic spectrum of complex tensors, the quantum functionals
  and interpolation.
\newblock {MSc} mathematics, master thesis, University of Amsterdam,
  Korteweg-de Vries Institute for Mathematics, August 2023.

\bibitem[vdBCL{\etalchar{+}}25]{MR4928469}
Maxim van~den Berg, Matthias Christandl, Vladimir Lysikov, Harold Nieuwboer,
  Michael Walter, and Jeroen Zuiddam.
\newblock Computing moment polytopes of tensors, with applications in algebraic
  complexity and quantum information.
\newblock In {\em S{TOC}'25---{P}roceedings of the 57th {A}nnual {ACM}
  {S}ymposium on {T}heory of {C}omputing}, pages 756--765. ACM, New York, 2025.
\newblock \href {https://doi.org/10.1145/3717823.3718221}
  {\path{doi:10.1145/3717823.3718221}}.

\bibitem[Vra21]{MR4357434}
P\'{e}ter Vrana.
\newblock Probabilistic refinement of the asymptotic spectrum of graphs.
\newblock {\em Combinatorica}, 41(6):873--904, 2021.
\newblock \href {https://doi.org/10.1007/s00493-020-4324-5}
  {\path{doi:10.1007/s00493-020-4324-5}}.

\bibitem[Vra25]{vrana2025errorexponentstripartitetobipartiteentanglement}
Péter Vrana.
\newblock Error exponents for tripartite-to-bipartite entanglement
  transformations, 2025.
\newblock \href {https://arxiv.org/abs/2507.13778} {\path{arXiv:2507.13778}}.

\bibitem[WZ26]{wigderson2022asymptotic}
Avi Wigderson and Jeroen Zuiddam.
\newblock Asymptotic spectra: Theory, applications, and extensions.
\newblock {\em Bull.\ Amer.\ Math.\ Soc.}, 63(2):199--331, 2026.
\newblock URL: \url{https://staff.fnwi.uva.nl/j.zuiddam/papers/convexity.pdf},
  \href {https://doi.org/10.1090/bull/1880} {\path{doi:10.1090/bull/1880}}.

\bibitem[Zui19]{MR4039606}
Jeroen Zuiddam.
\newblock The asymptotic spectrum of graphs and the {S}hannon capacity.
\newblock {\em Combinatorica}, 39(5):1173--1184, 2019.
\newblock \href {https://doi.org/10.1007/s00493-019-3992-5}
  {\path{doi:10.1007/s00493-019-3992-5}}.

\end{thebibliography}
\newcommand{\etalchar}[1]{$^{#1}$}

\end{document}